%% file: Bachelor-Arbeit_allgemein.tex
\documentclass[
headsepline, 
12pt, 
a4paper, 
parskip=on,
liststotoc] 
{scrreprt}	

\usepackage[utf8]{inputenc} 
\usepackage{footmisc}						
\usepackage[automark]{scrlayer-scrpage} 
\usepackage[latin,english]{babel}
\usepackage[T1]{fontenc}
\usepackage{amssymb}
\usepackage{extarrows}
\usepackage{esvect}
\usepackage{lmodern} 
\usepackage{textcmds}
\usepackage{dsfont}
\usepackage[left=30mm,right=25mm,bottom=25mm,top=30mm]{geometry}
\renewcommand{\arraystretch}{1.2} 

\usepackage[onehalfspacing]{setspace} 

\usepackage{graphicx}	
\usepackage{xspace}

\usepackage{textpos} 

\usepackage[colorlinks,bookmarks]{hyperref} 

\usepackage{tabularx}	

\usepackage{supertabular} 

\usepackage{longtable} 

\usepackage{ragged2e}	
	\newcolumntype{L}[1]{>{\RaggedRight\arraybackslash\hspace{0pt}}p{#1}} 	
	\newcolumntype{R}[1]{>{\RaggedLeft\arraybackslash\hspace{0pt}}p{#1}} 		
	\newcolumntype{C}[1]{>{\centering\arraybackslash\hspace{0pt}}p{#1}} 		

\usepackage{endnotes} 

\usepackage{hyperref}	
\hypersetup{colorlinks = true, linkcolor  = black, urlcolor=black, linkcolor=black}

\usepackage{booktabs} 
\usepackage{multirow} 

\usepackage[binary-units = true]{siunitx}
\DeclareSIUnit\px{px}
\usepackage{amsmath} 

\usepackage{float} 
\usepackage{placeins} 
\usepackage{makecell} 

\usepackage[center]{caption} 
\usepackage{appendix} 

\usepackage{pdfpages} 

\usepackage[version=4]{mhchem}

\usepackage{lscape} 

\usepackage{rotating} 

\usepackage{upgreek} 

\usepackage{wrapfig} 

\usepackage{icomma}

\makeatletter
\begingroup
\catcode`\_=\active
\protected\gdef_{\@ifnextchar|\subtextit\subtextup }
\endgroup
\def\subtextit|#1|{\sb{#1}}
\def\subtextup#1{\sb{\mathrm{#1}}}
\AtBeginDocument{\catcode`\_=12 \mathcode`\_=32768}
\makeatother

\RequirePackage{ifthen}

\newcommand*{\Cpp}{C\ensuremath{++}\xspace}
\usepackage{courier}
\usepackage{cancel}
\usepackage{chngcntr}
\counterwithout{footnote}{chapter}
\usepackage{pgfplots}
\usepackage{listings} 
\lstdefinestyle{Matlab}{
  frame=L,
  language=Matlab,
  showstringspaces=false,
}
\lstdefinestyle{c}{
  frame=L,
  language=C++,
  showstringspaces=false,
}
\begin{document}

\setcounter{tocdepth}{3}	
\setcounter{secnumdepth}{3}	

\hypersetup{
pdftitle={Evaluation of the Breit-Hartree contribution to the total energy of open atomic shells},
pdfsubject={Bachelorarbeit},
pdfauthor={Benjamin Buchholz},
citecolor=black}

\pagestyle{scrheadings} 
\clearscrheadings	
\clearscrplain		
\rohead{\pagemark}	
\lohead{\headmark}	

\setlength\abovedisplayshortskip{0,55cm}
\setlength\belowdisplayshortskip{0,55cm}
\setlength\abovedisplayskip{0,55cm}
\setlength\belowdisplayskip{0,55cm}

\include{Titel}

\addchap*{Eidesstattliche Erklärung}
\label{cap:Erklaerung}

Hiermit erkläre ich, Benjamin Buchholz, die vorliegende Arbeit selbstständig angefertigt zu haben. Die Erstellung erfolgte ohne das unerlaubte Zutun Dritter. Alle Hilfs-mittel, die für die Erstellung der vorliegenden Arbeit benutzt wurden, befinden sich ausschließlich im Literaturverzeichnis. Alles, was aus anderen Arbeiten unverändert oder mit Abänderungen übernommen wurde, ist kenntlich gemacht.

Die Arbeit wurde bisher keiner anderen Prüfungsbehörde vorgelegt.

\vspace{2em}
Hamburg, 2. Dezember 2021

\vspace{1em}

\rule[-1cm]{5cm}{0.1mm}\\[1em]
Benjamin Buchholz

\clearpage

\pagenumbering{Roman} 

\phantomsection  

\addcontentsline{toc}{chapter}{Abstract}

\setcounter{page}{6}

\begin{minipage}{\linewidth}
\vspace*{-4mm}
\begin{abstract}
\textbf{Abstract}

In this work the \textit{Breit-Hartree} interaction, as the lowest order relativistic correction to the \textit{Coulomb} interaction, is extensively analyzed in the framework of relativistic Density Functional Theory. Its relation to the magnetostatic dipole-dipole interaction is recapitulated, and its contribution to the total energy of the ground state of an atom or ion is investigated analytically and numerically. Specifically, an atom or ion is treated as a hollow sphere in zeroth order with a magnetization density solely generated by the spin density of open atomic shells.\hspace{1.5mm}An analytical solution is derived for a radially dependent magnetization density within a spherical volume and implemented in \Cpp and \textsc{Matlab}. The \textit{Breit-Hartree} contribution is calculated for an $\textup{Mn}^{2+}$ and a $\textup{Gd}^{3+}$ ion and compared with the second order \textit{M\o{}ller-Plesset} correlation energy correction. Additionally, the result for the $\textup{Gd}^{3+}$ ion is discussed against the backdrop of experimental data and an improvement in experimental data prediction is shown. Moreover, the applicability of the \textit{Breit-Hartree} contribution for atoms, ions and solid states is presented and a suggestion for further calculations of this correction for atoms and ions is submitted.    
\end{abstract}
\begin{abstract}
\textbf{Kurzzusammenfassung}

In dieser Arbeit wird die \textit{Breit-Hartree}-Wechselwirkung als relativistische Korrektur niedrigster Ordnung zur \textit{Coulomb}-Wechselwirkung im Rahmen der relativistischen Dichtefunktionaltheorie ausführlich analysiert.\hspace{1.5mm}Ihr Zusammenhang mit der magnetostatischen Dipol-Dipol-Wechselwirkung wird demonstriert und ihr Beitrag zur Gesamtenergie des Grundzustands eines Atoms oder Ions wird analytisch und numerisch untersucht.\hspace{1.5mm}Insbesondere wird ein Atom oder Ion in nullter Ordnung als Hohlkugel mit einer Magnetisierungsdichte behandelt, die allein durch die Spindichte nicht vollbesetzter Atomschalen erzeugt wird. Für eine radial abhängige Magnetisierungsdichte innerhalb eines Kugelvolumens wird eine analytische Lösung abgeleitet und in \Cpp und \textsc{Matlab} implementiert. Der \textit{Breit-Hartree}-Beitrag wird für ein $\textup{Mn}^{2+}$- und ein $\textup{Gd}^{3+}$- Ion berechnet und mit der Korrelationsenergiekorrektur in zweiter Ordnung der \textit{M\o{}ller-Plesset}-Störungstheorie verglichen. Zusätzlich wird das Ergebnis für das $\textup{Gd}^{3+}$-Ion vor dem Hintergrund experimenteller Daten diskutiert und eine Verbesserung der experimentellen Datenvorhersage aufgezeigt. Zudem wird die Anwendbarkeit des \textit{Breit-Hartree}-Beitrags für Atome, Ionen und Festkörper dargelegt und ein Vorschlag für weitere Berechnungen dieser Korrektur für Atome und Ionen unterbreitet.
\end{abstract}
\end{minipage} 

\phantomsection

\addcontentsline{toc}{chapter}{Acknowledgement}

\addchap*{Acknowledgement}

In preparation and throughout the process of writing of this thesis I have received support, assistance and motivation, without which I would not have been able to cope with a subject beyond the study of mechanical engineering.
\\\\
I would first like to thank my supervisor, PD Dr. Manuel Richter (department head of the research group for Numerical Simulation in Solide State Physics, Institute for Theoretical Solide State Physics, Leibniz Institute for Solide State and Materials Research Dresden), whose expertise and experience was crucial within the process of analyzing and understanding the integral equation representing the \textit{Breit-Hartree} interaction. His insightful feedback lead me into a clear path for the evaluation. Furthermore, I am thankful for the serval disussions which helped me extending my knowledge and were able to motivate me.
\\\\
In addition, I would like to acknowledge my first examiner Prof. Dr.-Ing. Denis Kramer, who supported the coorperation with the Leibniz Institute for Solide State and Materials Research, was always open for questions and dedicated to supporting.
\\\\
Finally, I am grateful for the motivation and support from my family, which helped me to never lose sight of my aims. Moreover, I would like to acknowledge Mr. Steiner's support with linguistic problems.

\pdfbookmark[chapter]{Contents}{contents}

\tableofcontents

\clearpage

\input{Formelzeichen}

\clearpage

\pagenumbering{arabic}	

\input{Introduction}

\input{Origin_of_the_Breit-Hartree_interaction_and_its_importance_in_condensed_matter_physics_V2}

\input{Evaluation_of_the_Breit-Hartree_contribution_to_the_total_energy_of_a_homogeneously_magnetized_sphere}

\input{Evaluation_of_the_Breit-Hartree_contribution_to_the_total_energy_of_a_spinpolarized_half-filled_atomic_shell}

\input{Comparison_and_discussion_in_the_context_of_experimental_information}

\input{Conclusion}

\phantomsection

\addcontentsline{toc}{chapter}{Bibliography}

\interlinepenalty=10000 

\bibliography{References}

\interlinepenalty=4000

\clearpage

\listoffigures

\appendix

\input{Appendix}

\newpage\thispagestyle{empty}\mbox{}\newpage

\end{document}

%% file: Titel.tex
\thispagestyle{empty}
\begin{titlepage}
	
\begin{textblock}{65}(106,-16)
\includegraphics[width=0.5\textwidth]{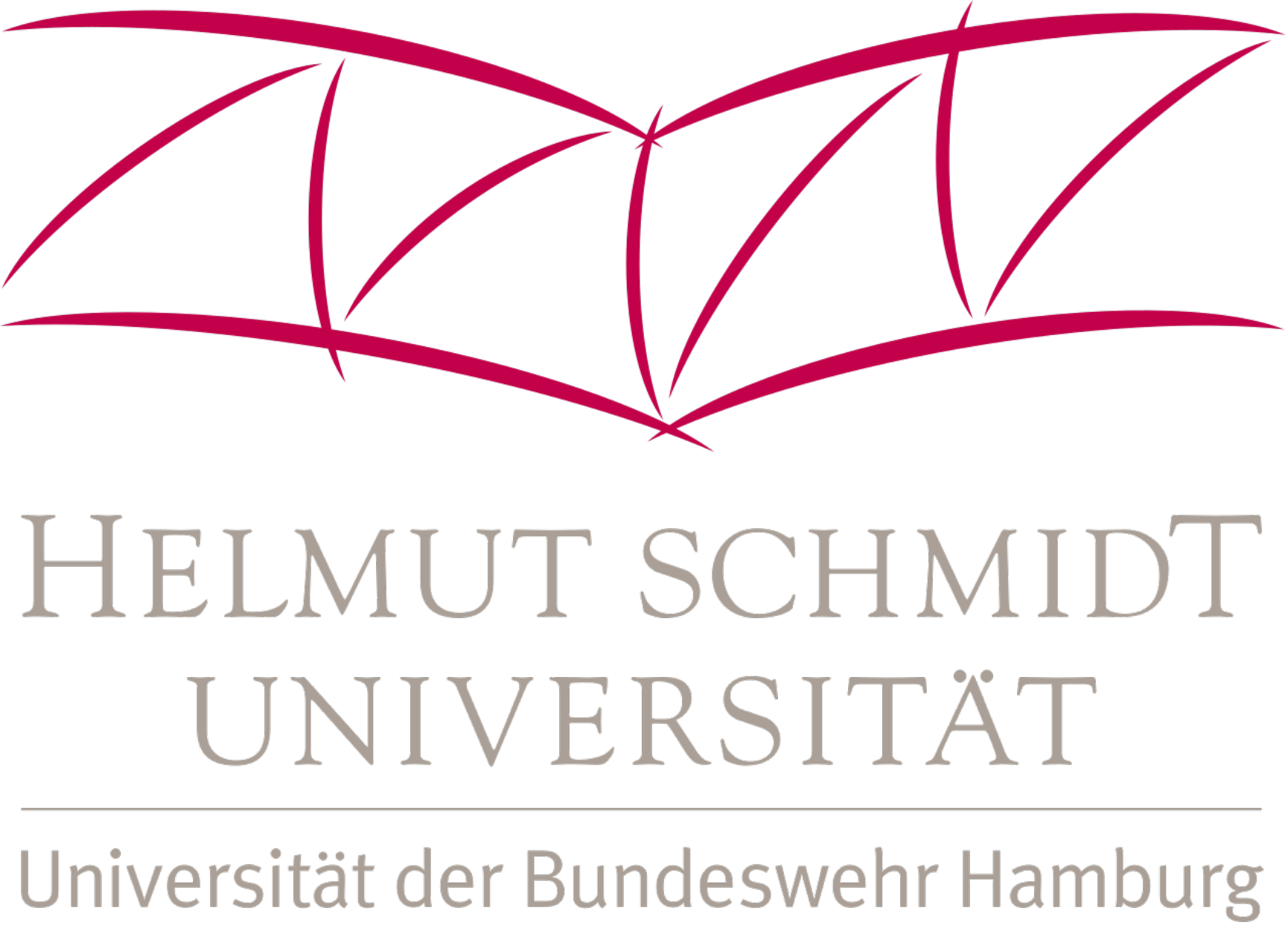}
\end{textblock}
					
\begin{textblock}{65}(95,14)
\includegraphics[width=0.85\textwidth]{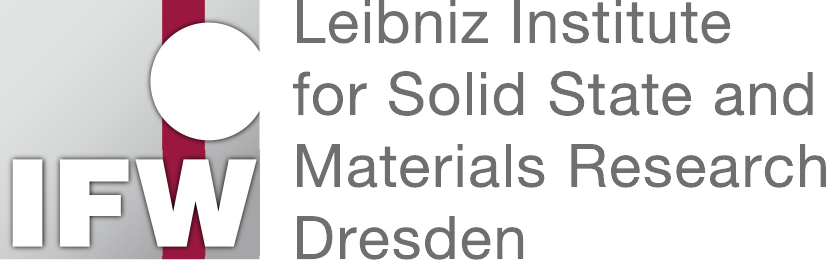}
\end{textblock}

\textcolor[RGB]{140,130,121}{	
\begin{textblock}{80}(0,-23)
\renewcommand{\baselinestretch}{1.1}\footnotesize
  \textsf{Helmut-Schmidt-University/ 
University of the \\German Federal Armed Forces Hamburg\\
  	Faculty of Mechanical Engineering\\ 
 	Institute for Computational Material Design\\
	Prof. Dr.-Ing. Denis Kramer}
\end{textblock}}

\textcolor[RGB]{140,130,121}{	
\begin{textblock}{80}(0,-5)
\renewcommand{\baselinestretch}{1.1}\footnotesize
  \textsf{Leibniz Institute for Solide State and \\Materials Research Dresden\\
Institute for Theoretical Solide State Physics\\
Numerical Simulation in Solide State Physics
\\PD Dr. Manuel Richter}
\end{textblock}}

\textcolor[RGB]{140,130,121}{
\begin{textblock}{65}(0,10)													
  {\large \textsf{\underline{~~~~~~~~~~~~~~~~~~~~~~~~~~~~~~~~~~~~~~~~~~~~~~~~~~~~~~~~~~~~~~~~~~~~~~~~~~~~~~~~~~~~~~~~~~~~~}}}            
\end{textblock}}
	
\begin{textblock}{65}(44,30)
	\begin{color}{black}													
  {\Huge \textbf{\textsf {Bachelor Thesis}}}            
	\end{color}
\end{textblock}																			
																				
\begin{textblock}{120}(0,58)										
    
\textcolor[RGB]{165,0,52}{				
\textbf{{\Large Benjamin Buchholz} \\[1cm] }}
\end{textblock}	
\begin{textblock}{157}(0,80)
\textcolor[RGB]{140,130,121}{
\textbf{{\Large Evaluation of the \textit{Breit-Hartree} contribution to the \\total energy
of open atomic shells \\[1.5cm] 
Absch\"atzung des \textit{Breit-Hartree} Beitrages zur Gesamtenergie von nicht voll besetzten Atomschalen}} \\ [2.0cm]
\begin{tabular}{@{}cl@{}}
{\Large Supervisors:} & {\Large Prof. Dr.-Ing. Denis Kramer}\\
&{\Large PD Dr. Manuel Richter}\\
\end{tabular}\\[2.0cm]
{\Large Hamburg, December 2nd, 2021}}

\end{textblock}

\begin{textblock}{100}(-30,135)												
 \includegraphics{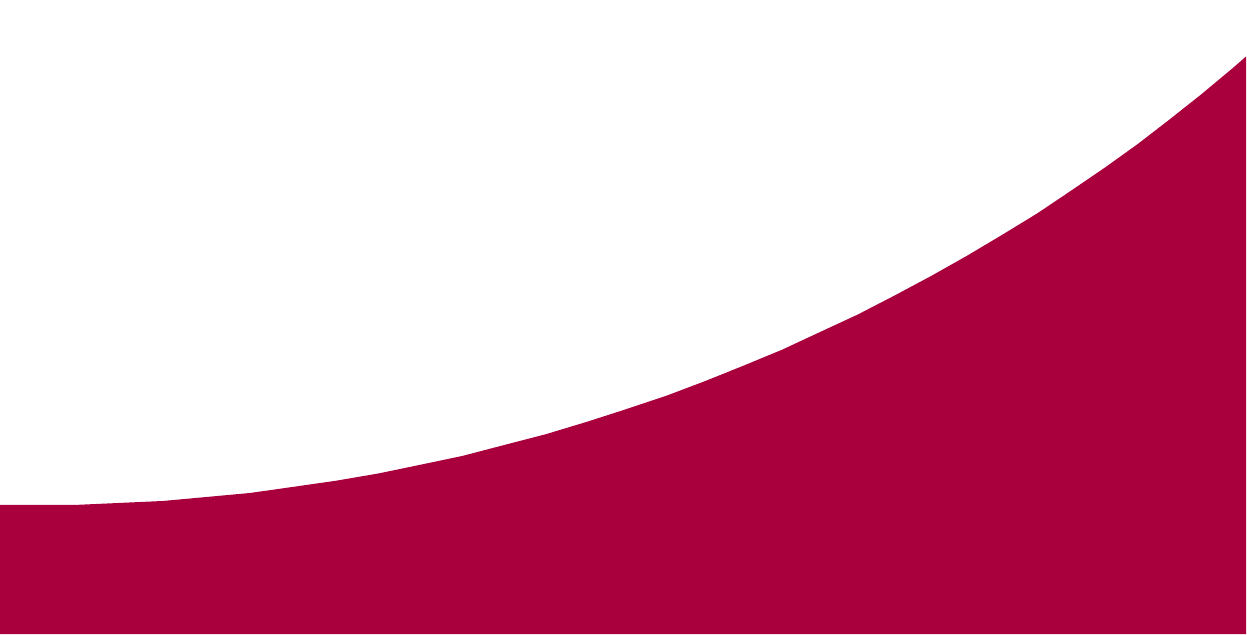}
\end{textblock}

\end{titlepage}
\newpage\thispagestyle{empty}\mbox{}\newpage 
\begin{titlepage}
\newgeometry{left=30mm,right=25mm,bottom=25mm,top=25mm}

\begin{center}
\begin{tabular}{p{\textwidth}}

\centering
\begin{minipage}[c]{0.4\textwidth}
\includegraphics[width=\textwidth]{IFW_Logo_EN.pdf}
\end{minipage}
\hspace{1cm}
\begin{minipage}[c]{0.2\textwidth}
\includegraphics[width=\textwidth]{Intellus-Partner-HSU.pdf}
\end{minipage}

\vspace{0.4cm}

\large{Helmut-Schmidt-University/ \\
University of the German Federal Armed Forces Hamburg}\\
\small{Faculty of Mechanical Engineering\\
Institute for Computational Material Design }

\vspace{0,7cm}

\LARGE{\textsc{
Evaluation of the \textit{Breit-Hartree} contribution to the total energy of open atomic shells}}

\vspace{0,7cm}

\textbf{\LARGE{Bachelor Thesis}}

\vspace{0,6cm}

\small{for the acquisition of the academic degree} 

\large{Bachelor of Science (B.Sc.)}

\vspace{0,4cm}

\small{submitted by}

\vspace{0,4cm}

\large{\textbf{Benjamin Buchholz}} \\
\small{born on June 24th, 1998 in Dresden }

\vspace{0,4cm}

supervised by

\vspace{0,4cm}

\large{\textbf{PD Dr. M. Richter}} 

\vspace{0,1cm}

Leibniz Institute for Solide State and Materials Research Dresden\\
\small{Institute for Theoretical Solide State Physics\\
Research group for Numerical Simulation in Solide State Physics}

\vspace{1.0cm}

\begin{tabular}{lllll}
\textbf{Student ID Number:} & & & 00893529\\ 
\textbf{Date of submission:} & & & December 2nd, 2021 \\
\textbf{first examiner:} & & & Prof. Dr.-Ing. D. Kramer\\
\textbf{second examiner:} & & & PD Dr. M. Richter \\
\end{tabular}

\end{tabular}
\end{center}
\end{titlepage}

\thispagestyle{empty}

\begin{titlepage}
\newgeometry{left=30mm,right=25mm,bottom=25mm,top=25mm}

\begin{center}
\begin{tabular}{p{\textwidth}}

\centering
\begin{minipage}[c]{0.4\textwidth}
\includegraphics[width=\textwidth]{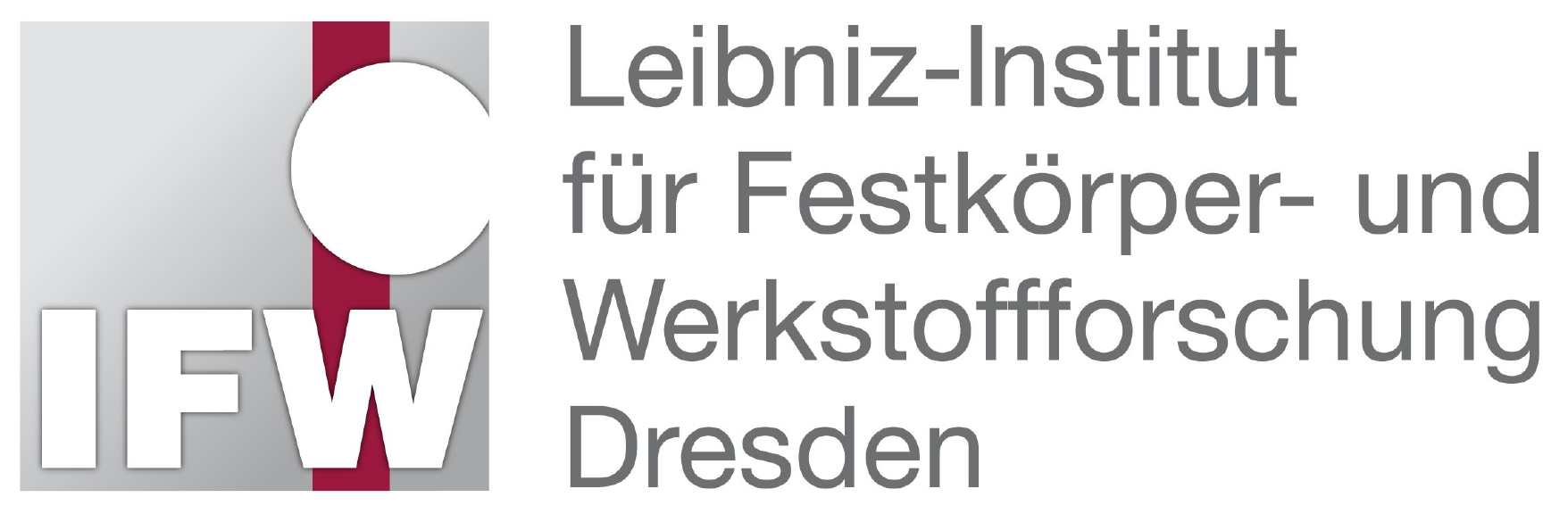}
\end{minipage}
\hspace{1cm}
\begin{minipage}[c]{0.2\textwidth}
\includegraphics[width=\textwidth]{Intellus-Partner-HSU.pdf}
\end{minipage}

\vspace{0.4cm}

\large{Helmut-Schmidt-Universit\"at/ \\
Universit\"at der Bundeswehr Hamburg}\\
\small{Fakult\"at Maschinenbau\\
Institut f\"ur Computational Material Design }

\vspace{0,7cm}

\LARGE{\textsc{
Absch\"atzung des \textit{Breit-Hartree}-\\Beitrages zur Gesamtenergie von nicht voll besetzten Atomschalen}}

\vspace{0,7cm}

\textbf{\LARGE{Bachelorarbeit}}

\vspace{0,6cm}

\small{zur Erlangung des akademischen Grades} 

\large{Bachelor of Science (B.Sc.)}

\vspace{0,4cm}

\small{vorgelegt von}

\vspace{0,4cm}

\large{\textbf{Benjamin Buchholz}} \\
\small{geboren am 24. Juni 1998 in Dresden}

\vspace{0,4cm}

betreut von

\vspace{0,4cm}

\large{\textbf{PD Dr. M. Richter}} 

\vspace{0,1cm}

Leibniz Institut  f\"ur Festk\"orper- und Werkstoffforschung Dresden\\
\small{Institut f\"ur Theoretische Fesk\"orperphysik\\
Forschungsgruppe f\"ur Numerische Simulation in der Fesk\"orperphysik}

\vspace{1.0cm}

\begin{tabular}{lllll}
\textbf{Martikelnummer:} & & & 00893529\\ 
\textbf{Abgabedatum:} & & & 2. Dezember 2021\\
\textbf{Erstgutachter:} & & & Prof. Dr.-Ing. D. Kramer\\
\textbf{Zweitgutachter:} & & & PD Dr. M. Richter \\
\end{tabular}

\end{tabular}
\end{center}
\end{titlepage}

\thispagestyle{empty}

%% file: Formelzeichen.tex
\addchap{Nomenclature}

\textbf{Latin letters}
\begin{table}[H]
\renewcommand{\arraystretch}{1}
\newcolumntype{s}{>{\hsize=.5\hsize}X}
\newcolumntype{S}{>{\hsize=.8\hsize}X}
\begin{tabularx}{\textwidth}{sSX}
\toprule
Symbol				&				Unit				& 				Denotation \\    
\midrule
$E$		&				\si{\joule}		&	energy \\
$L$\,=\,$\vert\vec{L}\,\vert$		&				\si{\kilogram\square\meter\per\second}		&	angular momentum \\
$n$		&				\si{\per\cubic\meter}		&	electron density \\
$s$		&				\si{\per\cubic\meter}		&	spin density \\
$k_{F}=(3\pi^{2}n)^\frac{1}{3}$		&				\si{\per\meter}		&	\textit{Fermi} wave vector \cite{ashcroft1976solid}\\
$M$\,=\,$\vert\vec{M}\,\vert$			&				\si{\ampere\per\meter}		&	magnetization density \\
$B$\,=\,$\vert\vec{B}\,\vert$		&				\si{\tesla}\,=\,\si{\kilogram\per\ampere\per\square\second}		&	magnetic flux density \\
$H$\,=\,$\vert\vec{H}\,\vert$		&				\si{\ampere\per\meter}		&	magnetic field strength \\
$D$\,=\,$\vert\vec{D}\,\vert$		&				\si{\ampere\second\per\square\meter}		&	electric displacement field \\
$j$=$\vert\vec{j}\,\vert$		&				\si{\ampere\per\square\meter}		&	non-relativistic current density \\
$N_{el}$&				-		&	number of electrons \\
$Z$ &				-		&	atomic or proton number  \\
\bottomrule						
\end{tabularx}
\end{table}

\textbf{Greek letters}
\begin{table}[H]
\renewcommand{\arraystretch}{1}
\newcolumntype{s}{>{\hsize=.5\hsize}X}
\newcolumntype{S}{>{\hsize=.8\hsize}X}
\begin{tabularx}{\textwidth}{sSX}
\toprule
Symbol				&				Unit				& 				Denotation \\      
\midrule

$\Omega$				&		\si{\cubic\meter}			&	volume \\

$\varepsilon_{i}$				&		\si{\joule}			&	eigen energy to the eigenstate $\psi_{i}$  \\

$\psi$				&		$\frac{1}{\sqrt{\textup{m}^{3}}}$			&	one particle wave function \\

$\Psi$				&		$\frac{1}{\sqrt{\textup{m}^{3}}}$			&	many particle wave function \\

$\sigma $				&		-		&	electron spin  \\

$\rho$ & \si{\joule\per\cubic\meter} & energy (contribution) density	\\

$\phi_{m}$		&				\si{\ampere}		&	magnetic scalar potential \\

$\rho_{m}$		&				\si{\ampere\per\square\meter}		&	effective magnetic charge density \\

$\pi$		&				-	&	ratio of a circle's circumference to its diameter \cite{WolframPI} \\
\bottomrule						
\end{tabularx}
\end{table}

\clearpage

\textbf{Physical constants}
\begin{table}[H]
\renewcommand{\arraystretch}{1.24}
\newcolumntype{s}{>{\hsize=.35\hsize}X}
\newcolumntype{S}{>{\hsize=.90\hsize}X}
\begin{tabularx}{\textwidth}{sSX}
\toprule
Symbol		&				Value and Unit				& 				Denotation \\    
\midrule

$c$  &				$2,99792458 \cdot 10^{8}$ \si{\meter\per\second}		&	speed of light in vacuum \cite{NISTc}\\

$e $ &				$1,602176634 \cdot 10^{-19}$ \si{\ampere\second}		&	elementary charge \cite{NISTe} \\

$m_{e}$  &				$9,1093837015 \cdot 10^{-31}$ \si{\kilogram}		&	electron mass \cite{NISTme}\\

$h $ &				$6,62607015 \cdot 10^{-34}$ \si{\joule\second}		&	\textit{Planck} constant \cite{NISTh}\\

$\hslash:=\frac{h}{2\pi}$  &			$	1,054571817 \cdot 10^{-34}$ \si{\joule\second}	&	reduced \textit{Planck} constant \cite{NISTredPlanck} \\

$\mu_{B} := \frac{e\hslash}{2m_{e}}$&				$9,2740100783\cdot 10^{-24}$ \si{\joule\per\tesla}		&	\textit{Bohr} magneton \cite{NISTmueB}\\

$\mu_{0} $&				$1,25663706212\cdot 10^{-6}$ \si{\newton\per\square\ampere}		&	vacuum magnetic permeability \cite{NISTmue0}  \\

$r_{B} $&				$5,29177210903\cdot 10^{-11}$ \si{\meter}		&	\textit{Bohr} radius \cite{NISTrBohr}  \\

\bottomrule						
\end{tabularx}
\end{table}
\clearpage

\textbf{Mathematical symbols and operators}
\begin{table}[H]
\renewcommand{\arraystretch}{0.75}
\newcolumntype{s}{>{\hsize=.3\hsize}X}
\newcolumntype{S}{>{\hsize=.98\hsize}X}
\begin{tabularx}{\textwidth}{sXS}
\toprule
Symbol				&				representation				& 				Denotation \\      
\midrule

$ vol\left(\Omega\right)$			&	 $  vol\left(\Omega\right) = \displaystyle{\int\limits_{\Omega}} d^{\,n}x 	$		&	volume of a $n$-dimensional set $\Omega$   \\\\

$ \bar{\psi}$			&	 $ \psi \bar{\psi} = \vert \psi \vert	^{2} \;\; ; \;\; \psi \in  \mathbb{C}	$		&	complex conjugate  \\\\

$ \psi^{\dagger}$			&	 $ \psi^{\dagger} := \left(\bar{\psi}\right)^{\top} \;\; ; \;\; \psi \in \mathbb{C}	$		&	 transposed conjugate \\\\

$ \vec{e}_{i}$			&	 $ \vec{e}_{i}:=\left( e_{j}\right);\, e_{j}=\begin{cases}1\;;\,j=i \\ 
0 \;;\,j\neq i 
\end{cases}; $		&	 cartesian unit vectors in $d$ dimensions \\
 &	 $j=\left\lbrace 1,2,...,d\right\rbrace;\;\vert\vec{e}_{i}\vert=1	$		&	  \\\\

$ \vec{\nabla}	$			&		$ \vec{\nabla}=\displaystyle{\sum\limits_{i=1}^{d}}\,\dfrac{\partial}{\partial x_{i}}\vec{e}_{i}$			&	\textit{Nabla}-Operator in $d$ cartesian dimensions \\\\

$ \Delta	$			&		$\Delta= \vec{\nabla}\cdot\vec{\nabla}=\displaystyle{\sum\limits_{i=1}^{d}}\,\dfrac{\partial^{2}}{\partial x_{i}^{2}} $			&	\textit{Laplace}-Operator in $d$ cartesian dimensions \\\\

$ \delta_{ij} $&	$\delta_{ij}= \begin{cases}1\;;\,i=j \\ 
0 \;;\,i\neq j 
\end{cases}$	&	\textit{Kronecker}-delta  \\\\

$ \underline{\sigma}_{0} $&	$\underline{\sigma}_{0}\equiv \underline{\mathds{1}} = \begin{pmatrix}
1 & 0  \\ 
0 & 1    \\ 
\end{pmatrix}$	&	\textit{Pauli}-matrices  \\\\

$ \underline{\sigma}_{x}$&	$\underline{\sigma}_{x}= \begin{pmatrix} 
0 & 1  \\ 
1 & 0    \\ 
\end{pmatrix}$	&	\textit{Pauli}-matrices $\cite{Eschrig1996}$ \\\\

$ \underline{\sigma}_{y}$&	$\underline{\sigma}_{y}=  \begin{pmatrix}
0 & -i  \\ 
i & 0    \\ 
\end{pmatrix}$	&	\textit{Pauli}-matrices $\cite{Eschrig1996}$ \\\\

$ \underline{\sigma}_{z}$ &	$\underline{\sigma}_{z}=  \begin{pmatrix} 
1 & 0  \\ 
0 & -1    \\ 
\end{pmatrix}$	&	\textit{Pauli}-matrices $\cite{Eschrig1996}$ \\\\

$ \underline{\alpha}^{i}$			&		$\underline{\alpha}^{i}=  \begin{pmatrix}
0 & \underline{\sigma}_{i}  \\ 
\underline{\sigma}_{i} & 0    \\ 
\end{pmatrix}  $			&	\textit{Dirac}-matrices  $\cite{Eschrig1996}$\\\\

$ \underline{\gamma}^{i}$			&		$\underline{\gamma}^{i}=  \begin{pmatrix} 
0 & \underline{\sigma_{i}}  \\ 
-\underline{\sigma_{i}} & 0  \\ 
\end{pmatrix}$	&	\textit{Dirac}-matrices $\cite{Eschrig1996}$\\\\

$ \underline{\beta}$			&		$\underline{\beta}:=\underline{\gamma}^{0} $	&	\textit{Dirac}-matrices $\cite{Eschrig1996}$\\\\

\bottomrule						
\end{tabularx}
\end{table}

\clearpage

\textbf{Mathematical symbols and operators in spherical coordinates}
\begin{table}[H]
\renewcommand{\arraystretch}{0.75}
\newcolumntype{s}{>{\hsize=.2\hsize}X}
\newcolumntype{S}{>{\hsize=0.8\hsize}X}
\begin{tabularx}{\textwidth}{sXS}
\toprule
Symbol				&				representation				& 				Denotation \\      
\midrule
$ r$			&	 $ r \in  \mathbb{R}^{+} :=	\left[0,\,\infty\right)	$		&	radial coordinate  \\\\

$ \theta$			&	 $ \theta \in  \left[0,\,\pi\right]	$		&	polar angle  \\\\

$ \varphi$			&	 $ \varphi \in  \left[0,\,2\pi\right)		$		&	 azimuthal angle  \\\\

$ \vec{e}_{r}$			&	 $  \vec{e}_{r}:=  \begin{pmatrix} 
\sin{\left(\theta\right)}\cos{\left(\varphi\right)}   \\ 
\sin{\left(\theta\right)}\sin{\left(\varphi\right) }   \\ 
 \cos{\left(\theta\right)}    \\
\end{pmatrix} 	$		&	radial basis vector \cite{Bartelmann_2015} \\\\

$ \vec{e}_{\theta}$			&	 $  \vec{e}_{\theta}:=  \begin{pmatrix}
\cos{\left(\theta\right)}\cos{\left(\varphi\right) }  \\ 
\cos{\left(\theta\right)}\sin{\left(\varphi\right)  }  \\ 
 -\sin{\left(\theta\right)}    \\
\end{pmatrix} 	$		&	polar basis vector \cite{Bartelmann_2015}\\\\

$ \vec{e}_{\varphi}$		&	 $  \vec{e}_{\varphi}:=  \begin{pmatrix}
-\sin{\left(\varphi\right)}   \\ 
\cos{\left(\varphi\right) }   \\ 
 0    \\
\end{pmatrix} 	$		&	azimuthal basis vector \cite{Bartelmann_2015}\\\\

$ \vec{r}$			&	 $ \vec{r}:=r\cdot\vec{e}_{r}\;;\;\;r:=\vert\vec{r}\,\vert	$		&	position vector  
\end{tabularx}
\end{table}
\begin{table}[H]
\renewcommand{\arraystretch}{0.75}
\newcolumntype{s}{>{\hsize=.12\hsize}X}
\newcolumntype{S}{>{\hsize=.65\hsize}X}
\begin{tabularx}{\textwidth}{sX}
$ d\Omega$			&	  	 volume element with \textit{Jacobian} determinant\\\\

 	&	 $ d\Omega:=d^{3}r=r^{2}\sin{\left(\theta\right)}\,dr\,d\theta \,d\varphi	$		  \\\\

$ \vec{\nabla}	$			 	&	\textit{Nabla}-Operator in spherical coordinates \cite{Bartelmann_2015} \\\\

 	&		$ \vec{\nabla}:=\vec{e}_{r}\cdot\dfrac{\partial}{\partial r}+\vec{e}_{\theta}\cdot\dfrac{1}{r}\dfrac{\partial}{\partial \theta}+\vec{e}_{\varphi}\cdot\dfrac{1}{r\sin{\left(\theta\right)}}\dfrac{\partial}{\partial \varphi}$				
\\\\
$ \vec{\nabla}\cdot \vec{f}	$			 	&	divergence of a vector field $\vec{f}$ in spherical coordinates \cite{Bartelmann_2015} \\\\

 	&		$ \vec{\nabla}\cdot \vec{f}=\dfrac{1}{r^{2}}\dfrac{\partial}{\partial r}\left(r^{2}f_{r}\right)+ \dfrac{1}{r \sin{\left(\theta\right)}}\dfrac{\partial}{\partial \theta}\left(\sin{\left(\theta\right)}f_{\theta}\right)+ \dfrac{1}{r\sin {\left(\theta\right)}}\dfrac{\partial\,f_{\varphi}}{\partial \varphi} $				
\\\\

$ \Delta	$			 	&	\textit{Laplace}-Operator in spherical coordinates \cite{Bartelmann_2015} \\\\

 	&		$\Delta:=\dfrac{1}{r^{2}}\dfrac{\partial}{\partial r}\left(r^{2}\dfrac{\partial}{\partial r}\right)+ \dfrac{1}{r^{2}\sin{\left(\theta\right)}}\dfrac{\partial}{\partial \theta}\left(\sin{\left(\theta\right)}\dfrac{\partial}{\partial \theta}\right)+ \dfrac{1}{r^{2}\sin^{2}{\left(\theta\right)}}\dfrac{\partial^{2}}{\partial \varphi^{2}} $			
\\\\
\bottomrule						
\end{tabularx}
\end{table}

\clearpage

\textbf{Indices and abbreviations}
\begin{table}[H]
\renewcommand{\arraystretch}{0.9}
\newcolumntype{S}{>{\hsize=0.45\hsize}X}
\begin{tabularx}{\textwidth}{SX}
\toprule
Symbol				&				meaning \\      
\midrule

el & electron \\

$\textup{N}_{\textup{el}}$ & total number of electrons \\

eig & \textit{eigen} \\

ext & external \\

eff & effective \\

n & self-consistent iteration over the electron density $n$ \\

1s & $1s$-subshell \\

3d & $3d$-subshell \\

4f & $4f$-subshell \\

H & \textit{Hartree} \\

BH & \textit{Breit-Hartree} \\

CH & \textit{Coulomb-Hartree} \\

XC & exchange and correlation \\

mo & model \\

nuc-el & nuclei-electron interaction \\

LL & \textit{Levy-Lieb} \\

B & \textit{Bohr} \\

$0$ & ground state \\

z & $z$-axis\\

c & contact \\

d & dipolar \\

dip & dipole \\

$\uparrow,\,\downarrow$ & spin up, spin down \\

m & magnetic \\

i & internal space \\

M & magnetized matter \\

e &	exterior space \\

R & full sphere with radius $R$\\

reg. & regularisation(s) \\

tot. & total \\

max & maximal \\

num. & numerical \\

analyt. & analytical \\

I.b.P. & integration by parts \\

hom & homogeneous solution \\

part & particular solution \\
 
LH & \textit{Lorentz}-\textit{Heaviside} unit system \\

GCGS & \textit{Gaussian} \textit{Centimetre–Gram–Second} system of units \\

SI &  International System of Units \\

\bottomrule						
\end{tabularx}
\end{table}
\begin{table}[H]
\renewcommand{\arraystretch}{0.9}
\newcolumntype{S}{>{\hsize=0.45\hsize}X}
\begin{tabularx}{\textwidth}{SX}     
\midrule
DFT & density functional theory \\

SDFT & spin density functional theory \\

LDA & local density approximation \\

GGA & generalized gradient approximation\\

LSDA & local spin density approximation \\

FPLO & full-potential local-orbital code\footnotemark[0]  \\

data & numerically calculated data \\

Ln & lanthanide \\

S & spin \\

L & oribtal \\

so & spin-orbit \\

C,MP2 & second order \textit{M\o{}ller-Plesset} correlation energy  \\

F & \textit{Fermi} \\

$\textup{Gd}^{3+}$ & $\textup{Gd}^{3+}$-ion \\

$\textup{Mn}^{2+}$ & $\textup{Mn}^{2+}$-ion \\

i.a. & inter alia \\

e.g. & exempli gratia \\
\bottomrule						
\end{tabularx}
\end{table}
 \footnotetext{https://www.fplo.de,
version dirac21.00-61}

%% file: Introduction.tex
\chapter{Introduction}
Relativistic effects are vital, for example, in the understanding of magnetism and highly resolved atomic or ionic spectra, and can be adequately included as relativistic corrections in the relativistic Spin Density Functional Theory (SDFT) \cite{ferenc}, \cite{Richter2001}, \cite{Jansen1988}, which e.g.\hspace{1mm}can be used to calculate the total ground state energy of atomic and ionic states. Consequently, precisely resolved spectroscopy measurements can be predicted theoretically, for instance for the trivalent ionic states of the lanthanides \cite{Melsen1994CalculationsOV}. However, there  are still deviations between the predicted and experimental values, which require further corrections \cite{Melsen1994CalculationsOV}.
\\
Hence, the analysis of the lowest order contribution to the \textit{Coulomb} energy, which is the quantum electrodynamical \textit{Breit} interaction, is motivated. The main part of this interaction, in the zero-frequency or non-retardation limit \cite{ferenc}, \cite{Chantler_2014} and especially in the \textit{Hartree} approximation, is a magnetic interaction, which is aquivalent to the magnetic dipole-dipole interaction \cite{Pellegrini_2020}.\hspace{1.5mm}It is also capable of explaining magnetic shape anisotropy and depends on the magnetization density originating from spin and orbital angular momentum density \cite{ComplexMaterials}. Hence, one assumes a non-zero contribution from this so-called \textit{Breit-Hartree} interaction for atomic or ionic states with open shells, which provide a high spin polarization.\hspace{1.5mm}Consequently, an evaluation of the \textit{Breit-Hartree} contribution to the total ground state energy of open atomic shells is interesting for the purpose of improvement of relativistic Spin Density Functional Theory calculations and the possible explanation of deviations from experimental results.
\\
Subsequently, the main aim of this thesis is the analytical and numerical calculation of the \textit{Breit-Hartree} contribution of a hollow sphere with a radially dependent magnetization density, as the zeroth order approximation of spin polarized atoms and ions. Moreover, and due to these restrictions, one even expects a possible solution to be generally applicable for spherical bulk volumes with an arbitrary, radially dependent magnetization densitiy. Additionally, the evaluation of the \textit{Breit-Hartree} contribution includes the exemplary application of a possible solution as well as the discussion of the magnitude of this first order correction to e.g. the $\textup{Mn}^{2+}$ and $\textup{Gd}^{3+}$ ion. Therefore, one has to investigate the origin of the \textit{Breit-Hartree} interaction and its connection to the magnetostatic dipole-dipole interaction in the framework of relativistic Density Functional Theory (DFT) to tie onto current research, referring to recent publications which e.g.\hspace{1mm}focused on the magnetic dipole-dipole interaction for structures at the nanoscale within the DFT concept \cite{Pellegrini_2020} or applied the \textit{Breit} interaction for bulk systems, for the purpose of the examination of magnetic anisotropy. Thus, the analysis of the intra-atomic \textit{Breit-Hartree} interaction ideally extends their research. 
  
\clearpage

%% file: Origin_of_the_Breit-Hartree_interaction_and_its_importance_in_condensed_matter_physics_V2.tex
\chapter{Origin of the \textit{Breit-Hartree} interaction and its importance in condensed matter physics}

\section{Preliminary considerations for atomic many-electron systems}

This work will consider the \textit{Breit-Hartree} contribution as a correction to the ground state total energy of an atomic or ionic state. Such a state is indeed a many-electron system, requiring the application of a quantum mechanical many-body theory, which one generally constructs for multiple nuclei, whereby the \textit{Breit-Hartree} contribution contributes per atom or ion. 

First, the  \textit{Born}-\textit{Oppenheimer} approximation is applied, which allows one to solve the many-electron \textit{Schrödinger} equation separately from the \textit{Schrödinger} equation for the nuclei. Furthermore, one considers a stationary system of $N_{nuc}$ atomic nuclei at rest and $N_{el}$ electrons, neglecting any motions of the nuclei. Thus this work focuses on the electronic structure in atomic systems. The electrons possess a spin $\sigma\in\left\lbrace\uparrow,\downarrow\right\rbrace$ and are described by the many-particle wave functions $\Psi_{\nu}\left(\vec{r}_{1},\sigma_{1};\dots;\vec{r}_{N_{el}},\sigma_{N_{el}}\right)$ to the energy values $E_{\nu}$, which are the solutions of the many-particle \textit{Schrödinger} equation \cite{RichterLecture}:
\vspace*{-3mm}\begin{equation}\label{SchrödingerManyElectron}
\hat{H}\;\Psi_{\nu}\left(\vec{r}_{1},\,\sigma_{1};\,\dots;\,\vec{r}_{N_{el}},\,\sigma_{N_{el}}\right) = E_{\nu}\;\Psi_{\nu}\left(\vec{r}_{1},\,\sigma_{1};\,\dots;\,\vec{r}_{N_{el}},\,\sigma_{N_{el}}\right)
\end{equation}
In (\ref{SchrödingerManyElectron}) $\hat{H}$ is the many-particle Hamilton operator, which can be decomposed into a kinetic part $\hat{T}$, a part for the electron-nuclei interaction $\hat{U}$ and the electron-electron interaction part $\hat{W}$. It is given by \cite{RichterLecture} in natural units and reads:
\begin{equation}\label{HamiltonManyElectron}
\hat{H} = -\dfrac{1}{2}\sum_{i=1}^{N_{el}} \Delta_{i} + \sum_{i=1}^{N_{el}} v_{ext}(\vec{r}_{i}) + \sum_{i\neq j}^{N_{el}} w(\vert \vec{r}_{i}-\vec{r}_{j}\vert) = \hat{T} + \hat{U} + \hat{W}   
\end{equation}
Where $\Delta_{i}$ is the \textit{Laplace} operator and $ v_{ext}(\vec{r}_{i},\,\sigma)$ is the potential of the external field generated by the nuclei acting on a particle at the position $\vec{r}_{i}$ with spin $\sigma$. For the case of the \textit{Coulomb} interaction, we can say that $w(\vert \vec{r}_{i}-\vec{r}_{j}\vert)=\vert \vec{r}_{i}-\vec{r}_{j}\vert^{-1}$. 
 
An analytic solution for (\ref{SchrödingerManyElectron}) can be found for $1 \leq N_{el} \leq 3$, with the result that for $N_{el}>3$ a solution has to be approximated \cite{RichterLecture}. In addition one can claim that in general for $N_{el}>100$, stationary states of the many-particle wave function $\psi_{\nu}$ cannot be resolved experimentally \cite{RichterLecture}. However, typical problems in solid state physics involve $N_{el}\gg 100$ electrons. Although a many-particle wave function cannot be calculated in such cases, one can still gain information about the system from its ground state. 
\\
The energy of the ground state $E_{0}$ of the many-particle \textit{Schrödinger} equation (\ref{SchrödingerManyElectron}) can be computed by means of the Density Functional Theory without the use of the corresponding many-particle wave function $\Psi_{0}$. In order to do this, all quantities can be expressed as functionals of the electronic density, which will be noted as $F\left[n\right]:=F\left[n\left(\vec{r}\,\right)\right]$ for a functional $F$ with functions $n\left(\vec{r}\,\right)$ as variables. 
\\
The total energy is the observable quantity of the hermitian \textit{Hamilton} operator. The expectation value of the total energy $E_{\nu}$ with respect to the state $\Psi_{\nu}$ is then defined using the \textit{Dirac} notation \cite{RichterLecture}:
\begin{equation} \label{Generalexpectationvalues}
E_{\nu} =\sum\limits_{\sigma}\int \Psi^{\dagger}_{\nu}\left(\vec{r},\,\sigma\right)\hat{H}\left(\vec{r},\,\sigma\right)\Psi_{\nu}\left(\vec{r},\,\sigma\right)\,d^{3}r=: \langle \Psi_{\nu} \vert \hat{H} \vert \Psi_{\nu} \rangle
\end{equation}
In this context, the equation (\ref{Generalexpectationvalues}) describes the energy $E_{0}$ of the ground state $\Psi_{0}$ for $\nu=0$. 

In conclusion, the DFT forms the theoretical framework for the calculations in this thesis, since its theoretical concept can also be used for the analysis of single atomic and ionic states if the nuclear potential is computed from only one nucleus.
\clearpage

\section{Density Functional Theory}

The Density Functional Theory (DFT) is a quantum mechanical approach to study the electronic structure and properties of the ground state of many-electron systems. Therefore it is also useful for the calculation of the ground state energy of atomic systems like single atoms or ions, due to the fact that $N_{el}> 3$ holds with only a few exceptions. 

\begin{figure}[H]
\begin{center}
\includegraphics[trim=5 5 5 -7,width=0.44\textwidth]{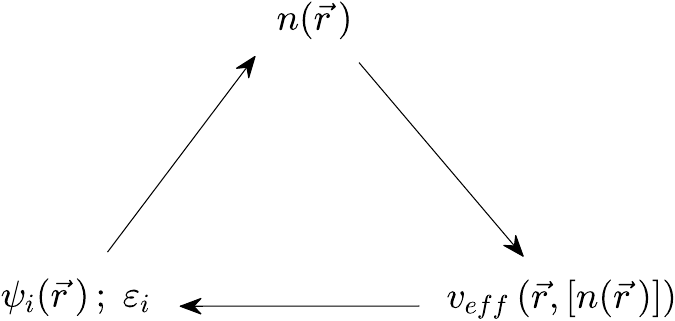}
\end{center}
\caption{Iteration cycle for the self-consistent solution density} \label{Iteration cycle}
\end{figure}

Within the concept of DFT all quantities are expressed as functionals of the electron density $n\left(\vec{r}\,\right)$ which is given by \cite{Eschrig1996},\cite{RichterLecture},\cite{Dreizler1985}:
\begin{equation}\label{DefinitionElecDens}
n\left(\vec{r}\,\right) = \sum\limits_{\sigma} \vert \Psi (\vec{r},\,\sigma) \vert^{2} = \sum\limits_{\sigma}\Psi^{\dagger}(\vec{r},\,\sigma)\Psi(\vec{r},\,\sigma)
\end{equation}
Consequently, the electron density for an atomic system with $N_{el}$ electrons must fulfill the following condition:
\vspace*{-3mm}\begin{equation}\label{ConditionElecDens}
\int\limits_{\mathbb{R}^{3}} n(\vec{r}\,) \, d^{3}r= N_{el}
\end{equation}
All Density Functional Theories are based on the \textit{Hohenberg-Kohn} theorems, which predict that the electron density of non-degenerated ground states determines the external potential $v_{ext}\left(\vec{r}\,\right)$ \cite{Dreizler1985}. As a consequence the \textit{Hamilton} operator $\hat{H}\left[v_{ext}\right]$ is fully defined \cite{Dreizler1985}. By definition, the ground state energy is equal to the minimum of the expectation value of the \textit{Hamilton} operator of all normalized, fermionic, many-particle wave functions $\Psi_{n}$ generating a particular density $n\left(\vec{r}\,\right)$ \cite{Eschrig1996},\cite{RichterLecture}:
\begin{equation}\label{groundstateenergy1}
E_{0}\left[v_{ext}\right]:=\underset{n}{\textup{inf}} \left\lbrace\underset{\Psi_{n}}{\textup{inf}}\left\lbrace \langle \Psi_{n} \vert \hat{H}\left[v_{ext}\right] \vert\Psi_{n} \rangle \right\rbrace\left|\,\int\limits_{\mathbb{R}^{3}} n(\vec{r}\,) \, d^{3}r= N_{el}\right.\right\rbrace
\end{equation}
In (\ref{groundstateenergy1}) the first infimum has to be computed with respect to the wave functions $\Psi_{n}$, while the second infimum has to be found with repect to the electron density $n$ and under the conditon of a physical density, which fullfills relation (\ref{ConditionElecDens}) for a given system with $N_{el}$ electrons. With the \textit{Hamiltonian} from (\ref{HamiltonManyElectron}) one can derive the \textit{Hohenberg-Kohn} variational principle containing the \textit{Levy-Lieb} functional $F_{LL}\left[n\right]$ starting from (\ref{groundstateenergy1}) and following the steps shown in \cite{RichterLecture}:
\begin{equation}\label{groundstateenergy2}
E_{0}\left[v_{ext}\right]=\underset{n}{\textup{inf}} \Biggl\{ \,\underbrace{\int\limits_{\mathbb{R}^{3}} n \cdot v_{ext}\,d^{3}r}_{=:E_{nuc-el}\left[n\right]}+ \underbrace{\underset{\Psi_{n}}{\textup{inf}}\left\lbrace\langle \Psi_{n} \vert \hat{T}+\hat{W} \vert \Psi_{n} \rangle \right\rbrace}_{=:F_{LL}\left[n\right]}\left|\,\int\limits_{\mathbb{R}^{3}} n(\vec{r}\,) \, d^{3}r= N_{el}\right.\Biggr\} 
\end{equation}
The \textit{Levy-Lieb} functional is then split up, in order to separately analyse the functionals contained in it \cite{RichterLecture}:
\begin{equation}\label{LevyLieb}
F_{LL}\left[n\right]=\underset{\Psi_{n}}{\textup{inf}}\left\lbrace\langle \Psi_{n} \vert \hat{T}+\hat{W} \vert \Psi_{n} \rangle \right\rbrace=:T_{mo}\left[n\right] + E_{H}\left[n\right] + E_{XC}\left[n\right]
\end{equation}
From (\ref{groundstateenergy2}) and (\ref{LevyLieb}) one finds all functionals, which determine the ground state energy:
\vspace*{-5mm}\begin{equation}\label{groundstateenergy3}
E_{0}\left[v_{ext}\right]=\underset{n}{\textup{inf}} \Biggl\{ E_{nuc-el}\left[n\right]+T_{mo}\left[n\right] + E_{H}\left[n\right] + E_{XC}\left[n\right]  \left|\,\int\limits_{\mathbb{R}^{3}} n(\vec{r}\,) \, d^{3}r= N_{el}\right.\Biggr\} 
\end{equation}
The occurring functional $E_{nuc}\left[n\right]$ represents the electrostatic interaction of the electron density with the external field of the nuclei, while $T_{mo}\left[n\right]\neq T\left[n\right]$ is defined as the kinetic energy of a \textit{model} system of non-interacting electrons, which fulfill an effective one particle \textit{Schrödinger} equation. The electrostatic self-interaction of the electron density is encoded in the functional $E_{H}\left[n\right]$, whereat the electrons are assumed to be uncorrelated using the \textit{Hartree} or \textit{mean field} treatment. All of the above contributions to the total energy of the ground state are exact at this point, as their mathematical representation is unambiguous. However, the functional $E_{XC}\left[n\right]$ cannot be written in closed mathematical form, meaning that it has to be approximated e.g. with expressions for model systems like the homogeneous electron gas, in order to allow practical use of the theory. In general it describes the exchange-correlation interaction, due to the \textit{Pauli} exclusion principle for fermions and the \textit{Coulomb} repulsion as well as all other contributions which are not included in the functionals $E_{H}\left[n\right]$ and $T_{mo}\left[n\right]$. It therefore contains additional potential and kinetic energy terms. 
\\
DFT-programs use the \textit{Hohenberg-Kohn} variational principle to compute the ground state energy of a given system with the help of an effective single-particle \textit{Schrödinger} equation (so-called \textit{Kohn}-\textit{Sham} equation \cite{RichterLecture}, \cite{Dreizler1985}):
\begin{equation}\label{Schrödinger Hohenberg Kohn Hartree}
\left(-\dfrac{\Delta}{2} + v_{eff}\left(\vec{r}, \left[n\right]\right)\right) \psi_{i}(\vec{r})= \varepsilon_{i}\psi_{i}(\vec{r})
\end{equation}
In order to conduct these computations, one needs an approach for the \textit{exchange-correlation} energy functional. Commonly used methods for the description of the exchange functional are the Local Density Approximation (LDA) and the Generalized Gradient Approximation (GGA) \cite{Richter2001}. Additionally, a starting set of $N_{eig}=N_{el}$ single particle eigenstates is needed to generate the initial density $n(\vec{r}\,)$:
\vspace*{-2mm}\begin{equation}\label{Densityiteration}
n(\vec{r}\,)= \sum\limits_{i=1}^{N_{eig}} \vert\psi_{i}(\vec{r}\,)\vert^{2} = \sum\limits_{i=1}^{N_{eig}} \psi_{i}^{\dagger}(\vec{r}\,)\psi_{i}(\vec{r}\,)
\end{equation}
Now an effective potential $v_{eff}\left(\vec{r}, \left[n\right]\right)$ \cite{Dreizler1985},\cite{RichterLecture}) is calculated from this inital density by means of a variation of the \textit{ansatz} functional $E_{XC}\left[n\right]$ with respect to the density:
\begin{equation}\label{VeffIteration}
v_{eff}\left(\vec{r}, \left[n\right]\right):= v_{ext} (\vec{r}\,) + v_{H} \left(\vec{r}, \left[n\right]\right)+ v_{XC}\left(\vec{r}, \left[n\right]\right)=v_{ext} (\vec{r}\,) + v_{H} \left(\vec{r}, \left[n\right]\right)+ \dfrac{\delta E_{XC}\left[n\right]}{\delta n} 
\end{equation}
The updated effective potential leads to new eigenenergies $\varepsilon_{i}$ and eigenstates $\psi_{i}$ and a new density can be calculated from (\ref{Densityiteration}). This is conducted as an iterative process (figure \ref{Iteration cycle}), which leads to an approximation of the ground state density as well as to its expection value of the total energy, if a predefined convergence criterion is reached.
\vspace*{-2mm}
All considered approximations ensure that an iteration towards a self-consistent solution of (\ref{Schrödinger Hohenberg Kohn Hartree}) will represent a valid approximation of the interactions within the electronic model system whereat the self-consitent density, eigenenergies and one particle eigenstates describe this model system. The dominant contributions included in this framework, for instance to the total energy of the system, are the kinetic energy of non-interacting electrons. Additionally, the \textit{Coulomb} interaction is considered in the \textit{mean-field} or \textit{Hartree} approximation, while other interaction effects like the \textit{Pauli}-exclusion principle are defined in the \textit{ansatz} for the exchange and correlation potential. The interaction with the nuclei is included with the external potential within the \textit{Born}-\textit{Oppenheimer} approximation. Further corrections can be made for instance by using a relativistic Density Functional Theory, in order to include relativistic contributions to the kinetic energy and to embrace spin-spin interactions of the dipole type. 
\clearpage

\section{Relativistic Spin Density Functional Theory}\label{sec:SDFT}

The generality of the \textit{Hohenberg}-\textit{Kohn} theory allows the investigation of relativistic effects within the relativistic Density Functional Theory. Thereunto, one replaces the effective single-particle \textit{Schrödinger} equation with an effective one-particle \textit{Dirac} equation, generating the \textit{Kohn-Sham-Dirac} equation \cite{Eschrig1996}. Additional relativistic corrections, for instance in a second order \textit{Taylor}-expansion in $\frac{1}{c}$, are then treated in the expression for the exchange and correlation functional. This allows oneself to understand fundamental properties of atoms, molecules and solids, which solely emanate from the relativistic theory. A popular example is the different color of gold in comparison to silver, due to the contraction of the circumference of the 1s-electron orbit. This is a consequence of the relativistic speed $v_{1s}>0,5\,c$ of electrons with orbit close to the nucleus. Other relativistic effects are the occupation anomaly within the actinides and the lanthanides \cite{Richter2001} and the fine as well as hyperfine structure splitting within atoms, leading to more diversified energy spectra.

The relativistic formulation of DFT, and thus the relativistic current Density Functional Theory derives the effective one-particle \textit{Kohn}-\textit{Sham}-\textit{Dirac} equation in the most general manner from a relativistic four-current density $J^{\mu}$ and its corresponding operator $\hat{j}^{\mu}$ \cite{Eschrig1996}:
\vspace*{-2mm}\begin{equation}
\left[ -ic\underline{\alpha} \vec{\nabla} + \underline{\beta} m_{0} c^{2} - e c \underline{\beta}\,\underline{\gamma}^{\mu} \left( A_{\mu} + a_{\mu} + a_{\mu}^{XC} \right)\right] \psi_{i} =\varepsilon_{i}  \psi_{i}
\end{equation}\vspace*{-11mm}
\begin{equation} \label{current}
 -ea_{\mu}^{XC} := \dfrac{\delta E_{XC}[J^{\mu}]}{\delta J^{\mu}} \;\;\;\;  ; \;\; \;\; J^{\mu} = c\sum_{i=1}^{N_{el}} \bar{\psi_{i}}\gamma^{\mu}\psi_{i}
\end{equation}
In (\ref{current}) $ J^{\mu} = \langle \Psi \vert \hat{j}^{\mu} \vert \Psi \rangle $ is the expectation value of the four-current density gained from the \textit{Kohn-Sham} orbitals $\psi_{i}$, $c$ is the speed of light in vacuum and $\gamma^{\mu}$ is a \textit{Dirac} matrix. Moreover, $A_{\mu}$ is the external four-vector potential, and the \textit{Hartree} four-potential $a_{\mu} $ as well as the exchange and correlation four-potential $a_{\mu}^{XC}$ originate from the gauge invariance of quantum electrodynamics, which implies the conservation of charge according to \textit{Noether's} theorem. In this context, $a_{\mu}^{XC}$ is defined by a variation of the exchange and correlation functional with respect to the variation of its functional variable $J^{\mu}$ \cite{Eschrig1996}. 

Typically, an approximated current Density Functional Theory is used in DFT-programs, whereas for the exchange and correlation functional the same expressions as for the nonrelativistic case are applied. Therefore, retardation effects within the electron-electron interaction are not included and the effective one-particle \textit{Kohn}-\textit{Sham}-\textit{Dirac} equation is solved by replacing the functional variable of the four-current with the electron density $n(\vec{r},\,\sigma)$ with respect to the spin orientation $\sigma\in\left\lbrace\uparrow,\downarrow\right\rbrace$. Such theories are often referred to as Spin Density Functional Theories (SDFT), and define the spin density as the difference between the spin-up and the spin-down density \cite{Ullrich_2019} with the help of (\ref{DefinitionElecDens}): 
\vspace*{-4mm}\begin{equation} \label{spindensity}
 s(\vec{r}\,):= n_{\uparrow}(\vec{r}\,)-n_{\downarrow}(\vec{r}\,)=n(\vec{r},\uparrow)-n(\vec{r},\downarrow)= \vert\Psi(\vec{r},\uparrow) \vert^{2}-\vert\Psi(\vec{r},\downarrow) \vert^{2}
\end{equation}
The spin density from (\ref{spindensity}) is then used to calculate the magnetization density along the $z$-axis, including the \textit{Bohr} magneton $\mu_{B}$ \cite{Ullrich_2019}:
\begin{equation} \label{magnetdensity}
 \vec{M}(\vec{r}\,)=-\mu_{B}\left( n_{\uparrow}(\vec{r}\,)-n_{\downarrow}(\vec{r}\,)\right)\vec{e}_{z}=-\mu_{B}s(\vec{r}\,)\vec{e}_{z}
\end{equation}
The magnetization density is calculated by means of the spin-projected electron densities according to (\ref{magnetdensity}) and can therefore be used for calculations within the framework of SDFT. Moreover, the functional equation (\ref{groundstateenergy3}) for the ground state energy is still valid if the correponding SDFT-functionals are used. Although, the main interest within DFT-research is to construct an \textit{ansatz} for the functional $E_{XC}\left[n\right]$, which represents the exchange and correlation between the electrons, the focus in SDFT is also themed on corrections to the representation of the functional $E_{H}\left[n\right]$, respectivly to the potential $v_{H} \left(\vec{r}, \left[n\right]\right)$ in (\ref{VeffIteration}). These corrections will eventually improve the approximation of the ground state energy as well as lead to better approaches for the exchange and correlation functional. Such a relativistic correction to the \textit{Coulomb} potential in the \textit{Hartree} treatment \cite{RichterLecture} and its corresponding functional $E_{CH}\left[n\right]$, which is the usual expression for $E_{H}\left[n\right]$, is the \textit{Breit-Hartree} interaction functional with respect to the magnetization density:
\vspace*{-1mm}\begin{equation} \label{E_Hcorrection}
E_{H}\left[n\right]\cong E_{CH}\left[n\right] + \Delta E_{\,BH}\left[\vec{M}\right]=\dfrac{1}{2}\int\dfrac{n\left(\vec{r}^{\,\prime}\right)n\left(\vec{r}\,\right)}{\vert \vec{r}-\vec{r}^{\,\prime}\vert}\,d^{3}rd^{3}r^{\,\prime}+ \Delta E_{\,BH}\left[\vec{M}\right]
\end{equation}
It is derived from the \textit{Breit} interaction and represents the lowest order relativistic correction to the \textit{Coulomb} interaction \cite{ComplexMaterials}, whereat it is a valid contribution for atomic states of atoms or ions with $Z\ll137$ \cite{Bethe}. Therefore an evaluation of the \textit{Breit} interaction in the \textit{mean-field} or \textit{Hartree} approximation and the use of its correction $\Delta E_{\,BH}\left[\vec{M}\right]$ to the total ground state energy of atomic systems is justified. Current applications of Density Functional Theory with estimations of the \textit{Breit} contribution only analyse bulk systems \cite{bornemann2010abinitio}, which motivates an analysis of the \textit{Breit-Hartree} correction to the total ground state energy of a single atom or ion. 
\clearpage

\section{\textit{Breit-Hartree} interaction}\label{BHINERACTION}

The virtual exchange particles responsible for electromagnetic interaction are photons, whose maximal propagation speed is the speed of light within vaccum $c=2,99792458  \cdot \, 10^{8} \; \textup{m}\, \textup{s}^{-1}$. Their finite speed leads to retardation effects pertaining the electron-electron interaction. Moreover, the relativistic \textit{Dirac} theory distinguishes electrons by reference to their spin, which represents the intrinsic angular momentum of an electron and leads to spin-spin interactions beyond those due to the Pauli principle, which are usually not included in nonrelativistic SDFT. However, they lead to corrections in relativistic SDFT for instance in the form of the subsequently analysed \textit{Breit-Hartree} contribution.
\\
In order to introduce the \textit{Breit} interaction one uses the framework of quantum electrodynamics as it was used by \textit{G. Breit} \cite{Breit_1929}, \cite{Breit_1932}, where relativistic corrections are generally defined by means of an operator $\hat{W}$ acting on a state $\Psi$. The contribution to the expectation value of the total energy $E$ is then given by the expectation value of this particular operator $W$, which is the matrix element $\langle \vec{\Psi}_{i} \vert \hat{W} \vert \vec{\Psi}_{j} \rangle$ for $i=j$, using standard \textit{Dirac} notation \cite{Wachter_2005}. Hence, for the energy contribution $W_{0}$ to the ground state $\vec{\Psi}_{0}$ one has to calculate the following six-dimensional integral, as $\hat{W}=\hat{W}\left(\vec{r},\,\vec{r}^{\,\prime}\right)$:
\begin{equation} \label{expectationvalue}
W_{0} = \langle \Psi_{0} \vert \hat{W} \vert \Psi_{0} \rangle=\int\limits_{\mathbb{R}^{3}\left(\vv{r}\right)} \int\limits_{\mathbb{R}^{3}\left(\vv{r}^{\,\prime}\right)}\Psi^{\dagger}_{0}\left(\vec{r}\,\right)\hat{W}\left(\vec{r},\,\vec{r}^{\,\,\prime}\right)\Psi_{0}\left(\vec{r}^{\,\,\prime}\right)\,d^{3}r\,d^{3}r^{\prime}
\end{equation}
Within the framework of relativistic Spin Density Functional Theory the modified electron-electron interaction (including the \textit{Coulomb} interaction) was derived by \textit{Jansen et al.} \cite{ComplexMaterials} and the kernel of the corresponding interaction operator $W_{\kappa\lambda\mu\nu}$ in lowest order of perturbation theory, assuming the summation over repeated indices reads:
\begin{equation} \label{WeeOperator}
W_{\kappa\lambda\mu\nu} = \dfrac{e^{2}}{\vert\vec{r}-\vec{r}^{\,\,\prime}\vert}\left(\delta_{\lambda\mu}\delta_{\kappa\nu}-\dfrac{1}{2}\left(\underline{\alpha}_{\lambda\mu}\,\underline{\alpha}_{\kappa\nu}+\dfrac{\left[\underline{\alpha}_{\lambda\mu}\left(\vec{r}^{\,\,\prime}-\vec{r}\,\right)\right]\bigl[ \underline{\alpha}_{\kappa\nu}\left(\vec{r}^{\,\,\prime}-\vec{r}\,\right) \bigr]}{\left(\vec{r}^{\,\,\prime}-\vec{r}\,\right)^{2}}\right)\right)
\end{equation}
In (\ref{WeeOperator}) the product of \textit{Kronecker} deltas constitutes the \textit{Coulomb} interaction, while the second term containing the \textit{Dirac} matrices is the \textit{Breit} operator, depicting the \textit{Breit} interaction. It is composed of a relativistic retardation correction (first term) to the \textit{Coulomb} interaction, due to the finite propagation speed of the virtual photons carrying the \textit{Coulomb} interaction and an expression (second term) corresponding to the magnetic interaction beetween the electron spins \cite{Chantler_2014}. Therefore the \textit{Breit} operator describes the interaction of the magnetic moments of two electrons conducted by the exchange of a single virtual photon \cite{Chantler_2014}, whose energy depends on their frequency and consequently classifies the \textit{Breit} operator as frequency-dependent too. Note that the \textit{Breit} interaction in this form only holds within the treatment of first order perturbation theory for a particluar atomic state \cite{Bethe} and is stated in the zero-frequency or non-retardation limit of the fully quantum electrodynamical interaction operator \cite{ferenc}, \cite{Chantler_2014}.
\\
As the leading relativistic correction the \textit{Breit} interaction shall be analysed in detail, by means of the previously mentioned \textit{Hartree} approximation. It was applied and written as an integral equation for the energy contribution to the total energy within a finite volume $\Omega\subset\mathbb{R}^{3}$ by \textit{H.J.F. Jansen} in 1988 \cite{Jansen1988}:
\begin{equation} \label{EnergieBH}
\begin{split}
\begin{gathered}
\Delta E_{\,BH}\left[\vec{M}\right] := \tilde{I}_{d} + \tilde{I}_{c}= \dfrac{1}{2} \int\limits_{\vv{r}\in\,\Omega}\int\limits_{\vv{r}^{\prime}\in\,\Omega} \dfrac{1}{\vert  \vec{r}-\vec{r}^{\,\,\prime} \vert^{3}}\left(\vec{M}\left(\vec{r}\,\right)\vec{M}\left(\vec{r}^{\,\,\prime}\right)\right.
\\[17pt]
\left.-3\cdot\dfrac{\left[ \left(\vec{r}-\vec{r}^{\,\,\prime}\right)\vec{M}\left(\vec{r}\,\right)\right]\left[ \left(\vec{r}-\vec{r}^{\,\,\prime}\right)\vec{M}\left(\vec{r}^{\,\,\prime}\right)\right]}{\vert  \vec{r}-\vec{r}^{\,\,\prime} \vert^{2}}\right)d^{\,3}rd^{\,3}r^{\,\prime}
-\dfrac{4\pi}{3} \int\limits_{\vv{r}\in\,\Omega} \vec{M}\left(\vec{r}\,\right)\vec{M}\left(\vec{r}\,\right)  d^{\,3}r
\end{gathered}
\end{split}
\end{equation}
Where $\vec{M}\left(\vec{r}\,\right)$ is the magnetization density as defined in (\ref{magnetdensity}) and the vectors $\vec{r},\,\vec{r}^{\,\,\prime}\in \Omega$ denote different position vectors within the considered volume. Note that in addition to the spin density the magnetization density contains orbital contributions from the whole electronic shell of the atom, which are also included in the \textit{Breit-Hartree} interaction \cite{Chantler_2014}. In this regard, the magnetization density generally constitutes the total density of momenta within the total electron shell of an atom with the exception of the nucleus and is therefore related to the orbital
angular momentum density and the spin density \cite{ComplexMaterials}. However, the focus of this thesis is the analysis of the \textit{Breit-Hartree} interaction in specific atomic systems, in which the orbital momenta can be neglected and the spin magnetization density is dominant. An example with a large spin magnetization is the $\textup{Gd}^{3+}$ ion, which has a half filled $4f$ shell containing $7$ $f$-electrons and no orbital momentum due to the spherical symmetry of the electron density (orbital angular momentum $L=0\,\textup{kg}\textup{m}^{2}\textup{s}^{-1}$). Based on the dependence of the magnetization density on the spin density given in (\ref{magnetdensity}), the \textit{Breit-Hartree} contribution is a functional of the spin density and therefore more relevant if the spin-up density $n_{\uparrow}\left(\vec{r}\,\right)$ and the spin-down density $n_{\downarrow}\left(\vec{r}\,\right)$ are strongly different. This for instance occurs in open atomic shells and is mainly responsible for a strong total atomic magnetic moment of an atom. In the limit of a vanishing spin density $s\left(\vec{r}\,\right)\rightarrow\,0$, the magnetization density strives towards zero, which leads to a zero contribution of the \textit{Breit-Hartree} interaction to the total energy of the volume $\Omega$ of the atomic system.
In addition, one can state from (\ref{EnergieBH}), that only if $\vec{r}\neq \vec{r}^{\,\,\prime}$ the integral $\tilde{I}_{d}$ has a finite integrand, assuming that the magnetization density is a continuous vector field within the finite volume $\Omega$. In addition, one notices that $\tilde{I}_{d}$ is the integral form of the second term in the \textit{Breit}-operator, which is the electron-electron interaction kernel (\ref{WeeOperator}) without the term for the \textit{Coulomb} interaction $\delta_{\lambda\mu}\delta_{\kappa\nu}$. It can be thought of as a representation of the magnetic interaction of the electron spins \cite{Pellegrini_2020}. Therefore $\tilde{I}_{d}$ is classified as the total energy contribution of the magnetic interaction within the spin density, which can also be declared as a dipolar interaction of the magnetization density with itself. Furthermore, it should remarked that the integral $\tilde{I}_{c}$ is only three-dimensional, in contrast to the six-dimensional integral $\tilde{I}_{d}$. The integrand of $\tilde{I}_{c}$ is a scalar product of $\vec{M}\left(\vec{r}\,\right)$ with itself and can be rewritten as:
\begin{equation} \label{IcIntegrand}
\vec{M}\left(\vec{r}\,\right)\cdot\vec{M}\left(\vec{r}\,\right) =\vert\vec{M}\left(\vec{r}\,\right)\vert\vert\vec{M}\left(\vec{r}\,\right)\vert\underbrace{\cos{\left(\measuredangle\left(\vec{M}\left(\vec{r}\,\right),\,\vec{M}\left(\vec{r}\,\right)\,\right)\right)}}_{=\,1}= \vert\vec{M}\left(\vec{r}\,\right)\vert^{2}
\end{equation}
(\ref{IcIntegrand}) demonstrates the exclusive dependence of $\tilde{I}_{c}$ on the magnitude of the magnetization density at one spacial point $\vec{r}$ in the volume. As a consequence, this allows the interpretation of $\tilde{I}_{c}$ as a contact interaction \cite{Pellegrini_2020}. Furthermore, a simular contact term can be recognizes from the magnetic field of a magnetic point dipole, by averaging over a finite volume (see also \cite{Bartelmann_2015}). This could be explained by the \textit{Hartree}-approximation of the \textit{Breit} interaction as a mean-field treatment for the many-electron system, averaging the interaction of the electrons/spins to an equivalent interaction of an electron/spin density. Therefore the contribution of \textit{Breit}-\textit{Hartree} interaction to the total energy of a finite volume is the nonrelativistic limit of the \textit{Breit} interaction and is furthermore equivalent to the potential energy originating from the classical magnetostatic dipole-dipole interaction within magnetized matter. However, the disadvantage of the \textit{Hartree} approximation and the zero-frequency limit is that the retardation corrections of the general \textit{Breit} interaction do not contribute to the \textit{Breit}-\textit{Hartree} correction \cite{PhysRevA.4.41}, \cite{bornemann2010abinitio}. This makes it a mainly magnetic interaction, emphasizing the magnetostatic interpretation, whereat the magnetic part is in any case the dominant contribution for atomic calculations as it was shown by \textit{Kozlov} for the \textit{Breit} interaction without the \textit{Hartree} approximation \cite{Kozlov}. Additionally, a high quality of the \textit{Hartree} approximation of the \textit{Breit} interaction respectively the classical interpretation as the magnetic dipole-dipole interaction was justified by \textit{S. Bornemann et al.} for a monolayer bulk system \cite{bornemann2010abinitio}. This further motivates the analysis of the \textit{Hartree} treatment of the \textit{Breit} interaction for single atomic and ionic states.
\\
The integral form of the \textit{Breit}-\textit{Hartree} interaction from (\ref{EnergieBH}) was written by \textit{H.J.F. Jansen} \cite{Jansen1988} applying the \textit{Gaussian} \textit{Centimetre–Gram–Second} system of units (GCGS) whereat \textit{Jansen} deviates from the \textit{Lorentz}-\textit{Heaviside} unit system as his basic framework for notation within his work \cite{Jansen1988}, according to the reference \cite{mandl2010quantum}. Another verification of the unit system of (\ref{EnergieBH}) is the work of \textit{Pellegrini et al.} \cite{Pellegrini_2020}, which identically stated $\tilde{I}_{d}$ as the \textit{Hartree} part of the magnetic dipole-dipole interaction and $\tilde{I}_{c}$ as the magnetic contact interaction, citing \textit{Jansen's} paper. In order to enable better comparability with current research results, one converts equation (\ref{EnergieBH}) into SI-units. This means that mechanical quantities are left unchanged, while the magnetization density in GCGS-units is converted, by the use of a conversion factor \cite{Bartelmann_2015}, \cite{FranzSchwabl2019} containing the vacuum magnetic permeability $\mu_{0}$ \cite{NISTmue0}:
\vspace*{-1mm}\begin{equation} \label{SI}
\left[\vec{M}\right]^{\textup{GCGS}}=\sqrt{\dfrac{\mu_{0}}{4\pi}}\left[\vec{M}\right]^{\textup{SI}}
\end{equation}
Consequently, the SI-conforming integral representation of the \textit{Breit-Hartree} interaction depends exclusively on SI-quantities and can be derived through (\ref{EnergieBH}), (\ref{IcIntegrand}) and (\ref{SI}):
\vspace*{-3mm}\begin{equation}\label{EnergieBHSI}
\begin{split}
\begin{gathered}
\Delta E_{\,BH}\left[\vec{M}\right] := I_{d} + I_{c}= \dfrac{\mu_{0}}{8\pi} \int\limits_{\vv{r}\in\,\Omega}\int\limits_{\vv{r}^{\prime}\in\,\Omega}\dfrac{1}{\vert  \vec{r}-\vec{r}^{\,\,\prime} \vert^{3}}\left(\vec{M}\left(\vec{r}\,\right)\vec{M}\left(\vec{r}^{\,\,\prime}\right)\right.
\\[17pt]     
\left.-3\cdot\dfrac{\left[ \left(\vec{r}-\vec{r}^{\,\,\prime}\right)\vec{M}\left(\vec{r}\,\right)\right]\left[ \left(\vec{r}-\vec{r}^{\,\,\prime}\right)\vec{M}\left(\vec{r}^{\,\,\prime}\right)\right]}{\vert  \vec{r}-\vec{r}^{\,\,\prime} \vert^{2}}\right)d^{\,3}rd^{\,3}r^{\,\prime}
-\dfrac{\mu_{0}}{3}\int\limits_{\vv{r}\in\,\Omega} \vert\vec{M}\left(\vec{r}\,\right)\vert^{2}   d^{\,3}r
\end{gathered}
\end{split}
\end{equation}
For the purpose of an analysis in greater detail, one can express (\ref{EnergieBHSI}) only in terms of the magnitudes and angles of the vector quantities in the integrand. First, one defines the following angles:
\vspace*{-2mm}\begin{equation} \label{angles}
\begin{split}
\begin{gathered}
\kappa:=\measuredangle\left(\vec{r},\,\vec{r}^{\,\,\prime}\right);\;\alpha:=\measuredangle\left(\vec{M}\left(\vec{r}\,\right),\,\vec{M}\left(\vec{r}^{\,\,\prime}\right)\right);\;
\\[5pt]
\beta:=\measuredangle\left(\vec{r}-\vec{r}^{\,\,\prime},\,\vec{M}\left(\vec{r}\,\right)\right);\;
\gamma:=\measuredangle\left(\vec{r}-\vec{r}^{\,\,\prime},\,\vec{M}\left(\vec{r}^{\,\,\prime}\right)\right)
\end{gathered}
\end{split}
\end{equation}
From (\ref{angles}) and the law of cosine applied to the expression $\vert \vec{r}-\vec{r}^{\,\,\prime}\vert$ one receives from (\ref{EnergieBHSI}):
\vspace*{-3mm}\begin{equation}\label{EnergieBHangles}
\begin{split}
\begin{gathered}
\Delta E_{\,BH}\left[\vec{M}\right] = \dfrac{\mu_{0}}{8\pi} \int\limits_{\vv{r}\in\,\Omega}\int\limits_{\vv{r}^{\prime}\in\,\Omega} \dfrac{1}{\vert  \vec{r}-\vec{r}^{\,\,\prime} \vert^{3}}\left(\vert\vec{M}\left(\vec{r}\,\right)\vert\vert\vec{M}\left(\vec{r}^{\,\,\prime}\right)\vert\cos{\left(\alpha\right)}\right.
\\[15pt]
\left.-3\cdot\dfrac{\vert\vec{M}\left(\vec{r}\,\right)\vert\vert\vec{M}\left(\vec{r}^{\,\,\prime}\right)\vert\cancel{\vert\vec{r}-\vec{r}^{\,\,\prime} \vert^{2}}\cos{\left(\beta\right)}\cos{\left(\gamma\right)}}{\cancel{\vert\vec{r}-\vec{r}^{\,\,\prime}\vert^{2}}}\right)d^{\,3}rd^{\,3}r^{\,\prime}
-\dfrac{\mu_{0}}{3}\int\limits_{\vv{r}\in\,\Omega} \vert\vec{M}\left(\vec{r}\,\right)\vert^{2}  d^{\,3}r
\\[15pt]
= \dfrac{\mu_{0}}{8\pi} \int\limits_{\vv{r}\in\,\Omega}\int\limits_{\vv{r}^{\prime}\in\,\Omega} \dfrac{\vert\vec{M}\left(\vec{r}\,\right)\vert\vert\vec{M}\left(\vec{r}^{\,\,\prime}\right)\vert\bigl[\cos{\left(\alpha\right)}-3\cos{\left(\beta\right)}\cos{\left(\gamma\right)}\bigr]}{\bigl[\vert \vec{r}\,\vert^{2}+\vert\vec{r}^{\,\,\prime} \vert^{2}-2\vert\vec{r}\,\vert\vert\vec{r}^{\,\,\prime} \vert\cos{\left(\kappa\right)}\bigr]^{\frac{3}{2}}}\,d^{\,3}rd^{\,3}r^{\,\prime}
\end{gathered}
\end{split}
\end{equation}
\begin{equation*}
\begin{split}
\begin{gathered}
-\dfrac{\mu_{0}}{3} \int\limits_{\vv{r}\in\,\Omega} \vert\vec{M}\left(\vec{r}\,\right)\vert^{2}  \,d^{\,3}r
\end{gathered}
\end{split}
\end{equation*}
From (\ref{EnergieBHangles}) one can verify that the value of the integral carries units of energy. On that account numerical factors are ignored and the unit of the magnetization density is derived from (\ref{magnetdensity}):
\vspace*{-3mm}\begin{equation} \label{units}
\begin{split}
\begin{gathered}
\underline{\underline{\left[\vert\vec{M}\vert\right] }}= \left[\mu_B\right]\left[s\left(\vec{r}\,\right)\right]\left[\vec{e}_{z}\,\right] = \dfrac{\textup{J}}{\textup{T}}\cdot \dfrac{1}{\textup{m}^{3}} \cdot  1= \dfrac{\cancel{\textup{kg}}  \cdot \cancel{\textup{m}^{2}} \cdot \cancel{\textup{s}^{-2}}}{\cancel{\textup{kg}} \cdot \textup{A}^{-1} \cdot \cancel{\textup{s}^{-2}}}\cdot \dfrac{1}{\textup{m}^{\cancel{3}}} =\underline{\underline{\dfrac{\textup{A}}{\textup{m}}}}
\\[15pt]
\underline{\underline{\left[\Delta E_{\,BH}\right] }}= \left[\mu_{0}\right]\left\lbrace\dfrac{\left[\vert\vec{M}\vert\right]^{2}\left[f\left(\alpha,\beta,\gamma\right)\right]}{\left[\vert\vec{r}\,\vert\right]^{2\cdot\frac{3}{2}}}\left[dr\right]^{3+3}+\left[\vert\vec{M}\vert\right]^{2}\left[dr\right]^{3} \right\rbrace
\\[15pt]
= \dfrac{\textup{N}}{\cancel{\textup{A}^2}} \left\lbrace\dfrac{\cancel{\textup{A}^{2}}\cdot \cancel{\textup{m}^{-2}}\cdot 1}{\cancel{\textup{m}^{3}}}  \cdot \textup{m}^{\cancel{6}}+\cancel{\textup{A}^{2}}\cdot \cancel{\textup{m}^{-2}}\cdot \textup{m}^{\cancel{3}} \right\rbrace = \textup{Nm} = \underline{\underline{\textup{J}}}
\end{gathered}
\end{split}
\end{equation}
Finally, it must be remarked that the inclusion of the \textit{Breit-Hartree} correction in the \textit{Hartree}-functional (\ref{E_Hcorrection}) allows oneself to solve equation (\ref{EnergieBHSI}) with self-consistency. However, one expects the relative order of this contribution to be very small, which leads to the valid treatment of the \textit{Breit-Hartree} interaction as a perturbation correction to the total energy of the ground state, because its influence on the self-consistent DFT iterations is assumed to be negligible.\hspace{2mm}Hence, the \textit{Breit-Hartree} contribution is adequately evaluated, by means of (\ref{EnergieBHSI}) and the insertion of an independently and self-consistently found magnetization density. Consequently, one ascertains the definition of the magnetization density according to (\ref{spindensity}) and (\ref{magnetdensity}), that the \textit{Breit-Hartree} correction can be found directly from a spin-projected electron density, which was independently computed by a relativistic SDFT program. 
\clearpage

\section{Importance of the \textit{Breit-Hartree} contribution in condensed matter physics}\label{Important}

The \textit{Breit} interaction operator was originally worked out by \textit{G. Breit}, in order to evaluate the retardation effects of the interaction between two electrons occurring from the relativistic Dirac equation within the framework of quantum electrodynamics \cite{Breit_1929}. As a consequence, he further developed an operator similar to (\ref{WeeOperator}), in which diagonal elements correspond to proper energy values, representing a perturbation energy contribution to the energy eigenvalues of the \textit{Dirac} equation \cite{Breit_1932}. This originates from the exchange of a virtual photon, conducting the spin-spin interaction \cite{Breit_1932} and is generally dependent on the frequency of the exchanged photon. 
\\
In conclusion, the \textit{Breit} correction is assumed to deliver a crucial contribution to the total energy of spin systems, even in the low-frequency limit, because the frequency-dependent correction only contributes $10\%$ to the frequency-independent contribution \cite{Chantler_2014}. Particularly, in combination with the \textit{Hartree} approximation performed by \textit{Jansen} \cite{Jansen1988}, the so-called \textit{Breit-Hartree}-contribution includes the perturbation energy, which arises from the dipole-dipole interaction of the electron spin and is therefore interesting for spin-polarized and magnetized systems. In this regard, this interaction has to be interpreted as the dipolar interaction of two spin densities $s\left(\vec{r}\,\right)$ and $s\left(\vec{r}^{\,\prime}\right)$ leading to the magnetization densities $\vec{M}\left(\vec{r}\,\right)$ and $\vec{M}\left(\vec{r}^{\,\prime}\right)$ \cite{Pellegrini_2020}. It should be remarked, that this work mainly focuses on magnetization densities, which only arise from spin densities, although in general the magnetization density is the total angular momentum density. This interpretation is valid if a Spin Density Functional Theory is applied to calculate the expectation value of the total ground state energy, allowing an insertion of this energy correction for instance into current relativistic SDFT-code. Hence, the \textit{Breit-Hartree} contribution should be considered for all atoms with a proton number significantly below $Z=137$, in order to improve the approximation of the total energy of single atoms. This restriction for the application of the \textit{Breit} interaction in the \textit{Hartree} approximation ensures validity and was discussed in detail by \textit{H.A. Bethe} and \textit{E.E. Slapeter} \cite{Bethe}. However, one does not cover retardation effects with the inclusion of the \textit{Breit-Hartree} correction \cite{bornemann2010abinitio}, \cite{PhysRevA.4.41}, which means that retardation effects would have to be contained in the functional $E_{XC}\left[n\right]$, which is unusual, due to the fact that the main approaches are taken from the non-relativistic DFT. Nevertheless, the explicit calculation of the \textit{Breit-Hartree} contribution for certain atomic and ionic systems is encouraged, because it adds to recent research \cite{Pellegrini_2020}, which was able to model and approximate the exchange correlation part of the \textit{Breit} interaction for a homogeneous electron gas \cite{Pellegrini_2020}. 
\\
Furthermore, one knows from section \ref{sec:SDFT} that the magnitude of the \textit{Breit-Hartree} contribution depends on the spin density and is only non-zero for atoms with open shells. Therefore, its evaluation is essential for calculations of parameters of atoms with a distinctive magnetic moment, for instance within the lanthanides or actinides, because the total magnetic moment can be decomposed into the magnetic moments of the nucleus and the atomic electron shell, whereby the latter carries the main part of the total magnetic moment. In addition, current research does not exclude the possiblity that this interaction could deliver better approximations for the spin polarization energy of atomic shells \cite{Melsen1994CalculationsOV}, especially for heavy ions with high spin densitys like the $\textup{Gd}^{3+}$ ion, where the \textit{Breit-Hartree} correction is expected to be significant. Moreover, one should mention that the \textit{Breit-Hartree} contribution is expected to be of a vanishing magnitude for closed shells, as they represent the limit of a zero-spin density (section \ref{sec:SDFT}). Furthermore, it is a fact that orbital polarization energy contributions have a lower order than pure spin polarization contributions for open atomic shells \cite{Melsen1994CalculationsOV} and in particular can be left aside for half-filled atomic shells. Hence, the focus of this thesis is the evaluation of the \textit{Breit-Hartree} contribution for highly spin-polarized atomic shells, which promise significant corrections to the total energy of their ground state.
\\
Finally, the importance of the \textit{Breit-Hartree} interaction was shown by \textit{Jansen} \cite{Jansen1988}, by extracting its role as the origin of magnetic shape anisotropy. This is mainly justified by the fact that the \textit{Breit-Hartree} contribution represents the lowest order correction related with the spin-spin interaction. Therefore it is the main correction to the \textit{Coulomb} potential and improves the description of magnetically dominated atomic systems by improving the effective potential.

In general the above discussion of the importance of the \textit{Breit-Hartree} contribution in condensed matter physics promises further improvement of existing relativistic SDFT-programs.\hspace{1.5mm}These are commonly used in solid state physics and provide precise material parameters used in several research fields such as 2D and nano materials.
\clearpage

%% file: Evaluation_of_the_Breit-Hartree_contribution_to_the_total_energy_of_a_homogeneously_magnetized_sphere.tex
\chapter{Evaluation of the \textit{Breit-Hartree} contribution to the total energy of a homogeneously magnetized hollow sphere}\label{BHContributionMConst}

\section{Analytical approach}\label{AnApSphere}

In order to analyse the \textit{Breit-Hartree} contribution to the total energy of open atomic shells one starts with the evaluation of the integral equation (\ref{EnergieBHSI}) for the most simple magnetization density possible. Therefore, one defines a constant vectorfield $\vec{M}\left(\vec{r}\,\right)= \vec{\textup{const.}}$ for the magnetization density within a finite volume $\Omega$. In zeroth order an atom can be approximated as a hollow sphere with an inner radius $R_{1}=0\,\textup{m}$ and an outer radius $R_{2}>0\,\textup{m}$, whereby inside this volume $\Omega$ the magnetization density is a constant vector, thus the product of a constant magnitude $M$ and a constant unit vector $\vec{e}:=\left(e_{x}\,e_{y}\,e_{z} \right)^{T}$ with $\vert \vec{e} \,\vert = \sqrt{{e_{x}}^{2}+{e_{y}}^{2}+{e_{z}}^{2}}= 1$. The magnitude $M:=-\vert \vec{M}\left(\vec{r}\,\right) \vert=-\mu_{B}s$ always conforms to the definition (\ref{magnetdensity}), if $\vec{e}=\vec{e}_{z}$ holds, whereat $s$ is the constant spin density with reference to $\Omega$. As it is always possible to choose a coordinate system, where the magnetization is constant along the $z$-axis, one can set $\vec{e}=\vec{e}_{z}$ without loss of generality. The magnetization density outside the sphere equals the zero vector. As a consequence the vector field of the magnetization density reads:
\begin{equation}\label{defM}
\vec{M}\left(\vec{r}\,\right)= \vec{\textup{const.}} = 
\begin{cases}
M\vec{e}_{z}\overset{(\ref{magnetdensity})}{=} -\mu_{B}s\vec{e}_{z}&;\;\; \forall \; \vec{r},\,\vec{r}^{\,\,\prime} \,\in\,\Omega
\\
\vec{0}&;\;\; \forall \; \vec{r},\,\vec{r}^{\,\,\prime} \,\notin\,\Omega
\end{cases}
\end{equation}
The integration volume $\Omega$ as a sphere is then naturally described in spherical coordintes $(r,\,\theta,\,\varphi)$ which leads to $\Omega:=r\,\in\,\left[R_{1},R_{2}\right]\times\theta\,\in\,\left[0,\pi\right]\times\varphi\,\in\,\left[0,2\pi\right)$. The boundaries of the radial coordinate are chosen for a hollow sphere with the full sphere included as a special case of $\Omega$ for $R_{1}=0\,\textup{m}$, enabling oneself to obtain more general results. 
\\
The transformed position vectors $\vec{r},\,\vec{r}^{\,\,\prime}\in \Omega$ are given by:
\begin{equation}\label{sphereCoordin}
\vec{r}= \begin{pmatrix}
r\sin{\left(\theta\right)}\cos{\left(\varphi\right)}\\
r\sin{\left(\theta\right)}\sin{\left(\varphi\right)}\\
r\cos{\left(\theta\right)}
\end{pmatrix} \;\;;\;\; \vec{r}^{\,\,\prime}= \begin{pmatrix}
r^{\prime}\sin{\left(\theta^{\,\prime}\right)}\cos{\left(\varphi^{\,\prime}\right)}\\
r^{\prime}\sin{\left(\theta^{\,\prime}\right)}\sin{\left(\varphi^{\,\prime}\right)}\\
r^{\prime}\cos{\left(\theta^{\,\prime}\right)}
\end{pmatrix} 
\end{equation}
The volume elements $d^{\,3}r$ and $d^{\,3}r^{\,\prime}$ occurring in (\ref{EnergieBH}) need to be transformed with the \textit{Jacobian} determinant:
\begin{equation}\label{volElem}
d^{\,3}r=r^{2}\sin{\left(\theta\right)}\,dr d\theta d\varphi\;\;;\;\;d^{\,3}r^{\,\prime}=r^{\prime\,2}\sin{\left(\theta^{\prime}\right)}\, dr^{\prime} d\theta^{\prime} d\varphi^{\prime}
\end{equation}
In order to conduct the calculation of the Breit-Hartree contribution to the total energy of the homogeneously magnetized sphere $\Omega$, the structure of the integral equation can be recalled as a sum of two separate integrals $I_{d}$ and $I_{c}$ from (\ref{EnergieBHSI}). 

First, the three-dimensional integral $I_{c}$ from (\ref{EnergieBHSI}) is evaluated, whereat it represents the contact interaction within the spin density at same spatial points within the homogeneously magnetized sphere \cite{Pellegrini_2020}. Using (\ref{defM}), (\ref{volElem}) and the integration volume $\Omega$ one calculates the result (\ref{Icv1A}),(\ref{Icv1AA}):
\begin{equation}\label{Icv1}
\begin{split}
\begin{gathered}
\underline{\underline{I_{c} =-\dfrac{4\pi}{9}\mu_{0}M^{2} \left(R_{2}^{3}-R_{1}^{3}\right)=-\dfrac{1}{3}\mu_{0}M^{2}  vol\left(\Omega\right)}}
\end{gathered}
\end{split}
\end{equation}
One regards (\ref{Icv1}) and extracts the dependencies of the contribution of the contact interaction to the total energy of a hollow sphere with the volume $vol\left(\Omega\right)$, by inserting the representation of $M$ according to (\ref{defM}):
\begin{equation}\label{Icv2}
\begin{split}
\begin{gathered}
I_{c}= -\dfrac{1}{3}\mu_{0}\left(-\mu_{B}s\right)^{2} vol\left(\Omega\right)=-\dfrac{1}{3}\mu_{0}\mu_{B}^{2}s^{2} vol\left(\Omega\right)
\end{gathered}
\end{split}
\end{equation} 
In (\ref{Icv2}) the magnitude of the spin density has a quadratic influence on $I_{c}=I_{c}\left(s, vol\left(\Omega\right)\right)$, while the energy correction due to the contact interaction is only linearly dependent on the volume.
\\
Second, the integral $I_{d}$ from (\ref{EnergieBHSI}), which represents the dipolar interaction, has to be calculated for $\vec{r},\,\vec{r}^{\,\prime}\in\Omega$. From (\ref{defM}) one can show that the integrand has the typical format of a dipole field, mainly with respect to the exponents of the denominators: 
\begin{equation*}\label{Idv1a}
\begin{split}
\begin{gathered}
I_{d}:=\dfrac{\mu_{0}}{8\pi} \int\limits_{\vv{r},\,\vv{r}^{\,\prime}\,\in\,\Omega} \dfrac{1}{\vert  \vec{r}-\vec{r}^{\,\,\prime} \vert^{3}}\left(\vec{M}\left(\vec{r}\,\right)\vec{M}\left(\vec{r}^{\,\,\prime}\,\right)\right.
\end{gathered}
\end{split}
\end{equation*}
\begin{equation}\label{Idv1}
\begin{split}
\begin{gathered}
\left.-3\cdot\dfrac{\left[ \left(\vec{r}-\vec{r}^{\,\,\prime}\right)\vec{M}\left(\vec{r}\,\right)\right]\left[ \left(\vec{r}-\vec{r}^{\,\,\prime}\right)\vec{M}\left(\vec{r}^{\,\,\prime}\,\right)\right]}{\vert \vec{r}-\vec{r}^{\,\,\prime} \vert^{2}}\right)d^{\,3}rd^{\,3}r^{\,\prime}
\\[16pt]
\overset{(\ref{defM})}{=}\dfrac{\mu_{0}}{8\pi}M^{2} \int\limits_{\vv{r},\,\vv{r}^{\,\prime}\,\in\,\Omega}\dfrac{\overbrace{\vec{e}_{z}\cdot\vec{e}_{z}}^{=\vert\vec{e}_{z}\vert^{2}=1}}{\vert  \vec{r}-\vec{r}^{\,\,\prime} \vert^{3}}-3\cdot\dfrac{\left[ \left(\vec{r}-\vec{r}^{\,\,\prime}\right)\vec{e}_{z}\right]\left[ \left(\vec{r}-\vec{r}^{\,\,\prime}\right)\vec{e}_{z}\right]}{\vert  \vec{r}-\vec{r}^{\,\,\prime} \vert^{5}} d^{\,3}rd^{\,3}r^{\,\prime}
\\[16pt]
=\dfrac{\mu_{0}}{2}M^{2} \int\limits_{\vv{r},\,\vv{r}^{\,\prime}\,\in\,\Omega} \dfrac{1}{\vert \vec{r}-\vec{r}^{\,\,\prime} \vert^{3}}-3\cdot\dfrac{\left[ \left(\vec{r}-\vec{r}^{\,\,\prime}\right)\vec{e}_{z}\right]^{2}}{\vert \vec{r}-\vec{r}^{\,\,\prime} \vert^{5}} \,d^{\,3}rd^{\,3}r^{\,\prime}
\end{gathered}
\end{split}
\end{equation} 
By the use of (\ref{volElem}) and (\ref{sphereCoordin}) the integral (\ref{Idv1}) can be written in spherical coordinates. For this purpose, a converted expression for the magnitude of the distance vector $\vert \vec{r}-\vec{r}^{\,\,\prime} \vert$ derived in the appendix (\ref{DistVecSphericalCoord3}) is used:  
\begin{equation}\label{Idv2}
\begin{split}
\begin{gathered} 
I_{d}\underset{(\ref{volElem}),(\ref{sphereCoordin})}{\overset{(\ref{DistVecSphericalCoord3})}{=}}\dfrac{\mu_{0}}{8\pi}M^{2} \int\limits_{\vv{r},\,\vv{r}^{\,\prime}\,\in\,\Omega} \dfrac{r^{2}r^{\prime\,2}\sin{\left(\theta\right)}\sin{\left(\theta^{\prime}\right)}}{\left[r^{2}+r^{\prime\,2} -2rr^{\prime}\left( \sin{\left(\theta\right)}\sin{\left(\theta^{\prime}\right)}\cos{\left(\varphi-\varphi^{\prime}\right)}+\cos{\left(\theta\right)}\cos{\left(\theta^{\prime}\right)} \right)\right]^{\frac{3}{2}}}
\\\\
-\dfrac{3\left[ r\cos{\left(\theta\right)}-r^{\prime}\cos{\left(\theta^{\prime}\right)}\right]^{2}r^{2}r^{\prime\,2}\sin{\left(\theta\right)}\sin{\left(\theta^{\prime}\right)}}{\left[r^{2}+r^{\prime\,2} -2rr^{\prime}\left( \sin{\left(\theta\right)}\sin{\left(\theta^{\prime}\right)}\cos{\left(\varphi-\varphi^{\prime}\right)}+\cos{\left(\theta\right)}\cos{\left(\theta^{\prime}\right)} \right)\right]^{\frac{5}{2}}}drdr^{\prime}d\theta  d\theta ^{\prime}d\varphi d\varphi^{\prime} 
\end{gathered}
\end{split}
\end{equation}
An analytical solution of (\ref{Idv2}) is assumed to not be possible by iterative integration with respect to the six coordinates.\hspace{1.5mm}The main reasons for this assumption are the complex angle term in the denominator, which contains all angle coordinates within a sum of two products and the high dimensionality of the integral alongside the singular character of the integrand. Therefore a transition to a numerical approach is necessary, in order to calculate the dipole interaction contribution of the \textit{Breit-Hartree} correction to the total energy of the homogeneously magnetized sphere $\Omega$.

\clearpage

\section{Numerical approach}\label{NumApSphere}

The integrand of (\ref{Idv2}) is singular for a vanishing denominator $\vert\vec{r}-\vec{r}^{\,\,\prime}\vert$, which is the magnitude of the distance vector. Consequently, a numerical success in calculating the six dimensional integral $I_{d}$ is unlikely, because the singularity occurs at each spatial point within the integration volume $\Omega$. This is why for for an initial analysis a crude regularisation, using the exclusion of undefined values from the evaluation of the integrand, is applied. Moreover, the high dimensionality of the problem aggravates convergence, which leaves two choices for the numerical integration technique, which will be discussed in the following subsections. 

\subsection{Multidimensional \textit{Monte Carlo }integration}\label{MCarlo}

\begin{figure}[h]
\begin{center}
\includegraphics[trim=135 12 134 10,clip,width=\textwidth]{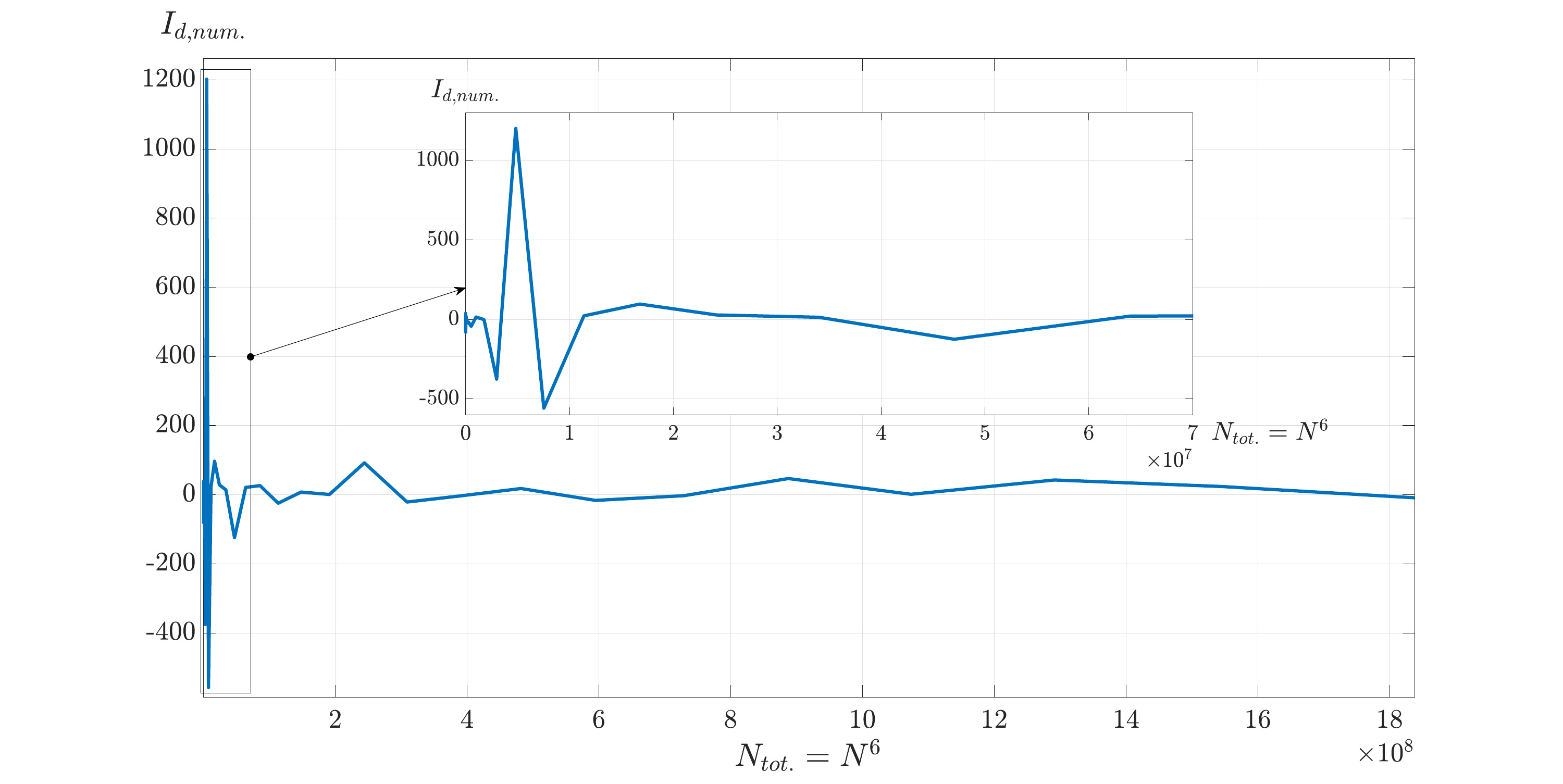}
\end{center}
\caption{Numerical results of the \textit{Monte Carlo} implementation (\Cpp) \\ for $I_{d,num.}$ with $M=1$ for a dimensionless full sphere with $R_{2}=R^{\prime}_{2}=1$} 
\label{Fig1MonteCarloDipol}
\end{figure}

First, one could construct a set of randomized points within the integration volume to conduct a \textit{Monte-Carlo} integration, which is commonly used for highly dimensional integrals. Certainly, the set must contain $N^{6}$ different random numbers with $N$ as the number of evaluations of the integrand per coordinate. In addition, the \textit{Monte-Carlo} integration needs many evaluations of the integrand, for instance of the order $10^{6}$ per coordinate to converge. Knowing these two facts one can assume that the total amount of function evaluations, which need the most runtime within \textit{Monte-Carlo} integrations, will extend $10^{36}$ in total. Based on this realization, one can imagine a long runtime, in order to achieve convergence. In combination with a singular integrand and the difficulty of constructing a sufficiently large and truly randomized as well as uniformly distributed set of sampling points within the sphere, one has to expect a low rate of convergence. Particularly, if the set of randomized numbers is too small and the pseudo-random number series in current programming libraries repeats itself, one might get incorrect results or reach a threshold where convergence is lost. Nevertheless, this option for the numerical calcualtion of $I_{d}$ is applied in order to compare to numerical approaches. This comparison could then be pathbreaking for further numerical analysis or motivate the implementation of a better random number generator, in order to improve the \textit{Monte Carlo} method. However, for the purpose of the first examination of the numeric behaviour of the integral $I_{d}$ from (\ref{Idv2}) a standard implementation in \Cpp is assumed to be sufficient.
\\
One first defines the approximation $\tilde{I}_{d,num.}$ of the integral $I_d$ with the \textit{Monte Carlo} integration as given in \cite{Quarteroni}, using the average of the integrand evaluations $\bar{f}$ and multiplying it with the integration volume $vol\left(\tilde{\Omega}\right)$. One regards $N$ random sampling points for each of the six coordinates in the integrand of (\ref{Idv2}), resulting in a total number of $N_{tot}=N^{6}$ evaluations of integrand. These function evaluations are then summed up and divided by $N_{tot}$ as well as multiplied by the hypervolume $\tilde{\Omega}:=r\,\in\,\left[R_{1},R_{2}\right]\times r^{\prime}\,\in\,\left[R_{1}^{\prime},R_{2}^{\prime}\right]\times\theta\,\in\,\left[0,\pi\right]\times\theta^{\prime}\,\in\,\left[0,\pi\right]\times\varphi\,\in\,\left[0,2\pi\right)\times\varphi^{\prime}\,\in\,\left[0,2\pi\right)$, which is a hypercube describing the six dimensional integration volume. Hence the following approximation holds:\vspace*{-1mm}
\begin{equation}\label{IdMonteCarlo1}
\begin{split}
\begin{gathered} 
I_{d}=\int\limits_{r=R_{1}}^{\textup{R}_{2}}\int\limits_{r^{\prime}=R_{1}^{\prime}}^{\textup{R}_{2}^{\prime}}\int\limits_{\theta=\theta_{1}}^{\theta_{2}}\int\limits_{\theta^{\prime}=\theta^{\prime}_{1}}^{\theta^{\prime}_{2}}\int\limits_{\varphi=\varphi_{1}}^{\varphi_{2}}\int\limits_{\varphi^{\prime}=\varphi^{\prime}_{1}}^{\varphi^{\prime}_{2}} f\left(r,r^{\prime},\theta,\theta^{\prime},\varphi,\varphi^{\prime}\right)\,d\varphi d\varphi^{\prime} d\theta d\theta^{\prime} dr dr^{\prime} 
\\[15pt]
\cong \underbrace{\left(\varphi_{2}-\varphi_{1}\right)\left(\varphi_{2}^{\prime}-\varphi_{1}^{\prime}\right)\left(\theta_{2}-\theta_{1}\right)\left(\theta_{2}^{\prime}-\theta_{1}^{\prime}\right)\left(R_{2}-R_{1}\right)\left(R_{2}^{\prime}-R_{1}^{\prime}\right)}_{=:\, vol\left(\tilde{\Omega}\right)}\cdot
\\[7pt]
\underbrace{\dfrac{1}{N_{tot}}\cdot\sum\limits_{i=1}^{N_{tot}} f\left(r_{i},r^{\prime}_{i},\theta_{i},\theta^{\prime}_{i},\varphi_{i},\varphi^{\prime}_{i}\right)}_{=:\bar{f}\left(r_{i},r^{\prime}_{i},\theta_{i},\theta^{\prime}_{i},\varphi_{i},\varphi^{\prime}_{i}\right)}=vol\left(\tilde{\Omega}\right)\cdot \bar{f}\left(r_{i},r^{\prime}_{i},\theta_{i},\theta^{\prime}_{i},\varphi_{i},\varphi^{\prime}_{i}\right) =:\tilde{I}_{d,num.}
\end{gathered}
\end{split}
\end{equation}\vspace*{-1mm}
In (\ref{IdMonteCarlo1}) the integrand is extracted from (\ref{Idv2}) and reads as follows:\vspace*{-1mm}
\begin{equation}\label{MCintegrandIdv2}
\begin{split}
\begin{gathered} 
f\left(r_{i},r^{\prime}_{i},\theta_{i},\theta^{\prime}_{i},\varphi_{i},\varphi^{\prime}_{i}\right):=\dfrac{\mu_{0}}{8\pi}M^{2} r_{i}^{2}r_{i}^{\prime\,2}\sin{\left(\theta_{i}\right)}\sin{\left(\theta_{i}^{\prime}\right)}\cdot
\\[9pt]
\left( \left[r_{i}^{2}+r_{i}^{\prime\,2} -2r_{i}r_{i}^{\prime}\left( \sin{\left(\theta_{i}\right)}\sin{\left(\theta_{i}^{\prime}\right)}\cos{\left(\varphi_{i}-\varphi_{i}^{\prime}\right)}+\cos{\left(\theta_{i}\right)}\cos{\left(\theta_{i}^{\prime}\right)} \right)\right]^{-\frac{3}{2}}\right.
\\[9pt]
\left.-\dfrac{3\left[ r_{i}\cos{\left(\theta_{i}\right)}-r_{i}^{\prime}\cos{\left(\theta_{i}^{\prime}\right)}\right]^{2}}{\left[r_{i}^{2}+r_{i}^{\prime\,2} -2r_{i}r_{i}^{\prime}\left( \sin{\left(\theta_{i}\right)}\sin{\left(\theta_{i}^{\prime}\right)}\cos{\left(\varphi_{i}-\varphi_{i}^{\prime}\right)}+\cos{\left(\theta_{i}\right)}\cos{\left(\theta_{i}^{\prime}\right)} \right)\right]^{\frac{5}{2}}}\right)
\end{gathered}
\end{split}
\end{equation}
This integrand is then calculated at $N_{tot}$ random values for the six integration variables, which are generated in their specific integration interval, using a generated random number $x_{i}\in\left[0,1\right]$:
\begin{equation}\label{MonteCarlo}
\begin{split}
\begin{gathered} 
r_{i}=R_{1}+ \left(R_{2}-R_{1}\right) \cdot x_{i}
\;\;;\;\;
r^{\prime}_{i}=R^{\prime}_{1} + \left(R^{\prime}_{2}-R^{\prime}_{1}\right) \cdot x_{i}
\\
\varphi_{i}=\varphi_{1} + \left(\varphi_{2}-\varphi_{1}\right) \cdot x_{i}
\;\;;\;\;
\varphi^{\prime}_{l}=\varphi^{\prime}_{1} +\left(\varphi^{\prime}_{2}-\varphi^{\prime}_{1}\right) \cdot x_{i}
\\
\theta_{i}=\theta_{1} + \left(\theta_{2}-\theta_{1}\right) \cdot x_{i}
\;\;;\;\;
\theta^{\prime}_{i}=\theta^{\prime}_{1} +\left(\theta^{\prime}_{2}-\theta^{\prime}_{1}\right) \cdot x_{i}
\end{gathered}
\end{split}
\end{equation}
Note that the distribution of the sampling points is uniform in a six dimensional hyper cube volume, representing the integration volume $\tilde{\Omega}$. As a consequence the sampling points are clustered along the center and the poles of the physical sphere $\Omega$.
\\
The \textit{Monte Carlo} method (\ref{IdMonteCarlo1}) is implemented in a \Cpp file and can be found in the digital appendix\footnote{montecarlo_integration[I_d_regularised].cpp}, whereat a regularisation was inserted. This rigorous regularisation sets the evaluation of the integrand to zero if it reaches singular values. The output of the \Cpp implementation for a constant, dimensionless magnetization density of $M=1$ along the $z$-axis within a dimensionless full sphere with $R_{1}=R^{\prime}_{1}=0$ and $R_{2}=R^{\prime}_{2}=1$ can be seen in (\ref{outMonteCarloId}). Whereby the boundaries of the magnetized physical sphere $\Omega:=r\,\in\,\left[0,1\right]\times\theta\,\in\,\left[0,\pi\right]\times\varphi\,\in\,\left[0,2\pi\right)$ were again choosen in favour of the numerical behaviour, allowing oneself to analyse physical radii in the dimension of an atom ($R_{2}\sim 1\si{\angstrom}$) or different magnitudes of the magnetization density $M$ by scaling the numerical result for the dimensionless full sphere ($R_{1}=0$) by the prefactor $a$, which contains physical and numerical constants from (\ref{MCintegrandIdv2}) as well as the correct SI-units. From (\ref{IdMonteCarlo1}) and (\ref{prefactor}) one gets:
\begin{equation}\label{aPrefactor}
\tilde{I}_{d,num.}= a\cdot I_{d,num.}=\dfrac{\mu_{0}}{8\pi}M^{2}R_{2}^{3}\cdot I_{d,num.}
\end{equation}
Consequently, the implemented integrand is $\frac{\mu_{0}M^{2}}{8\pi}f(r_{i},r^{\prime}_{i},\theta_{i},\theta^{\prime}_{i},\varphi_{i},\varphi^{\prime}_{i})$ with the original integrand from (\ref{MCintegrandIdv2}).
The used random number generator provided by the \Cpp library \verb|random| generates random numbers within the interval $\left[0,N_{tot.,max.}\right]$. In this context, $N_{tot.,max.}$\footnote{$N_{tot.,max.}=18446744073709551615$} (\ref{outMonteCarloId}) is the maximum total number of function evaluations for this implementation. The random numbers generated by the $64$bit \textit{Mersenne-Twister} engine \cite{mersenne} are then uniformly distributed over the interval $\left[0,1\right]$. From this distribution and from (\ref{MonteCarlo}), one can calculate random numbers within each integration variable interval, in order to carry out the \textit{Monte Carlo} integration, whereupon a different set of random values is used for the approximation of the integral for each number of sampling points $N_{tot.}$. The maximum number of sampling points per integration variable, which is equivalent to the number of evaluations of the integrand per integration variable, was set to $N_{max.}=35$, in order to estimate the behaviour of the implementation concerning the integral from (\ref{Idv2}) and to reduce the runtime. The results shown in appendix (\ref{outMonteCarloId}) are plotted over the total number of sampling points and can be seen in figure \ref{Fig1MonteCarloDipol}, which was generated with \textsc{Matlab}\footnote{figure_output_montecarlo_integration[I_d_regularised].fig}.
\noindent\linebreak \noindent 
The output in (\ref{outMonteCarloId}) shows that regularisations are not necessary if the working precision and the number of different random values $N_{tot.,max.}$ are sufficient. This can be explained by the implemented random distribution of sampling points, which cause undefined numeric values only with a probability close to zero. However, one cannot exclude the occurrence of values, which have to be regularized, because the working precision and the number of different random values $N_{tot.,max.}$ could still lead to singular function evaluations with a non-zero probability. In contrast, the \textit{Gauss–Legendre} quadrature evaluates the integrand at a fixed distribution of sampling points, originating from the roots of the \textit{Legendre} polynomials, which leads to more regularisations. This can be seen in figure \ref{Fig1GaussDipol} and is further discussed in the following section. 
\\
From the diagram in figure \ref{Fig1MonteCarloDipol} one can state that no true convergence is reached. Particularly, an unpredictable behaviour in the form of peaks for more than $8\cdot10^{8}$ evaluations of the integrand is observed. Moreover, one notices a general oscillatory behaviour with distinct peaks in the magnified miniature diagram. Therefore, one can only assume that the  \textit{Monte Carlo} approach will compute converging results for higher sampling point numbers combined with a problem-specific random number distribution.
\\
The reason for this unusual behaviour originates from the certain set of random sampling points generated by the pseudo random number library in \Cpp. In addition, the convergence is interrupted, if many sampling points evaluate the integrand near the singularities, leading to extremely high contributions to the total approximation, and the oscillations and peaks in $I_{d,num.}$. To understand this, one remembers the representation $\vert\vec{r}-\vec{r}^{\,\prime}\vert$ of the denominator of the integrand in (\ref{Idv2}) and concludes that the integrand indeed has singularities at each point in the sphere $\Omega$ for $\vec{r}\cong\vec{r}^{\,\prime}$. Therefore the working precision or number representation and the generation of the sampling points have strong influence on the oscillatory behaviour and the need of regularisations. Furthermore, the mentioned clustering of the sampling points due to their uniform distribution in the six dimensional hyper cube integration volume might harm the convergence behaviour. This could be prevented by a coordinate tranformation, which generates a uniform distribution of the sampling points within the sphere $\Omega$.

In conclusion, one should mention the runtime duration as a limiting factor for more sampling points and a higher precision. In general, one can estimate that the duration of computation is mainly determined by the number of function evaluations, resulting in an upper bound for the runtime of $\mathcal{O}\left(N_{tot.}\right)$. The \textit{Monte Carlo} integration intrinsically requires many function evaluations $N_{tot.}>10^{8}$ (see also section \ref{Benchmark}), due to the fact that its convergence is coupled to the law of large numbers. This means that the described behaviour could be irrelevant for $N_{tot.}>35^{6}$. Additionally, it should be noted that this analysis is only valid for the presented implementation.   

\subsection{Multidimensional \textit{Gauss–Legendre} integration}\label{Gauss}

Second, the multidimensional \textit{Gauss–Legendre} quadrature shall be discussed in the view of a first estimation of the dipole interaction contribution integral. On the one hand, this quadrature is only applicable for multidimensional integrals with finite integration boundaries, requiring an integration in spherical coordinates, whereby it is not predestined for singular integrands.\hspace{1.5mm}On the other hand and in contrast to the \textit{Monte-Carlo} integration, the rate of convergence of the \textit{Gauss–Legendre} quadrature is higher, leading to a possibly lower number of calculations of the integrand per coordinate. Because of this, the total amount of function evaluations and thus the runtime duration should be smaller compared to the \textit{Monte-Carlo} approach. Additionally, the set of sampling points and weights needed for this quadrature can be calculated in advance with high precision. Based on this analysis the six dimensional \textit{Gauss–Legendre} quadrature will be implemented and applied to the integral $I_d$ from (\ref{Idv2}) in \Cpp.
\\
One defines the approximation $\tilde{I}_{d,num.}$ of the integral $I_d$ with the \textit{Gauss–Legendre} quadrature for $N$ sampling points per coordinate. In that case, one has to include an affine transformation from the standard interval $\left[-1,1\right]$ of the \textit{Gauss–Legendre} quadrature to the interval of every coordinate in the integration volume $\Omega$ \cite{schwarz2009numerische}: 
\begin{equation}\label{IdGauss1}
\begin{split}
\begin{gathered} 
I_{d}=\int\limits_{r=R_{1}}^{\textup{R}_{2}}\int\limits_{r^{\prime}=R_{1}^{\prime}}^{\textup{R}_{2}^{\prime}}\int\limits_{\theta=\theta_{1}}^{\theta_{2}}\int\limits_{\theta^{\prime}=\theta^{\prime}_{1}}^{\theta^{\prime}_{2}}\int\limits_{\varphi=\varphi_{1}}^{\varphi_{2}}\int\limits_{\varphi^{\prime}=\varphi^{\prime}_{1}}^{\varphi^{\prime}_{2}} f\left(r,r^{\prime},\theta,\theta^{\prime},\varphi,\varphi^{\prime}\right)\,d\varphi d\varphi^{\prime} d\theta d\theta^{\prime} dr dr^{\prime} 
\\[19pt]
I_{d}\cong \frac{\left(\varphi_{2}-\varphi_{1}\right)}{2}\frac{\left(\varphi_{2}^{\prime}-\varphi_{1}^{\prime}\right)}{2}\frac{\left(\theta_{2}-\theta_{1}\right)}{2}
\frac{\left(\theta_{2}^{\prime}-\theta_{1}^{\prime}\right)}{2}\frac{\left(R_{2}-R_{1}\right)}{2}
\frac{\left(R_{2}^{\prime}-R_{1}^{\prime}\right)}{2}\cdot
\\[15pt]
\sum\limits_{i,j,k,l,m,n=1}^{N} w_{i}w_{j}w_{k}w_{l}w_{m}w_{n} \cdot f\left(r_{i},r^{\prime}_{j},\theta_{k},\theta^{\prime}_{l},\varphi_{m},\varphi^{\prime}_{n}\right)=:\tilde{I}_{d,num.}
\end{gathered}
\end{split}
\end{equation}
In (\ref{IdGauss1}) the weights $w_{\ast}$ are the standard \textit{Gauss–Legendre}-weights, while the sampling points are transformed \textit{Gauss–Legendre}-points $x_{\ast}$ with $\ast=i,j,k,l,m,n$, which are equivalent to the roots of the \textit{Legendre}-polynomials \cite{schwarz2009numerische}:
\begin{equation}\label{Legendre}
\begin{split}
\begin{gathered} 
r_{i}=\frac{R_{1}+R_{2}}{2} + \frac{R_{2}-R_{1}}{2} \cdot x_{i}
\;\;;\;\;
r^{\prime}_{j}=\frac{R^{\prime}_{1}+R^{\prime}_{2}}{2} + \frac{R^{\prime}_{2}-R^{\prime}_{1}}{2} \cdot x_{j}
\\[12pt]
\varphi_{k}=\frac{\varphi_{1}+\varphi_{2}}{2} + \frac{\varphi_{2}-\varphi_{1}}{2} \cdot x_{k}
\;\;;\;\;
\varphi^{\prime}_{l}=\frac{\varphi^{\prime}_{1}+\varphi^{\prime}_{2}}{2} + \frac{\varphi^{\prime}_{2}-\varphi^{\prime}_{1}}{2} \cdot x_{l}
\\[12pt]
\theta_{m}=\frac{\theta_{1}+\theta_{2}}{2} + \frac{\theta_{2}-\theta_{1}}{2} \cdot x_{m}
\;\;;\;\;
\theta^{\prime}_{n}=\frac{\theta^{\prime}_{1}+\theta^{\prime}_{2}}{2} + \frac{\theta^{\prime}_{2}-\theta^{\prime}_{1}}{2} \cdot x_{n}
\end{gathered}
\end{split}
\end{equation}
The sets of standard sampling points $x_{\ast}$ and weights $w_{\ast}$ on the interval $\left[-1,1\right]$ for the \textit{Gauss–}\textit{Legendre} quadrature are calculated separately with two modified \textsc{Mathematica} scripts from \cite{Legendre}, which can be found in the digital appendix\footnote{mathematica_gauss_legendre_weights(n_max=200_precision_44_digit).nb}$^{;}$\footnote{mathematica_gauss_legendre_roots(n_max=200_precision_44_digit).nb}. This allows oneself to generate two \small{\verb|.txt|} files\footnote{gauss_legendre_weights(n_max=200_precision_44_digit).txt}$^{;}$\footnote{gauss_legendre_roots(n_max=200_precision_44_digit).txt} with $N_{max.}=200$ sampling points and weights with an internal precision of $44$ digits. Those files are then read and streamed to two memory arrays allowing a maximum precision of $18$ digits for the data type $long\,double$ in \Cpp. Hence, the sums over the six indices from (\ref{IdGauss1}) can be conducted with the use of $for$-$loops$, while the integrand $f\left(r_{i},r^{\prime}_{j},\theta_{k},\theta^{\prime}_{l},\varphi_{m},\varphi^{\prime}_{n}\right)$ is evaluated at the transformed sampling points (\ref{Legendre}) in each step. The integrand according to (\ref{Idv2}) is therefore given by:
\begin{equation*}
f\left(r_{i},r^{\prime}_{j},\theta_{k},\theta^{\prime}_{l},\varphi_{m},\varphi^{\prime}_{n}\right):=\dfrac{\mu_{0}}{8\pi}M^{2} r_{i}^{2}r_{j}^{\prime\,2}\sin{\left(\theta_{k}\right)}\sin{\left(\theta_{l}^{\prime}\right)}\cdot
\end{equation*}
\begin{equation}\label{integrandIdv2}
\left( \left[r_{i}^{2}+r_{j}^{\prime\,2} -2r_{i}r_{j}^{\prime}\left( \sin{\left(\theta_{k}\right)}\sin{\left(\theta_{l}^{\prime}\right)}\cos{\left(\varphi_{m}-\varphi_{n}^{\prime}\right)}+\cos{\left(\theta_{k}\right)}\cos{\left(\theta_{l}^{\prime}\right)} \right)\right]^{-\frac{3}{2}}\right.
\end{equation}
\begin{equation*}
\left.-\dfrac{3\left[ r_{i}\cos{\left(\theta_{k}\right)}-r_{j}^{\prime}\cos{\left(\theta_{l}^{\prime}\right)}\right]^{2}}{\left[r_{i}^{2}+r_{j}^{\prime\,2} -2r_{i}r_{j}^{\prime}\left( \sin{\left(\theta_{k}\right)}\sin{\left(\theta_{l}^{\prime}\right)}\cos{\left(\varphi_{m}-\varphi_{n}^{\prime}\right)}+\cos{\left(\theta_{k}\right)}\cos{\left(\theta_{l}^{\prime}\right)} \right)\right]^{\frac{5}{2}}}\right)
\end{equation*}

\begin{figure}[h]
\begin{center}
\includegraphics[trim=58 10 100 5,clip,width=\textwidth]{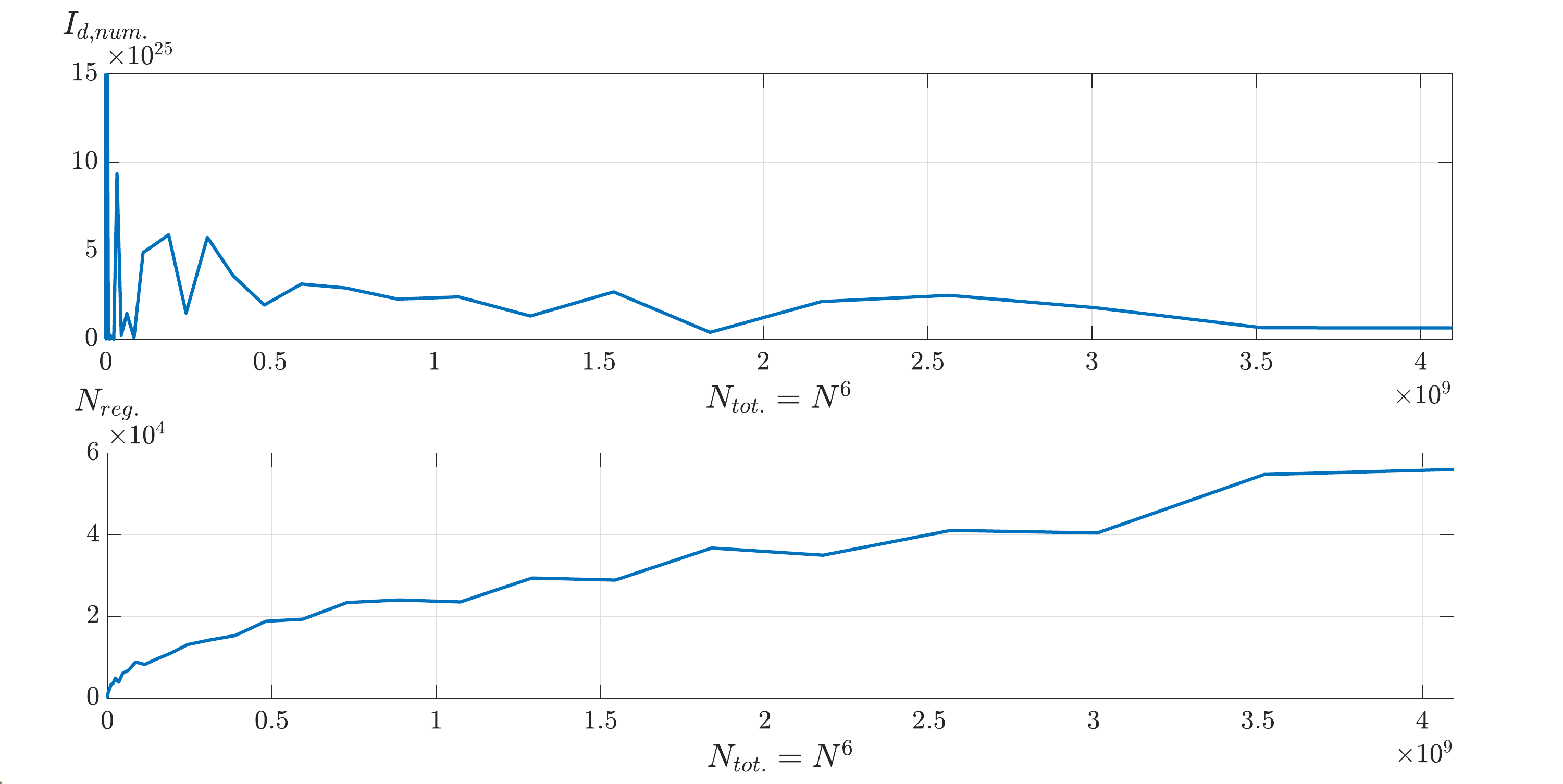}
\end{center}
\caption{Numerical results of the \textit{Gauss-Legendre} implementation (\Cpp) \\ for $I_{d,num.}$ with $M=1$ for a dimensionless full sphere with $R_{2}=R^{\prime}_{2}=1$} 
\label{Fig1GaussDipol}
\end{figure}

The \Cpp implementation of the \textit{Gauss–Legendre} integration can be found in the digital appendix\footnote{gauss_legendre_integration[I_d_regularised].cpp}, whereat a regularisation was inserted. This rigorous regularisation sets the evaluation of the integrand to zero if it reaches singular values.  The output of the \Cpp implementation of the \textit{Gauss–Legendre} integration for a constant, dimensionless magnetization density of $M=1$ along the $z$-axis within a dimensionless full sphere with $R_{1}=R^{\prime}_{1}=0$ and $R_{2}=R^{\prime}_{2}=1$ can be seen in (\ref{outGaussLegendreId}), whereby the boundaries of the sphere $\Omega:=r\,\in\,\left[0,1\right]\times\theta\,\in\,\left[0,\pi\right]\times\varphi\,\in\,\left[0,2\pi\right)$ were choosen in favour of the numerical behaviour. Furthermore, one is still able to analyse physical radii in the dimension of an atom ($R_{2}\sim 1\si{\angstrom}$) or different magnitudes of the magnetization density $M$ by scaling the numerical result for the dimensionless full sphere ($R_{1}=0$) by the factor $a$ from (\ref{aPrefactor}), whereat the derivation of the scaling factor was given in (\ref{prefactor}) and the implemented integrand is $\frac{\mu_{0}M^{2}}{8\pi}f(r_{i},r^{\prime}_{j},\theta_{k},\theta^{\prime}_{l},\varphi_{m},\varphi^{\prime}_{n})$ with the original integrand from (\ref{integrandIdv2}). Moreover, the number of regularizations needed for a calculation was counted to analyse its effect and influence onto the numerically received value of the integration. For the purpose of an analysis of the numerical behaviour of the implementation concerning the integral from (\ref{Idv2}) the maximum number of sampling points per integration variable, which is equivalent to the number of evaluations of the integrand per integration variable, was set to $N_{max.}=40$. The numerical results shown in the appendix (\ref{outGaussLegendreId}) are displayed over the total number of sampling points used for calculation and can be seen in figure \ref{Fig1GaussDipol}, which was computerized with \textsc{Matlab}\footnote{figure_output_gauss_legendre_integration[I_d_regularised].fig}.

In figure \ref{Fig1GaussDipol} one notices a power law relation between the number of regularisations $N_{reg.}$ and the total number of sampling points $N_{tot.}$. Thus the increase of $N_{reg.}$ is of order $6\cdot 10^{4}$. Furthermore, several peaks in the numerically calculated integral values can be observed even above $10^{9}$ sampling points. Additionally, an irregular oscillation of the numerical results occurs, while the values seem to tend against a value of order $6\cdot 10^{24}$ (\ref{outGaussLegendreId}). This trend is mainly determined by a falling slope which is higher for small numbers of function evaluations and decreases for higher values of $N_{tot.}$. Furthermore a stagnation occurs at $N_{tot.}>2.5\cdot 10^{9}$. 
\\
An interpretation of the numeric behaviour in figure \ref{Fig1GaussDipol} is that the singular integrand is responsible for the oscillatory values of $I_{d,num.}$, while the regularisations have an influence on the results, due to their higher contribution at low numbers of sampling points. One can assume that the high number of regularisations is correlated with the symmetry of the sampling points, which originate from the \textit{Legendre} polynomials, in combination with the denominator of the integrand $\vert\vec{r}-\vec{r}^{\,\prime}\vert$ in (\ref{Idv2}), which contains all integration variables and tends to be zero within the working precision, if the inserted varibles within a function evaluation are close together. This is the case for the \textit{Gauss–Legendre} quadrature for the integration variables with ($r^{\prime},\,\varphi^{\prime},\,\theta^{\prime}$) and without prime ($r,\,\varphi,\,\theta$), which are subtracted within the integrand. For increasing values of $N_{tot.}$ the relative contribution of the regularisations becomes smaller with respect to the total number of integrand evaluations, as it can be estimated with the help of the second diagram in figure \ref{Fig1GaussDipol}. Consequently, this leads to the stagnation of the numeric values above $N_{tot.}>3.5\cdot10^{9}$. 
\\
As a reason for the occurrence of the peaks one can imagine that the \textit{Gauss-Legendre} quadrature for those values of $N_{tot.}$ are unfavorable for the integrand from (\ref{integrandIdv2}). This would underestimate the integral in combination with the crude regularisation (addition of a zero at singular integrand evaluations), especially if many sampling points are far away from the singularity at each spatial point.
\\
Due to the above discussion of the numerical behaviour of the \textit{Gauss-Legendre} integration one cannot expect converging results, since even the power of ten oscillates for high numbers of sampling points, whereat the regularisation is determining for small values of $N_{tot.}$, but still has an underestimating influence on the results. The high order of $I_{d,num.}$ in comparison to the \textit{Monte-Carlo} integration could be explained by the fixed distribution of the \textit{Gauss-Legendre}-quadrature on the integration intervals, which might lead to many calculations of the intgrand near its singularities. This would explain the overestimation of the integral. Furthermore, one can imagine that the choice of a certain quadrature rule or a particular set of random points is itself a regularisation, because the set of sampling points can cause different numbers of singular function evaluations for evaluations of the integrand near the singularities. This consequently results in different numerical results with higher order for more evaluations near the singularities. 

\subsection{Problem-specific comparison of the numerical methods on the basis of a benchmark problem}\label{Benchmark}

\begin{figure}[h]
\begin{center}
\includegraphics[trim=132 10 132 0,clip,width=\textwidth]{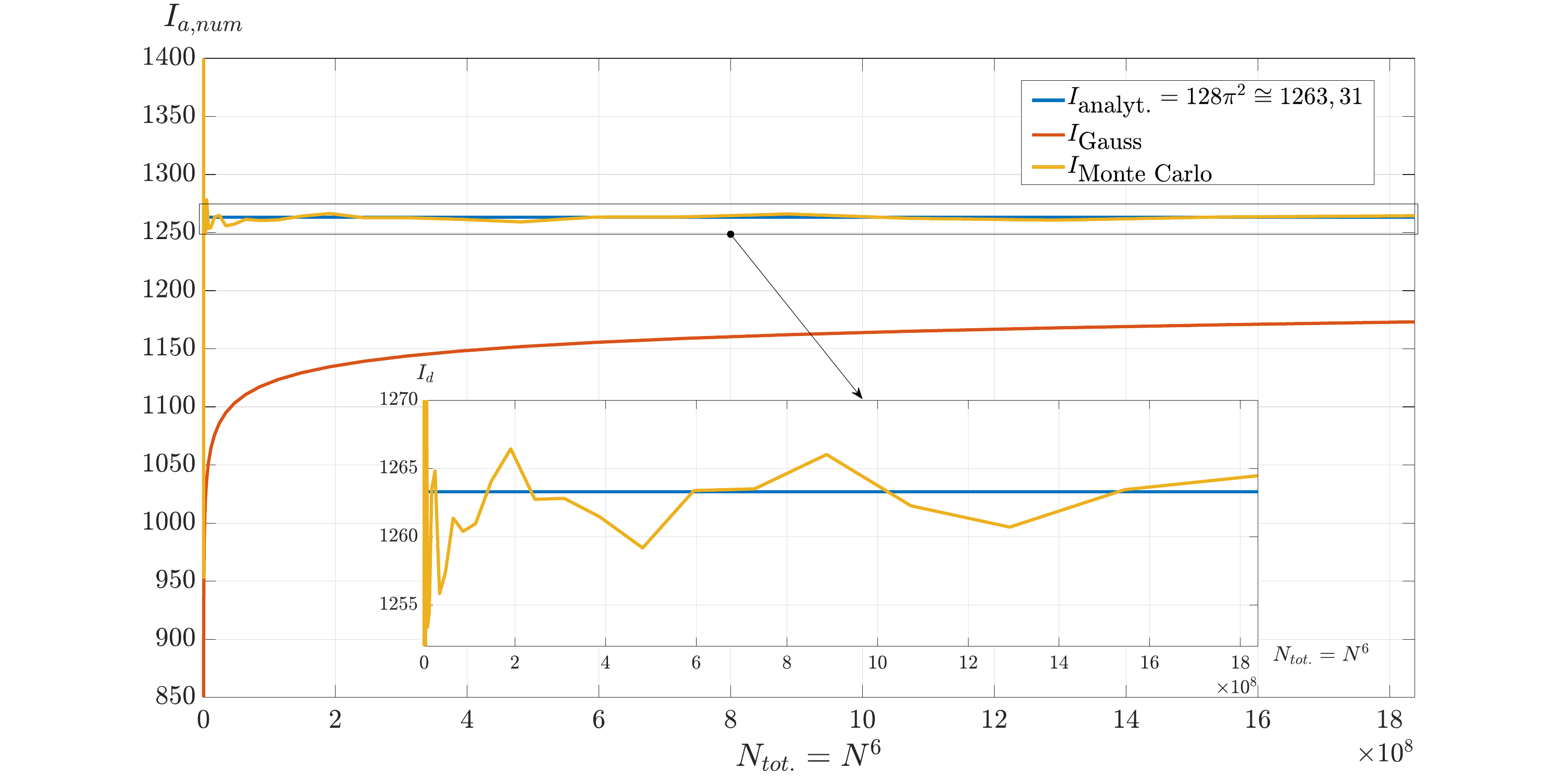}
\end{center}
\caption{Numerical results $I_{\textup{Gauss}}$ and $I_{\textup{Monte Carlo}}$ of the \textit{Gauss-Legendre} and the \textit{Monte Carlo} implementation (\Cpp) for the benchmark problem $I_{a}$} 
\label{FigBenchmark}
\end{figure}

Both numerical methods were constructed with similar boundary conditions and the same integrand. Similarly many different numbers of function evaluations were analysed. However, the comparison of the interpretations of the numerical behaviour given in the sections \ref{MCarlo} and \ref{Gauss} is only possible if one can show the characteristics of both implementations by calculating a benchmark problem. Nevertheless, this still allows for certain numerical behaviours to be assigned to the integrand of the \textit{Breit-Hartree} problem and excludes implementation errors.
\\
As a benchmark problem, one may use an analytically solvable integral $I_{a}$ in spherical coordinates with the same integration boundaries as defined for the integration volume $\tilde{\Omega}$. It is weakly singular in all integration variables and its integrand reads:
\begin{equation}\label{benchmark1}
f\left(r,r^{\prime},\theta,\theta^{\prime},\varphi,\varphi^{\prime}\right):=\dfrac{1}{\sqrt{r\cdot r^{\prime}\cdot\theta\cdot\theta^{\prime}\cdot\varphi\cdot\varphi^{\prime}}}
\end{equation}
Hence, the analytical solution, which is approximated by the numerical result $I_{a,num.}$, is stated in general form without inserting any values of the integration boundaries:
\begin{equation}\label{benchmark2}
\begin{split}
\begin{gathered} 
I_{a}=\int\limits_{r=R_{1}}^{\textup{R}_{2}}\int\limits_{r^{\prime}=R_{1}^{\prime}}^{\textup{R}_{2}^{\prime}}\int\limits_{\theta=\theta_{1}}^{\theta_{2}}\int\limits_{\theta^{\prime}=\theta^{\prime}_{1}}^{\theta^{\prime}_{2}}\int\limits_{\varphi=\varphi_{1}}^{\varphi_{2}}\int\limits_{\varphi^{\prime}=\varphi^{\prime}_{1}}^{\varphi^{\prime}_{2}} f\left(r,r^{\prime},\theta,\theta^{\prime},\varphi,\varphi^{\prime}\right)\,d\varphi d\varphi^{\prime} d\theta d\theta^{\prime} dr dr^{\prime} 
\\[12pt]
\overset{(\ref{benchmark1})}{=}2^{6}\cdot\left(\sqrt{R_{2}}-\sqrt{R_{1}}\right)\left(\sqrt{R_{2}^{\prime}}-\sqrt{R_{1}^{\prime}}\right)\left(\sqrt{\varphi_{2}}-\sqrt{\varphi_{1}}\right)\left(\sqrt{\varphi^{\prime}_{2}}-\sqrt{\varphi^{\prime}_{1}}\right)
\\[12pt]
\cdot\left(\sqrt{\theta_{2}}-\sqrt{\theta_{1}}\right)\left(\sqrt{\theta^{\prime}_{2}}-\sqrt{\theta^{\prime}_{1}}\right)\cong I_{a,num.}
\end{gathered}
\end{split}
\end{equation}
The implementations for the \textit{Monte Carlo} and the \textit{Gauss-Legendre} method in \Cpp can be found in the digital appendix\footnote{montecarlo_integration[benchmark_regularised].cpp}$^{;}$\footnote{gauss_legendre_integration[benchmark_regularised].cpp}, while the output for the \textit{Gauss-Legendre} quadrature ($N_{max.}=50$) and the \textit{Monte Carlo} integration ($N_{max.}=35$) is displayed in (\ref{outGaussLegendreBenchmark}) and (\ref{outMonteCarloBenchmark}). In this regard, the implementations of the two numerical methods made use of the same relations for the approximation of the integral $I_{a,num}$, as used for the \textit{Breit-Hartree} integral $I_{d}$. The integration results for $N_{tot.}\in\left[1,35^{6}\right]$ are shown in the \textsc{Matlab}-generated\footnote{figure_output_benchmark_integration_regularised.fig} figure \ref{FigBenchmark}.
The first diagram in this figure shows the plot of the computed values of $I_{a,num.}$ for the two numerical integration methods as well as the analytical solution for increasing numbers of sampling points. One notices that both methods underestimate the true value, whereat the \textit{Gauss-Legendre} quadrature converges noticeably slower than the \textit{Monte Carlo} method. Nevertheless, the \textit{Gauss-Legendre} approach converges continually (\ref{outGaussLegendreBenchmark}) in comparison to the oscillatory behaviour of the \textit{Monte Carlo} integration (\ref{outMonteCarloBenchmark}), which can be observed in detail in the magnified miniature plot. The output files (\ref{outMonteCarloBenchmark}) and (\ref{outGaussLegendreBenchmark}) display the needed regularisations, in order to approximate the singular integral $I_{a}$. Furthermore, the \textit{Monte Carlo} method and the \textit{Gauss-Legendre} integration does not need any regularisations.
Hereinafter, the phenomena occurring in the benchmark problem are interpreted in order to extract specific characteristics of the numerical methods respectively their implementations and the different integrands of the integrals $I_{d}$ and $I_{a}$. 
\\ 
First, the importance and number of regularisations is clearly dependent on the integrand, because the \textit{Monte Carlo} method requires no regularisations for the integral $I_{d}$, while for the integral $I_{a}$ both integration techniques do not produce undefined values. From this one deduces that for the problem from (\ref{Idv2}) the preferred numerical approach, in terms of needed regularisations, is the \textit{Monte Carlo} method. However, one has to consider the distribution of pseudo random numbers, which can lead to more regularisations if the denominator of the singular integrand $I_{d}$ is zero within the working precision. Therefore, an implementation should at least use \textit{double} precision and make use of an established random number set for the \textit{Monte Carlo} method, as it is provided by the \textit{Mersenne-Twister} algorithm, in order to reduce the number of regularisations.
\\
Second, one analyses the convergence of the two numerical methods, leading to the conclusion that in the benchmark problem the \textit{Monte Carlo} integration converges faster and oscillatory, while the \textit{Gauss-Legendre} integration approaches a stationary result very slowly but continually. The reason for this is the high dimensionality of the integrals, which slows down the rate of convergence for the \textit{Gauss-Legendre} integration and explains the widely used \textit{Monte Carlo} technique for high-dimensional integration. In additon, one can now understand the high order of the numerical results of $I_{d}$ produced by the \textit{Gauss-Legendre} method (figure \ref{Fig1GaussDipol}). The magnitude of these values is  decreasing and the approximations for small values of $N_{tot.}$ are arbitrarily large, which is specific to the integrand of $I_{d}$ after comparison with the benchmark problem. Therefore the true solution of the problem from (\ref{Idv2}) seems to be very small, as it is quickly estimated by the \textit{Monte Carlo} integration, similarly to the computation of the integral $I_{a}$. With regard to the \textit{Hartree-Breit} integral $I_{d}$, one can consequently expect faster results  from the \textit{Monte Carlo} approach (with regard to convergence) with less evaluations of the integrand or with less number of sampling points and therefore lower runtime duration. However, the convergence is strongly dependent on the set of random sampling points and will improve with increasing $N_{tot.}$. One again emphazises, that the random number generating function and its probability distribution within the \Cpp implementation is key for a precise calculation, due to its influence on the oscillatory convergence and the necessity of regularisations. As a consequence, one has to investigate the behaviour of the \textit{Monte Carlo} integration with respect to convergence for every implementation separately. The statement of increasing convergence with larger numbers of sampling points also holds for the \textit{Gauss-Legendre} integration, whereat convergence is very likely for extremly high values of $N_{tot.}$, negatively implying an infinite runtime duration. 
\\
Third, the deviation from the analytical value of the integral $I_{a}$ should be mentioned, which is larger for the \textit{Gauss} integration, although it slowly decreases. For the \textit{Monte Carlo} integration one recognizes in (\ref{outMonteCarloBenchmark}) that the error seems to remain constant for $N_{tot.}\cong10^{9}$. This could be explained by the uniform distribution of the random numbers, which generate a fast convergence towards a deviated approximate solution, whose deviation from the true value is highly dependent on the particular set of random points. This is typical for the \textit{Monte Carlo} integration and requires a problem specific distribution of the random numbers for the approximation of the integral, which should make use of at least as many different random numbers as needed for the highest value of $N_{tot.}$. 

In conclusion, one should focus on improvements of the \textit{Monte-Carlo} method in order to receive results within finite runtime and under the certainty, that crude regularisations as presented in section \ref{MCarlo} become negligible or obsolete, which was previously discussed for the \textit{Hartree-Breit} integral $I_{d}$ and the benchmark integral $I_{a}$. However, one has to construct a set of $N_{tot.,max.}\geq N_{max.}^{6}$ different pseudo random sampling points in a certain manner, so that the deviation from the true solution is negligible, sufficient convergence is reached fast and the number of regularisations is small $N_{reg.}\ll N_{tot.}$. On that point, one can use the derived scaling factor from (\ref{prefactor}), in order to carry out the computation with an arbitrary choosable boundary $R_{2}$, in favor with numerical precision. In addition one should include a transformation for the sampling points, which generates a uniform distribution within the sphere and therefore excludes the sampling point clustering. Last, the random sampling points should improve the convergence behavior, whereat a convergence criterion has to be established. An additional verification for the correctness of the order of the numerical results can be found in the appendix in subsection \ref{CHECKMonteCarlo}.
\\ 
Furthermore, one has to remember that all numerical characteristics were derived for $N_{tot.}=N^{6}$ function evaluations, meaning that the shown behaviour for the given implementations is only an excerpt of the true numerical behaviour. Additionally, it should be mentioned that all computations were carried out in \textit{double} precision in \Cpp on a \textsc{Windows} user system, whereat the runtime durations can be found at the end of the output files in (\ref{NumericalCalulations}). 
\clearpage

\section{Analytical calculation of the \textit{Breit-Hartree} contribution of a test dipole}\label{AnalyticalTestDipol}

In view of the fact that besides the analytical iterative integration (section \ref{AnApSphere}), two different numerical methods (section \ref{NumApSphere}) were not capable of solving the integral equation from (\ref{EnergieBHSI}) for a homogeneously magnetized sphere, the problem can be simplified. Thereunto one considers a test dipole at the spatial point $\vec{a}\in\mathbb{R}^{3}$ in the constant spin density $s$ within the sphere. On that point, one uses the interpretation of the \textit{Breit-Hartree} interaction as an interaction between two spin densities $s\left(\vec{r}\,\right)$ and $s\left(\vec{r}^{\,\prime}\right)$ along the $z$-direction from section \ref{Important} with the spin density $s\left(\vec{r}\,\right)$ reduced to one test dipole generating the test dipole magnetization density $\vec{M}^{\,dip}\left(\vec{r},\,\vec{a}\,\right)$. This is represented by the insertion of a three dimensional delta distribution $\delta\left(\vec{r}-\vec{a}\right)$ into the definiton of the magnetization density $\vec{M}\left(\vec{r}\,\right)$, while the magnetization density $\vec{M}\left(\vec{r}^{\,\prime}\right)$ is left unchanged. On this occasion, it shall be emphasised, that the magnetization densities $\vec{M}\left(\vec{r}\,\right)$ and $\vec{M}\left(\vec{r}^{\,\prime}\right)$ originate from the same electron or spin density, although they are interpreted as two different virtual densities interacting with each other. This is because the position vectors $\vec{r}$ and $\vec{r}^{\,\prime}$ at which the virtual densities are evaluated are always different, meaning $\vec{r}\neq\vec{r}^{\,\prime}$ while the contact interaction integral $I_{d}$ from (\ref{EnergieBHSI}) covers the case $\vec{r}=\vec{r}^{\,\prime}$. Consequently and by means of (\ref{magnetdensity}) and (\ref{defM}) one gets for a constant magnitude of the magnetization density $M=\mu_{B}s=\textup{const.}$:
\begin{equation}\label{defMtestdipol}
\begin{split}
\begin{gathered}
\vec{M}^{\,dip}\left(\vec{r},\,\vec{a}\,\right) = 
\begin{cases}
M\delta\left(\vec{r}-\vec{a}\right)\vec{e}_{z}= -\mu_{B}s\delta\left(\vec{r}-\vec{a}\right)\vec{e}_{z}&;\;\; \forall \; \vec{r}\,\in\,\Omega
\\
\vec{0}&;\;\; \forall \; \vec{r} \,\notin\,\Omega
\end{cases}
\\[12pt]
\vec{M}\left(\vec{r}^{\,\prime}\right) = 
\begin{cases}
M\vec{e}_{z}= -\mu_{B}s\vec{e}_{z}&;\;\; \forall \; \vec{r}^{\,\,\prime} \,\in\,\Omega
\\
\vec{0}&;\;\; \forall \; \vec{r}^{\,\,\prime} \,\notin\,\Omega
\end{cases}
\end{gathered}
\end{split}
\end{equation}
Hence, the total \textit{Breit-Hartree} contribution can be calculated by an integration of the energy contribution density $\varrho_{BH}\left(\vec{a}\,\right):=\Delta E_{BH}^{\,dip}\left(\vec{a}\,\right)$ of a dipole at the spatial point $\vec{a}$ over the spherical volume $\Omega:=r\,\in\,\left[R_{1},R_{2}\right]\times\theta\,\in\,\left[0,\pi\right]\times\varphi\,\in\,\left[0,2\pi\right)$ with respect to the integration variables $d\Omega=d^{3}a$:
\begin{equation}\label{totalEBHfromdipolContri1}
\Delta E_{BH} = \int\limits_{\Omega}\varrho_{BH}\left(\vec{a}\,\right)\,d\Omega = \int\limits_{\Omega} \varrho_{BH}\left(\vec{a}\,\right) \,d^{3}a
\end{equation}
In the next step one has to derive the function $\Delta E_{BH}^{\,dip}\left(\vec{a}\,\right)$. From (\ref{EnergieBHSI}) one remembers the decompostion $\varrho_{BH}\left(\vec{a}\,\right)=\Delta E_{BH}^{\,dip}\left(\vec{a}\,\right) := I^{\,dip}_{d} + I^{\,dip}_{c}$ of the integral equation into the dipole and the contact interaction, enabling a separate evaluation, whereat $I^{\,dip}_{d}$ and $I^{\,dip}_{c}$ can be interpreted as the contact interaction contribution density and the dipole interaction contribution density. With the help of (\ref{totalEBHfromdipolContri1}) one finds:
\begin{equation}\label{totalEBHfromdipolContri2}
\Delta E_{BH} = \int\limits_{\Omega} I^{\,dip}_{d} + I^{\,dip}_{c} \,d\Omega = \int\limits_{\Omega} I^{\,dip}_{d} \,d\Omega +  \int\limits_{\Omega} I^{\,dip}_{c} \,d\Omega =:\Delta E_{BH}^{\,d}+\Delta E_{BH}^{\,c}
\end{equation}
First one inserts the definition of the magnetization density (\ref{defMtestdipol}) into the representation of $I^{\,dip}_{c}$ as it is stated in (\ref{EnergieBHSI}) and calculates the total contact contribution $\Delta E_{BH}^{\,c}$ with the help of (\ref{totalEBHfromdipolContri2}) and $vol\left(\Omega\right)=\frac{4\pi}{3}\left(R_{2}^{3}-R_{1}^{3}\right)$:
\begin{equation}\label{IcDIP1}
\begin{split}
\begin{gathered}
\underline{\underline{\Delta E_{BH}^{\,c}}}\overset{(\ref{totalEBHfromdipolContri2})}{=}\int\limits_{\Omega} I^{\,dip}_{c} \,d\Omega \overset{(\ref{EnergieBHSI})}{=} -\dfrac{1}{3}\mu_{0}\int\limits_{\Omega}\int\limits_{\Omega} \vert\vec{M}^{\,dip}\left(\vec{r},\,\vec{a}\,\right)\vert^{2}  d^{\,3}r \,d\Omega
\\\\
\overset{(\ref{defMtestdipol})}{=}-\dfrac{1}{3}\mu_{0}\int\limits_{\Omega}\int\limits_{\Omega} \vert M\delta\left(\vec{r}-\vec{a}\right)\vec{e}_{z}\vert^{2}  d^{\,3}r \,d\Omega
= -\dfrac{1}{3}\mu_{0}M^{2}\int\limits_{\Omega}\int\limits_{\Omega} \underbrace{\vert \delta\left(\vec{r}-\vec{a}\right)\vert^{2}}_{=\,\delta\left(\vec{r}-\vec{a}\right)}\underbrace{\vert\vec{e}_{z}\vert^{2}}_{=\,1}  d^{\,3}r\,d\Omega 
\\\\
= -\dfrac{1}{3}\mu_{0}M^{2}\int\limits_{\Omega}\underbrace{\int\limits_{\Omega} \delta\left(\vec{r}-\vec{a}\right)d^{\,3}r }_{=\,1}\,d\Omega
=-\dfrac{1}{3}\mu_{0}M^{2} \underbrace{\int\limits_{\Omega}\,d\Omega}_{=\,vol\left(\Omega\right)}=\underline{\underline{-\dfrac{4\pi}{9}\mu_{0}M^{2}\left(R_{2}^{3}-R_{1}^{3}\right)}}
\end{gathered}
\end{split}
\end{equation}  
In (\ref{IcDIP1}) the following characteristic of the \textit{Dirac} delta distribution was used \cite{Bartelmann_2015}:
\begin{equation}\label{delta1}
\int \delta\left(\vec{r}-\vec{a}\,\right)\,d^3r = 1
\end{equation}
Note that the result in (\ref{IcDIP1}) is identical to the result from (\ref{Icv1}) and (\ref{Icv2}). This proves that the total \textit{Breit-Hartree} contribution for the volume $\Omega$ can be calculated by an integral over an energy contribution density as it was given in (\ref{totalEBHfromdipolContri1}). Also one notices, that a permutation of $\vec{r}$ and $\vec{r}^{\,\prime}$ has no effect on the final results, because it is equivalent to a renaming of the considered position vectors, which can be verified from (\ref{EnergieBHSI}).
\\
Second, one applies the magnetization densities from (\ref{defMtestdipol}) in $I^{\,dip}_{d}$ as it is formulated in (\ref{EnergieBHSI}):
\begin{equation}\label{IdDIP1}
\begin{split}
\begin{gathered}
I^{\,dip}_{d}:=\dfrac{\mu_{0}}{8\pi} \int\limits_{\vv{r},\,\vv{r}^{\,\prime}\,\in\,\Omega} \dfrac{1}{\vert  \vec{r}-\vec{r}^{\,\,\prime} \vert^{3}}\left(\vec{M}^{\,dip}\left(\vec{r},\,\vec{a}\,\right)\vec{M}\left(\vec{r}^{\,\,\prime}\,\right)\right.
\\[15pt]
\left.-3\cdot\dfrac{\left[ \left(\vec{r}-\vec{r}^{\,\,\prime}\right)\vec{M}^{\,dip}\left(\vec{r},\,\vec{a}\,\right)\right]\left[ \left(\vec{r}-\vec{r}^{\,\,\prime}\right)\vec{M}\left(\vec{r}^{\,\,\prime}\,\right)\right]}{\vert  \vec{r}-\vec{r}^{\,\,\prime} \vert^{2}}\right)d^{\,3}rd^{\,3}r^{\,\prime}
\\[10pt]
=\dfrac{\mu_{0}}{8\pi} \int\limits_{\vv{r},\,\vv{r}^{\,\prime}\,\in\,\Omega} \dfrac{1}{\vert  \vec{r}-\vec{r}^{\,\,\prime} \vert^{3}}\left(M^{2}\delta\left(\vec{r}-\vec{a}\right)\overbrace{\vec{e}_{z}\cdot\vec{e}_{z}}^{=\vert\vec{e}_{z}\vert^{2}=1}\right.
\\[15pt]
\left.-3\cdot\dfrac{\left[ \left(\vec{r}-\vec{r}^{\,\prime}\right)M \delta\left(\vec{r}-\vec{a}\right)\vec{e}_{z}\right]\left[ \left(\vec{r}-\vec{r}^{\,\prime}\right)M \vec{e}_{z}\,\right]}{\vert  \vec{r}-\vec{r}^{\,\,\prime} \vert^{2}}\right)\,d^{\,3}rd^{\,3}r^{\,\prime}
\\[15pt]
=\dfrac{\mu_{0}}{8\pi}M^{2} \int\limits_{\vv{r},\,\vv{r}^{\,\prime}\,\in\,\Omega}\delta\left(\vec{r}-\vec{a}\right)\left(\dfrac{1}{\vert \vec{r}-\vec{r}^{\,\,\prime} \vert^{3}}-3\cdot\dfrac{\left[ \left(\vec{r}-\vec{r}^{\,\,\prime}\right)\vec{e}_{z}\right]\left[ \left(\vec{r}-\vec{r}^{\,\,\prime}\right)\vec{e}_{z}\right]}{\vert \vec{r}-\vec{r}^{\,\,\prime} \vert^{5}}\right) \,d^{\,3}rd^{\,3}r^{\,\prime}
\end{gathered}
\end{split}
\end{equation} 
The following relation holds for the integration over a delta distribution \cite{Bartelmann_2015}:
\begin{equation}\label{deltarule}
\int \delta\left(\vec{r}-\vec{a}\right)f\left(\vec{r}\,\right)\,d^3r = f\left(\vec{a}\,\right)
\end{equation}
From (\ref{IdDIP1}) and (\ref{deltarule}) one gets: 
\begin{equation}\label{IdDIP2}
\begin{split}
\begin{gathered}
I^{\,dip}_{d}=\dfrac{\mu_{0}}{8\pi}M^{2} \int\limits_{\vv{r}^{\,\prime}\,\in\,\Omega}\dfrac{1}{\vert  \vec{a}-\vec{r}^{\,\,\prime} \vert^{3}}-3\cdot\dfrac{\left[ \left(\vec{a}-\vec{r}^{\,\,\prime}\right)\vec{e}_{z}\right]^{2}}{\vert  \vec{a}-\vec{r}^{\,\,\prime} \vert^{5}} \,d^{\,3}r^{\,\prime}
\end{gathered}
\end{split}
\end{equation} 
The integral (\ref{IdDIP2}) is generally of the same complexity as (\ref{EnergieBHSI}). But for a test dipole on the connecting line between south and north pole the integration becomes analytically solvable. Therefore one focuses on a test dipole at the position vector $\vec{a}:=z\vec{e}_{z}$ with $z\in\left\lbrace \left[-R_{2},-R_{1}\right] \cup \left[R_{1},R_{2}\right]\right\rbrace$ for the hollow sphere $\Omega$. Consequently, one finds from (\ref{IdDIP2}), (\ref{DistVecZ1}) and by a transformation into spherical coordinates with the help of (\ref{sphereCoordin}) and (\ref{volElem}):
\begin{equation}\label{IdDIP3}
\begin{split}
\begin{gathered}
I^{\,dip}_{d}=\dfrac{\mu_{0}}{8\pi}M^{2} \int\limits_{\varphi^{\prime}=0}^{2\pi}d\varphi^{\prime}\int\limits_{\theta^{\prime}=0}^{\pi}\int\limits_{r^{\prime}=R_{1}}^{\textup{R}_{2}}\left(\dfrac{1}{\left(z^{2}-2zr^{\prime} \cos{\left(\theta^{\prime}\right)}+r^{\prime\,2} \right)^{\frac{3}{2}}}\right.
\\[15pt]
\left.-\dfrac{3\left(z-r^{\,\,\prime}\cos{\left(\theta^{\prime}\right)}\right)^{2}}{\left(z^{2}-2zr^{\prime} \cos{\left(\theta^{\prime}\right)}+r^{\prime\,2} \right)^{\frac{5}{2}}} \right)r^{\prime\,2}\sin{\left(\theta^{\prime}\right)}\, dr^{\prime} d\theta^{\prime} 
\\[15pt]
=\dfrac{\mu_{0}M^{2} \cancel{2\pi}}{4\,\cancel{2\pi}}\int\limits_{r^{\prime}=R_{1}}^{\textup{R}_{2}}\int\limits_{\theta^{\prime}=0}^{\pi}\left(\dfrac{z^{2}-2zr^{\prime} \cos{\left(\theta^{\prime}\right)}+r^{\prime\,2}-3\left(z-r^{\,\,\prime}\cos{\left(\theta^{\prime}\right)}\right)^{2}}{\left(z^{2}-2zr^{\prime} \cos{\left(\theta^{\prime}\right)}+r^{\prime\,2} \right)^{\frac{5}{2}}} \right)r^{\prime\,2}\sin{\left(\theta^{\prime}\right)}\, d\theta^{\prime}dr^{\prime} 
\end{gathered}
\end{split}
\end{equation} 
Note that in (\ref{IdDIP3}) \textit{Fubini's} theorem was used to change the order of integration. This allows to first carry out the $\varphi^{\prime}$-integration, which gives a prefactor of $2\pi$ due to the fact that the integrand does not depend on $\varphi^{\prime}$. 

The remaining two dimensional integral (\ref{IdDIP3}) is further solved by the introduction of a substitution for the basis of the power in the denominator:
\begin{equation}\label{IDIPSubU}
\begin{split}
\begin{gathered}
u=z^{2}-2zr^{\prime} \cos{\left(\theta^{\prime}\right)}+r^{\prime\,2} \,;\,\, d\theta^{\prime}=\dfrac{du}{2zr^{\prime} \sin{\left(\theta^{\prime}\right)}}\,;\,\,u\left(0\right)=\left(z-r^{\prime}\right)^{2}\,;\,\,u\left(\pi\right)=\left(z+r^{\prime}\right)^{2}
\end{gathered}
\end{split}
\end{equation} 
The substitution (\ref{IDIPSubU}) reduces the integrand of (\ref{IdDIP3}) to a sum of integrable powers of $u$. Thus the $u$-integration can be completed, leaving an integral over the radial coordinate $r^{\prime}$. This is explicitly shown in (\ref{uIntIDIP}) and results in:
\begin{equation}\label{IdDIP4}
\begin{split}
\begin{gathered}
I^{\,dip}_{d}=-\dfrac{3}{16z^{3}}\mu_{0}M^{2}\int\limits_{r^{\prime}=R_{1}}^{\textup{R}_{2}}r^{\prime}\left(\vert z+r^{\prime}\vert-\vert z-r^{\prime}\vert-\dfrac{2\left(z^{2}-3r^{\prime\,2}\right)\left(\vert z-r^{\,\prime}\vert-\vert z+r^{\prime}\vert\right)}{3\vert z+r^{\prime}\vert\vert z-r^{\prime}\vert}\right.
\end{gathered}
\end{split}
\end{equation}
\begin{equation*} 
\begin{split}
\begin{gathered}
\left.-\dfrac{\left(z^{2}-r^{\prime\,2}\right)^{2}\left(\vert z-r^{\prime}\vert^{3}-\vert z+r^{\prime}\vert^{3}\right)}{3\vert z+r^{\prime}\vert^{3}\vert z-r^{\prime}\vert^{3}}\right)\,dr^{\,\prime}
\end{gathered}
\end{split}
\end{equation*} 
In (\ref{IdDIP4}) one observes two moduli, hence a case analysis is required, in order to integrate with respect to $r^{\prime}$. One rembers that the test dipole has to be inside the magnetized hollow sphere $\Omega$, meaning $z\in\left\lbrace \left[-R_{2},-R_{1}\right] \cup \left[R_{1},R_{2}\right]\right\rbrace$. Moreover, in accordance with the allowed interval for $z$, one can deduce from (\ref{IdDIP4}) that $I^{\,dip}_{d}=0$ is true for all $z$ inside the hollow space or outside of the sphere. This is because the magnitude of the magnetization density is a prefactor in (\ref{IdDIP4}) and is only non-zero, according to its definition in (\ref{defM}), if the test dipole is within the volume $\Omega$. Furthermore, in (\ref{pointsymm1}) one has shown that the integrand $f(z)$ of (\ref{IdDIP4}) is point-symmetric with regard to the point of origin $z=0$, which means $f(-z)=-f(z)$. Therefore the different cases for the moduli as well as the $r^{\prime}$-integration can be done for $z\in\left[R_{1},R_{2}\right]$, generating the solution $I^{\,dip}_{d}\left(z\in\left[R_{1},R_{2}\right]\right)$. The solution for $z\in\left[-R_{2},-R_{1}\right]$ is then given by:
\begin{equation}\label{IdDIP5-+}
I^{\,dip}_{d\,-}:=I^{\,dip}_{d}\left(z\in\left[-R_{2},-R_{1}\right]\right)=-I^{\,dip}_{d}\left(z\in\left[R_{1},R_{2}\right]\right)=:-I^{\,dip}_{d\,+}
\end{equation}
For the case analysis one gets:
\begin{equation}\label{caseanalysis}
\vert z+r^{\prime}\vert = z+r^{\prime} \;\;;\;\; \forall \,z\,\in\,\left[R_{1},R_{2}\right]
\;\;;\;\;
\vert z-r^{\prime}\vert =
\begin{cases}
z-r^{\prime}&;\;\; \forall \,z\,\in\,\left[R_{1},z\right]
\\
-\left(z-r^{\prime}\right)&;\;\; \forall \,z\,\in\,\left[z,R_{2}\right]
\end{cases}
\end{equation}
For the purpose of evaluating the integral (\ref{IdDIP4}) let the integrand $f\left(\vert z+r^{\prime}\vert,\,\vert z-r^{\prime}\vert\right)$. Then $I^{\,dip}_{d\,+}$ can be calculated under the use of (\ref{IdDIP4}) and (\ref{caseanalysis}) with the following representation:
\begin{equation}\label{IdDIP6}
\begin{split}
\begin{gathered}
I^{\,dip}_{d\,+}=-\dfrac{3}{16z^{3}}\mu_{0}M^{2}\left\lbrace \int\limits_{r^{\prime}=\textup{R}_{1}}^{\textup{z}} f\left( z+r^{\prime},\, z-r^{\prime}\right) \,dr^{\,\prime}+\int\limits_{r^{\prime}=\textup{z}}^{\textup{R}_{2}} f\left( z+r^{\prime},\,-\left( z-r^{\prime}\right)\right) \,dr^{\,\prime} \right\rbrace
\\[12pt]
=:-\dfrac{3}{16z^{3}}\mu_{0}M^{2}\left\lbrace I_{r^{\,\prime},\,1}+I_{r^{\,\prime},\,2} \right\rbrace
\end{gathered}
\end{split}
\end{equation}

The procedure of solution for the integrals $I_{r^{\prime},\,1}$ and $I_{r^{\prime},\,2}$ can be found in (\ref{rPrimeIntIDIP}). Eventually, their values from (\ref{1rPrime1IdDIP}) and (\ref{1rPrime2IdDIP}) are inserted into equation (\ref{IdDIP6}), which gives the following result for $z\in\left[R_{1},R_{2}\right]$:
\begin{equation}\label{IdDIP7}
I^{\,dip}_{d\,+}\underset{(\ref{IdDIP6})}{\overset{(\ref{1rPrime1IdDIP}),(\ref{1rPrime2IdDIP})}{=}}-\dfrac{3}{16z^{3}}\mu_{0}M^{2}\left\lbrace \dfrac{16}{9}\left(z^{3}-R_{1}^{3}\right)+0 \right\rbrace=-\dfrac{1}{3}\mu_{0}M^{2}\Biggl( 1-\biggl( \dfrac{R_{1}}{z}\biggr)^{3}\Biggr)
\end{equation}
For negative values of $z$ namely $z\in\left[-R_{2},-R_{1}\right]$ one has to insert the boundary $\vert z\vert$ into the case analysis. Consequently, one finds from (\ref{IdDIP5-+}), (\ref{caseanalysis}) and (\ref{IdDIP6}):
\begin{equation*}
I^{\,dip}_{d\,-}\underset{(\ref{IdDIP6})}{\overset{(\ref{caseanalysis}), (\ref{IdDIP5-+})}{=}}\dfrac{3}{16z^{3}}\mu_{0}M^{2}\left\lbrace \int\limits_{r^{\prime}=\textup{R}_{1}}^{\vert\textup{z}\vert} f\left( z+r^{\prime},\, z-r^{\prime}\right) \,dr^{\,\prime}\right.
\end{equation*}
\begin{equation}\label{IdDIP8}
\left.+\int\limits_{r^{\prime}=\vert\textup{z}\vert}^{\textup{R}_{2}} f\left( z+r^{\prime},\,-\left( z-r^{\prime}\right)\right) \,dr^{\,\prime} \right\rbrace
=\dfrac{3}{16z^{3}}\mu_{0}M^{2}\left\lbrace I_{r^{\,\prime},\,1}\left(\vert z\vert\right)+I_{r^{\,\prime},\,2}\left(\vert z\vert\right) \right\rbrace 
\end{equation}
Further evaluation of (\ref{IdDIP8}) is then carried out with the help of (\ref{1rPrime1IdDIP}) and (\ref{1rPrime2IdDIP}), whereat $z=-\vert z \vert$ holds for $z\in\left[-R_{2},-R_{1}\right]$ :
\begin{equation}\label{IdDIP9}
I^{\,dip}_{d\,-}\underset{(\ref{IdDIP8})}{\overset{(\ref{1rPrime1IdDIP}),(\ref{1rPrime2IdDIP})}{=}}-\dfrac{3}{16\vert z\vert^{3}}\mu_{0}M^{2}\left\lbrace \dfrac{16}{9}\left(\vert z\vert^{3}-R_{1}^{3}\right)+0 \right\rbrace=-\dfrac{1}{3}\mu_{0}M^{2}\Biggl( 1-\biggl( \dfrac{R_{1}}{\vert z\vert}\biggr)^{3}\Biggr)
\end{equation}
By comparison of (\ref{IdDIP7}) and (\ref{IdDIP9}) one notice that $I^{\,dip}_{d} = I^{\,dip}_{d\,-}$ holds for all possible values of $z$ within the hollow sphere. The total dipolar interaction contribution $
\Delta E_{BH}^{\,d}$ is then given by (\ref{totalEBHfromdipolContri2}) under the use of (\ref{IdDIP9}):
\begin{equation}\label{DELTAEBHd1}
\begin{split}
\begin{gathered}
\Delta E_{BH}^{\,d}\overset{(\ref{totalEBHfromdipolContri2})}{=}\int\limits_{\Omega} I^{\,dip}_{d\,-} \,d\Omega\overset{(\ref{IdDIP9})}{=}-\dfrac{1}{3}\mu_{0}M^{2}\int\limits_{\Omega}  1-\biggl( \dfrac{R_{1}}{\vert z\vert}\biggr)^{3} \,d\Omega
\\[11pt]
=-\dfrac{1}{3}\mu_{0}M^{2}\left\lbrace\int\limits_{\Omega} d\Omega-R_{1}^{3}\int\limits_{\Omega}\vert z\vert^{-3} \,d\Omega\right\rbrace
=-\dfrac{1}{3}\mu_{0}M^{2}\left\lbrace vol\left(\Omega\right)-R_{1}^{3}\int\limits_{\Omega}\vert z\vert^{-3} \,d\Omega\right\rbrace
\end{gathered}
\end{split}
\end{equation}
In (\ref{DELTAEBHd1}) the last integral gives an infinte contribution as it can be seen in (\ref{InfDIPzaxis}). However, this was to be expected, because the test dipole was defined to be located along the $z$-axis of the hollow sphere with $z\in\left\lbrace \left[-R_{2},-R_{1}\right] \cup \left[R_{1},R_{2}\right]\right\rbrace$. As a consequence, one can imagine that divergent terms occur, since they were not canceled by contributions from other positions within the volume $\Omega$ and outside the $z$-axis. Therefore the general application of (\ref{totalEBHfromdipolContri2}) was not justified and the introduction of a restriction is necessary, which enforces only finite or physical corrections. This is only possible by an exclusive consideration of a full sphere with $R_{1}=0$, which results in:
\begin{equation}\label{DELTAEBHd2}
\underline{\underline{\Delta E_{BH}^{\,d}}}\overset{(\ref{DELTAEBHd1})}{=}-\dfrac{1}{3}\mu_{0}M^{2} vol\left(\Omega\right) = -\dfrac{1}{3}\mu_{0}M^{2}\dfrac{4\pi}{3}R_{2}^{3} =\underline{\underline{-\dfrac{4\pi}{9}\mu_{0}M^{2} R_{2}^{3}}}
\end{equation}
From (\ref{totalEBHfromdipolContri2}), (\ref{IcDIP1}) and (\ref{DELTAEBHd2}) one can estimate an analytical expression for the \textit{Breit-Hartree} contribution of a test dipole on the $z$-axis of a homogeneously magnetized full sphere with radius $R_{2}$ and $R_{1}=0$:
\begin{equation}\label{DELTAEBHDIP}
\begin{split}
\begin{gathered}
\underline{\underline{\Delta E_{BH}}}\overset{(\ref{totalEBHfromdipolContri2})}{=}\Delta E_{BH}^{\,d}+\Delta E_{BH}^{\,c} = -\dfrac{4\pi}{9}\mu_{0}M^{2} R_{2}^{3}-\dfrac{4\pi}{9}\mu_{0}M^{2}\left(R_{2}^{3}-\cancel{R_{1}^{3}}^{\,0}\right) 
\\[14pt]
  =\underline{\underline{ -\dfrac{8\pi}{9}\mu_{0}M^{2} R_{2}^{3}}}
\end{gathered}
\end{split}
\end{equation}

In general, it is noticeable from (\ref{DELTAEBHd1}) for the full sphere ($R_{1}=0$) that the dipole interaction part $\Delta E_{BH}^{\,d}$ of the \textit{Breit-Hartree} contribution is constant and does not depend on the position of the test dipole on the $z$-axis. In addition, one finds the same characteristics for the contact interaction contribution $\Delta E_{BH}^{\,c}$ by regarding (\ref{IcDIP1}), which leads to the understanding that the \textit{Breit-Hartree} contribution for the test dipol approximation applied to a full sphere includes equal-sized terms from the contact and the dipole-dipole interaction. Finally, one expresses the solution (\ref{DELTAEBHDIP}) with the spin density representation of the magnetization density according to (\ref{defM}), if $vol\left(\Omega\right)$ is the volume of a full sphere with radius $R_2$:
\begin{equation}\label{DELTAEBHDIP2}
\begin{split}
\begin{gathered}
\Delta E_{BH}=-\dfrac{2}{3}\mu_{0}M^{2}vol\left(\Omega\right) 
\overset{(\ref{defM})}{=}-\dfrac{2}{3}\mu_{0}\mu_{B}^{2}s^{2}vol\left(\Omega\right) 
\end{gathered}
\end{split}
\end{equation}
In (\ref{DELTAEBHDIP2}) one recognizes the same proportionalities to the volume and the square of the spin density as they were found for the contact interaction in (\ref{Icv2}). 

In conclusion, one has shown that the test dipole approach leads to an identical result for the contact term (\ref{Icv2}) as from the integral representation (\ref{EnergieBHSI}) given by \textit{Jansen}. Moreover, an analytic expression for the \textit{Breit-Hartree} contribution for a full sphere was found (\ref{DELTAEBHDIP}), which contains terms of equal size from the dipolar and the contact interaction. Hereinafter, one uses these results for the full sphere for the verification of further solutions for the \textit{Breit-Hartree} contribution. However, the restriction to the $z$-axis and the inadmissible volume integration must always be taken into account.

\clearpage

\section{Magnetostatic approach and solution for a homogeneously magnetized hollow sphere}\label{MagnetostatikSolution}

The problem of a homogeneously magnetized hollow sphere with the magnetization density defined in (\ref{defM}) and the finite volume $\Omega$ is treated stationary. Therefore the classical theory of electrodynamics as well as more accurately the theory of magnetostatics holds, allowing oneself to apply a different analytical approach. This is valid, because the \textit{Breit-Hartree} contribution is equivalent to the potential energy arising from the dipole-dipole and contact interaction within magnetized matter, which was discussed in section \ref{BHINERACTION}. Thus one has to evaluate the potential interaction energy from the integral over the potential energy density generated by the magnetization density $\vec{M}(\vec{r}\,)$ interacting with the magnetic flux density $\vec{B}(\vec{r}\,)$ caused by itself \cite{TUDortmund}, \cite{Cohen-Tannoudji}: \vspace{3.5mm}
\begin{equation}\label{E_WW_DIP}
\Delta E_{BH}=-\int\limits_{\Omega} \vec{M}\left(\vec{r}\,\right)\cdot\vec{B}\left(\vec{r}\,\right)\,d\Omega
\end{equation}
In (\ref{E_WW_DIP}) the magnetic flux density $\vec{B}(\vec{r}\,)$ is interpreted as an external field with the same magnitude as the field generated by the magnetization density. The magnetostatic representation (\ref{E_WW_DIP}) is even consistent with the integral expression of the \textit{Breit-Hartree} interaction (\ref{EnergieBHSI}) stated by \textit{Jansen}, which can be understood in detail in \cite{Cohen-Tannoudji}, where \textit{C. Cohen-Tannoudji} derived the integrals for the dipole-dipole and contact interaction in SI-units as well as with the publication \cite{Pellegrini_2020} from \textit{Pellegrini et al.}, which shows the connection between (\ref{EnergieBHSI}) and the magnetostatic energy (\ref{E_WW_DIP}). 
\\
In absence of free current densities and for the stationary case \textit{Ampère's} law is a homogeneous, vector valued differential equation and reads in SI-units \cite{Bartelmann_2015}:
\begin{equation}\label{AmpereLaw}
\vec{\nabla}\times\vec{H}=\vec{j}+\dfrac{\partial\vec{D}}{\partial t}= \vec{\,0} 
\end{equation}
Consequently, the magnetic field $\vec{H}$ is non-rotational and has the general solution:
\begin{equation}\label{magneticfield}
\vec{H}\left(\vec{r}\,\right)=-\vec{\nabla}\cdot\phi_{m}\left(\vec{r}\,\right) \;\;;\; \vec{j}=\vec{\,0}\,;\;\; \vec{r}\in\Omega
\end{equation}
Whereat $\phi_{m}\left(\vec{r}\,\right)$ is the magnetic scalar potential, which is an unambiguously determined function of the position vector $\vec{r}$, if the volume $\Omega$ is simply connected \cite{Bartelmann_2015}, which is the case for a hollow sphere $\Omega\in\mathbb{R}^{3}$. Now one combines the material equation $\vec{B}=\mu_{0}\left(\vec{H}+\vec{M}\right)$ with the requirement that the magnetic flux density is source-free thus $\vec{\nabla}\cdot\vec{B}=0$ and inserts (\ref{magneticfield}):
\begin{equation}\label{LaplacGl1}
\begin{split}
\begin{gathered}
\vec{\nabla}\cdot\vec{B}=\mu_{0}\cdot\vec{\nabla}\cdot\left(\vec{H}+\vec{M}\right)=0 \;\;\;\longleftrightarrow\;\;\; -\vec{\nabla}\cdot\vec{H}=\vec{\nabla}\cdot\vec{M}
\\[10pt]
\xlongleftrightarrow{\text{(\ref{magneticfield})}} \;\;\; \underline{\underline{\Delta\phi_{m}\left(\vec{r}\,\right) =\vec{\nabla}\cdot\vec{M}=:\rho_{m}\left(\vec{r}\,\right) }}
\end{gathered}
\end{split}
\end{equation}
In (\ref{LaplacGl1}) one receives the \textit{Poisson} equation for the effective magnetic surface charge density $\rho_{m}\left(\vec{r}\,\right)$, whereby it must be emphasized that $\rho_{m}$ is a density of virtually separated magnetic dipole charges, excluding any interpretation as magnetic monopoles. For the constant magnetization density from (\ref{defM}) one finds from (\ref{DivM(r)2}) that $\rho_{m}= 0$ holds in the whole space with the exclusion of the surface of the sphere. Hence, the \textit{Poisson} equation becomes the \textit{Laplace} equation $\Delta\phi_{m}\left(\vec{r}\,\right) = 0$. However, the magnetization density is not continuous, requiring a piecewise-defined \textit{ansatz} for $\phi_{m}\left(\vec{r}\,\right)$ in spherical coordinates (\ref{sphereCoordin}) with regard to the $\varphi$-independence of the magnetic potential, due to the spherical symmetry of the problem \cite{Bartelmann_2015}:
\begin{equation}\label{phimAnsatz1}
\phi_{m}\left(\vec{r}\,\right)= 
\begin{cases}
\phi_{i}\left(\vec{r}\,\right)=\phi_{i}\left(r,\,\theta\right)&;\;\;\; \forall \; \vec{r}\,\in\,\left\lbrace\vec{r}\,\notin\,\Omega\,\vert\,0\leq r\leq R_{1}\right\rbrace
\\
\phi_{M}\left(\vec{r}\,\right)=\phi_{M}\left(r,\,\theta\right)&;\;\;\; \forall \; \vec{r}\,\in\,\Omega
\\
\phi_{e}\left(\vec{r}\,\right)=\phi_{e}\left(r,\,\theta\right)&;\;\;\; \forall \;\vec{r}\,\in\,\left\lbrace\vec{r}\,\notin\,\Omega\,\vert\,R_{2}\leq r<\infty \right\rbrace
\end{cases}
\end{equation}
The general solution of the \textit{Laplace} equation for a $\varphi$-independent potential $\phi_{m}\left(r,\,\theta\right)$, containing the \textit{Legendre} polynomials $\textup{P}_{l}\left(\cos{\left(\theta\right)}\right)$, is well known and can be found in the relevant literature \cite{Bartelmann_2015}:
\begin{equation}\label{LaplaceSoultiongenrale}
\phi_{m}\left(r,\,\theta\right)= \sum\limits_{l=1}^{\infty}\left(A_{l}r^{l}+B_{l}r^{-\left(l+1\right)}\right)\textup{P}_{l}\left(\cos{\left(\theta\right)}\right)
\end{equation}
In order to evaluate the piecewise-defined potential from (\ref{phimAnsatz1}) one uses the general solution from (\ref{LaplaceSoultiongenrale}). In addition, one modifies the general contribution, so that the magnetic potential is physically regular, meaning that it vanishes at an infinite distance from the origin and is finite at the origin. Equivalently, one demands the validity of $\lim\limits_{r\rightarrow \infty}{\phi_{m}\left(r,\,\theta\right)}=0$ and $\lim\limits_{r\rightarrow 0}{\phi_{m}\left(r,\,\theta\right)}\neq \pm\infty$, leading to:
\begin{equation}\label{phimAnsatz2}
\phi_{m}\left(\vec{r}\,\right)= 
\begin{cases}
\phi_{i}\left(r,\,\theta\right)=\sum\limits_{l=1}^{\infty}A_{l}r^{l}\textup{P}_{l}\left(\cos{\left(\theta\right)}\right)
\\
\phi_{M}\left(r,\,\theta\right)=\sum\limits_{l=1}^{\infty}\left(B_{l}r^{l}+C_{l}r^{-\left(l+1\right)}\right)\textup{P}_{l}\left(\cos{\left(\theta\right)}\right)
\\
\phi_{e}\left(r,\,\theta\right)=\sum\limits_{l=1}^{\infty} D_{l}r^{-\left(l+1\right)}\textup{P}_{l}\left(\cos{\left(\theta\right)}\right)
\end{cases}
\end{equation}
There is a unique solution for the four constant coefficients in (\ref{phimAnsatz2}), if one deduces four equations from four different boundary conditions. These conditions are received by the requirement of continuous components $B_{r}$, $H_{\varphi}$ and $H_{\theta}$ of the magnetic flux density $\vec{B}$ and the magnetic field $\vec{H}$ \cite{Bartelmann_2015}, which ensures closed field lines for both vector fields as well as solenoidality of the $\vec{B}$-field, whereat those conditions are necessary, because the magnetization density is discontinuous at the bounding surfaces of $\Omega$ according to (\ref{defM}).
By means of the material equation and the definition of the magnetization density (\ref{defM}), one finds that the requirement is fulfilled, if the radial component of $\vec{B}$ is continuous and the magnetic potential has the same function values for $r=R_{1}$ and $r=R_{2}$ and every possible value of the polar angle $\theta$. Hence, the boundary conditions are:
\begin{equation}\label{LaplacBoundary}
\begin{split}
\begin{gathered}
B_{r,i}\left(R_{1},\,\theta\right)\overset{!}{=}B_{r,M}\left(R_{1},\,\theta\right)\;\;;\;\; B_{r,M}\left(R_{2},\,\theta\right)\overset{!}{=}B_{r,e}\left(R_{2},\,\theta\right)
\\[5pt]
\phi_{i}\left(R_{1},\,\theta\right)\overset{!}{=}\phi_{M}\left(R_{1},\,\theta\right)\;\;;\;\;\phi_{M}\left(R_{2},\,\theta\right)\overset{!}{=}\phi_{e}\left(R_{2},\,\theta\right)
\end{gathered}
\end{split}
\end{equation}
From (\ref{magneticfield}), (\ref{phimAnsatz2}), (\ref{LaplacBoundary}) and the material equation for $\vec{B}$ one determines the constants, which is carried out in (\ref{AppendixConstCoeffHollowSphereConst}) and shows that the solution contains only contributions with $l=1$ from the \textit{Legendre} polynomials. The constants are then inserted into (\ref{phimAnsatz2}), keeping only terms with $l=1$:
\begin{equation}\label{phimSolutionHollowSphere}
\underline{\underline{\phi_{m}\left(\vec{r}\,\right)= 
\begin{cases}
\phi_{i}\left(r,\,\theta\right)= 0 \,\textup{A}
\\
\phi_{M}\left(r,\,\theta\right)=\dfrac{M}{3}\left(r-\dfrac{R_{1}^{3}}{r^{2}}\right)\cos{\left(\theta\right)}
\\
\phi_{e}\left(r,\,\theta\right)=\dfrac{M}{3} \left(R_{2}^{3}-R_{1}^{3}\right)\dfrac{\cos{\left(\theta\right)}}{r^{2}}=\dfrac{\vec{m}\cdot\vec{r}}{4\pi r^{\,3}}
\end{cases}}}
\end{equation}
The solution for the magnetic potential of a homogeneously magnetized hollow sphere exhibits, that within the internal space the potential is zero and therefore field-free. Moreover, one finds the potential for a full sphere with radius $R$, by setting $R_{1}=0 \textup{m}$:
\begin{equation}\label{phimSolutionFullSphere}
\underline{\underline{\phi_{m,R}\left(\vec{r}\,\right)= 
\begin{cases}
\phi_{M,R}\left(r,\,\theta\right)=\dfrac{M}{3}r\cos{\left(\theta\right)}
\\\\
\phi_{e,R}\left(r,\,\theta\right)=\dfrac{M}{3} R^{3}\,\dfrac{\cos{\left(\theta\right)}}{r^{2}}=\dfrac{\vec{m}_{R}\cdot\vec{r}}{4\pi r^{\,3}}
\end{cases}}}
\end{equation}
In (\ref{phimSolutionFullSphere}) the potential in the exterior space $\phi_{e}$ can be interpreted as the potential of a point dipole with the magnetic moment $\vec{m}_{R}=\frac{4\pi}{3}R^{3}M\vec{e}_{z}$ for the full sphere, while in (\ref{phimSolutionHollowSphere}) the magnetic moment is $\vec{m}=\frac{4\pi}{3}(R_{2}^{3}-R_{1}^{3})M\vec{e}_{z}$ for the hollow sphere \cite{FranzSchwabl2019}. In addition, one notices, that the solution for the hollow sphere is simply given by the subtraction of the potentials $\phi_{m,R_{2}}$ and $\phi_{m,R_{1}}$ of two full spheres with radii $R_{1}$ and $R_{2}$:
\begin{equation}\label{phimSolutionInterpet}
\phi_{m}\left(\vec{r}\,\right)= 
\begin{cases}
\phi_{i}\left(r,\,\theta\right)=\phi_{M,R_{2}}-\phi_{M,R_{1}} = 0 \,\textup{A}
\\
\phi_{M}\left(r,\,\theta\right)=\phi_{M,R_{2}}-\phi_{e,R_{1}}=\dfrac{M}{3}\left(r-\dfrac{R_{1}^{3}}{r^{2}}\right)\cos{\left(\theta\right)}
\\
\phi_{e}\left(r,\,\theta\right)=\phi_{e,R_{2}}-\phi_{e,R_{1}}=\dfrac{M}{3} \left(R_{2}^{3}-R_{1}^{3}\right)\dfrac{\cos{\left(\theta\right)}}{r^{2}}
\end{cases}
\end{equation}
Hereinafter, one can evaluate the \textit{Breit-Hartree} contribution to the total energy of a homogeneously magnetized hollow sphere $\Omega$. For this, one uses the material equation, (\ref{magneticfield}) and the magnetic potential from (\ref{phimSolutionHollowSphere}), for the purpose of the calculation of the magnetic flux density $\vec{B}$. This is explicitly shown in (\ref{BfieldhollowSphereConst1App}), where only the magnetized volume is regarded, because the magnitude $M$ of the magnetization density is zero for the internal and exterior space, generating no contribution to the \textit{Breit-Hartree} correction:
\begin{equation}\label{BfieldhollowSphereConst1}
\vec{B}_{M}\left(r,\,\theta\right)= \mu_{0}\dfrac{M}{3}\left(2\vec{e}_{z}-\left(\dfrac{R_{1}}{r}\right)^{3}\left(2\cos{\left(\theta\right)}\vec{e}_{r}+\sin{\left(\theta\right)}\vec{e}_{\theta}\right)\right)
\end{equation}
In (\ref{BfieldhollowSphereConst1}) $\vec{e}_{r}$ and $\vec{e}_{\theta}$ are basis vectors in spherical coordinates, while $\vec{e}_{z}$ is a cartesian basis vector. Moreover, one observes two parts in the representation of the $\vec{B}$-field ($\vec{e}_{z}+\vec{e}_{z}=2\vec{e}_{z}$), which correspond to a magnetic contact and a magnetic dipole-dipole term, whereat the second summand only exists. This structure is typical for a magnetic point dipole \cite{Bartelmann_2015} and shows that the contact and dipole-dipole interactions are included in a similar way as it can be found in the integral expression (\ref{EnergieBHSI}) from \textit{Jansen}.
\\
Eventually, one inserts (\ref{BfieldhollowSphereConst1}) and (\ref{defM}) into (\ref{E_WW_DIP}), receiving the \textit{Breit-Hartree} contribution, by following the evaluation in (\ref{EHBHollowSphereMConstApp}):
\begin{equation}\label{EHBHollowSphereMConst}
\underline{\underline{\Delta E_{BH}  =-\dfrac{8\pi}{9}\mu_{0}M^{2}\left(R_{2}^{3}-R_{1}^{3}\right)=  -\dfrac{2}{3}\mu_{0}M^{2}vol\left(\Omega\right)}}
\end{equation}
One ascertains that the estimation for the \textit{Breit-Hartree} correction for a test dipole on the $z$-axis from (\ref{DELTAEBHDIP}) and (\ref{DELTAEBHDIP2}) is equal to the solution in (\ref{EHBHollowSphereMConst}) for the consideration of a full sphere. Hence, the finite expression in (\ref{EHBHollowSphereMConst}) justifies the explanation from section \ref{AnalyticalTestDipol}, that the infinite integral in (\ref{DELTAEBHd1}) vanishes, if the test dipole is not restricted to the $z$-axis. As a consequence, the \textit{Breit-Hartree} contribution to the total energy of a homogeneously magnetized hollow sphere $\Omega$ is generally given by (\ref{EHBHollowSphereMConst}) and is directly proportional to the spherical volume $\Omega$ and the square of the magnetization or spin density:
\begin{equation}\label{SpinEHBHollowSphereMConst}
\underline{\underline{\Delta E_{BH}\, \overset{(\ref{EHBHollowSphereMConst})}{=}\, -\dfrac{2}{3}\mu_{0}M^{2}vol\left(\Omega\right)\,\overset{(\ref{defM})}{=}\, -\dfrac{2}{3}\mu_{0}\mu_{B}^{2}s^{2}vol\left(\Omega\right)}}
\end{equation}
Let $E_{BH,max.}$ be the maximum possible energy contribution, which is reached for a stretched ellipsoid
with an infinite aspect ratio \cite{manuelconversation} and $\zeta$ the demagnetization factor, then one can rewrite the expression (\ref{EHBHollowSphereMConst}):
\begin{equation}\label{Demagnetfactor}
\Delta E_{BH} \overset{(\ref{EHBHollowSphereMConst})}{=} -\dfrac{2}{3}\mu_{0}M^{2}vol\left(\Omega\right)\overset{(\ref{defM})}{=} \left(1-\dfrac{1}{3}\right)\left(-\mu_{0}\mu_{B}^{2}s^{2}vol\left(\Omega\right)\right) =:\left(1-\zeta\right)E_{BH,max.}
\end{equation} 
From (\ref{Demagnetfactor}) one notices the demagnetization factor $\zeta=\frac{1}{3}$, which is characteristic for spherical volumes \cite{Bartelmann_2015}. The occurrence of this factor is explained by the fact, that the dipole-dipole interaction for colinear spins along the $z$-axis is attractive and leads to the negative maximum potential energy $E_{BH,max.}$ for a constant magnitude $s$ and a given volume $\Omega$, while the dipolar interaction orthogonal to the $z$-axis is repulsive and leads to a positive demagnetization contribution to the potential interaction energy. In conclusion, one receives a \textit{Breit-Hartree} contribution, which is lower than the maximum possible interaction energy, whereat the reduction factor is the previously mentioned, geometry-dependent demagnetization factor. Hence, it is shown that the magnetic shape anisotropy is considered in the \textit{Breit-Hartree} interaction, which was previously mentioned by \textit{Jansen} \cite{Jansen1988}, \cite{ComplexMaterials}.
\clearpage

%% file: Evaluation_of_the_Breit-Hartree_contribution_to_the_total_energy_of_a_spinpolarized_half-filled_atomic_shell.tex
\chapter{Evaluation of the \textit{Breit-Hartree} contribution to the total energy of a half-filled, spin-polarized atomic shell at the ground state}

The \textit{Breit-Hartree} contribution is proportional to the square of the spin density for a constant magnetization density as it was shown in section \ref{MagnetostatikSolution}, using the classical theory of magnetostatics. One therefore assumes a similar proportionality for a radially dependent magnetization density respectively spin density, due to the fact that a constant magnetization density is a special case of a magnetization density with radial dependence. In additon, a radial spin density can be calculated with current fully relativistic DFT-programms for different atoms and ions, which would allow an application of \textit{Breit-Hartree} correction to different atomic and ionic states. Consequently, it is justified to use the same magnetostatic approach, in order to derive an analytical expression for the \textit{Breit-Hartree} contribution to the total energy of a hollow sphere $\Omega$ with a radially dependent magnetization density. In this context, $\Omega=r\,\in\,\left[R_{1},R_{2}\right]\times\theta\,\in\,\left[0,\pi\right]\times\varphi\,\in\,\left[0,2\pi\right)$ is an approximation in zeroth order for a spin-polarized, half-filled atomic shell as it occurs e.g. in the $\textup{Gd}^{3+}$ ion, where the $4f$-shell is half-filled and generates a radially dependent spin density $s(r)$. As discussed in section \ref{Important} one expects significant correction contributions to the total energy of the $\textup{Gd}^{3+}$ ion, because the orbital angular momentum of the $4f$-shell is zero and the spin density predominantly contributes to the magnetization density. Moreover, only the spin density is included in the definition for the magnetization density within this thesis as it was stated in (\ref{magnetdensity}). 

\section{Magnetostatic approach and solution for a hollow sphere with a radially dependent magnetization density}\label{MagnetostatikSolutionHollow}

First, the magnetization density along the $z$-axis is defined for the hollow sphere $\Omega$, under the use of (\ref{magnetdensity}):
\begin{equation}\label{defM(r)}
\vec{M}\left(\vec{r}\,\right)= \vec{M}\left(r\right) = 
\begin{cases}
M(r)\vec{e}_{z}\overset{(\ref{magnetdensity})}{=} -\mu_{B}s(r)\vec{e}_{z}&;\;\; \forall \; \vec{r}\,\in\,\Omega
\\
\vec{0}&;\;\; \forall \; \vec{r} \,\notin\,\Omega
\end{cases}
\end{equation}
Second, one remarks that the equations (\ref{E_WW_DIP}), (\ref{AmpereLaw}) and (\ref{magneticfield}) still hold, due to the stationary consideration of the volume $\Omega$ under the consideration of the absence of free currents. As a consequence the \textit{Poisson} equation from (\ref{LaplacGl1}) applies to the whole space. For the case of the radially dependent magnetization density from (\ref{defM(r)}), one finds the effective magnetic charge density, by evaluating the divergence of the magnetization density in spherical coordinates (see (\ref{DivM(r)1})):
\begin{equation}\label{rhom(r)}
\rho_{m}\left(\vec{r}\,\right)\overset{(\ref{LaplacGl1})}{=}\vec{\nabla}\cdot \vec{M} \,\underset{(\ref{defM(r)})}{\overset{(\ref{DivM(r)1})}{=}} \,
\begin{cases}
\dfrac{d\,M(r)}{dr}\cos{\left(\theta\right)}&;\;\; \forall \; \vec{r}\,\in\,\Omega
\\
0&;\;\; \forall \; \vec{r} \,\notin\,\Omega
\end{cases}
\end{equation}
Hence, it is sufficient to solve the \textit{Laplace} equation for the internal and exterior space, while for the magnetized volume the \textit{Poisson} equation does not simplify. Based on that, one can use the \textit{ansatz} from (\ref{phimAnsatz2}) for a physically regular magnetic potential inside and outside the hollow sphere, which means $\lim\limits_{r\rightarrow \infty}{\phi_{m}\left(r,\,\theta\right)}=0$ and $\lim\limits_{r\rightarrow 0}{\phi_{m}\left(r,\,\theta\right)}\neq \pm\infty$ is ensured. Moreover, one remembers the $\varphi$-independence of the potential, due to the spherical symmetry, which allows the introduction of the product \textit{ansatz} $\phi_{m}\left(\vec{r}\,\right)=\phi_{m}\left(r,\,\theta\right)=f\left(r\right)q\left( \theta\right)$. In addition, one knows the solution of the magnetic potential for the magnetized volume from (\ref{phimSolutionHollowSphere}), which has the general structure $\phi_{M}\left(r,\,\theta\right)=f\left(r\right)\cos{\left(\theta\right)}$ and is the special case of radially dependent magnetization density with $M(r)=M=\textup{const.}$. This justifies that the $\theta$-dependence of the solution is the same as in (\ref{phimSolutionHollowSphere}), meaning that only terms with $l=1$ need to be kept. This can also be found from the complete \textit{separation ansatz} $\phi_{M}\left(r,\,\varphi,\,\theta\right)=f(r)p(\varphi)q(\theta)$ of the \textit{Laplace} operator in spherical coodinates, which can be reviewed in \cite{Arens}, whereat the $\theta$-dependence originates from the special \textit{Legendre} differential equation for the special case of a $\varphi$-independency ($p(\varphi)=1$) and the still valid boundary conditions from (\ref{LaplacBoundary}), which eventually lead to $l=1$ and $q(\theta)=\cos{(\theta)}$, due to the fact that the \textit{Legendre} polynomials $\textup{P}_{l}(\cos{(\theta)})$ are the solutions of the special \textit{Legendre} differential equation \cite{Arens}. Ultimately, it follows that the \textit{ansatz} for the magnetic potential with $\textup{P}_{1}\left(\cos{\left(\theta\right)}\right)=\cos{\left(\theta\right)}$, $A:=A_{1}$ and $D:=D_{1}$ is given as follows:
\begin{equation}\label{MAnsatz1}
\phi_{m}\left(\vec{r}\,\right)= 
\begin{cases}
\phi_{i}\left(r,\,\theta\right)\,\underset{(\ref{phimAnsatz2})}{\overset{l=1}{=}}\,A\,r  \cos{\left(\theta\right)} &;\;\;\; \forall \; \vec{r}\,\in\,\left\lbrace\vec{r}\,\notin\,\Omega\,\vert\,0\leq r\leq R_{1}\right\rbrace
\\
\phi_{M}\left(r,\,\theta\right)=f\left(r\right)\cos{\left(\theta\right)}&;\;\;\; \forall \; \vec{r}\,\in\,\Omega
\\
\phi_{e}\left(r,\,\theta\right)\,\underset{(\ref{phimAnsatz2})}{\overset{l=1}{=}}\,D\,r^{-2} \cos{\left(\theta\right)} &;\;\;\; \forall \;\vec{r}\,\in\,\left\lbrace\vec{r}\,\notin\,\Omega\,\vert\,R_{2}\leq r<\infty \right\rbrace
\end{cases}
\end{equation}
Third, one has to solve the \textit{Poisson} equation for $\phi_{M}\left(r,\,\theta\right)$ in spherical coordinates. Because of the $\varphi$-independence any derivative with respect to $\varphi$ drops. By inserting (\ref{MAnsatz1}) one can derive the determining differential equation for $f(r)$, which is presented in detail in (\ref{AppPoissonGL1}):
\begin{equation}\label{PoissonGL1}
\begin{split}
\begin{gathered}
\Delta\,\phi_{M}\left(r,\,\theta\right)= \vec{\nabla}\cdot \vec{M} \overset{(\ref{rhom(r)})}{=}\dfrac{d\,M(r)}{dr}\cos{\left(\theta\right)}
\\\\
\longleftrightarrow\;\;\dfrac{1}{r^{2}}\dfrac{\partial}{\partial r}\left(r^{2}\dfrac{\partial\phi_{M}\left(r,\,\theta\right)}{\partial r}\right)+ \dfrac{1}{r^{2}\sin{\left(\theta\right)}}\dfrac{\partial}{\partial \theta}\left(\sin{\left(\theta\right)}\dfrac{\partial\phi_{M}\left(r,\,\theta\right)}{\partial \theta}\right) =\dfrac{d M(r)}{dr}\cos{\left(\theta\right)}
\end{gathered}
\end{split}
\end{equation}
\begin{equation}\label{EULERDGL}
\begin{split}
\begin{gathered}
\xlongleftrightarrow[(\ref{AppPoissonGL1})]{(\ref{MAnsatz1})}\;\; r^{2}\,  \dfrac{d^{\,2} f(r)}{d r^{2}}+2r \,\dfrac{d f(r)}{d r}   -2 f(r) =r^{2} \,\dfrac{d M(r)}{dr} 
\end{gathered}
\end{split}
\end{equation}
The result (\ref{EULERDGL}) is an inhomogeneous \textit{Euler} differential equation for the function $f(r)$, which is explicitly solved in (\ref{AppSolutionPoisson}), whereby the homogeneous equation is converted into an ordinary linear differential equation of the second order with constant coefficients, with the aid of the substitution shown in (\ref{AppSubstitutionEuler}). The homogeneous solution to this differential equation can then be generated from the characteristic polynomial and is used, by means of backsubstitution and with the aid of the variation of constants, to obtain a particular solution for the inhomogeneity. Hence, the solution $f(r)$ of the inhomogeneous \textit{Euler} differential equation is assembled from the homogeneous solution $f_{hom}(r)$ and the particular solution $f_{part}(r)$ and reads according to (\ref{Appf(r)solution}): 
\begin{equation}\label{f(r)solution}
f(r)= \tilde{C}_{1}r+\dfrac{\tilde{C}_{2}}{r^{2}}-\dfrac{rM(R_{1})}{3} +\dfrac{R_{1}^{3}M(R_{1})}{3r^{2}}+\dfrac{1}{r^{2}}\int\limits^{r}_{R_{1}} \tilde{r}^{\,2}M(\tilde{r})\,d\tilde{r}
\end{equation}
Consequently, one can calculate the constants $\tilde{C}_{1}$ and $\tilde{C}_{2}$ (see also (\ref{DConstantC1tilde}) and (\ref{AConstantC2tilde})), with the help of the boundary conditons from (\ref{LaplacBoundary}), which also apply to a radially magnetized, hollow sphere. Thus, the complete solution for $f(r)$ from (\ref{AppUPDATEf(r)solution}) is:
\begin{equation}\label{Completef(r)solution}
\underline{f(r)=\dfrac{1}{r^{2}}\int\limits^{r}_{R_{1}} \tilde{r}^{\,2}M(\tilde{r})\,d\tilde{r}}
\end{equation}
The verification of (\ref{Completef(r)solution}) is proven, by means of the inhomogeneous \textit{Euler} differential equation (\ref{EULERDGL}) and can be found in (\ref{AppCheckf(r)}).
Subsequently, the magnetic potential for the magnetized volume is assembled in (\ref{AppSolutionPoisson}), by means of the \textit{ansatz} from (\ref{MAnsatz1}) and the coefficients $A$ and $D$ from (\ref{AConstant}) and (\ref{DConstant}) and reads according to (\ref{AppFullPotentialHollowM(r)}):
\begin{equation}\label{PotentialHollowM(r)}
\underline{\underline{\phi_{m}\left(\vec{r}\,\right)= 
\begin{cases}
\phi_{i}\left(r,\,\theta\right)\;\;\underset{(\ref{AConstant})}{\overset{(\ref{MAnsatz1})}{=}}\;\;0\,\textup{A} &;\;\;\; \forall \; \vec{r}\,\in\,\left\lbrace\vec{r}\,\notin\,\Omega\,\vert\,0\leq r\leq R_{1}\right\rbrace
\\
\phi_{M}\left(r,\,\theta\right)\;\underset{}{\overset{(\ref{AppfPoissonPhiMsolution})}{=}}\;\dfrac{\cos{\left(\theta\right)}}{r^{2}}\displaystyle{\int\limits^{r}_{R_{1}}} \tilde{r}^{\,2}M(\tilde{r})\,d\tilde{r}&;\;\;\; \forall \; \vec{r}\,\in\,\Omega
\\
\phi_{e}\left(r,\,\theta\right)\;\;\underset{(\ref{DConstant})}{\overset{(\ref{MAnsatz1})}{=}}\;\;\dfrac{\cos{\left(\theta\right)}}{r^{2}}\displaystyle{\int\limits^{\textup{R}_{2}}_{R_{1}}} \tilde{r}^{\,2}M(\tilde{r})\,d\tilde{r}&;\;\;\; \forall \;\vec{r}\,\in\,\left\lbrace\vec{r}\,\notin\,\Omega\,\vert\,R_{2}\leq r<\infty \right\rbrace
\end{cases}}}
\end{equation}
Similarly to the case of a constant magnetization density (\ref{phimSolutionHollowSphere}), one notices a field-free internal space, because the potential $\phi_{i}$ is zero. Moreover, the potential $\phi_{e}$ for the exterior space can still be interpreted as the potential of a magnetic point dipole $\phi_{e}=\frac{\vec{m}\cdot\vec{r}}{4\pi r^{\,3}}$ \cite{FranzSchwabl2019}, whereat the magnetic moment originates from the difference of the electron numbers with spin-up and spin-down, concerning the volume $\Omega$. This can be derived from (\ref{defM(r)}), (\ref{spindensity}) and (\ref{ConditionElecDens}), by showing that the integral for the magnetic moment is the reduced volume integral over the spin density:
\begin{equation}\label{magneticmomentSpinupdown}
\begin{split}
\begin{gathered}
\vec{m}=4\pi\int\limits^{\textup{R}_{2}}_{R_{1}} \tilde{r}^{\,2}M(\tilde{r})\,d\tilde{r}\,\vec{e}_{z} \overset{(\ref{defM(r)})}{=}-\mu_{B}\cdot4\pi\int\limits^{\textup{R}_{2}}_{R_{1}} \tilde{r}^{\,2}s(\tilde{r})\,d\tilde{r}\,\vec{e}_{z}=- \mu_{B} \int\limits_{\Omega} s(\tilde{r})\,d\Omega\,\vec{e}_{z} 
\\[5pt]
\overset{(\ref{spindensity})}{=}-\mu_{B}\int\limits_{\Omega} n_{\uparrow}(r)-n_{\downarrow}(r)\,d\Omega\,\vec{e}_{z}\overset{(\ref{ConditionElecDens})}{=}-\mu_{B}\left( N_{el,\uparrow} -N_{el,\downarrow}\right)\vec{e}_{z}
\end{gathered}
\end{split}
\end{equation}
In addition, one finds the potential for a full sphere with radius $R$, by setting $R_{1}=0\,\textup{m}$ in (\ref{PotentialHollowM(r)}).
Hence, one realizes, that the solution for the hollow sphere is given, analogously to the case with constant magnetization density from (\ref{phimSolutionInterpet}), by the subtraction of the potentials $\phi_{m,R_{2}}$ and $\phi_{m,R_{1}}$ of two full spheres with radii $R_{1}$ and $R_{2}$, which is explicitly done in (\ref{AppphimHollowM(r)Interpet}). As a result, it is a fact, that the point dipole representation, for the potential $\phi_{M}$ inside $\Omega$, is not permissible, because the functional representation of the magnetization density is unknown and the upper integration boundary is not constant. 
Furthermore, one can show that the special case of constant magnetization density (\ref{phimSolutionHollowSphere}) is contained in the solution for a radially dependent magnetization density (\ref{PotentialHollowM(r)}). This was suggested in the preceding text and can be appreciated in (\ref{AppM(r)containsM}).
\\
With the solution for the magnetic potential, one is able to evaluate the \textit{Breit-Hartree} contribution to the total energy of a hollow sphere $\Omega$ with a radially dependent magnetization density. Thereto, the same procedure from section \ref{MagnetostatikSolution} is applied, utilizing the material equation $\vec{B}=\mu_{0}\left(\vec{H}+\vec{M}\right)$, (\ref{magneticfield}) and the magnetic potential from (\ref{PotentialHollowM(r)}), for the purpose of the calculation of the magnetic flux density $\vec{B}$. This is done in (\ref{BfieldhollowSphereM(r)1App}), where only the magnetized volume is regarded, because the magnitude $M(r)$ of the magnetization density is zero (see also (\ref{defM(r)})) for the internal and exterior space, generating no contribution to the \textit{Breit-Hartree} correction:
\begin{equation}\label{BfieldhollowSphereM(r)1}
\vec{B}_{M}\left(r,\,\theta\right)=\mu_{0}\left( M(r)\left(\vec{e}_{z}-\vec{e}_{r}\cos{\left(\theta\right)}\right)+\dfrac{1}{r^{3}} \int\limits^{r}_{R_{1}} \tilde{r}^{\,2}M(\tilde{r})\,d\tilde{r}\left(2\cos{\left(\theta\right)}\vec{e}_{r}+ \sin{\left(\theta\right)}\vec{e}_{\theta}\right)\right) 
\end{equation}
In (\ref{BfieldhollowSphereM(r)1}) one again notices that the $\vec{B}$-field consists of multiple parts, which are assumed to be connected with the magnetic dipole-dipole interaction and the contact interaction. Hence, the general structure of the \textit{Breit-Hartree} interaction in its integral expression (\ref{EnergieBHSI}) follows intrinsically from the magnetostatic potential interaction energy as it was generally shown in \cite{Cohen-Tannoudji}.
Subsequently, one again calculates the potential dipole-dipole interaction energy from (\ref{E_WW_DIP}) from the magnetic flux density (\ref{BfieldhollowSphereM(r)1}) and the magnetization density (\ref{defM(r)}). This leads to the \textit{Breit-Hartree} correction to the total energy of the ground state of the hollow sphere, by following the evaluation in (\ref{EHBHollowSphereM(r)App2}):
\begin{equation}\label{EHBHollowSphereM(r)}
\underline{\underline{\Delta E_{BH}= -\dfrac{8\pi}{3} \mu_{0}\int\limits_{r=R_{1}}^{\textup{R}_{2}} r^{2}\left[M(r)\right]^{2}dr }}
\end{equation}
From (\ref{EHBHollowSphereM(r)}) one can again show in (\ref{SpecialcaseEHBM(r)toMConst}), that the result for the \textit{Breit-Hartree} contribution for the hollow sphere with a radially dependent magnetization density represents the general form of the result from (\ref{EHBHollowSphereMConst}) for a constant magnetization density and therefore includes all previously obtained expressions for the \textit{Breit-Hartree} correction and is completely verified.
\\
The expression in (\ref{EHBHollowSphereM(r)}) depends on the square of the magnetization or spin density, which can be seen by rewriting $M(r)$ with the help of (\ref{defM(r)}): 
\begin{equation}\label{EHBHollowSpheres(r)}
\underline{\underline{\Delta E_{BH}= -\dfrac{8\pi}{3} \mu_{0} \mu_{B}^{2}\int\limits_{r=R_{1}}^{\textup{R}_{2}} r^{2}\left[s(r)\right]^{2}dr = -\dfrac{2}{3} \mu_{0} \mu_{B}^{2}\int\limits_{\Omega} \left[s(r)\right]^{2}d\Omega }}
\end{equation}
In (\ref{EHBHollowSpheres(r)}) one has written the result as a integral over the volume $\Omega$, which leads to a the same demagnetisation factor for a spherical volume $\zeta=\frac{1}{3}$ \cite{Bartelmann_2015}, presumed one follows the procedure from (\ref{Demagnetfactor}).
 
In conclusion, the expression (\ref{EHBHollowSphereM(r)}) is the general calculation rule of the \textit{Breit-Hartree} contribution for atomic and ionic states in zeroth order approximation, meaning the treatment of these states as stationary systems with a finite spherical volumes and radially dependent magnetization densities. On that account, the radially dependent values of the full magnetization density, including for instance spin-orbit contribution, has to be known, which goes beyond the spin density part of the magnetization density as defined in (\ref{magnetdensity}), which are the focus of this work. Nevertheless, the generality of (\ref{EHBHollowSphereM(r)}) would allow future calculations, including differently calculated magnetization densities with isotroic, radially dependent magnitude, which might be adjusted to the treated atom or ion. However, within this thesis one analyses the \textit{Breit-Hartree} contribution to the total ground state energy of open atomic or ionic shells like half-filled valence shells, which can be treated validly with the magnetization density from (\ref{magnetdensity}) and the equation (\ref{EHBHollowSpheres(r)}) for the \textit{Breit-Hartree} correction. In addition, one remarks  that the solution is valid for a mircoscopic system like an ion as well as for a macroscopic system like a magnetized full or hollow sphere.

\clearpage

\section{Numerical calculation of the \textit{Breit-Hartree} contribution for a $\textup{Gd}^{3+}$ ion}\label{Gd3+E_HB}

The result (\ref{EHBHollowSpheres(r)}) allows the calculation of the \textit{Breit-Hartree} contribution from a radially dependent spin density $s(r)$ and represents a potential energy contribution, which originates from the magnetic dipole-dipole interaction of this spin density with itself \cite{Pellegrini_2020}. In order to investigate the magnitude of this correction to the total energy of atomic and ionic ground states one starts with the $3+$ ionic state of Godolinium. This ion from the series of the lanthanides has the electron configuration $_{64}\textup{Gd}^{3+}:\left[\textup{Xe}\right]4f^{\,7}5d^{\,0}6s^{\,0}$ \cite{Tafelwerk}, thus a highly spin polarized, half-filled $4f$-subshell, which leads to the expectation of a significant contribution of the \textit{Breit-Hartree} correction. Moreover, mainly the spin polarization contributes to the magnetization density of the $\textup{Gd}^{3+}$ ion, justifying the investigation within the framework of this thesis, in conformity with the definition of the magnetization density from (\ref{magnetdensity}). Therefore, the use of the spin denity representation (\ref{EHBHollowSpheres(r)}) is verified, instead of the expression (\ref{EHBHollowSphereM(r)}) containing the magnetization density. Subsequently, this makes the solution (\ref{EHBHollowSpheres(r)}) most applicable for atomic and ionic states with an orbital angular momentum of $L=0\,\textup{kg}\textup{m}^{2}\textup{s}^{-1}$, which is the case for a half-filled valence shell as it can be found in the $\textup{Gd}^{3+}$ ion. 
 
The spin density of the $\textup{Gd}^{3+}$ ion is dominated by the $4f$-valence shell and was calculated, using the \textit{Dirac} module of the Full-Potential Local-Orbital code (FPLO)\footnote{\label{note0}https://www.fplo.de,
version dirac21.00-61} in its full relativistic mode \cite{manuelconversation}. The code
solves the \textit{Kohn-Sham-Dirac} equations for a free ion with fixed occupation numbers and the exchange-correlation potential was approximated with the \textit{Perdew-Burke-Ernzerhof} implementation of the generalized gradient approximation \cite{manuelconversation}, \cite{PhysRevLett.77.3865}. Furthermore, a finite nucleus with constant charge density was assumed to avoid density singularities at the origin $r \rightarrow 0$ \cite{manuelconversation}. The produced \small{\verb|.txt|}-file\footnote{\label{Gd3+data}Gd3+ion_spin_density.txt} contains a mesh of $2000$ values for the radius $r_{data}\in[0,\,30\,r_{B}]$ in \textit{Bohr} radii $r_{B}$ and the assigned numerical values for the spin density $s_{data}=r^{2}s(r)$ in units of $r_{B}^{2}r_{B}^{-3}=r_{B}^{-1}$. 
\\
The calculation of the \textit{Breit-Hartree} contribution is then carried out, by means of a \textit{Gauss-Legendre} integration within a \Cpp implementation\footnote{\label{note1}EHB_gauss_legendre_integration[Gd3+_ion].cpp} shown in the digital appendix. For this, the numerical data\footref{Gd3+data} from the FPLO code is read to a memory array and converted into SI-units:
\begin{equation}\label{Gd3+Ion_E_HBSIUNITS}
r_{SI}=r_{data}\cdot r_{B}\;\;\;;\;\;\;s_{SI}=\dfrac{s_{data}}{r^{2}_{data}\cdot r_{B}^{3}}
\end{equation}
Hence, one can derive a general, numerically calculable relation, by starting from (\ref{EHBHollowSpheres(r)}) and by means of the \textit{Gauss-Legendre} quadrature as stated in section \ref{Gauss} similar to (\ref{IdGauss1}). With the SI conforming data values of the spin density of the analyzed ionic or atomic state, one finds with the aid of \cite{schwarz2009numerische}:
\begin{equation}\label{Gd3+Ion_E_HBGAUSS}
\underline{\underline{\Delta E_{BH} }}\overset{(\ref{EHBHollowSpheres(r)})}{=} -\dfrac{8\pi}{3} \mu_{0} \mu_{B}^{2}\int\limits_{r=R_{1}}^{\textup{R}_{2}} r^{2}\left[s(r)\right]^{2}dr \overset{(\ref{IdGauss1})}{\approx}\underline{\underline{ -\mu_{0} \mu_{B}^{2}\dfrac{8\pi}{3} \cdot\dfrac{R_{2}-R_{1}}{2}\sum\limits_{i=1}^{N_{tot.}} w_{i}\cdot r_{i}^{2}\left[s_{i}(r_{i})\right]^{2} }}
\end{equation}
In (\ref{Gd3+Ion_E_HBGAUSS}) the weights $w_{i}$ are the standard \textit{Gauss–Legendre}-weights, while the sampling points $r_{i}$ are transformed \textit{Gauss–Legendre}-points $x_{i}$, whereat the $x_{i}$ are equivalent to the roots of the \textit{Legendre}-polynomials \cite{schwarz2009numerische}:
\begin{equation}\label{GD3+Legendre} 
r_{i}=\frac{R_{1}+R_{2}}{2} + \frac{R_{2}-R_{1}}{2} \cdot x_{i}
\end{equation}
Where $R_{1}$ and $R_{2}$ are chosen as the minimum and maximum value of the mesh for the radial data $r_{data}$, in order to include all contributions of the spin density distribution, which appears quadratic and therefore positive in the integrand in (\ref{Gd3+Ion_E_HBGAUSS}). 
The necessary sets of standard sampling points $x_{i}$ and weights $w_{i}$ on the interval $\left[-1,1\right]$ for the \textit{Gauss}-\textit{Legendre} quadrature are calculated separately with two modified \textsc{Mathematica} scripts from \cite{Legendre}, which can be found in the digital appendix\footnote{\label{note2}mathematica_gauss_legendre_weights(n_max=200_precision_44_digit).nb}$^{;}$\footnote{\label{note3}mathematica_gauss_legendre_roots(n_max=200_precision_44_digit).nb}. This allows to generate two \small{\verb|.txt|} files\footnote{\label{note4}gauss_legendre_weights(n_max=200_precision_44_digit).txt}$^{;}$\footnote{\label{note5}gauss_legendre_roots(n_max=200_precision_44_digit).txt} with $N_{max.}=200$ sampling points and weights with an internal precision of $44$ digits, which are used to evaluate the numerical equation (\ref{Gd3+Ion_E_HBGAUSS}). Eventually, the values $s_{i}(r_{i})$ have to be linearly interpolated from the given data values. Hence the values $r_{1,i}<r_{i}$ and $r_{2,i}>r_{i}$ are searched in the mesh for the radius for each iteration $i\in[1,N_{tot.}]$, so that the corresponding data values $s_{data}$ are $s_{1,i}:=s_{data}(r_{1,i})$ and $s_{2,i}:=s_{data}(r_{2,i})$, leading to the interpolation relation:
\begin{equation}\label{Gd3+Ion_E_HBs(r)Interpolation}
s_{i}(r_{i})=s_{1,i}+\dfrac{s_{2,i}-s_{1,i}}{r_{2,i}-r_{1,i}}\cdot (r_{i}-r_{1,i})
\end{equation}
In addition, one examines the specific spin density data of the ion with the help of (\ref{magneticmomentSpinupdown}) or (\ref{spindensity}) and (\ref{ConditionElecDens}), because the difference of the number of electrons with spin-up and spin-down for the $\textup{Gd}^{3+}$ ion are exactly the seven $4f$-electrons of the half-filled valence shell, which leads to the following conditon by use of the \textit{Gauss-Legendre} quadrature:
\begin{equation}\label{Gd3+Ion_E_HBCONDITION4f}
\begin{split}
\begin{gathered}
N_{el}^{\,4f}:=N^{\,Gd3+}_{el,\uparrow} -N^{\,Gd3+}_{el,\downarrow}\overset{(\ref{ConditionElecDens})}{=}\int\limits_{\Omega} n_{\uparrow}(r)-n_{\downarrow}(r)\,d\Omega
\overset{(\ref{spindensity})}{=}4\pi\int\limits_{r=R_{1}}^{\textup{R}_{2}} r^{2}s(r)\,dr\overset{!}{=} 7
\\
\scalebox{2}{\rotatebox[origin=c]{180}{$\Lsh$}}
\;\;
N_{el}^{\,4f}\approx 4\pi\cdot\dfrac{R_{2}-R_{1}}{2}\sum\limits_{i=1}^{N_{tot.}} w_{i}\cdot r_{i}^{2}s_{i}(r_{i})\overset{!}{\approx} 7
\end{gathered}
\end{split}
\end{equation}
The condition (\ref{Gd3+Ion_E_HBCONDITION4f}) is checked in parallel to the evaluation of the \textit{Breit-Hartree} contribution and additionally a second implementation\footnote{\label{note6}Breit_Hartree_contribution_Gd_ion.m} in \textsc{Matlab} is used for the purpose of verification. 
\\
The output file\footnote{output_gauss_legendre_integration[Gd3+_ion].txt} of the \Cpp implementation\footref{note1} and the output\footnote{output_MATLAB_integration[Gd3+_ion].txt} of the \textsc{Matlab} implementation\footref{note2} generate equal results and exhibit that the given data for the spin density fulfills the condition (\ref{Gd3+Ion_E_HBCONDITION4f}), which can be seen in (\ref{CppGD3+Appendix}) and (\ref{MATLABEHBAppendix}). Whereby the difference to the integer number of seven $4f$-electrons likely originates from small, positron-like components of the \textit{Kohn-Sham} wave function, due to the high nuclear charge \cite{manuelconversation}. Therefore, the final result for the \textit{Breit-Hartree} contribution to the total energy of the ground state of a $\textup{Gd}^{3+}$ ion has the following magnitude, whereat the conversion factor from joules to electron volts is the inverse of the elementary charge \cite{NISTe}:
\begin{equation}\label{Gd3+Ion_E_HB}
\underline{\underline{\Delta E_{BH}^{\textup{\,Gd}^{3+}}\approx -4.647\cdot 10^{-21} \textup{J} = -0.029 \,\textup{eV}}}
\end{equation}
On that account, one remarks that the result only slightly decreases in the fourth decimal place\footnote{output_gauss_legendre_integration[Gd3+_ion_pointlike_nucleus].txt}$^{;}$\footnote{output_MATLAB_integration[Gd3+_ion_pointlike_nucleus].txt}, if the spin density is calculated with a pointlike nucleus\footnote{Gd3+_spin_density_pointlike_nucleus.txt} with constant charge density and the same configurations for the FPLO code. Therefore the given result holds identically for a finite and a pointlike nucleus approximation. However, one should always treat the nucleus as finite and not pointlike object, due to the fact that this is the more realistic treatment.
\\  
The ground state total energy of the $3+$ ionic state of gadolinium for a finite nucleus is $E_{0}^{\,\textup{\,Gd}^{3+}}\approx -306894.91651 \,\textup{eV}$ according to the FPLO calculation \cite{manuelconversation}, which classifies the relative order of the \textit{Breit-Hartree} correction as very small with: 
\begin{equation}\label{Gd3+Ion_E_HB_relativeorder}
\dfrac{\Delta E_{BH}^{\textup{\,Gd}^{3+}}}{E_{0}^{\,\textup{\,Gd}^{3+}}}\cdot 100\%\approx 9.45\cdot10^{-6}\,\%
\end{equation} 
The corrected total ground state energy of the $\textup{Gd}^{3+}$ ion, therefore changes in the second decimal place:
\begin{equation}\label{Gd3+Ion_E_0_corrected}
\underline{\underline{\tilde{E}_{0}^{\,\textup{\,Gd}^{3+}}\approx E_{0}^{\,\textup{\,Gd}^{3+}}+\Delta E_{BH}^{\textup{\,Gd}^{3+}}\approx -306894.94551 \,\textup{eV}}}
\end{equation}
In addition, it should be remarked, that the total energy of the groundstate is given as the deviation from the high relativistic rest energy of the whole system, which leads to its negative value.
\clearpage

\section{Numerical calculation of the \textit{Breit-Hartree} contribution for an $\textup{Mn}^{2+}$ ion}\label{Mn2+E_HB}

In order to further analyse the \textit{Breit-Hartree} contribution, one applies the procedure presented in section \ref{Gd3+E_HB} to the $2+$ ionic state of manganese. This has mainly two reasons, whereat the first reason results from the electron configuration $_{25}\textup{Mn}^{2+}:\left[\textup{Ar}\right]3d^{\,5}4s^{\,0}$ \cite{Tafelwerk} of this light-transition-metal ion. The $3d$-valence shell is half-filled with five electrons, leading to a lower spin density compared to the $\textup{Gd}^{3+}$ ion. Hence, the expected \textit{Breit-Hartree} contribution should be lower. The second reason is the comparibility, because the valence shells of $\textup{Mn}^{2+}$ and $\textup{Gd}^{3+}$ are both half-occupied and have an equal orbital momentum of $L=0\,\textup{kg}\textup{m}^{2}\textup{s}^{-1}$.  
 
One follows the same procedure as presented in section \ref{Gd3+E_HB}, under the use of the 
spin density of the $\textup{Mn}^{2+}$ ion, mainly originating from the $3d$-subshell. Hence, the spin density was computed again with the \textit{Dirac} modul of the Full-Potential Local-Orbital code (FPLO)\footref{note0} in its full relativistic mode \cite{manuelconversation}. The same computation specifications as for the $\textup{Gd}^{3+}$ ion were used, i.a.\hspace{1mm}including the assumption of a finite nucleus with constant charge density \cite{manuelconversation}. The output \small{\verb|.txt|}-file\footnote{\label{note7}Mn2+ion_spin_density.txt} then contains a mesh of $2000$ values for the radius $r_{data}\in[0,\,30\,r_{B}]$ in \textit{Bohr} radii $r_{B}$ and the assigned numerical values for the spin density $s_{data}=r^{2}s(r)$ in units of $r_{B}^{2}r_{B}^{-3}=r_{B}^{-1}$. Thus, one converts the given data into SI-units with the help of (\ref{Gd3+Ion_E_HBSIUNITS}).
\\
The calculation of the \textit{Breit-Hartree} contribution is then carried out, by means of a \textit{Gauss-Legendre} integration within a similar \Cpp implementation\footnote{\label{note8}EHB_gauss_legendre_integration[Mn2+_ion].cpp} shown in the digital appendix, which basically reads the numerical data for the $2+$ ionic state of manganese from the output file\footref{note7} of the FPLO code. The numerical calculation
is then based on the general relation (\ref{Gd3+Ion_E_HBGAUSS}), whereat the boundaries of the integration interval $R_{1}$ and $R_{2}$ are chosen as the minimum and maximum value of the mesh for the radial data $r_{data}$, in order to  include all contributions of the spin density distribution. 
In addition, the \Cpp implementation\footref{note8} makes use of the same sets of standard sampling points $x_{i}$ and weights $w_{i}$ on the interval $\left[-1,1\right]$ for the \textit{Gauss}-\textit{Legendre} quadrature, stored in the two \small{\verb|.txt|} files\footref{note4}$^{;}$\footref{note5}. Eventually, the same linear interpolation from (\ref{Gd3+Ion_E_HBs(r)Interpolation}) is applied for the given data values of the spin density. 
Additionally, one evaluates the adapted condition (\ref{Gd3+Ion_E_HBCONDITION4f}) for the five $3d$ electrons of the $\textup{Mn}^{2+}$ ion, on account of ensuring that the data for the spin density is physically valid. Similar to (\ref{Gd3+Ion_E_HBCONDITION4f}) and by means of the \textit{Gauss-Legendre} quadrature the condition reads:
\begin{equation}\label{Mn2+Ion_E_HBCONDITION4f}
\begin{split}
\begin{gathered}
N_{el}^{\,3d}:=N^{\,\textup{Mn2+}}_{el,\uparrow} -N^{\,\textup{Mn2+}}_{el,\downarrow}\overset{(\ref{ConditionElecDens})}{=}\int\limits_{\Omega} n_{\uparrow}(r)-n_{\downarrow}(r)\,d\Omega
\overset{(\ref{spindensity})}{=}4\pi\int\limits_{r=R_{1}}^{\textup{R}_{2}} r^{2}s(r)\,dr\overset{!}{=} 5
\\
\scalebox{2}{\rotatebox[origin=c]{180}{$\Lsh$}}
\;\;
N_{el}^{\,3d}\approx 4\pi\cdot\dfrac{R_{2}-R_{1}}{2}\sum\limits_{i=1}^{N_{tot.}} w_{i}\cdot r_{i}^{2}s_{i}(r_{i})\overset{!}{\approx} 5
\end{gathered}
\end{split}
\end{equation}
In contrast to section \ref{Mn2+E_HB}, the condition (\ref{Mn2+Ion_E_HBCONDITION4f}) is then checked in parallel to the computation of the \textit{Breit-Hartree} contribution within the implementation\footref{note8}. Moreover, a redundant implementation\footnote{\label{note9}Breit_Hartree_contribution_Mn_ion.m} in \textsc{Matlab} serves the purpose of verification of the results. 
\\
The output file\footnote{output_gauss_legendre_integration[Mn2+_ion].txt} of the \Cpp implementation\footref{note8} and the output\footnote{output_MATLAB_integration[Mn2+_ion]} of the \textsc{Matlab} implementation\footref{note9} generate identical results for the condition and the \textit{Breit-Hartree} correction, proving that the given data for the spin density fulfills the condition (\ref{Mn2+Ion_E_HBCONDITION4f}), which can be seen in (\ref{CppMN2+Appendix}) and (\ref{MATLABEHBAppendix}).  Therefore, the final result for the \textit{Breit-Hartree} contribution to the total energy of the ground state of an $\textup{Mn}^{2+}$ ion is verified and has the following value, whereat the conversion factor from joules to electron volts is the inverse of the elementary charge \cite{NISTe}:
\begin{equation}\label{Mn2+Ion_E_HB}
\underline{\underline{\Delta E_{BH}^{\,\textup{Mn}^{2+}}\approx -9.824\cdot 10^{-22} \textup{J} = -0.006\, \textup{eV}}}
\end{equation}
The ground state total energy of the $2+$ ionic state of manganese for a finite nucleus is $E_{0}^{\,\textup{\,Mn}^{2+}}\approx -31501.26865  \,\textup{eV}$ according to the FPLO calculation \cite{manuelconversation}, which leads to the relative order of the \textit{Breit-Hartree} correction for the $2+$ ionic state of manganese: 
\begin{equation}\label{Mn2+Ion_E_HB_relativeorder}
\dfrac{\Delta E_{BH}^{\textup{\,Mn}^{2+}}}{E_{0}^{\,\textup{\,Mn}^{2+}}}\cdot 100\%\approx 1.90\cdot10^{-5}\,\%
\end{equation} 
The corrected total energy of the ground state of the $\textup{Mn}^{2+}$ ion, therefore changes in the second decimal place:
\begin{equation}\label{Mn2+Ion_E_0_corrected}
\underline{\underline{\tilde{E}_{0}^{\,\textup{\,Mn}^{2+}}\approx E_{0}^{\,\textup{\,Mn}^{2+}}+\Delta E_{BH}^{\textup{\,Mn}^{2+}}\approx -31501.27465 \,\textup{eV}}}
\end{equation}
Again, it should be noted, that the total energy of the ground state is given as the deviation from the high relativistic rest energy of the whole system, which leads to its negative value.

\clearpage

%% file: Comparison_and_discussion_in_the_context_of_experimental_information.tex
\chapter{Comparison and discussion in the context of experimental information}\label{Ende}

In order to discuss the application of the \textit{Breit-Hartree} contribution to the trivalent state of gadolinium and the divalent state of manganese, one first compares the results from sections \ref{Gd3+E_HB} and \ref{Mn2+E_HB}. 

\begin{figure}[h]
\begin{center}
\includegraphics[trim=140 17 140 5,clip,width=\textwidth]{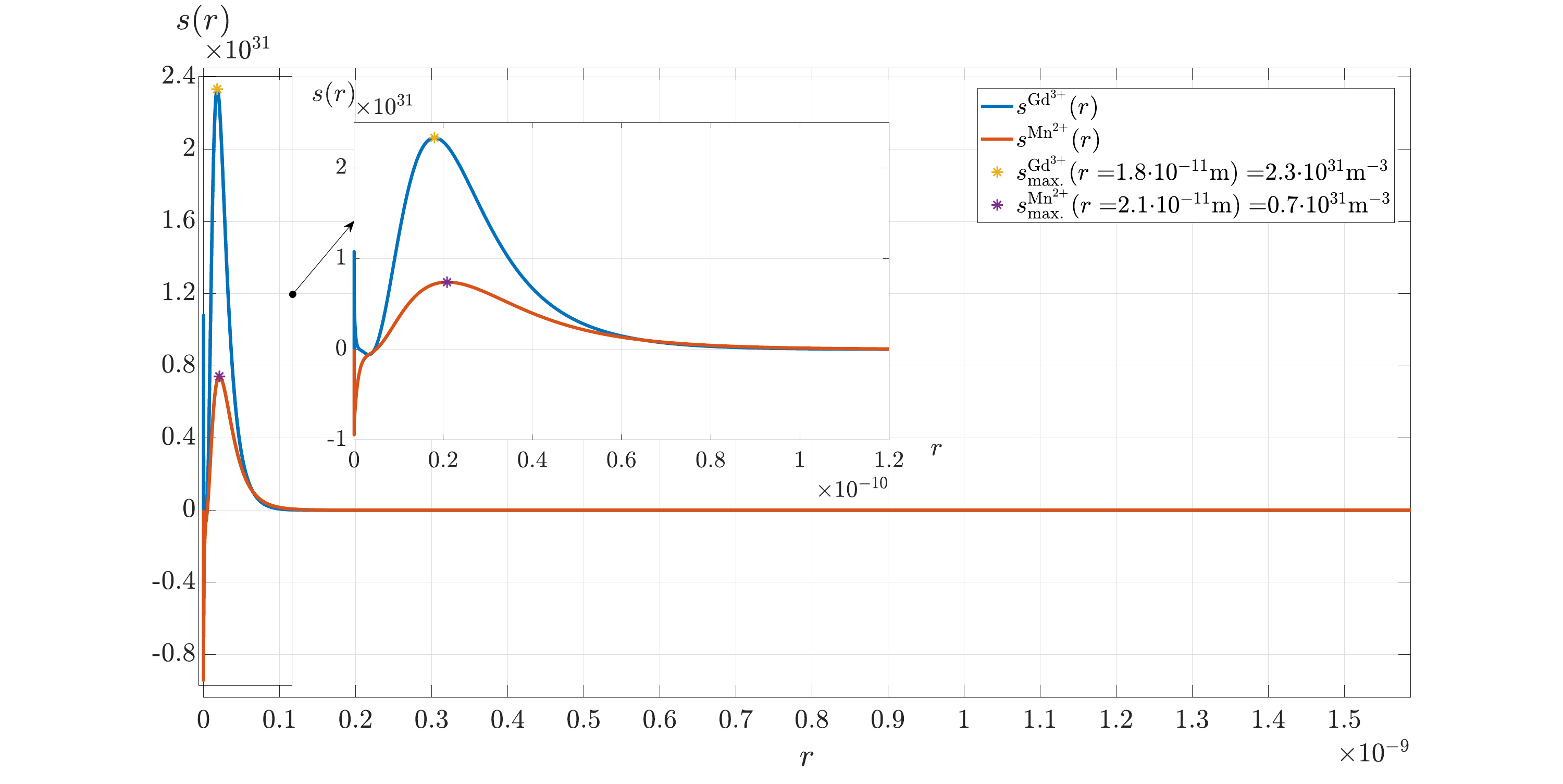}
\end{center}
\caption{Spin densities in $\textup{m}^{-3}$ of the $\textup{Gd}^{3+}$ and the $\textup{Mn}^{2+}$ ion plotted over the radius in $\textup{m}$}
\label{FigSpindensities}
\end{figure}

On that point, one notices from (\ref{Gd3+Ion_E_HB}) and (\ref{Mn2+Ion_E_HB}), that the \textit{Breit-Hartree} correction for the $\textup{Mn}^{2+}$ ion is approximatly $4.8$ times the contribution of the $\textup{Gd}^{3+}$ ion. In addition, one finds through (\ref{Gd3+Ion_E_HB_relativeorder}) and (\ref{Mn2+Ion_E_HB_relativeorder}) that the relative order of the correction is $10^{-5}$ and the relative order for the $\textup{Mn}^{2+}$ ion is two times bigger than for the $\textup{Gd}^{3+}$ ion. Therefore and despite the fact, that both ions have a highly spin polarized valence shell, the \textit{Breit-Hartree} contribution has a small relative order. Based on this, one has to further analyze the significance of this correction to the total energy in section \ref{mollerplesset}. Nevertheless, one recognizes that the minor relative order of the correction justifies its non-self-consistent calculation. Hence, the treatment as a final perturbation correction to the ground state total energy is valid, confirming the assumption from the last paragraph in section \ref{BHINERACTION}. 
\\
Hereinafter, the results of the comparison for the two ions shall be analysed with the help of their spin density and their electron configurations. On order to enable comparison, one displays the spin densities in $1\textup{m}^{-3}$ of both ions over the radius in $1\textup{m}$, using the data\footref{Gd3+data}$^{;}$\footref{note7} from the FPLO code and a \textsc{Matlab} script\footnote{\label{Matlab_s(r)}figure_output_spin_densities_Gd_Mn_ion.m} for the conversion into SI-units and the generation of the figure \ref{FigSpindensities}.
\\
Consequently, the higher spin density of the $\textup{Gd}^{3+}$ ion can be seen, which originates from the seven $4f$-electrons, while the spin densitiy for the manganese ion is lower, because there are only five unpaired electrons in the $3d$ valence shell. This explains the $4.8$ times larger \textit{Breit-Hartree} contribution for the $\textup{Gd}^{3+}$ ion and verifies the prediction that the \textit{Breit-Hartree} contribution is large for high spin densities as they are generated by open shells with many unpaired electrons. However, the difference in the relative order of the contribution is simply caused by the larger, total number of electrons for the gadolinium ion. This can be shown by estimating the relative order from the quotient of the number of unpaired electrons, which represent the origin of the \textit{Breit-Hartree} correction and the total number of electrons. Hence, it follows from (\ref{Gd3+Ion_E_HB_relativeorder}) and (\ref{Mn2+Ion_E_HB_relativeorder}):
\vspace*{-0.5mm}\begin{equation}\label{relOrderHalb}
\underline{\dfrac{N_{el,Mn^{2+}}^{\,3d}}{N_{el,Mn^{2+}}^{\,\textup{tot.}}}\div\dfrac{N_{el,Gd^{3+}}^{\,4f}}{N_{el,Gd^{3+}}^{\,\textup{tot.}}}}=\dfrac{5}{25}\div\dfrac{7}{64}\approx \underline{ 2 } \approx 1.90\cdot10^{-5}\div 9.45\cdot10^{-6}= \underline{\dfrac{\Delta E_{BH}^{\textup{\,Mn}^{2+}}}{E_{0}^{\,\textup{\,Mn}^{2+}}}\div\dfrac{\Delta E_{BH}^{\textup{\,Gd}^{3+}}}{E_{0}^{\,\textup{\,Gd}^{3+}}}}
\end{equation}
However, one is not able to derive general assertions concerning the magnitude of the \textit{Breit-Hartree} contribution for atoms or ions. Indeed it is preferable to conduct further investigations, using the given \Cpp and \textsc{Matlab} program for different spin densities, in order to analyse correlations with the type of the spin polarized subshell, the difference beetween atomic and ionic states and the scaling of the magnitude with the proton number $Z$.
\clearpage

\section{Analysis of the \textit{Breit-Hartree} contribution for the $\textup{Gd}^{3+}$ ion in the context of experimental data}\label{experimentGd3+}

The fifteen elements of the lanthanides are characterized by filling of the $4f$-subshell and have the general electron configuration of $\left[\textup{Xe}\right]4f^{\,x+1}5d^{\,0}6s^{2}$ for their atomic ground state with $x\,\in\,\left\lbrace x\,\in\,\mathbb{Z}\,\vert\,0\leq x\leq 14\right\rbrace$, with the exception of $\textup{La}$ ($x=0$), $\textup{Ce}$ ($x=1$), $\textup{Gd}$ ($x=7$) and $\textup{Lu}$ ($x=14$), which share the configuration $\left[\textup{Xe}\right]4f^{\,x}5d^{\,1}6s^{2}$ \cite{Melsen1994CalculationsOV}, \cite{Tafelwerk}, highlighting that the subshells sorted from the lowest to the highest energy would be $6s$, $4f$ followed by $5d$. 

\begin{figure}[h]
\hspace*{11mm}\includegraphics[trim=300 410 48 53,clip,width=0.7\textwidth,height=14cm]{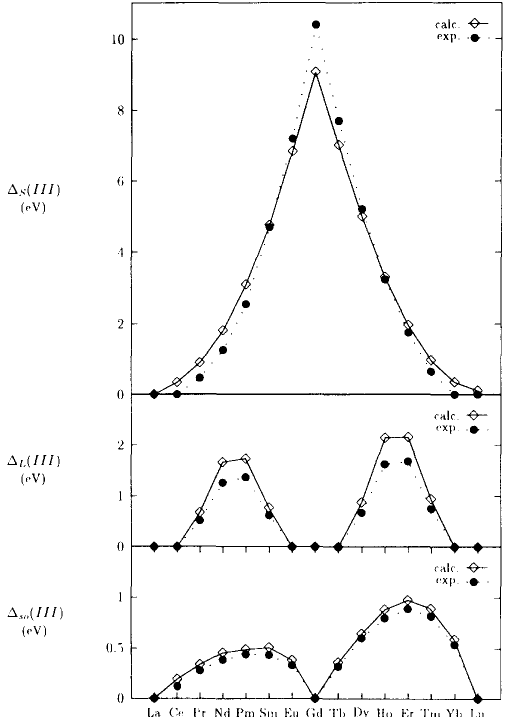}
\caption{Theoretical (diamonds with solid lines) and
experimental values (circles with dotted lines) of
the spin polarization energy $\Delta_{\textup{S}}(III)$,
orbital polarization energy $\Delta_{\textup{L}}(III)$ and spin-orbit coupling energy $\Delta_{\textup{so}}(III)$ contributions of the $3+$ ionic state
of the lanthanides \cite{Melsen1994CalculationsOV} in $\textup{eV}$}
\label{FigMehlsen}
\end{figure}

The lanthanides show several anomalies like 	
the lanthanide contraction as well as their discontinous oxidation states \cite{Melsen1994CalculationsOV}. This means that all lanthanides $\textup{Ln}$ share the trivalent ionic state $\textup{Ln}^{3+}:\,\left[\textup{Xe}\right]4f^{\,x}5d^{\,0}6s^{0}$. However, they are only partly stable in the divalent ionic state $\textup{Ln}^{2+}:\,\left[\textup{Xe}\right]4f^{\,x+1}5d^{\,0}6s^{2}$ for $\textup{Sm}$ ($x=5$), $\textup{Eu}$ ($x=6$), $\textup{Tm}$ ($x=12$), $\textup{Yb}$ ($x=13$) and the $4+$ state $\textup{Ln}^{4+}:\,\left[\textup{Xe}\right]4f^{\,x-1}5d^{\,0}6s^{0}$ for $\textup{Ce}$, $\textup{Pr}$ ($x=2$), $\textup{Nd}$ ($x=3$), $\textup{Tb}$ ($x=8$), $\textup{Dy}$ ($x=9$) \cite{Chempie}. These valence stabilites were i.a.\hspace{1mm}analysed in regard of the di- and trivalent states by \textit{J. Melsen} \cite{Melsen1994CalculationsOV} as well as by \textit{Richter} \cite{Richter2001}, \cite{RichterLecture2}. For the application of the \textit{Breit-Hartree} correction, calculated in section \ref{Gd3+E_HB}, one follows the work of \textit{Melsen} \cite{Melsen1994CalculationsOV}. He calculated the total energy of the $3+$ ionic ground states, including polarization energy contributions of the $\textup{Ln}^{3+}$ ions, by means of fully relativistic SDFT calculations and under the use of the local spin density approximation for the exchange and correlation functional. Furthermore, he spereately calculated the spin polarization energy $\Delta_{\textup{S}}(III)$, the
orbital polarization energy $\Delta_{\textup{L}}(III)$ as well as the spin-orbit coupling energy $\Delta_{\textup{so}}(III)$ contribution to the total energy and compared them to experimental data \cite{Melsen1994CalculationsOV}. Finally, \textit{Melsen} found that the spin-pairing energy contributes the most, due to the different occupation of the $4f$-valence shell and is at its maximum for the $\textup{Gd}^{3+}$ ion with its seven $4f$-electrons \cite{Melsen1994CalculationsOV}. Moreover, he remarked a deviation from the experimental data, which can be seen in figure \ref{FigMehlsen}, as compared to the calulated polarization energy contributions \cite{Melsen1994CalculationsOV}. Note that in figure \ref{FigMehlsen} all energies are displayed unsigned. 
For the $3+$ ionic state of gadolinium one can estimate a deviation of $1.2\,\textup{eV}$ beetween the experimental and the theoretical spin polarization energy, with the help of figure \ref{FigMehlsen}. 
\\
This gap can be partially explained by the \textit{Breit-Hartree} contribution, whose applicability to the $\textup{Gd}^{3+}$ ion was already discussed in section \ref{Gd3+E_HB}, including the calculation of the magnitude $\Delta E_{BH}^{\textup{\,Gd}^{3+}}\approx -0.029 \,\textup{eV}$ of the correction from (\ref{Gd3+Ion_E_HB}). The absolute value of this contribution corrects the spin polarization energy calculated by \textit{Melsen} towards the direction of the experimentally gained value. Subsequently, one can postulate that the \textit{Breit-Hartree} contribution, as a perturbation correction to the total ground state energy, explains $2.42\,\%$ of the error discoverd in \cite{Melsen1994CalculationsOV}. 
\newpage
\vspace*{-4mm}
\clearpage

\section{Applicability of the \textit{Breit-Hartree} contribution}\label{applicability}

\subsection{Comparison of the \textit{Breit-Hartree} contribution with the second order \textit{M\o{}ller-Plesset} correlation energy correction}\label{mollerplesset}

In order to properly discuss the applicability of the \textit{Breit-Hartree} contribution, one compares this contribution with a different energy correction. This comparison subsequently enables the classification of the \textit{Breit-Hartree} correction with regard to its magnitude. 

One orients oneself with the recent research \cite{Pellegrini_2020} from \textit{Pellegrini et al.}, stating a second order correlation energy contribution for a dipolar homogeneous electron gas for the framework of SDFT in combination with the Local Spin Density Approximation (LSDA). This correction arises from the second order \textit{M\o{}ller-Plesset} (MP2) correlation energy per electron \cite{Pellegrini_2020} and depends on the \textit{Fermi} wave vector $k_{F}(n)=(3\pi^{2}n)^\frac{1}{3}$ \cite{ashcroft1976solid}, which is itself a function of the electron density $n$. One focuses on a radially dependent electron denisty $n=n(r)$, simplifying the calculation and the comparison with the results for the \textit{Breit-Hartree} contribution for the $\textup{Gd}^{3+}$ and $\textup{Mn}^{2+}$ ion and formulates the MP2-correlation energy contribution according to \cite{Pellegrini_2020} in GCGS-units:
\vspace*{-1.5mm}\begin{equation}\label{MP2Plesset1}
E_{C,MP2}[n(r)]=-\dfrac{43-46\ln{(2)}}{525}\dfrac{e^{4}\hslash^{2}\left(k_{F}(n(r))\right)^{4}}{2m_{e}^{3}c^{4}\pi^{2}}
\end{equation}    
Thus, one finds the MP2-correlation energy contribution to the total energy of an atom or ion approximated as a hollow sphere $\Omega$, by inserting the definition of the \textit{Fermi} wave vector $k_{F}[n]$ and by integration over the volume $\Omega$ in spherical coordinates (see also (\ref{AppPlesset2})):
\begin{equation}\label{MP2Plesset2}
\Delta E_{C,MP2}=-\dfrac{43-46\ln{(2)}}{525}\dfrac{2\cdot 3^{\frac{4}{3}}\pi^{\frac{1}{3}}e^{4}\hslash^{2}}{m_{e}^{3}c^{4}}\int\limits_{\textup{R}_{1}}^{\textup{R}_{2}}r^{2}\left(n(r)\right)^{\frac{4}{3}}\,dr
\end{equation}
From (\ref{MP2Plesset2}) one can evaluate the MP2-correlation energy contribution for the $\textup{Gd}^{3+}$  and the $\textup{Mn}^{2+}$ ion, using the radially dependent data for the electron density\footref{gdMp2}$^{;}$\footref{mnMp2}, generated by the FPLO-code\footref{note0} in the same configuration as in the sections \ref{Gd3+E_HB} and \ref{Mn2+E_HB}.
\\
Subsequently, the necessary integration and conversion into SI-units is implemented redundantly in \Cpp\footnote{\label{CppMP2ImplementationGd3+}MP2_gauss_legendre_integration[Gd3+_ion].cpp}$^{;}$\footnote{\label{CppMP2ImplementationMn2+}MP2_gauss_legendre_integration[Mn2+_ion].cpp} and \textsc{Matlab}\footnote{\label{MATLABMP2ImplementationGd3+}MP2_correlation_energy_contribution_Gd_ion.m}$^{;}$\footnote{\label{MATLABMP2ImplementationMn2+}MP2_correlation_energy_contribution_Mn_ion.m} and can be understood in the appendix (\ref{AppMP2}). 
The results follow from the output files\footref{R1MP2}$^{;}$\footref{R2MP2}$^{;}$\footref{R3MP2}$^{;}$\footref{R4MP2} and are displayed in the appendix (\ref{CppGD3+AppendixMP2}, \ref{CppMN2+AppendixMP2}, \ref{MATLABMP2AppendixMP2}). Hence, the calculated MP2-correlation energy contribution for the $\textup{Gd}^{3+}$ ion is:
\begin{equation}\label{MP2PlessetGd3+}
\underline{\underline{\Delta E_{C,MP2}^{\textup{Gd}^{3+}}=-5.734\cdot10^{-50}\,\textup{J}=-3.580\cdot10^{-31}\,\textup{eV}}}
\end{equation}
Whereas one finds for the $\textup{Mn}^{2+}$ ion:\vspace*{-2mm}
\begin{equation}\label{MP2PlessetMn2+}
\underline{\underline{\Delta E_{C,MP2}^{\textup{Mn}^{2+}}=-1.156\cdot10^{-50}\,\textup{J}=-7.215\cdot10^{-32}\,\textup{eV}}}
\end{equation}
Consequently, it can be postulated, that the \textit{Breit-Hartree} contribution is approximately $28$ orders of magnitude higher than the second order \textit{M\o{}ller-Plesset} (MP2) correlation energy contribution. Moreover, one notices that the MP2-correction energy for the $\textup{Gd}^{3+}$ ion is five times larger than the contribution to the total energy of a $\textup{Mn}^{2+}$ ion, which is nearly the same factor between those ions as it was discovered at the beginning of this chapter \ref{Ende}. The much bigger order of the the \textit{Breit-Hartree} correction could be explained by the fact that it is a first order correction to the $Hartree$ energy functional, while the MP2 correlation energy corrects the exchange and correlation functional (compare (\ref{groundstateenergy3}) and (\ref{E_Hcorrection})) and is found in second order \textit{M\o{}ller-Plesset} perturbation theory. Due to the fact that the main energy contributions result from the $Hartree$ energy functional, it seems plausible that a correction to the correlation energy is smaller, although one rather assumes that this correction is not applicable to atoms or ions.

In conclusion, one can classify the \textit{Breit-Hartree} contribution as more significant than the second order \textit{M\o{}ller-Plesset} (MP2) correlation energy contribution. Hence, it should always be applied with a higher priority than the MP2 correction, except for the case of a vanishing spin density, where the \textit{Breit-Hartree} contribution would be zero, due to the fact that the atomic shell is empty or fully occupied.
\\
Note, that in the appendix (\ref{ConstAppMP2}) the calculation for a constant electron density, which neglects the radial dependency of the electron density, resulted in a slightly smaller \textit{M\o{}ller-Plesset} correlation energy contribution of the same order as the results in (\ref{MP2PlessetGd3+}) and (\ref{MP2PlessetMn2+}). This emphasizes the importance of the consideration of the radial dependency of the electron density for the MP2 correlation energy contribution, showing that the approximation of atoms and ions as hollow spheres with a \textit{radially-dependent} magnetization density is also crucial for the calculation of the \textit{Breit-Hartree} contribution. Nevertheless, the MP2 correlation energy contribution is negligibly small for the $\textup{Gd}^{3+}$ and the $\textup{Mn}^{2+}$ ion, which raises doubts about its applicability to single ions. The reason for this is that this contribution was stated by \textit{Pellegrini et al.} \cite{Pellegrini_2020} for the dipolar homogeneous electron gas, which is a model for bulk systems of solid states and might fail for ions approximated as a finite volume.

\subsection{Future applications of the \textit{Breit-Hartree} correction}\label{future}

The correction to the total energy of a stationary ground state of an atomic or ionic system, due to the \textit{Breit} interaction in the \textit{Hartree} approximation was generally stated by the integral equation (\ref{EnergieBHSI}) in SI-units. It depends on a general magnetization density $\vec{M}(\vec{r}\,)$, whereat a treatment of the \textit{Breit-Hartree} contribution as a perturbation correction to the total ground state energy is sufficiently valid, referring to the beginning of chapter \ref{Ende}. Consequently, it can be found from a self-consistently computed magnetization density from relativistic SDFT programs. In this context, it is applicable for proton numbers significantly below $Z=137$ as it was mentioned in section \ref{Important} (see also \cite{Bethe}). Therefore, one predicts correct results for the \textit{Breit-Hartree} contribution for nuclei with sufficiently large decay times, mainly excluding large nuclei with $Z>137$, where the strong nuclear potential causes the creation of electron-positron pairs \cite{Jansen1988}. Furthermore, the interpretation as the magnetostatic dipole-dipole interaction as discussed in the sections \ref{BHINERACTION}, \ref{Important} and \ref{MagnetostatikSolution} allows the application of the expression (\ref{EnergieBHSI}) for finite bulk volumes within a stationary state and a given magnetization density. However, the integral representation (\ref{EnergieBHSI}) is not always useful to calculate the \textit{Breit-Hartree} contribution, because the integrand is singular in every spatial point and therefore causes difficulties in the convergence behaviour of the numerical integration techniques (compare section \ref{NumApSphere}). Hence, one summarizes the analytically gained solutions for certain cases and describes possible procedures to receive further solutions for the remaining cases.
 
\begin{figure}[H]
\begin{center}
\includegraphics[trim=5 252 60 190,clip,width=\textwidth]{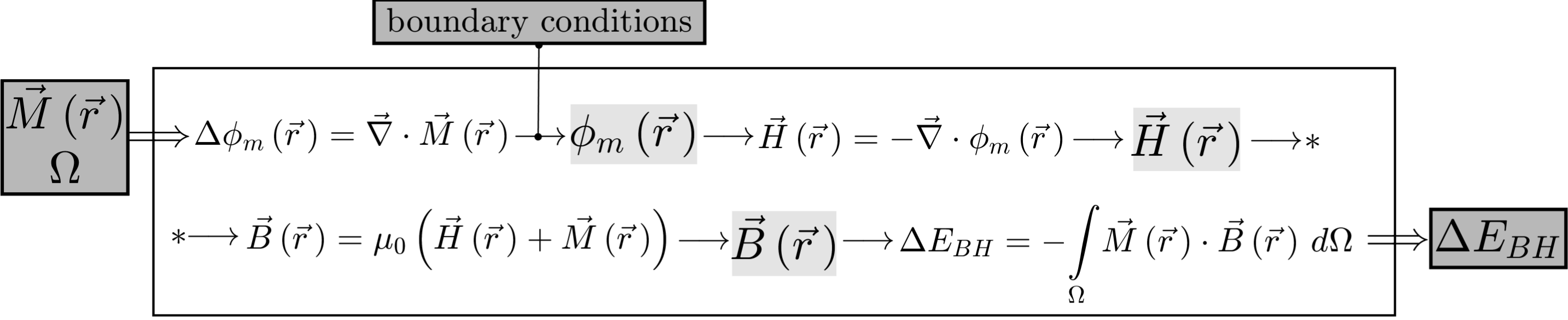}
\end{center}
\caption{Calculation procedure for the \textit{Breit-Hartree} contribution $\Delta E_{BH}$, under the use of the magnetostatic approach}
\label{FigMagnetApproach}
\end{figure}

In this work, a general, analytic expression for the \textit{Breit-Hartree} contribution was found for a radially dependent magnetization density inside a spherical volume $\Omega$ and reads in SI-units according to (\ref{EHBHollowSphereM(r)}):
\begin{equation} 
\underline{\underline{\Delta E_{BH}= -\dfrac{8\pi}{3} \mu_{0}\int\limits_{r=R_{1}}^{\textup{R}_{2}} r^{2}\left[M(r)\right]^{2}dr }}
\end{equation}
It allows the calculation of the \textit{Breit-Hartree} correction for spherical volumes and a given magnetization density $M(r)$, whereby the case of a constant magnetization density is also covered. This means that bulk and atomic systems can be investigated, as long as the magnetizatin density is known. This includes calculations with a differently defined magnetization density, which at best include the total density of momenta within the total electron shell of an atom or ion with exception of the nucleus. 
\\
Moreover, an implementation and application of this solution for single atomic and ionic states was presented in section \ref{Gd3+E_HB}, whereat the magentization density entirely arises from the spin density. Hence, the \textit{Breit-Hartree} contribution for this case scales with the square of the spin density and is therefore significant for open atomic shells. Especially for valence shells with a half- or nearly half-occupation an analysis of the \textit{Breit-Hartree} correction is recommended. In this context, one should examine the magnitude of this correction of the ground state energy with respect to the proton number $Z\,\in\,[1,80]$ as well as with regard to the type of  the partially occupied valence shell ($s$-, $p$-, $d$-, $f$-subshell for $Z\leq80$). In addition, these examinations should be carried out for atoms and ions separately, in order to compare the behaviour of the \textit{Breit-Hartree} contribution for atomic and ionic states. Preferably, one starts with atoms and ions with half-filled valence shells, providing an orbital angular momentum of $L=0\,\textup{kg}\textup{m}^{2}\textup{s}^{-1}$. For such candidates the magnetization density can be calculated from the spin density, using (\ref{spindensity}) and (\ref{magnetdensity}) and the \textit{Breit-Hartree} contribution can be evaluated with the expression from (\ref{EHBHollowSpheres(r)}) in SI-units:
\vspace*{-1mm}\begin{equation}
\underline{\underline{\Delta E_{BH}= -\dfrac{8\pi}{3} \mu_{0} \mu_{B}^{2}\int\limits_{r=R_{1}}^{\textup{R}_{2}} r^{2}\left[s(r)\right]^{2}dr }}
\end{equation}
A list of interesting candidates, as a starting point for possible future calculations, can be found in the digital appendix\footnote{\label{candidates}applications_Breit-Hartee_contribution_atomic_and_ionic_states.txt}.   
\\
The remaining cases for the application of the \textit{Breit-Hartree} contribution deal with a general magnetization density $\vec{M}(\vec{r}\,)$. Consequently, they require the evaluation of the integral from (\ref{EnergieBHSI}), which is not analytically solvable for a sphere according to section \ref{AnApSphere} and is assumed to be exclusively solvable numerically for general volumes. In this context, the \textit{Monte Carlo} integration is the most promising method compared to other numerical integration techniques, which was examined in section \ref{NumApSphere} and additionally verified in the appendix in subsection \ref{CHECKMonteCarlo}. One remarks, that for the numerical integration the angle representation (\ref{EnergieBHangles}) of (\ref{EnergieBHSI}) should be used in combination with a problem specific set of at least $10^{9}$ random sampling points. For non-spherical volumes the use of cartesian coordinates is recommended, while for spherical volumes one conveniently chooses spherical coordinates.  
\\
In addition, the magnetostatic approach is possible, whereat one first solves the \textit{Poisson}-equation from (\ref{LaplacGl1}) numerically for the magnetic potential under consideration of certain boundary conditions. After that one calculates the field of the magnetic flux density similar to (\ref{BfieldhollowSphereConst1App}) or (\ref{BfieldhollowSphereM(r)0App}) and the magnetic potential dipole-dipole interaction energy from (\ref{E_WW_DIP}) by the use of numerical differentiation and integration. The received interaction energy is then equal to the \textit{Breit-Hartree} contribution as explicitly shown in the references \cite{Pellegrini_2020}, \cite{Cohen-Tannoudji}. A numerical implementation would therefore follow the calculation scheme displayed in the \textsc{Matlab} generated figure\footnote{\label{scheme}figure_magnetostatic_approach.fig} \ref{FigMagnetApproach}.
\vspace{2.5mm}
 \vspace*{-1.5mm}
It must be remarked, that the representation of the differential operators in cartesian or spherical coordinates should be choosen with regard to the shape of the volume. Consequently, the previously discussed cases are completely numerically computable, allowing future calculations of the \textit{Breit-Hartree} contribution for arbitrary magnetization densities and volumes.

\clearpage

%% file: Conclusion.tex
\chapter{Conclusion}

The \textit{Breit-Hartree} interaction was introduced as a correction to the \textit{Hartree} energy functional within the general framework of relativistic Spin Density Functional Theory and for the purpose of the calculation of the total energy of the ground state of many-electron systems. On that account, its origin from the zero-frequency limit of the quantum electrodynamical \textit{Breit} operator in the \textit{Hartree} approximation was investigated. Based on that, one expounded the interpretation of the \textit{Breit-Hartree} interaction as the magnetostatic dipole-dipole interaction with a dependency on the magnetization density. 
In addition, the connection between the integral representation (\ref{EnergieBHSI}) of the \textit{Breit-Hartree} interaction, stated by \textit{Jansen}, to the expression (\ref{E_WW_DIP}) for the magnetostatic potential energy \cite{Pellegrini_2020}, \cite{Cohen-Tannoudji}, caused by a magnetization density in a finite volume interacting with its self-induced field of the magnetic flux density, is shown .

First, the integral equation (\ref{EnergieBHSI}) was converted into SI-units and stated in the angle representation (\ref{EnergieBHangles}), whereat one discovered its analytically unsolvability even for the most simple case of a constant magnetization density in a spherical volume. Subsequently, the case of a constant magnetization density was further analyzed numerically with \Cpp implementations of a \textit{Monte Carlo} and a \textit{Gauss-Legendre} integration technique, revealing a singular behaviour of the integrand. Although no numerical solution was generated, the behavior of the integrand and the two integration methods could be assessed. Thus, the superiority of the \textit{Monte Carlo} method was determined and recommendations for future calculations were derived. 
\\
Second, an analytical solution for a test dipole on the $z$-axis of a full sphere with a constant magnetization density was derived, showing proportionalites of the \textit{Breit-Hartree} contribution to the square of the magnetization density and the volume of the sphere.\hspace{1.5mm}This was then used to verify the results for the magnetostatic approach from figure \ref{FigMagnetApproach}. Consequently, one was able to apply this approach for a constant and a radially dependent magnetization density within a hollow sphere, whereat the general solution (\ref{EHBHollowSphereM(r)}) covers both cases and fulfills the main aim of this work. With regard to the \textit{Breit-Hartree} contribution to the total ground state energy of open atomic shells as the main interest of this thesis, the magnetization density was defined (compare (\ref{spindensity}) and (\ref{magnetdensity})) as entirely originating from the spin angular momentum density, which led from (\ref{EHBHollowSphereM(r)}) to the equation (\ref{EHBHollowSpheres(r)}).
\\
Third, the gained calculation rule (\ref{EHBHollowSpheres(r)}) was implemented redundantly in \Cpp and \textsc{Matlab}, in order to evaluate the \textit{Breit-Hartree} correction exemplarily for the $\textup{Mn}^{2+}$ and the $\textup{Gd}^{3+}$ ion. The two programs equally compute a contribution of $-0.029\,\textup{eV}$ for the $\textup{Gd}^{3+}$ ion and of $-0.006\,\textup{eV}$ for the $\textup{Mn}^{2+}$ ion, whereat the results were compared against the background of the respective spin densities. 
\\
Fourth, one has discussed the \textit{Breit-Hartree} contribution in the context of experimental data, whereat it corrects the total ground state energy of the trivalent gadolinium ion or the spin polarization energy as a part of the total energy. In this connection, it was possible to explain $2.42\%$ of the deviation of the DFT-calculated result from the experimental value \cite{Melsen1994CalculationsOV}.
\\
Fifth, the second order \textit{M\o{}ller-Plesset} correlation energy correction was implemented, calculated and compared with the \textit{Breit-Hartree} contribution for the $\textup{Mn}^{2+}$ and the  $\textup{Gd}^{3+}$ ion, for the purpose of analyzing the applicability of the \textit{Breit-Hartree} correction. As a consequence, the significance of the \textit{Breit-Hartree} interaction was shown, because it generates a $28$ orders of magnitude larger contribution than the \textit{M\o{}ller-Plesset} correction. 

Eventually, this allowed to classify the \textit{Breit-Hartree} contribution as an important correction of the ground state total energy of non-radioactive atomic and ionic states with $Z\leq80$, whereat the solution (\ref{EHBHollowSphereM(r)}) can be used for a known magnetization density. Additionally, atoms and ions for future investigations of the behaviour of the magnitude of the \textit{Breit-Hartree} contribution calculated from this solution were stated. Moreover, the previously assumed generality of the equation (\ref{EHBHollowSphereM(r)}) holds and would even enable evaluations of this solution for spherical bulk systems with a known, radially-dependent magnetization density. In addition, one has discussed that future computations of the \textit{Breit-Hartree} contribution for general, inhomogeneous magnetization densities for atoms, ions and bulk system are possible. In this context, the first possibility is the evaluation of the integral representation (\ref{EnergieBHSI}) with the \textit{Monte Carlo} integration, whereas the second is the more recommended, numerical implementation of the magnetostatic approach (figure \ref{FigMagnetApproach}) which is . 
\clearpage

%% file: Appendix.tex
\chapter{Appendix}

\section{Analytical calculations}

\subsection{Evaluation of the integral $I_{c}$ from (\ref{Icv1})}

The three-dimensional integral $I_{c}$ from (\ref{EnergieBHSI}) is evaluated, whereat it represents the contact interaction within the spin density at same spatial points within the homogeneously magnetized sphere \cite{Pellegrini_2020}. Using (\ref{defM}), (\ref{volElem}) and the integration volume $\Omega$, one can calculate:
\begin{equation}\label{Icv1A}
\begin{split}
\begin{gathered}
\underline{\underline{I_{c}}} := -\dfrac{\mu_{0}}{3} \int\limits_{\Omega} \vert\vec{M}\left(\vec{r}\,\right)\vert^{2}  d^{\,3}r 
\underset{(\ref{volElem})}{\overset{(\ref{defM}),(\ref{sphereCoordin})}{=}}-\dfrac{\mu_{0}}{3} \int\limits_{r=R_{1}}^{\textup{R}_{2}}\int\limits_{\theta=0}^{\pi}\int\limits_{\varphi=0}^{2\pi} M^{2} \cdot r^{2}\sin{\left(\theta\right)}\,d\varphi d\theta dr
\\\\
= -\dfrac{\mu_{0}}{3} M^{2}\int\limits_{r=R_{1}}^{\textup{R}_{2}}r^{2}dr\int\limits_{\varphi=0}^{2\pi}d\varphi\int\limits_{\theta=0}^{\pi}  \sin{\left(\theta\right)}d\theta 
\\\\
= -\dfrac{\mu_{0}}{3} M^{2}\left[\dfrac{r^{3}}{3}\right]_{R_{1}}^{\textup{R}_{2}}\biggl[\varphi\biggr]_{0}^{2\pi}\biggl[-\cos{\left(\theta\right)}\biggr]_{0}^{\pi} 
\\\\
= -\dfrac{\mu_{0}}{3} 
M^{2}\cdot\dfrac{R_{2}^{3}-R_{1}^{3}}{3}\cdot2\pi\cdot2
 = \underline{\underline{-\dfrac{4\pi}{9}\mu_{0}M^{2} \left(R_{2}^{3}-R_{1}^{3}\right)}}
\end{gathered}
\end{split}
\end{equation}  

Note that in (\ref{Icv1}) \textit{Fubini's} theorem for the change of the order of integration holds, due to the coordinate-independent boundaries of the integration volume $\Omega:=r\,\in\,\left[R_{1},R_{2}\right]\times\theta\,\in\,\left[0,\pi\right]\times\varphi\,\in\,\left[0,2\pi\right)$.
 
It further holds for a hollow sphere with the integration volume $\Omega=\frac{4\pi}{3}\left(R_{2}^{3}-R_{1}^{3}\right)$:
\begin{equation}\label{Icv1AA}
\begin{split}
\begin{gathered}
\underline{\underline{I_{c} =-\dfrac{4\pi}{9}\mu_{0}M^{2} \left(R_{2}^{3}-R_{1}^{3}\right)=-\dfrac{1}{3}\mu_{0}M^{2}  vol\left(\Omega\right)}}
\end{gathered}
\end{split}
\end{equation} 

\clearpage

\subsection{Magnitude of the distance vector $\vert \vec{r}-\vec{r}^{\,\,\prime} \vert$ and $\vert  z\vec{e}_{z}-\vec{r}^{\,\,\prime} \vert$ in spherical coordinates}

The magnitude of the distance vector $\vert  \vec{r}-\vec{r}^{\,\,\prime} \vert $ in spherical coordinates is derived by means of (\ref{sphereCoordin}) and the law of cosine with $\kappa:=\measuredangle\left(\vec{r},\,\vec{r}^{\,\,\prime}\,\right)$ from (\ref{angles}):

\begin{equation}\label{DistVecSphericalCoord1}
\vert \vec{r}-\vec{r}^{\,\,\prime} \vert\, = \sqrt{\vert \vec{r}\,\vert^{2}+\vert\vec{r}^{\,\,\prime} \vert^{2}-2\vert\vec{r}\,\vert\vert\vec{r}^{\,\,\prime} \vert\cos{\left(\kappa\right)}}=\sqrt{r^{2}+r^{\prime\,2} -2rr^{\prime} \cos{\left(\kappa\right)}}
\end{equation}

The term $\cos{\left(\kappa\right)}$ is contained in the scalar product of $\vec{r}$ and $\vec{r}^{\,\,\prime}$, leading to the following expression with the help of a addition theorem for sine and cosine:
\begin{equation}\label{DistVecSphericalCoord2}
\begin{split}
\begin{gathered}
\cos{\left(\kappa\right)}=\dfrac{ \vec{r}\cdot\vec{r}^{\,\,\prime} }{ \vert\vec{r}\,\vert\vert\vec{r}^{\,\,\prime}\vert } =\dfrac{ \vec{r}\cdot\vec{r}^{\,\,\prime} }{ rr^{\prime} } 
\\\\
= \dfrac{ \cancel{rr^{\prime}}\left( \sin{\left(\theta\right)}\cos{\left(\varphi\right)}\sin{\left(\theta^{\prime}\right)}\cos{\left(\varphi^{\prime}\right)}+\sin{\left(\theta\right)}\sin{\left(\varphi\right)}\sin{\left(\theta^{\prime}\right)}\sin{\left(\varphi^{\prime}\right)}+\cos{\left(\theta\right)}\cos{\left(\theta^{\prime}\right)} \right)}{\cancel{rr^{\prime}}}
\\\\
=  \sin{\left(\theta\right)}\sin{\left(\theta^{\prime}\right)}\underbrace{\left[\cos{\left(\varphi\right)}\cos{\left(\varphi^{\prime}\right)}+\sin{\left(\varphi\right)}\sin{\left(\varphi^{\prime}\right)}\right]}_{=\cos{\left(\varphi-\varphi^{\prime}\right)}}+\cos{\left(\theta\right)}\cos{\left(\theta^{\prime}\right)}
\\\\
=  \sin{\left(\theta\right)}\sin{\left(\theta^{\prime}\right)}\cos{\left(\varphi-\varphi^{\prime}\right)}+\cos{\left(\theta\right)}\cos{\left(\theta^{\prime}\right)}
\end{gathered}
\end{split}
\end{equation}

By inserting (\ref{DistVecSphericalCoord1}) into (\ref{DistVecSphericalCoord2}) one gets:

\begin{equation}\label{DistVecSphericalCoord3}
\underline{\underline{\vert  \vec{r}-\vec{r}^{\,\,\prime} \vert \,= \sqrt{r^{2}+r^{\prime\,2} -2rr^{\prime}\left( \sin{\left(\theta\right)}\sin{\left(\theta^{\prime}\right)}\cos{\left(\varphi-\varphi^{\prime}\right)}+\cos{\left(\theta\right)}\cos{\left(\theta^{\prime}\right)} \right) }}}
\end{equation}
\\
The magnitude of the distance vector $\vert z\vec{e}_{z}-\vec{r}^{\,\,\prime} \vert $ in spherical coordinates is derived by means of (\ref{sphereCoordin}) and the law of cosine with $\kappa:=\measuredangle\left(\vec{e}_{z},\,\vec{r}^{\,\,\prime}\,\right)$ from (\ref{angles}). In spherical coordinates the polar axis is the $z$-axis and therefore the angle $\kappa$ is identical to the polar angle $\theta^{\prime}$. Thus, one finds: 
\begin{equation}\label{DistVecZ1}
\begin{split}
\begin{gathered} 
\underline{\underline{\vert z\vec{e}_{z}-\vec{r}^{\,\,\prime} \vert\, }}= \sqrt{\vert z\vec{e}_{z}\,\vert^{2}+\vert\vec{r}^{\,\,\prime} \vert^{2}-2\vert z\vec{e}_{z}\,\vert\vert\vec{r}^{\,\,\prime} \vert\cos{\left(\theta^{\prime}\right)}}
\\\\
=\sqrt{\vert z\vert^{2}\vert \vec{e}_{z}\vert^{2}+\vert\vec{r}^{\,\,\prime} \vert^{2}-2\vert z\vert\vert \vec{e}_{z}\vert\vert\vec{r}^{\,\,\prime} \vert\cos{\left(\theta^{\prime}\right)}}=\underline{\underline{\sqrt{z^{2}-2zr^{\prime} \cos{\left(\theta^{\prime}\right)}+r^{\prime\,2} }}}
\end{gathered}
\end{split}
\end{equation}
\clearpage

\subsection{Derivation of the scaling factor $a$ for the numerical results of the integral $I_{d}$}

The integral $I_{d}$ (\ref{Idv2}) shall be calculated numerically for a dimensionless magnetization density magnitude of $M=1$ over a homogeneously magnetized sphere with dimensionless radial coordinates $r,\,r^{\prime}\in \left[0,1\right]$ leading to the result $I_{d}\approx\tilde{I}_{d,\,num.}=:a\cdot I_{d,\,num.}$. The magnetization density is constant within the sphere, allowing the calculation of the value of $\tilde{I}_{d,\,num}$ for an arbitrary magnitude of the magnetization density by multiplying the numerical result $I_{d,\,num.}$ by the prefactor $M^{2}$, as it arises from (\ref{Idv2}).
\\
One can determine the scaling factor occurring from the following substitutions for $r$ and $r^{\prime}$, which change the boundaries of those variables from an arbitrary full sphere with $r,\,r^{\prime}\in \left[0,R_{2}\right]$ to the numerical boundaries used in the computation of $I_{d,\,num.}$:

\begin{equation}\label{SubId}
\tilde{r}=\dfrac{r}{R_{2}}\;\;;\;\; dr=R_{2}d\tilde{r}\;\;;\;\;
\tilde{r}^{\prime}=\dfrac{r^{\prime}}{R_{2}}\;\;;\;\; dr^{\prime}=R_{2}d\tilde{r}^{\prime}
\end{equation}

The prefactor follows from the insertion of the substitutions (\ref{SubId}) into (\ref{Idv2}):
\begin{equation}\label{insertSubIdv2}
\begin{split}
\begin{gathered} 
I_{d}\approx\tilde{I}_{d,num.}\overset{(\ref{SubId})}{=}\dfrac{\mu_{0}}{8\pi}M^{2} \int \biggl\{
\\\\
\dfrac{1}{\left(R_{2}^{2}\right)^{\frac{3}{2}}\left[\tilde{r}^{\,2}+\tilde{r}^{\,\prime\,2} -2\tilde{r}\tilde{r}^{\,\prime}\left( \sin{\left(\theta\right)}\sin{\left(\theta^{\prime}\right)}\cos{\left(\varphi-\varphi^{\prime}\right)}+\cos{\left(\theta\right)}\cos{\left(\theta^{\prime}\right)} \right)\right]^{\frac{3}{2}}}
\\\\
-\dfrac{3R_{2}^{2}\left[ \tilde{r}\cos{\left(\theta\right)}-\tilde{r}^{\,\prime}\cos{\left(\theta^{\prime}\right)}\right]^{2}}{\left(R_{2}^{2}\right)^{\frac{5}{2}}\left[\tilde{r}^{2}+\tilde{r}^{\,\prime\,2} -2\tilde{r}\tilde{r}^{\,\prime}\left( \sin{\left(\theta\right)}\sin{\left(\theta^{\prime}\right)}\cos{\left(\varphi-\varphi^{\prime}\right)}+\cos{\left(\theta\right)}\cos{\left(\theta^{\prime}\right)} \right)\right]^{\frac{5}{2}}}
\\\\
\biggr\}\cdot R_{2}^{2}\cdot\tilde{r}^{\,2}\cdot R_{2}^{2}\cdot\tilde{r}^{\,\prime\,2}\sin{\left(\theta\right)}\sin{\left(\theta^{\prime}\right)}R_{2}^{2}d\tilde{r}d\tilde{r}^{\,\prime}d\theta  
d\theta^{\prime}d\varphi d\varphi^{\prime} 
\\\\
=\dfrac{\mu_{0}}{8\pi}\cdot M^{2} \cdot R_{2}^{3} \cdot \underbrace{f\left(M=1,\,R_{1}=0,\,R_{2}=1,\,\tilde{r},\,\tilde{r}^{\,\prime},\,\theta,\,\theta^{\,\prime},\,\varphi,\,\varphi^{\,\prime}\right)}_{=: \,I_{d,\,num.}}
\end{gathered}
\end{split}
\end{equation}

One assembles the complete prefactor $a$ with the help of (\ref{insertSubIdv2}), for the purpose of scaling the value $I_{d,\,num.}$ onto a full sphere of outer radius $R_2$ and a constant magnetization density along the $z$-axis with the arbitrary magnitude $M$ :

\begin{equation}\label{prefactor}
\underline{\underline{I_{d}\approx\tilde{I}_{d,\,num.}=\dfrac{\mu_{0}}{8\pi}\cdot M^{2}\cdot R_{2}^{3}\cdot I_{d,\,num.}=a\cdot I_{d,\,num.} }}
\end{equation}

\clearpage

\subsection{Application of the substitution (\ref{IDIPSubU}) to the integral (\ref{IdDIP3}) and integration with respect to $u$}\label{uIntIDIP}

The integral (\ref{IdDIP3}) reads:
\begin{equation}\label{copyIdDIP3}
\begin{split}
\begin{gathered}
I^{\,dip}_{d}
=\dfrac{\mu_{0}}{4}M^{2}\int\limits_{r^{\,\prime}=R_{1}}^{\textup{R}_{2}}\int\limits_{\theta^{\prime}=0}^{\pi}\left(\dfrac{z^{2}-2zr^{\,\prime} \cos{\left(\theta^{\prime}\right)}+r^{\,\prime\,2}-3\left(z-r^{\,\,\prime}\cos{\left(\theta^{\prime}\right)}\right)^{2}}{\left(z^{2}-2zr^{\,\prime} \cos{\left(\theta^{\prime}\right)}+r^{\,\prime\,2} \right)^{\frac{5}{2}}} \right)r^{\,\prime\,2}\sin{\left(\theta^{\prime}\right)} d\theta^{\prime}dr^{\prime} 
\end{gathered}
\end{split}
\end{equation} 

In order to express the integrand with the substituted variable $u$ one extends (\ref{IDIPSubU}):
\begin{equation}\label{extIDIPSubU}
\begin{split}
\begin{gathered}
u=z^{2}-2zr^{\,\prime} \cos{\left(\theta^{\prime}\right)}+r^{\prime\,2} \,;\,\, d\theta^{\prime}=\dfrac{du}{2zr^{\,\prime} \sin{\left(\theta^{\prime}\right)}}\,;\,\,
\\[14pt]
u\left(0\right)=\left(z-r^{\prime}\right)^{2}\,;\,\,u\left(\pi\right)=\left(z+r^{\prime}\right)^{2}\,;\,\,
\\[14pt]
\left(z-r^{\prime}\cos{\left(\theta^{\prime}\right)}\right)^{2}=\dfrac{1}{4z^{2}}\left(u^{2}+2u\left(z^{2}-r^{\prime\,2}\right)+\left(z^{2}-r^{\prime\,2}\right)^{2}\right)
\end{gathered}
\end{split}
\end{equation} 

Thus one can insert (\ref{extIDIPSubU}) into (\ref{copyIdDIP3}), rearrange the integrand and integrate with respect to the new integration variable $u$:

\begin{equation*}\label{IdDIPSUBU0}
\begin{split}
\begin{gathered}
I^{\,dip}_{d}
=\dfrac{\mu_{0}}{4}M^{2}\int\limits_{r^{\prime}=R_{1}}^{\textup{R}_{2}}\int\limits_{u=\left(z-r^{\prime}\right)^{2}}^{\left(z+r^{\prime}\right)^{2}}\left(\dfrac{u-\frac{3}{4z^{2}}\left(u^{2}+2u\left(z^{2}-r^{\prime\,2}\right)+\left(z^{2}-r^{\prime\,2}\right)^{2}\right)}{u^{\frac{5}{2}}} \right) \dfrac{r^{\prime\,\cancel{2}}\cancel{\sin{\left(\theta^{\prime}\right)}}}{2z\cancel{r^{\prime}} \cancel{\sin{\left(\theta^{\prime}\right)}}}dudr^{\prime} 
\end{gathered}
\end{split}
\end{equation*}
\begin{equation}\label{IdDIPSUBU1}
\begin{split}
\begin{gathered}
=-\dfrac{3}{32z^{3}}\mu_{0}M^{2}\int\limits_{r^{\prime}=R_{1}}^{\textup{R}_{2}}r^{\prime}\int\limits_{u=\left(z-r^{\prime}\right)^{2}}^{\left(z+r^{\prime}\right)^{2}}u^{-\frac{1}{2}}+\dfrac{2}{3}\left(z^{2}-3r^{\prime\,2}\right)u^{-\frac{3}{2}} +\left(z^{2}-r^{\prime\,2}\right)^{2}u^{-\frac{5}{2}} \,dudr^{\prime} 
\\[14pt]
=-\dfrac{3}{\cancel{2}\cdot16z^{3}}\mu_{0}M^{2}\int\limits_{r^{\,\prime}=R_{1}}^{\textup{R}_{2}}r^{\prime}\left[\cancel{2}\sqrt{u}-\dfrac{\cancel{2}\cdot2}{3\sqrt{u}}\left(z^{2}-3r^{\prime\,2}\right) -\dfrac{\cancel{2}}{3u^{-\frac{3}{2}}}\left(z^{2}-r^{\prime\,2}\right)^{2}\right]^{\left(z+r^{\prime}\right)^{2}}_{\left(z-r^{\prime}\right)^{2}} \,dr^{\,\prime} 
\\[14pt]
=-\dfrac{3}{16z^{3}}\mu_{0}M^{2}\int\limits_{r^{\prime}=R_{1}}^{\textup{R}_{2}}r^{\prime}\left[\vert z+r^{\prime}\vert-\vert z-r^{\prime}\vert-\dfrac{2}{3}\left(z^{2}-3r^{\prime\,2}\right)\left(\dfrac{1}{\vert z+r^{\prime}\vert}-\dfrac{1}{\vert z-r^{\prime}\vert}\right) \right.
\\[14pt]
\left.-\dfrac{1}{3}\left(z^{2}-r^{\prime\,2}\right)^{2}\left(\dfrac{1}{\vert z+r^{\prime}\vert^{3}}-\dfrac{1}{\vert z-r^{\prime}\vert^{3}}\right)\right]\,dr^{\prime} 
\end{gathered}
\end{split}
\end{equation}

The remaining integral (\ref{IdDIPSUBU1}) has to be evaluated with respect to $r^{\prime}$, under the consideration of the different cases for the moduli $\vert z+r^{\prime}\vert$ and $\vert z-r^{\prime}\vert$. Its rewritten representation reads:
\begin{equation}\label{IdDIPSUBU2}
\begin{split}
\begin{gathered}
\underline{\underline{I^{\,dip}_{d}=-\dfrac{3}{16z^{3}}\mu_{0}M^{2}\int\limits_{r^{\prime}=R_{1}}^{\textup{R}_{2}}r^{\prime}\left(\vert z+r^{\prime}\vert-\vert z-r^{\prime}\vert-\dfrac{2\left(z^{2}-3r^{\prime\,2}\right)\left(\vert z-r^{\,\prime}\vert-\vert z+r^{\prime}\vert\right)}{3\vert z+r^{\prime}\vert\vert z-r^{\prime}\vert}\right.}}
\\[14pt]
\underline{\underline{\left.-\dfrac{\left(z^{2}-r^{\prime\,2}\right)^{2}\left(\vert z-r^{\prime}\vert^{3}-\vert z+r^{\prime}\vert^{3}\right)}{3\vert z+r^{\prime}\vert^{3}\vert z-r^{\prime}\vert^{3}}\right)\,dr^{\,\prime}}} 
\end{gathered}
\end{split}
\end{equation} 

\clearpage

\subsection{Point symmetry of the integrand of (\ref{IdDIP4}) as a function of $z$ with regard to $z=0$}\label{pointsymm0}

The integrand $f(z)$ is extracted from (\ref{IdDIP4}):
\begin{equation}\label{pointsymm1}
\begin{split}
\begin{gathered}
f(z):=r^{\prime}\left(\vert z+r^{\prime}\vert-\vert z-r^{\prime}\vert-\dfrac{2\left(z^{2}-3r^{\prime\,2}\right)\left(\vert z-r^{\,\prime}\vert-\vert z+r^{\prime}\vert\right)}{3\vert z+r^{\prime}\vert\vert z-r^{\prime}\vert}\right.
\\\\
\left.-\dfrac{\left(z^{2}-r^{\prime\,2}\right)^{2}\left(\vert z-r^{\prime}\vert^{3}-\vert z+r^{\prime}\vert^{3}\right)}{3\vert z+r^{\prime}\vert^{3}\vert z-r^{\prime}\vert^{3}}\right)
\end{gathered}
\end{split}
\end{equation} 

The point symmetry of the integrand is shown by calculating $f(-z)$:
\begin{equation}\label{pointsymm2}
\begin{split}
\begin{gathered}
\underline{\underline{f(-z)}}=r^{\prime}\left(\vert -z+r^{\prime}\vert-\vert -z-r^{\prime}\vert-\dfrac{2\left((-z)^{2}-3r^{\prime\,2}\right)\left(\vert -z-r^{\,\prime}\vert-\vert -z+r^{\prime}\vert\right)}{3\vert -z+r^{\prime}\vert\vert -z-r^{\prime}\vert}\right.
\\\\
\left.-\dfrac{\left((-z)^{2}-r^{\prime\,2}\right)^{2}\left(\vert -z-r^{\prime}\vert^{3}-\vert -z+r^{\prime}\vert^{3}\right)}{3\vert -z+r^{\prime}\vert^{3}\vert -z-r^{\prime}\vert^{3}}\right)
\\\\
=r^{\prime}\left(\vert z-r^{\prime}\vert-\vert z+r^{\prime}\vert-\dfrac{2\left(z^{2}-3r^{\prime\,2}\right)\left(\vert z+r^{\,\prime}\vert-\vert z-r^{\prime}\vert\right)}{3\vert z-r^{\prime}\vert\vert z+r^{\prime}\vert}\right.
\\\\
\left.-\dfrac{\left(z^{2}-r^{\prime\,2}\right)^{2}\left(\vert z+r^{\prime}\vert^{3}-\vert z-r^{\prime}\vert^{3}\right)}{3\vert z-r^{\prime}\vert^{3}\vert z+r^{\prime}\vert^{3}}\right)= \underline{\underline{-f(z)}}
\end{gathered}
\end{split}
\end{equation} 

\clearpage

\subsection{Procedure of finding a solution for the integrals $I_{r^{\prime},\,1}$ and $I_{r^{\prime},\,2}$ } \label{rPrimeIntIDIP}

First one states the integral $I_{r^{\,\prime},\,1}$ with the help of (\ref{IdDIP6}) and the integrand $f\left(\vert z+r^{\prime}\vert,\,\vert z-r^{\prime}\vert\right)$ for $z\in\left[R_{1},z\right]$ from (\ref{IdDIP4}), which was already written down in (\ref{pointsymm1}):
\begin{equation}\label{1rPrime1IdDIP}
\begin{split}
\begin{gathered}
\underline{\underline{I_{r^{\,\prime},\,1}}}\overset{(\ref{IdDIP6})}{=}\int\limits_{r^{\prime}=\textup{R}_{1}}^{\textup{z}} f\left( z+r^{\prime},\, z-r^{\prime}\right) \,dr^{\,\prime}
\overset{(\ref{IdDIP4})}{=}\int\limits_{r^{\prime}=\textup{R}_{1}}^{\textup{z}}r^{\prime}\left( z+r^{\prime} -  \left(z-r^{\prime}\right)\right.
\\[14pt]
\left.-\dfrac{2\left(z^{2}-3r^{\prime\,2}\right)\left(  z-r^{\,\prime} - \left( z+r^{\prime}\right) \right)}{3\left( z+r^{\prime}\right)\left( z-r^{\prime}\right)}\right.
\left.-\dfrac{\left(z^{2}-r^{\prime\,2}\right)^{2}\left(\left( z-r^{\prime}\right)^{3}-\left( z+r^{\prime}\right)^{3}\right)}{3\left( z+r^{\prime}\right)^{3}\left( z-r^{\prime}\right)^{3}}\right)\,dr^{\,\prime}
\\[14pt]
=\int\limits_{r^{\prime}=\textup{R}_{1}}^{\textup{z}}r^{\prime}\left( 2r^{\prime}+\dfrac{4r^{\prime}\left(z^{2}-3r^{\prime\,2}\right)}{3\left(z^{2}-r^{\prime\,2}\right)}+\dfrac{\cancel{\left(z^{2}-r^{\prime\,2}\right)^{2}}2r^{\prime}\left(r^{\prime\,2}+3z^{2}\right)}{3\left(z^{2}-r^{\prime\,2}\right)^{\cancel{3}}}\right)\,dr^{\,\prime}
\\[14pt]
=\int\limits_{r^{\prime}=\textup{R}_{1}}^{\textup{z}}r^{\prime\,2}\left( \dfrac{6\left(z^{2}-r^{\prime\,2}\right)+4\left(z^{2}-3r^{\prime\,2}\right)+2\left(r^{\prime\,2}+3z^{2}\right)}{3\left(z^{2}-r^{\prime\,2}\right)}\right)\,dr^{\,\prime}
\\[14pt]
=\int\limits_{r^{\prime}=\textup{R}_{1}}^{\textup{z}}r^{\prime\,2}\left( \dfrac{16\cancel{\left(z^{2}-r^{\prime\,2}\right)}}{3\cancel{\left(z^{2}-r^{\prime\,2}\right)}}\right)\,dr^{\,\prime}=\dfrac{16}{3}\int\limits_{r^{\prime}=\textup{R}_{1}}^{\textup{z}}r^{\prime\,2}\,dr^{\,\prime}=\dfrac{16}{3}\left[\dfrac{r^{\prime\,3}}{3}\right]_{\textup{R}_{1}}^{\textup{z}}=\underline{\underline{\dfrac{16}{9}\left(z^{3}-R_{1}^{3}\right)}}
\end{gathered}
\end{split}
\end{equation}

Second, the integral $I_{r^{\,\prime},\,2}$ was evaluated. It is given by (\ref{IdDIP6}) and the integrand $f\left(\vert z+r^{\prime}\vert,\,\vert z-r^{\prime}\vert\right)$ for $z\in\left[z,R_{2}\right]$ from (\ref{IdDIP4}):
\begin{equation*}
\begin{split}
\begin{gathered}
\underline{\underline{I_{r^{\,\prime},\,2}}}\overset{(\ref{IdDIP6})}{=}\int\limits_{r^{\prime}=\textup{R}_{1}}^{\textup{z}} f\left( z+r^{\prime},\, -\left(z-r^{\prime}\right)\right) \,dr^{\,\prime}
\overset{(\ref{IdDIP4})}{=}\int\limits_{r^{\prime}=\textup{R}_{1}}^{\textup{z}}r^{\prime}\left( z+r^{\prime} +  \left(z-r^{\prime}\right)\right.
\\[14pt]
\left.-\dfrac{2\left(z^{2}-3r^{\prime\,2}\right)\left( -\left( z-r^{\,\prime}\right) - \left( z+r^{\prime}\right) \right)}{-3\left( z+r^{\prime}\right)\left( z-r^{\prime}\right)}\right.
\left.-\dfrac{\left(z^{2}-r^{\prime\,2}\right)^{2}\left(-\left( z-r^{\prime}\right)^{3}-\left( z+r^{\prime}\right)^{3}\right)}{-3\left( z+r^{\prime}\right)^{3}\left( z-r^{\prime}\right)^{3}}\right)\,dr^{\,\prime}
\\[14pt]
=\int\limits_{r^{\prime}=\textup{R}_{1}}^{\textup{z}}r^{\prime}\left( 2z-\dfrac{4z\left(z^{2}-3r^{\prime\,2}\right)}{3\left(z^{2}-r^{\prime\,2}\right)}-\dfrac{\cancel{\left(z^{2}-r^{\prime\,2}\right)^{2}}2z\left(z^{2}+3r^{\prime\,2}\right)}{3\left(z^{2}-r^{\prime\,2}\right)^{\cancel{3}}}\right)dr^{\,\prime}
\\[14pt]
=\int\limits_{r^{\prime}=\textup{R}_{1}}^{\textup{z}}zr^{\prime}\left( \dfrac{6\left(z^{2}-r^{\prime\,2}\right)-4\left(z^{2}-3r^{\prime\,2}\right)-2\left(z^{2}+3r^{\prime\,2}\right)}{3\left(z^{2}-r^{\prime\,2}\right)}\right)dr^{\,\prime}
\end{gathered}
\end{split}
\end{equation*}
\begin{equation}\label{1rPrime2IdDIP}
\begin{split}
\begin{gathered}
=\int\limits_{r^{\prime}=\textup{R}_{1}}^{\textup{z}}zr^{\prime}\left( \dfrac{0}{3\left(z^{2}-r^{\prime\,2}\right)}\right)dr^{\,\prime}=\underline{\underline{0}}
\end{gathered}
\end{split}
\end{equation}

\clearpage

\subsection{Infinite contribution of the second integral in (\ref{DELTAEBHd1})} \label{InfDIPzaxis}

The second integral from (\ref{DELTAEBHd1}) reads:
\begin{equation}\label{InfDIP1}
I_{\ast}:=-\dfrac{1}{3}\mu_{0}M^{2}\left(-R_{1}^{3}\right)\int\limits_{\Omega}\vert z\vert^{-3} \,d\Omega
\end{equation}

It is evaluated in spherical coordinates with $z=r\cos{\left(\theta\right)}$, $r>0$ and through (\ref{sphereCoordin}) and (\ref{volElem}) one obtains:
\begin{equation}\label{InfDIP1}
\begin{split}
\begin{gathered}
\underline{\underline{I_{\ast}}}=\dfrac{1}{3}\mu_{0}M^{2}R_{1}^{3}\int\limits_{r=\textup{R}_{1}}^{\textup{R}_{2}}\int\limits_{\varphi=0}^{2\pi}\int\limits_{\theta=0}^{\pi}\vert r\cos{\left(\theta\right)}\vert^{-3}r^{2}\sin{\left(\theta\right)} \,d\theta d\varphi dr
\\\\
=\dfrac{1}{3}\mu_{0}M^{2}R_{1}^{3}\int\limits_{\varphi=0}^{2\pi}d\varphi \int\limits_{r=\textup{R}_{1}}^{\textup{R}_{2}} \underbrace{\dfrac{r^{2}}{\vert r\vert^{3}}}_{=\,\frac{1}{r}} dr\int\limits_{\theta=0}^{\pi}\dfrac{\sin{\left(\theta\right)}}{\vert \cos{\left(\theta\right)}\vert^{3}} d\theta 
\\\\
=\dfrac{1}{3}\mu_{0}M^{2}R_{1}^{3}\cdot2\pi \int\limits_{r=\textup{R}_{1}}^{\textup{R}_{2}} \dfrac{1}{r}\, dr\left(\,\int\limits_{\theta=0}^{\frac{\pi}{2}}\dfrac{\sin{\left(\theta\right)}}{\left(\cos{\left(\theta\right)}\right)^{3}}\, d\theta +\int\limits_{\theta=\frac{\pi}{2}}^{\pi}\dfrac{\sin{\left(\theta\right)}}{-\left(\cos{\left(\theta\right)}\right)^{3}} \,d\theta\right)
\\\\
=\dfrac{\pi}{3}\mu_{0}M^{2}R_{1}^{3}\cdot\cancel{2}\ln{\left|\dfrac{R_{2}}{R_{1}}\right|} \left(\left[\dfrac{1}{\cancel{2}\left(\cos{\left(\theta\right)}\right)^{2}} \right]_{0}^{\frac{\pi}{2}} -\left[\dfrac{1}{\cancel{2}\left(\cos{\left(\theta\right)}\right)^{2}} \right]_{\frac{\pi}{2}}^{\pi}\right)
\\\\
=\dfrac{\pi}{3}\mu_{0}M^{2}R_{1}^{3}\ln{\left|\dfrac{R_{2}}{R_{1}}\right|} \left(\dfrac{1}{0}-\dfrac{1}{1} -\left(\dfrac{1}{\left(-1\right)^{2}}-\dfrac{1}{0}\right) \right)
\\\\
=\dfrac{\pi}{3}\mu_{0}M^{2}R_{1}^{3}\ln{\left|\dfrac{R_{2}}{R_{1}}\right|} \left(\infty-1 -1+\infty \right) = \underline{\underline{\infty}}
\end{gathered}
\end{split}
\end{equation}

\clearpage

\subsection{Divergence of the magnetization density} \label{DivMApp}

The divergence of the magnetization density $\vec{M}\left(\vec{r}\,\right)=M\left(r\right)\vec{e}_{z}$ according to (\ref{defM(r)}) is evaluated in spherical coordinates:
\begin{equation*} 
\begin{split}
\begin{gathered}
\underline{\underline{\vec{\nabla}\cdot \vec{M}}}=\dfrac{1}{r^{2}}\dfrac{\partial}{\partial r}\left(r^{2}M_{r}\right)+ \dfrac{1}{r \sin{\left(\theta\right)}}\dfrac{\partial}{\partial \theta}\left(\sin{\left(\theta\right)}M_{\theta}\right)+ \dfrac{1}{r\sin {\left(\theta\right)}}\dfrac{\partial\,M_{\varphi}}{\partial \varphi} 
\\\\
= \dfrac{1}{r^{2}}\dfrac{\partial}{\partial r}\left(r^{2}M\left(r\right)\underbrace{\vec{e}_{z}\cdot\vec{e}{r}}_{=\cos{\left(\theta\right)}}\right)+ \dfrac{1}{r \sin{\left(\theta\right)}}\dfrac{\partial}{\partial \theta}\left(\sin{\left(\theta\right)}M\left(r\right)\underbrace{\vec{e}_{z}\cdot\vec{e}_{\theta}}_{=-\sin{\left(\theta\right)}}\right)+ \dfrac{1}{r\sin {\left(\theta\right)}}\dfrac{\partial\,M\left(r\right)\overbrace{\vec{e}_{z}\cdot\vec{e}_{\varphi}}^{=\,0}}{\partial \varphi} 
\end{gathered}
\end{split}
\end{equation*}
\begin{equation}\label{DivM(r)1}
\begin{split}
\begin{gathered}
=\dfrac{\cos{\left(\theta\right)}}{r^{2}}\underbrace{\dfrac{d}{d r}\left(r^{2}M\left(r\right)\right)}_{=2rM(r)+r^{2}\frac{dM(r)}{dr}}- \dfrac{M\left(r\right)}{r \sin{\left(\theta\right)}}\underbrace{\dfrac{d}{d \theta}\left(\sin^{2}{\left(\theta\right)} \right)}_{=2\sin{\left(\theta\right)}\cos{\left(\theta\right)}}
\\\\
=\left(\dfrac{2M(r)}{r}+\dfrac{d\,M(r)}{dr}-\dfrac{2M(r)\cancel{\sin{\left(\theta\right)}}}{r\cancel{\sin{\left(\theta\right)}}}\right)\cos{\left(\theta\right)}=\underline{\underline{\dfrac{d\,M(r)}{dr}\cos{\left(\theta\right)}}}
\end{gathered}
\end{split}
\end{equation}
\\
For the special case of a constant magnetization density $M(r)=M=\textup{const.}$ one acquires from (\ref{DivM(r)1}):
\begin{equation}\label{DivM(r)2}
\begin{split}
\begin{gathered}
\underline{\underline{\vec{\nabla}\cdot \vec{M}}}= \underbrace{\dfrac{d\,M }{dr}}_{=0}\cos{\left(\theta\right)} =\underline{\underline{0}}
\end{gathered}
\end{split}
\end{equation}
\clearpage

\subsection{Determination of the coefficients of the magnetic potential for a homogeneously magnetized hollow sphere} \label{AppendixConstCoeffHollowSphereConst}

One recapitulates the boundary conditions from (\ref{LaplacBoundary}):
\begin{equation}\label{LaplacBoundaryApp1}
\begin{split}
\begin{gathered}
B_{r,i}\left(R_{1},\,\theta\right)\overset{!}{=}B_{r,M}\left(R_{1},\,\theta\right)\;\;;\;\; B_{r,M}\left(R_{2},\,\theta\right)\overset{!}{=}B_{r,e}\left(R_{2},\,\theta\right)
\\
\phi_{i}\left(R_{1},\,\theta\right)\overset{!}{=}\phi_{M}\left(R_{1},\,\theta\right)\;\;;\;\;\phi_{M}\left(R_{2},\,\theta\right)\overset{!}{=}\phi_{e}\left(R_{2},\,\theta\right)
\end{gathered}
\end{split}
\end{equation}
First, one rewrites the conditons for the radial component of the magnetic flux density $\vec{B}$, by means of its the material equation as well as with the definition of the magnetization density $\vec{M}$ according to (\ref{defM}) and $\vec{e}_{z}\cdot\vec{e}_{r}=\cos{\left(\theta\right)}$:
\begin{equation}\label{LaplacBoundaryApp2}
\begin{split}
\begin{gathered}
B_{r,i}\left(R_{1},\,\theta\right)\overset{!}{=}B_{r,M}\left(R_{1},\,\theta\right)
\\
\longleftrightarrow\;\; \cancel{\mu_{0}}\left(\vec{H}_{r,i}\left(R_{1},\,\theta\right)+\vec{M}_{r,i}\left(R_{1},\,\theta\right)\right)=\cancel{\mu_{0}}\left(\vec{H}_{r,M}\left(R_{1},\,\theta\right)+\vec{M}_{r,M}\left(R_{1},\,\theta\right)\right)
\\
\xlongleftrightarrow{\text{(\ref{defM})}}\;\; \vec{H}_{r,i}\left(R_{1},\,\theta\right)+0\cdot\vec{e}_{z}\cdot\vec{e}_{r}=\vec{H}_{r,M}\left(R_{1},\,\theta\right)+M\cdot\vec{e}_{z}\cdot\vec{e}_{r}
\\
\longleftrightarrow\;\; \vec{H}_{r,i}\left(R_{1},\,\theta\right) =\vec{H}_{r,M}\left(R_{1},\,\theta\right)+M\cos{\left(\theta\right)}
\\\\
B_{r,M}\left(R_{2},\,\theta\right)\overset{!}{=}B_{r,e}\left(R_{2},\,\theta\right)\\
\longleftrightarrow\;\; \cancel{\mu_{0}}\left(\vec{H}_{r,M}\left(R_{2},\,\theta\right)+\vec{M}_{r,M}\left(R_{2},\,\theta\right)\right)=\cancel{\mu_{0}}\left(\vec{H}_{r,e}\left(R_{2},\,\theta\right)+\vec{M}_{r,e}\left(R_{2},\,\theta\right)\right)
\\
\xlongleftrightarrow{\text{(\ref{defM})}}\;\; \vec{H}_{r,M}\left(R_{2},\,\theta\right)+M\cdot\vec{e}_{z}\cdot\vec{e}_{r}=\vec{H}_{r,e}\left(R_{2},\,\theta\right)+0\cdot\vec{e}_{z}\cdot\vec{e}_{r}
\\
\longleftrightarrow\;\; \vec{H}_{r,M}\left(R_{2},\,\theta\right)+M\cos{\left(\theta\right)}=\vec{H}_{r,e}\left(R_{2},\,\theta\right)
\end{gathered}
\end{split}
\end{equation}
In (\ref{LaplacBoundaryApp2}) the representation for the magnetic field $\vec{H}$ from (\ref{magneticfield}) is inserted and its radial component is given by the \textit{Nabla}-operator in spherical coordinates. Subsequently, one applies the \textit{ansatz} for $\phi_{m}$ from (\ref{phimAnsatz2}), observing that with $\textup{P}_{1}\left(\cos{\left(\theta\right)}\right)=\cos{\left(\theta\right)}$ \cite{Bartelmann_2015} the conditions are fulfilled. Therefore one only keeps terms with $l=1$ in the infinite sum and gets: 
\begin{equation*}\label{LaplacBoundaryApp3}
\begin{split}
\begin{gathered}
\vec{H}_{r,i}\left(R_{1},\,\theta\right) =\vec{H}_{r,M}\left(R_{1},\,\theta\right)+M\cos{\left(\theta\right)}
\;\;\xlongleftrightarrow{\text{(\ref{magneticfield})}}\;\; -\dfrac{\partial \phi_{i}}{\partial r}\bigg\vert_{R_{1}}=-\dfrac{\partial \phi_{M}}{\partial r}\bigg\vert_{R_{1}}+M\cos{\left(\theta\right)}
\\
\xlongleftrightarrow{\text{(\ref{phimAnsatz2})}}\;\; -\sum\limits_{l=1}^{\infty}A_{l}lR_{1}^{l-1}\textup{P}_{l}\left(\cos{\left(\theta\right)}\right)=-\sum\limits_{l=1}^{\infty}\left(B_{l}lR_{1}^{l-1}-C_{l}\dfrac{l+1}{R_{1}^{l+2}}\right)\textup{P}_{l}\left(\cos{\left(\theta\right)}\right)+M\cos{\left(\theta\right)}
\\
\xlongleftrightarrow{l=1}\;\; - A_{1} \cancel{\cos{\left(\theta\right)}} = \left(-B_{1}+C_{1}\dfrac{2}{R_{1}^{3}}\right) \cancel{\cos{\left(\theta\right)}} +M\cancel{\cos{\left(\theta\right)}}
\\
\longleftrightarrow\;\; A_{1} =B_{1}-C_{1}\dfrac{2}{R_{1}^{3}} - M 
\\\\
\vec{H}_{r,M}\left(R_{2},\,\theta\right)+M\cos{\left(\theta\right)}=\vec{H}_{r,e}\left(R_{2},\,\theta\right)\;\;
\xlongleftrightarrow{\text{(\ref{magneticfield})}}\;\; -\dfrac{\partial \phi_{M}}{\partial r}\bigg\vert_{R_{2}}+M\cos{\left(\theta\right)}=-\dfrac{\partial \phi_{e}}{\partial r}\bigg\vert_{R_{2}}
\end{gathered}
\end{split}
\end{equation*}
\begin{equation*}\label{LaplacBoundaryApp4}
\begin{split}
\begin{gathered}
\xlongleftrightarrow{\text{(\ref{phimAnsatz2})}}\;\; -\sum\limits_{l=1}^{\infty}\left(B_{l}lR_{2}^{l-1}-C_{l}\dfrac{l+1}{R_{2}^{l+2}}\right)\textup{P}_{l}\left(\cos{\left(\theta\right)}\right)+M\cos{\left(\theta\right)}=-\sum\limits_{l=1}^{\infty}-D_{l}\dfrac{l+1}{R_{2}^{l+2}}\textup{P}_{l}\left(\cos{\left(\theta\right)}\right)
\end{gathered}
\end{split}
\end{equation*}
\begin{equation}\label{LaplacBoundaryApp5}
\begin{split}
\begin{gathered}
\xlongleftrightarrow{l=1}\;\;  \left(-B_{1} +C_{l}\dfrac{2}{R_{2}^{3}}\right)\cancel{\cos{\left(\theta\right)}}+M\cancel{\cos{\left(\theta\right)}}=D_{l}\dfrac{2}{R_{2}^{3}}\cancel{\cos{\left(\theta\right)}}
\\
\longleftrightarrow\;\; B_{1}=  C_{l}\dfrac{2}{R_{2}^{3}}+M-D_{l}\dfrac{2}{R_{2}^{3}}
\end{gathered}
\end{split}
\end{equation}
In (\ref{LaplacBoundaryApp5}) the division by $\cos{\left(\theta\right)}\neq0$ is permitted and both equations are always true for $\cos{\left(\theta\right)}=0$. This is also valid for the following calculation.
\\
Second, the remaining boundary conditions in (\ref{LaplacBoundaryApp1}) are evaluated for $l=1$ and $\textup{P}_{1}\left(\cos{\left(\theta\right)}\right)=\cos{\left(\theta\right)}$ \cite{Bartelmann_2015} with the \textit{ansatz} from (\ref{phimAnsatz2}):
\begin{equation}\label{LaplacBoundaryApp6}
\begin{split}
\begin{gathered}
\phi_{i}\left(R_{1},\,\theta\right)\overset{!}{=}\phi_{M}\left(R_{1},\,\theta\right) \;\;\xlongleftrightarrow[\text{(\ref{LaplacBoundaryApp1})}]{l=1} \;\;A_{1}R_{1}\cancel{\cos{\left(\theta\right)}}=\left(B_{1}R_{1}+\dfrac{C_{1}}{R_{1}^{2}}\right)\cancel{\cos{\left(\theta\right)}}
\\
\longleftrightarrow\;\;A_{1} = B_{1} +\dfrac{C_{1}}{R_{1}^{3}} 
\\\\
\phi_{M}\left(R_{2},\,\theta\right)\overset{!}{=}\phi_{e}\left(R_{2},\,\theta\right)\;\;\xlongleftrightarrow[\text{(\ref{LaplacBoundaryApp1})}]{l=1} \;\;\left(B_{1}R_{2}+\dfrac{C_{1}}{R_{2}^{2}}\right)\cancel{\cos{\left(\theta\right)}}=\dfrac{D_{1}}{R_{2}^{2}}\cancel{\cos{\left(\theta\right)}}
\\
\longleftrightarrow\;\; B_{1}=-\dfrac{C_{1}}{R_{2}^{3}} +\dfrac{D_{1}}{R_{2}^{3}}
\end{gathered}
\end{split}
\end{equation}
One combines the relations containing $R_{1}$ from (\ref{LaplacBoundaryApp5}) and (\ref{LaplacBoundaryApp6}):
\begin{equation}\label{LaplacBoundaryApp7}
\begin{split}
\begin{gathered}
B_{1} +\dfrac{C_{1}}{R_{1}^{3}} =B_{1}-C_{1}\dfrac{2}{R_{1}^{3}} - M  \;\;\longleftrightarrow\;\; \underline{C_{1}=-\dfrac{M}{3}R_{1}^{3}}
\end{gathered}
\end{split}
\end{equation}
Consequently, one proceeds with the relations containing $R_{2}$ from (\ref{LaplacBoundaryApp5}) and (\ref{LaplacBoundaryApp6}), with the help of (\ref{LaplacBoundaryApp7}):
\begin{equation}\label{LaplacBoundaryApp8}
\begin{split}
\begin{gathered}
 C_{l}\dfrac{2}{R_{2}^{3}}+M-D_{l}\dfrac{2}{R_{2}^{3}} = -\dfrac{C_{1}}{R_{2}^{3}} +\dfrac{D_{1}}{R_{2}^{3}}\;\;\xlongleftrightarrow{\text{(\ref{LaplacBoundaryApp7})}}\;\;  \underline{D_{1}=\dfrac{M}{3}\left(R_{2}^{3}-R_{1}^{3}\right)}
\end{gathered}
\end{split}
\end{equation}
Subsequently, one receives $B_{1}$ from (\ref{LaplacBoundaryApp6}) with the help of (\ref{LaplacBoundaryApp7}) and (\ref{LaplacBoundaryApp8}):
\begin{equation}\label{LaplacBoundaryApp9}
\begin{split}
\begin{gathered}
\underline{B_{1}}=-\dfrac{C_{1}}{R_{2}^{3}} +\dfrac{D_{1}}{R_{2}^{3}}\;\underset{(\ref{LaplacBoundaryApp8})}{\overset{(\ref{LaplacBoundaryApp7})}{=}}\;\dfrac{MR_{1}^{3}}{3R_{2}^{3}} +\dfrac{M}{3}\left(1-\dfrac{R_{1}^{3}}{R_{2}^{3}}\right)=\underline{\dfrac{M}{3}}
\end{gathered}
\end{split}
\end{equation}
Eventually, one finds $A_{1}$ from (\ref{LaplacBoundaryApp6}) together with (\ref{LaplacBoundaryApp7}) and (\ref{LaplacBoundaryApp9}):
\begin{equation}\label{LaplacBoundaryApp10}
\begin{split}
\begin{gathered}
\underline{A_{1}}= B_{1} +\dfrac{C_{1}}{R_{1}^{3}} \;\underset{(\ref{LaplacBoundaryApp9})}{\overset{(\ref{LaplacBoundaryApp7})}{=}}\;\dfrac{M}{3} -\dfrac{MR_{1}^{3}}{3R_{1}^{3}}=\underline{0}
\end{gathered}
\end{split}
\end{equation}
The constants from (\ref{LaplacBoundaryApp7}), (\ref{LaplacBoundaryApp8}), (\ref{LaplacBoundaryApp9}) and (\ref{LaplacBoundaryApp10}) are then inserted into (\ref{phimAnsatz2}), keeping only terms with $l=1$ and receiving the solution for the magnetic potential of a homogeneously magnetized hollow sphere:
\begin{equation}\label{phimSolutionHollowSphereApp}
\underline{\underline{\phi_{m}\left(\vec{r}\,\right)= 
\begin{cases}
\phi_{i}\left(r,\,\theta\right)=A_{1}r\cos{\left(\theta\right)} \;\overset{(\ref{LaplacBoundaryApp10})}{=}\;0 \,\textup{A}
\\\\
\phi_{M}\left(r,\,\theta\right)=\left(B_{1}r+\dfrac{C_{1}}{r^{2}}\right)\cos{\left(\theta\right)}\;\underset{(\ref{LaplacBoundaryApp9})}{\overset{(\ref{LaplacBoundaryApp7})}{=}} \;\dfrac{M}{3}\left(r-\dfrac{R_{1}^{3}}{r^{2}}\right)\cos{\left(\theta\right)}
\\\\
\phi_{e}\left(r,\,\theta\right)=D_{1}\dfrac{\cos{\left(\theta\right)}}{r^{2}}\;\overset{(\ref{LaplacBoundaryApp8})}{=}\;\dfrac{M}{3} \left(R_{2}^{3}-R_{1}^{3}\right)\dfrac{\cos{\left(\theta\right)}}{r^{2}}
\end{cases}}}
\end{equation}

\clearpage

\subsection{Evaluation of the \textit{Breit-Hartree} contribution for a homogeneously magnetized hollow sphere $\Omega$} \label{AppendixEHBHollowSphereConst}

Here the material equation, (\ref{magneticfield}), (\ref{defM}) and the magnetic potential from (\ref{phimSolutionHollowSphere}) are used for the purpose of the calculation of the magnetic flux density $\vec{B}$. In this context, only the magnetized volume is regarded, because the magnitude $M$ of the magnetization density is zero for the internal and exterior space, generating no contribution to the $\vec{B}$ -field:
\begin{equation}\label{BfieldhollowSphereConst1App}
\begin{split}
\begin{gathered}
\underline{\underline{\vec{B}_{M}\left(r,\,\theta\right)}}= \mu_{0} \left( \vec{H}_{M}+\vec{M}_{M}\right)\underset{(\ref{defM})}{\overset{(\ref{magneticfield})}{=}}\mu_{0} \left( -\vec{\nabla}\phi_{M}+M\vec{e}_{z}\right)
\\[14pt]
\overset{(\ref{phimSolutionHollowSphere})}{=}  \mu_{0}\left( \dfrac{M}{3}\left(-\underbrace{\vec{\nabla}(r\cos{\left(\theta\right)})}_{=\vv{\nabla}z=\vv{e}_{z}\frac{d}{dz}z=\vv{e}_{z}}+R_{1}^{3}\cdot\underbrace{\vec{\nabla}(r^{-2}\cos{\left(\theta\right)})}_{=\vv{e}_{r}\cos{\left(\theta\right)}\frac{d}{dr}r^{-2}+\vv{e}_{\theta}\frac{1}{r^{3}}\frac{d}{d\theta}\cos{\left(\theta\right)}}\right)+M\vec{e}_{z}\right)
\\\\
=\mu_{0} \dfrac{M}{3}\left(-\vec{e}_{z}-\left(\dfrac{R_{1}}{r}\right)^{3}\left[2\cos{\left(\theta\right)}\vec{e}_{r}+\sin{\left(\theta\right)}\vec{e}_{\theta}\right]+3\vec{e}_{z}\right)
\\[14pt]
\underline{\underline{=\mu_{0} \dfrac{M}{3}\left(2\vec{e}_{z}-\left(\dfrac{R_{1}}{r}\right)^{3}\left[2\cos{\left(\theta\right)}\vec{e}_{r}+\sin{\left(\theta\right)}\vec{e}_{\theta}\right]\right)}}
\end{gathered}
\end{split}
\end{equation}

Eventually, one inserts (\ref{BfieldhollowSphereConst1App}) and (\ref{defM}) into (\ref{E_WW_DIP}), receiving the \textit{Breit-Hartree} contribution, by conducting the integration over the volume of the hollow sphere. In this context, \textit{Fubini's} theorem is applicable and one has to insert the volume element in spherical coordinates $d\Omega=r^{2}\sin{\left(\theta\right)}dr d\theta d\varphi$, including the \textit{Jacobian} determinant:
\begin{equation*}\label{EHBHollowSphereMConstApp0}
\begin{split}
\begin{gathered}
\underline{\underline{\Delta E_{BH}}}\overset{(\ref{E_WW_DIP})}{=}  -\int\limits_{\Omega} \vec{M}_{M}\cdot\vec{B}_{M}\,d\Omega
\end{gathered}
\end{split}
\end{equation*}
\begin{equation}\label{EHBHollowSphereMConstApp}
\begin{split}
\begin{gathered}
\underset{(\ref{defM})}{\overset{(\ref{BfieldhollowSphereConst1App})}{=}}-\mu_{0} \dfrac{M^{2}}{3}\int\limits_{\Omega} \left(2\underbrace{\vec{e}_{z}\cdot\vec{e}_{z}}_{=\vert\vv{e}_{z}\vert=1}-\left(\dfrac{R_{1}}{r}\right)^{3}\left[2\cos{\left(\theta\right)}\underbrace{\vec{e}_{z}\cdot\vec{e}_{r}}_{=\cos{\left(\theta\right)}}+\sin{\left(\theta\right)}\underbrace{\vec{e}_{z}\cdot\vec{e}_{\theta}}_{=-\sin{\left(\theta\right)}}\right]\right)d\Omega
\\[13pt]
=-\mu_{0} \dfrac{M^{2}}{3}\left[2\underbrace{\int\limits_{\Omega}d\Omega}_{=vol\left(\Omega\right)}- \int\limits_{\varphi=0}^{2\pi}d\varphi\int\limits_{r=R_{1}}^{\textup{R}_{2}}\left(\dfrac{R_{1}}{r}\right)^{3}r^{2}dr\underbrace{\int\limits_{\theta=0}^{\pi}\left(2\cos^{2}{\left(\theta\right)}-\sin^{2}{\left(\theta\right)} \right)\sin{\left(\theta\right)} d\theta}_{=\left[\cos^{2}{\left(\theta\right)}\sin{\left(\theta\right)}\right]_{0}^{\pi}=\,0}\right]
\\[16pt]
=-\mu_{0} \dfrac{M^{2}}{3}\left[2\cdot vol\left(\Omega\right)-0\right]=-\dfrac{2}{3}\mu_{0}M^{2}vol\left(\Omega\right) =\underline{\underline{ -\dfrac{8\pi}{9}\mu_{0}M^{2}\left(R_{2}^{3}-R_{1}^{3}\right)}}
\end{gathered}
\end{split}
\end{equation}

\clearpage

\subsection{Solution of the \textit{Poisson} equation for a hollow sphere with radially dependent magnetization density}\label{AppSolutionPoisson}

The \textit{Poisson} equation from (\ref{PoissonGL1}) has to be solved for the magnetized volume $\Omega=r\,\in\,\left[R_{1},R_{2}\right]\times\theta\,\in\,\left[0,\pi\right]\times\varphi\,\in\,\left[0,2\pi\right)$, leading to the magnetic potential $\phi_{M}\left(r,\,\theta\right)$. Therefore the \textit{Laplace} operator is written in spherical coordinates, dropping any derivative with respect to $\varphi$, because of the $\varphi$ -independence of the magnetic potential. Subsequently, the \textit{ansatz} $\phi_{M}\left(r,\,\theta\right)=f(r)\cos{\left(\theta\right)}$ from (\ref{MAnsatz1}) is inserted, which leads to an inhomogeneous \textit{Euler} differential equation for the function $f(r)$:
\begin{equation}\label{AppPoissonGL1}
\begin{split}
\begin{gathered}
\Delta\,\phi_{M}\left(r,\,\theta\right)= \vec{\nabla}\cdot \vec{M} \overset{(\ref{rhom(r)})}{=}\dfrac{d\,M(r)}{dr}\cos{\left(\theta\right)}
\\\\
\longleftrightarrow\;\;\dfrac{1}{r^{2}}\dfrac{\partial}{\partial r}\left(r^{2}\dfrac{\partial\phi_{M}\left(r,\,\theta\right)}{\partial r}\right)+ \dfrac{1}{r^{2}\sin{\left(\theta\right)}}\dfrac{\partial}{\partial \theta}\left(\sin{\left(\theta\right)}\dfrac{\partial\phi_{M}\left(r,\,\theta\right)}{\partial \theta}\right) =\dfrac{d M(r)}{dr}\cos{\left(\theta\right)}
\\\\
\xlongleftrightarrow{(\ref{MAnsatz1})}\;\;\dfrac{\cos{\left(\theta\right)}}{r^{2}}\underbrace{\dfrac{d}{d r}\left(r^{2}\dfrac{d\, f\left(r\right)}{d r}\right)}_{=2r\frac{d\, f(r)}{d r}+r^{2}\frac{d^{2}\, f\left(r\right)}{d r^{2}}}+ \dfrac{ f(r)}{r^{2}\sin{\left(\theta\right)}}\dfrac{d}{d \theta}\left(\sin{\left(\theta\right)}\underbrace{\dfrac{d\cos{\left(\theta\right)}}{d \theta}}_{=-\sin{\left(\theta\right)}}\right) =\dfrac{d M(r)}{dr}\cos{\left(\theta\right)}
\\\\
\longleftrightarrow\;\; \cos{\left(\theta\right)} \left(\dfrac{2\cancel{r}}{r^{\cancel{2}}}\dfrac{d\, f(r)}{d r}+ \dfrac{\cancel{r^{2}}}{\cancel{r^{2}}}\dfrac{d^{2}\, f(r)}{d r^{2}}\right)-\dfrac{ f\left(r\right)}{r^{2}\sin{\left(\theta\right)}}\underbrace{\dfrac{d}{d \theta}\left(\sin^{2}{\left(\theta\right)}\right)}_{=2\sin{\left(\theta\right)}\cos{\left(\theta\right)}} =\dfrac{d M(r)}{dr}\cos{\left(\theta\right)}
\\\\
\longleftrightarrow\;\; \cancel{\cos{\left(\theta\right)}} \left(\dfrac{2}{r}\dfrac{d\, f(r)}{d r}+ \dfrac{d^{2}\, f(r)}{d r^{2}}\right)-\dfrac{ 2f\left(r\right)\cancel{\sin{\left(\theta\right)}}\cancel{\cos{\left(\theta\right)}}}{r^{2}\cancel{\sin{\left(\theta\right)}}}  =\dfrac{d M(r)}{dr}\cancel{\cos{\left(\theta\right)}}
\\\\
\longleftrightarrow\;\;  \dfrac{d^{2}\, f(r)}{d r^{2}}+\dfrac{2}{r}\dfrac{d\, f(r)}{d r}   -\dfrac{ 2  }{r^{2} } f(r)=\dfrac{d M(r)}{dr} 
\\\\
\longleftrightarrow\;\; r^{2}\, \dfrac{d^{2}\, f(r)}{d r^{2}}+2r\,\dfrac{d\, f(r)}{d r} -2 f(r) =r^{2}\,\dfrac{d M(r)}{dr} \;\;\longleftrightarrow\;\;r^{2} f^{\,\prime\prime}+2rf^{\,\prime} -2 f  =r^{2}M^{\,\prime}
\end{gathered}
\end{split}
\end{equation}
In order to solve the \textit{Euler} differential equation the following substitution is introduced for $r>0$:
\begin{equation}\label{AppSubstitutionEuler}
\begin{split}
\begin{gathered}
r=e^{z}\,;\;\;z=\ln{(r)}\,;\;\;\dfrac{dr}{dz}=e^{z}=r\,;\;\;\dfrac{df}{dz}=\dfrac{df}{dr}\dfrac{dr}{dz}=f^{\,\prime}r\,;
\\\\
\dfrac{d^{\,2}f}{dz^{\,2}}=\dfrac{d}{dz}\left(f^{\,\prime}r\right)=\dfrac{d\left(f^{\,\prime}r\right)}{dr}\dfrac{dr}{dz}=\left(f^{\,\prime\prime}r+f^{\,\prime}\right)r=r^{2}f^{\,\prime\prime}+rf^{\,\prime}
\end{gathered}
\end{split}
\end{equation}
With the substitution (\ref{AppSubstitutionEuler}) the homogeneous \textit{Euler} differential equation from (\ref{AppPoissonGL1}) becomes an ordinary linear differential equation with constant coefficients:
\begin{equation}\label{AppODE1}
\begin{split}
\begin{gathered}
\xlongrightarrow{(\ref{AppPoissonGL1})}\;\;\;r^{2} f^{\,\prime\prime}+rf^{\,\prime}+rf^{\,\prime} -2 f  =0\;\;\;\xlongleftrightarrow{(\ref{AppSubstitutionEuler})}\;\;\;\dfrac{d^{\,2}f}{dz^{\,2}}+\dfrac{df}{dz}-2f=0
\end{gathered}
\end{split}
\end{equation}
(\ref{AppODE1}) is solved with the help of its characterictic second order polynomial $P_{2}(\lambda)$, which leads to the homogeneous solution $f_{hom}(r)$, using back substitution according to (\ref{AppSubstitutionEuler}):
\begin{equation}\label{AppODE2}
\begin{split}
\begin{gathered}
\xlongrightarrow{(\ref{AppODE1})}\;\;\;P_{2}(\lambda)=\lambda^{2}+\lambda-2=0\;\;\;\longrightarrow\;\;\;\lambda_{1}=1\,;\;\lambda_{2}=-2
\\
\scalebox{2}{\rotatebox[origin=c]{180}{$\Lsh$}}\;\;\;f_{hom}(r)=\tilde{C}_{1}e^{\lambda_{1}z}+\tilde{C}_{2}e^{\lambda_{2}z}=\tilde{C}_{1}e^{z}+\tilde{C}_{2}e^{-2z}
\\
\xlongrightarrow{(\ref{AppSubstitutionEuler})}\;\;\;\underline{f_{hom}(r)=\tilde{C}_{1}r+\tilde{C}_{2}r^{-2}}
\end{gathered}
\end{split}
\end{equation}

Applying the principle of variation of constants one can obtain a particular solution $f_{part}(r)$ for the inhomogeneity. Therefore, one makes the \textit{ansatz} for the variation of constants, calculates the required derivatives and inserts them into the inhomogeneous differential equation from (\ref{AppPoissonGL1}), while demanding that all derivatives of the varied constants vanish:
\begin{equation}\label{AppODE3}
\begin{split}
\begin{gathered}
f_{part}(r)=C_{1}(r)r+C_{2}(r)r^{-2}\,;\;\;f^{\,\prime}_{part}(r)=C^{\,\prime}_{1}r+C_{1}+C^{\,\prime}_{2}r^{-2}-2r^{-3}C_{2}
\\
\scalebox{2}{\rotatebox[origin=c]{180}{$\Lsh$}}
\;\;\underline{C^{\,\prime}_{1}r +C^{\,\prime}_{2}r^{-2}\overset{!}{=}0}\; \;\;\longrightarrow \;\;\;f^{\,\prime}_{part}(r)= C_{1} -2r^{-3}C_{2}
\,;
\\
f^{\,\prime\prime}_{part}(r)= C^{\,\prime}_{1} +6r^{-4}C_{2}-2r^{-3}C^{\,\prime}_{2}
\\\\
\xlongrightarrow{(\ref{AppPoissonGL1})}\;\;r^{2} f_{part}^{\,\prime\prime}+2rf_{part}^{\,\prime} -2 f_{part}  =r^{2}M^{\,\prime}
\\
\longleftrightarrow\;\;C^{\,\prime}_{1} +6r^{-4}C_{2}-2r^{-3}C^{\,\prime}_{2}+\dfrac{2\cancel{r}}{r^{\cancel{2}}} \left(C_{1} -2r^{-3}C_{2}\right)-\dfrac{2}{r^{2}} \left(C_{1} r+C_{2} r^{-2}\right) =\dfrac{\cancel{r^{2}}}{\cancel{r^{2}}}M^{\,\prime}
\\
\longleftrightarrow\;\;C^{\,\prime}_{1}-2r^{-3}C^{\,\prime}_{2} +\underbrace{6r^{-4}C_{2}+2r^{-1}C_{1} -4r^{-4}C_{2} -4r^{-1} C_{1}-2 r^{-4}C_{2} }_{=0} = M^{\,\prime}
\\
\scalebox{2}{\rotatebox[origin=c]{180}{$\Lsh$}}
\;\;\underline{C^{\,\prime}_{1}-2r^{-3}C^{\,\prime}_{2}=M^{\,\prime}}
\end{gathered}
\end{split}
\end{equation}
The two underlined conditions determine the functions $C^{\,\prime}_{1}(r)$ and $C^{\,\prime}_{2}(r)$ unambiguously, so that the particular solution can be found from the \textit{ansatz} in (\ref{AppODE3}), whereat the system of linear equations has to be solved by multiplying it with the inverse matrix followed by  integration with respect to $r$ from $R_{1}$ to $r$, in order to receive $C_{1}(r)$ and $C_{2}(r)$:
\begin{equation*}\label{AppODE4a}
\begin{split}
\begin{gathered}
\begin{pmatrix}
r&r^{-2}\\
1&-2r^{-3}
\end{pmatrix}
\begin{pmatrix}
C^{\,\prime}_{1}(r)\\
C^{\,\prime}_{2}(r)
\end{pmatrix}=
\begin{pmatrix}
0\\
M^{\,\prime}(r)
\end{pmatrix}
\end{gathered}
\end{split}
\end{equation*}
\begin{equation*}\label{AppODE4b}
\begin{split}
\begin{gathered}
\scalebox{2}{\rotatebox[origin=c]{180}{$\Lsh$}}
\;\;\begin{pmatrix}
C^{\,\prime}_{1}(r)\\
C^{\,\prime}_{2}(r)
\end{pmatrix}=
-\dfrac{r^{2}}{3}\begin{pmatrix}
-2r^{-3}&-r^{-2}\\
-1&r
\end{pmatrix}
\begin{pmatrix}
0\\
M^{\,\prime}(r)
\end{pmatrix}=
\dfrac{1}{3}\begin{pmatrix}
M^{\,\prime}(r)\\
-r^{3}M^{\,\prime}(r)
\end{pmatrix}
\end{gathered}
\end{split}
\end{equation*}
\begin{equation}\label{AppODE5}
\begin{split}
\begin{gathered}
C_{1}(r)=\dfrac{1}{3}\int\limits^{r}_{R_{1}} M^{\,\prime}(\tilde{r}) \,d\tilde{r}=\dfrac{1}{3}\left(M(r)-M(R_{1})\right)
\\
C_{2}(r)=-\dfrac{1}{3}\int\limits^{r}_{R_{1}} \tilde{r}^{\,3}M^{\,\prime}(\tilde{r})\,d\tilde{r}\;\overset{I.b.P.}{=}\;-\dfrac{1}{3}\left(r^{3}M(r)-R_{1}^{3}M(R_{1})-3\int\limits^{r}_{R_{1}} \tilde{r}^{\,2}M(\tilde{r})\,d\tilde{r}\right)
\\\\
\xlongrightarrow{(\ref{AppODE3})}\;\;f_{part}(r)=C_{1}(r)r+C_{2}(r)r^{-2}
\\
\underline{f_{part}(r)}=\dfrac{r}{3}\left(M(r)-M(R_{1})\right)-\dfrac{1}{3r^{2}}\left(r^{3}M(r)-R_{1}^{3}M(R_{1})-3\int\limits^{r}_{R_{1}} \tilde{r}^{\,2}M(\tilde{r})\,d\tilde{r}\right)
\\
=\dfrac{1}{3}\left(rM(r)-rM(R_{1})-rM(r)+\dfrac{R_{1}^{3}M(R_{1})}{r^{2}}+\dfrac{3}{r^{2}}\int\limits^{r}_{R_{1}} \tilde{r}^{\,2}M(\tilde{r})\,d\tilde{r}\right)
\\
=\underline{-\dfrac{rM(R_{1})}{3} +\dfrac{R_{1}^{3}M(R_{1})}{3r^{2}}+\dfrac{1}{r^{2}}\int\limits^{r}_{R_{1}} \tilde{r}^{\,2}M(\tilde{r})\,d\tilde{r} }
\end{gathered}
\end{split}
\end{equation}

Note that for $C_{2}(r)$ an integration by parts was carried out.
The solution of the inhomogeneous \textit{Euler} differential equation (\ref{AppPoissonGL1}) then results from the sum of the homogeneous and the particular solution from (\ref{AppODE2}) and (\ref{AppODE5}):
\begin{equation}\label{Appf(r)solution}
\begin{split}
\begin{gathered}
f(r)=f_{hom}(r)+f_{part}(r)
\\
\scalebox{2}{\rotatebox[origin=c]{180}{$\Lsh$}}
\;\;\underline{f(r)=\;\underset{(\ref{AppODE5})}{\overset{(\ref{AppODE2})}{=}}\;\tilde{C}_{1}r+\dfrac{\tilde{C}_{2}}{r^{2}}-\dfrac{rM(R_{1})}{3} +\dfrac{R_{1}^{3}M(R_{1})}{3r^{2}}+\dfrac{1}{r^{2}}\int\limits^{r}_{R_{1}} \tilde{r}^{\,2}M(\tilde{r})\,d\tilde{r}}
\end{gathered}
\end{split}
\end{equation}
Its correctness can be ensured, by inserting the solution (\ref{Appf(r)solution}) into the \textit{Euler} differential equation (\ref{AppPoissonGL1}), generating a true statement. 
Thereupon, the constants $\tilde{C}_{1}$ and $\tilde{C}_{2}$ are calculated, by means of the boundary conditions from (\ref{LaplacBoundaryApp1}), which are transformed into following set of equations, following the derivation in (\ref{LaplacBoundaryApp2}) and (\ref{LaplacBoundaryApp5}):
\begin{equation}\label{PoissonBoundaryApp1}
\begin{split}
\begin{gathered}
-\dfrac{\partial \phi_{i}}{\partial r}\bigg\vert_{R_{1}}\overset{!}{=}-\dfrac{\partial \phi_{M}}{\partial r}\bigg\vert_{R_{1}}+M\cos{\left(\theta\right)}
\;\;;\;\;
-\dfrac{\partial \phi_{M}}{\partial r}\bigg\vert_{R_{2}}+M\cos{\left(\theta\right)}\overset{!}{=}-\dfrac{\partial \phi_{e}}{\partial r}\bigg\vert_{R_{2}}
\\\\
\phi_{i}\left(R_{1},\,\theta\right)\overset{!}{=}\phi_{M}\left(R_{1},\,\theta\right) \;\;;\;\;
\phi_{M}\left(R_{2},\,\theta\right)\overset{!}{=}\phi_{e}\left(R_{2},\,\theta\right)
\end{gathered}
\end{split}
\end{equation}
The set of equations from (\ref{PoissonBoundaryApp1}) is further simplified by inserting the \textit{ansatz} for the magnetic potential from (\ref{MAnsatz1}), where the equations are trivially fulfilled for $\cos{\left(\theta\right)}=0$ and the division by $\cos{\left(\theta\right)}$ is valid for $\cos{\left(\theta\right)}\neq0$:
\begin{equation*}\label{PoissonBoundaryApp2a}
\begin{split}
\begin{gathered}
-A\cancel{\cos{\left(\theta\right)}}=-\dfrac{d\,f (r)}{d r}\bigg\vert_{R_{1}}+M(r)\cancel{\cos{\left(\theta\right)}}\;\longrightarrow\;A=\dfrac{d\,f (r)}{d r}\bigg\vert_{R_{1}}-M(R_{1})
\\
-\dfrac{d\,f (r)}{d r}\bigg\vert_{R_{2}}\cancel{\cos{\left(\theta\right)}}+M(R_{1})\cancel{\cos{\left(\theta\right)}}=\dfrac{2D}{r^{3}}\cancel{\cos{\left(\theta\right)}}
\end{gathered}
\end{split}
\end{equation*}
\begin{equation}\label{PoissonBoundaryApp2}
\begin{split}
\begin{gathered}
\scalebox{2}{\rotatebox[origin=c]{180}{$\Lsh$}}
\;\;
D =\dfrac{R_{2}^{3}}{2}\left(-\dfrac{d\,f (r)}{d r}\bigg\vert_{R_{2}} +M(R_{2}) \right)
\end{gathered}
\end{split}
\end{equation}
\begin{equation}\label{PoissonBoundaryApp3}
\begin{split}
\begin{gathered}
AR_{1}\cancel{\cos{\left(\theta\right)}}=f(R_{1})\cancel{\cos{\left(\theta\right)}} \;\longrightarrow\;A=\dfrac{f(R_{1})}{R_{1}} 
\\
f(R_{2})\cancel{\cos{\left(\theta\right)}}=\dfrac{D}{R_{2}^{2}}\cancel{\cos{\left(\theta\right)}}\;\longrightarrow\;D=f(R_{2})R_{2}^{2}
\end{gathered}
\end{split}
\end{equation}
The representations for $A$ and $D$ from (\ref{PoissonBoundaryApp2}) and (\ref{PoissonBoundaryApp3}) are equated and (\ref{Appf(r)solution}) is inserted, whereat for the derivative of an integral the  fundamental theorem of calculus applies. First, the constant $\tilde{C}_{1}$ is obtained from the equated representations for $A$: 
\begin{equation}\label{AConstantC2tilde}
\begin{split}
\begin{gathered}
A\,\overset{(\ref{PoissonBoundaryApp2})}{=}\,\dfrac{d\,f (r)}{d r}\bigg\vert_{R_{1}}-M(R_{1})=\dfrac{f(R_{1})}{R_{1}} \,\overset{(\ref{PoissonBoundaryApp3})}{=}\,A
\\
\scalebox{2}{\rotatebox[origin=c]{180}{$\Lsh$}}
\;\; \tilde{C}_{1}-\dfrac{2\tilde{C}_{2}}{R_{1}^{3}}\underbrace{-\dfrac{M(R_{1})}{3} -\dfrac{2\cancel{R_{1}^{3}}M(R_{1})}{3\cancel{R_{1}^{3}}}}_{=\,-M(R_{1})}-\dfrac{2}{R_{1}^{3}}\underbrace{\int\limits^{\textup{R}_{1}}_{R_{1}} \tilde{r}^{\,2}M(\tilde{r})\,d\tilde{r}}_{=\,0}+\underbrace{\dfrac{1}{\cancel{R_{1}^{2}}}\cancel{R_{1}^{2}}M(R_{1})-M(R_{1})}_{=\,0}
\\
=\tilde{C}_{1}+\dfrac{\tilde{C}_{2}}{R_{1}^{3}}\underbrace{-\dfrac{M(R_{1})}{3} +\dfrac{\cancel{R_{1}^{3}}M(R_{1})}{3\cancel{R_{1}^{3}}}}_{=\,0}+\dfrac{1}{R_{1}^{3}}\underbrace{\int\limits^{\textup{R}_{1}}_{R_{1}} \tilde{r}^{\,2}M(\tilde{r})\,d\tilde{r}}_{=\,0} 
\\
\longleftrightarrow\;\; \tilde{C}_{1}-\dfrac{2\tilde{C}_{2}}{R_{1}^{3}}-M(R_{1})=\tilde{C}_{1}+\dfrac{\tilde{C}_{2}}{R_{1}^{3}}\;\; \longrightarrow\;\;\underline{ \tilde{C}_{2} = -\dfrac{M(R_{1})R_{1}^{3}}{3}}
\end{gathered}
\end{split}
\end{equation}
Second, the constant $\tilde{C}_{2}$ is received from the equated representations for $D$: 
\begin{equation}\label{DConstantC1tilde}
\begin{split}
\begin{gathered}
D\,\overset{(\ref{PoissonBoundaryApp2})}{=}\,f(R_{2})R_{2}^{2}=\dfrac{R_{2}^{3}}{2}\left(-\dfrac{d\,f (r)}{d r}\bigg\vert_{R_{2}} +M(R_{2}) \right)\,\overset{(\ref{PoissonBoundaryApp3})}{=}\,D
\\
\scalebox{2}{\rotatebox[origin=c]{180}{$\Lsh$}}
\;\; \dfrac{2}{R_{2}}f(R_{2})-M(R_{2})  =-\dfrac{d\,f (r)}{d r}\bigg\vert_{R_{2}} 
\\
\scalebox{2}{\rotatebox[origin=c]{180}{$\Lsh$}}
\;\; 2\tilde{C}_{1}+\dfrac{2\tilde{C}_{2}}{R_{2}^{3}}-\dfrac{2M(R_{1})}{3} +\dfrac{2R_{1}^{3}M(R_{1})}{3R_{2}^{3}}+\dfrac{2}{R_{2}^{3}}\int\limits^{\textup{R}_{2}}_{R_{1}} \tilde{r}^{\,2}M(\tilde{r})\,d\tilde{r} -M(R_{2}) 
\\
=-\left(\tilde{C}_{1}-\dfrac{2\tilde{C}_{2}}{R_{2}^{3}}-\dfrac{M(R_{1})}{3} -\dfrac{2R_{1}^{3}M(R_{1})}{3R_{2}^{3}}-\dfrac{2}{R_{2}^{3}}\int\limits^{\textup{R}_{2}}_{R_{1}} \tilde{r}^{\,2}M(\tilde{r})\,d\tilde{r}+\dfrac{1}{\cancel{R_{2}^{2}}} \cancel{R_{2}^{2}}M(R_{2})\right) 
\\
\longleftrightarrow\;\;2\tilde{C}_{1} -\dfrac{2M(R_{1})}{3}  
=-\tilde{C}_{1} +\dfrac{M(R_{1})}{3} \;\; \longrightarrow\;\;\underline{ \tilde{C}_{1} =\dfrac{M(R_{1}) }{3}}  
\end{gathered}
\end{split}
\end{equation}
Third, one evaluates the constant $A$, with the help of the relations in (\ref{AConstantC2tilde}) and (\ref{DConstantC1tilde}):
\begin{equation}\label{AConstant}
\begin{split}
\begin{gathered}
\underline{ A}\,\overset{(\ref{AConstantC2tilde})}{=}\,\dfrac{f(R_{1})}{R_{1}} =\tilde{C}_{1}+\dfrac{\tilde{C}_{2}}{R_{1}^{3}}\,\underset{(\ref{DConstantC1tilde})}{\overset{(\ref{AConstantC2tilde})}{=}}\,\underbrace{\dfrac{M(R_{1})}{3}-\dfrac{M(R_{1})\cancel{R_{1}^{3}}}{3\cancel{R_{1}^{3}}}}_{=\,0} =\underline{ 0}
\end{gathered}
\end{split}
\end{equation}
Fourth, the constant $D$ is found from (\ref{DConstantC1tilde}), (\ref{Appf(r)solution}) and (\ref{AConstantC2tilde}):
\begin{equation}\label{DConstant}
\begin{split}
\begin{gathered}
D\;\overset{(\ref{DConstantC1tilde})}{=}\;f(R_{2})R_{2}^{2}
\\
\overset{(\ref{Appf(r)solution})}{=}\tilde{C}_{1}R_{2}^{3}+\dfrac{\tilde{C}_{2}\cancel{R_{2}^{2}}}{\cancel{R_{2}^{2}}}-\dfrac{R_{2}^{3}M(R_{1})}{3} +\dfrac{R_{1}^{3}M(R_{1})\cancel{R_{2}^{2}}}{3\cancel{R_{2}^{2}}}+\dfrac{\cancel{R_{2}^{2}}}{\cancel{R_{2}^{2}}}\int\limits^{\textup{R}_{2}}_{R_{1}} \tilde{r}^{\,2}M(\tilde{r})\,d\tilde{r}
\\
\underset{(\ref{DConstantC1tilde})}{\overset{(\ref{AConstantC2tilde})}{=}}\,\underbrace{\dfrac{M(R_{1}) }{3}R_{2}^{3}-\dfrac{M(R_{1})R_{1}^{3}}{3}-\dfrac{R_{2}^{3}M(R_{1})}{3} +\dfrac{R_{1}^{3}M(R_{1})}{3}}_{=\,0}+\int\limits^{\textup{R}_{2}}_{R_{1}} \tilde{r}^{\,2}M(\tilde{r})\,d\tilde{r}
\\
\scalebox{2}{\rotatebox[origin=c]{180}{$\Lsh$}}
\;\;\underline{D=\int\limits^{\textup{R}_{2}}_{R_{1}} \tilde{r}^{\,2}M(\tilde{r})\,d\tilde{r}}
\end{gathered}
\end{split}
\end{equation}
One updates the solution $f(r)$ from (\ref{Appf(r)solution}) with the expressions for the constants $\tilde{C}_{1}$ and $\tilde{C}_{2}$ from (\ref{DConstantC1tilde}) and (\ref{AConstantC2tilde}):
\begin{equation}\label{AppUPDATEf(r)solution}
\begin{split}
\begin{gathered}
f(r)=\dfrac{rM(R_{1})}{3}-\dfrac{R_{1}^{3}M(R_{1})}{3r^{2}}-\dfrac{rM(R_{1})}{3} +\dfrac{R_{1}^{3}M(R_{1})}{3r^{2}}+\dfrac{1}{r^{2}}\int\limits^{r}_{R_{1}} \tilde{r}^{\,2}M(\tilde{r})\,d\tilde{r}
\\
\longleftrightarrow\;\; \underline{f(r)=\dfrac{1}{r^{2}}\int\limits^{r}_{R_{1}} \tilde{r}^{\,2}M(\tilde{r})\,d\tilde{r}}
\end{gathered}
\end{split}
\end{equation}
The correctness of the updated solution (\ref{AppUPDATEf(r)solution}) is verified, with the aid of the inhomogeneous \textit{Euler} differential equation (\ref{AppPoissonGL1}):
\begin{equation}\label{AppCheckf(r)}
\begin{split}
\begin{gathered}
\xlongrightarrow{(\ref{AppPoissonGL1})}\;\;\dfrac{d^{2}\, f(r)}{d r^{2}}+\dfrac{2}{r}\dfrac{d\, f(r)}{d r}   -\dfrac{ 2  }{r^{2} } f(r)=\dfrac{d M(r)}{dr} 
\\
\scalebox{2}{\rotatebox[origin=c]{180}{$\Lsh$}}
\;\;\dfrac{6}{r^{4}}\int\limits^{r}_{R_{1}} \tilde{r}^{\,2}M(\tilde{r})\,d\tilde{r}-\dfrac{2}{r^{\cancel{3}}}\cancel{r^{2}}M(r)+\dfrac{d M(r)}{dr} +\dfrac{2}{r}\left(-\dfrac{2}{r^{3}}\int\limits^{r}_{R_{1}} \tilde{r}^{\,2}M(\tilde{r})\,d\tilde{r}+\dfrac{1}{\cancel{r^{2}}}\cancel{r^{2}}M(r)\right) 
\\
 -\dfrac{ 2  }{r^{2} } \dfrac{1}{r^{2}}\int\limits^{r}_{R_{1}} \tilde{r}^{\,2}M(\tilde{r})\,d\tilde{r}=\dfrac{d M(r)}{dr} 
\\
\longleftrightarrow\;\;\dfrac{d M(r)}{dr}+\underbrace{\left(\dfrac{6}{r^{4}} -\dfrac{4}{r^{4}}- \dfrac{2}{r^{4}}\right)\int\limits^{r}_{R_{1}} \tilde{r}^{\,2}M(\tilde{r})\,d\tilde{r}-\dfrac{2M(r)}{r}+\dfrac{2M(r)}{r}  }_{=\,0}=\dfrac{d M(r)}{dr} \;\;\;\;\;\;\blacksquare
\end{gathered}
\end{split}
\end{equation}
Eventually, the solution $\phi_{M}\left(r,\,\theta\right)$ of the \textit{Poisson} equation for a hollow sphere with radially dependent magnetization density is given by (\ref{MAnsatz1}) and (\ref{AppUPDATEf(r)solution}):
\begin{equation}\label{AppfPoissonPhiMsolution}
\underline{\phi_{M}\left(r,\,\theta\right)}=f(r)\cos{\left(\theta\right)}=\underline{\dfrac{\cos{\left(\theta\right)}}{r^{2}}\int\limits^{r}_{R_{1}} \tilde{r}^{\,2}M(\tilde{r})\,d\tilde{r}}
\end{equation}
The complete magnetic potential satisfies the \textit{Laplace} equation for the exterior of
the hollow sphere as well as its internal space, and the \textit{Poisson} equation within the magnetized volume $\Omega$ of the hollow sphere, so that the one finds with the help of (\ref{MAnsatz1}), (\ref{AConstant}), (\ref{DConstant}) and (\ref{AppfPoissonPhiMsolution}):
\begin{equation}\label{AppFullPotentialHollowM(r)}
\underline{\underline{\phi_{m}\left(\vec{r}\,\right)= 
\begin{cases}
\phi_{i}\left(r,\,\theta\right)\;\;\underset{(\ref{AConstant})}{\overset{(\ref{MAnsatz1})}{=}}\;\;0\,\textup{A} &;\;\;\; \forall \; \vec{r}\,\in\,\left\lbrace\vec{r}\,\notin\,\Omega\,\vert\,0\leq r\leq R_{1}\right\rbrace
\\
\phi_{M}\left(r,\,\theta\right)\;\underset{}{\overset{(\ref{AppfPoissonPhiMsolution})}{=}}\;\dfrac{\cos{\left(\theta\right)}}{r^{2}}\displaystyle{\int\limits^{r}_{R_{1}}} \tilde{r}^{\,2}M(\tilde{r})\,d\tilde{r}&;\;\;\; \forall \; \vec{r}\,\in\,\Omega
\\
\phi_{e}\left(r,\,\theta\right)\;\;\underset{(\ref{DConstant})}{\overset{(\ref{MAnsatz1})}{=}}\;\;\dfrac{\cos{\left(\theta\right)}}{r^{2}}\displaystyle{\int\limits^{\textup{R}_{2}}_{R_{1}}} \tilde{r}^{\,2}M(\tilde{r})\,d\tilde{r}&;\;\;\; \forall \;\vec{r}\,\in\,\left\lbrace\vec{r}\,\notin\,\Omega\,\vert\,R_{2}\leq r<\infty \right\rbrace
\end{cases}}}
\end{equation}

The potential of a full sphere with $R_{1}=0\textup{m}$ and $R:=R_{2}$ reads:
\begin{equation}\label{AppFullPotentialfullsphereM(r)}
\phi_{m,R}\left(\vec{r}\,\right)= 
\begin{cases}
\phi_{M,R}\left(r,\,\theta\right)\;\underset{}{\overset{(\ref{AppFullPotentialHollowM(r)})}{=}}\;\dfrac{\cos{\left(\theta\right)}}{r^{2}}\displaystyle{\int\limits^{r}_{0}} \tilde{r}^{\,2}M(\tilde{r})\,d\tilde{r}&;\;\;\; \forall \; \vec{r}\,\in\,\Omega
\\
\phi_{e,R}\left(r,\,\theta\right)\;\underset{}{\overset{(\ref{AppFullPotentialHollowM(r)})}{=}}\;\;\dfrac{\cos{\left(\theta\right)}}{r^{2}}\displaystyle{\int\limits^{R}_{0}} \tilde{r}^{\,2}M(\tilde{r})\,d\tilde{r}&;\;\;\; \forall \;\vec{r}\,\notin\,\Omega
\end{cases}
\end{equation}
The solution for the hollow sphere is given, analogously to the case with constant magnetization density from (\ref{phimSolutionInterpet}), by the subtraction of the potentials $\phi_{m,R_{2}}$ and $\phi_{m,R_{1}}$ of two full spheres with radii $R_{1}$ and $R_{2}$. 
\begin{equation}\label{AppphimHollowM(r)Interpet}
\begin{split}
\begin{gathered}
\phi_{m}\left(\vec{r}\,\right)= 
\begin{cases}
\phi_{i}\left(r,\,\theta\right)=\phi_{M,R_{2}}-\phi_{M,R_{1}} = 0 \,\textup{A}
\\
\phi_{M}\left(r,\,\theta\right)=\phi_{M,R_{2}}-\phi_{e,R_{1}}=\dfrac{\cos{\left(\theta\right)}}{r^{2}}\left(\displaystyle{\int\limits^{r}_{0}} \tilde{r}^{\,2}M(\tilde{r})\,d\tilde{r}-\displaystyle{\int\limits^{\textup{R}_{1}}_{0}} \tilde{r}^{\,2}M(\tilde{r})\,d\tilde{r}\right)
\\
\phi_{e}\left(r,\,\theta\right)=\phi_{e,R_{2}}-\phi_{e,R_{1}}=\dfrac{\cos{\left(\theta\right)}}{r^{2}}\left(\displaystyle{\int\limits^{\textup{R}_{2}}_{0}} \tilde{r}^{\,2}M(\tilde{r})\,d\tilde{r}-\displaystyle{\int\limits^{\textup{R}_{1}}_{0}} \tilde{r}^{\,2}M(\tilde{r})\,d\tilde{r}\right)
\end{cases}
\\\\
= 
\begin{cases}
\phi_{i}\left(r,\,\theta\right)\;=\;0\,\textup{A} 
\\
\phi_{M}\left(r,\,\theta\right)=\dfrac{\cos{\left(\theta\right)}}{r^{2}}\left(\displaystyle{\int\limits^{\textup{R}_{1}}_{0}} \tilde{r}^{\,2}M(\tilde{r})\,d\tilde{r}+\displaystyle{\int\limits^{r}_{0}} \tilde{r}^{\,2}M(\tilde{r})\,d\tilde{r}\right)
=\displaystyle{\int\limits_{R_{1}}^{r}} \tilde{r}^{\,2}M(\tilde{r})\,d\tilde{r}
\\
\phi_{e}\left(r,\,\theta\right)=\dfrac{\cos{\left(\theta\right)}}{r^{2}}\left(\displaystyle{\int\limits^{\textup{R}_{1}}_{0}} \tilde{r}^{\,2}M(\tilde{r})\,d\tilde{r}+\displaystyle{\int\limits^{\textup{R}_{2}}_{0}} \tilde{r}^{\,2}M(\tilde{r})\,d\tilde{r}\right)=\displaystyle{\int\limits_{R_{1}}^{\textup{R}_{2}}} \tilde{r}^{\,2}M(\tilde{r})\,d\tilde{r}
\end{cases}
\end{gathered}
\end{split}
\end{equation}

One can show that the special case of constant magnetization density (\ref{phimSolutionHollowSphere}) is contained in the solution for a radially dependent magnetization density (\ref{AppFullPotentialfullsphereM(r)}), by inserting $M(r)=M=\textup{const.}$ into (\ref{phimSolutionHollowSphere}):
\begin{equation}\label{AppM(r)containsM}
\begin{split}
\begin{gathered}
\phi_{m}\left(\vec{r}\,\right)= 
\begin{cases}
\phi_{i}\left(r,\,\theta\right)\;=\;0\,\textup{A}  
\\ 
\phi_{M}\left(r,\,\theta\right)=\dfrac{M\cos{\left(\theta\right)}}{r^{2}}\displaystyle{\int\limits^{r}_{R_{1}}} \tilde{r}^{\,2} \,d\tilde{r}=\dfrac{M\cos{\left(\theta\right)}}{3r^{2}}\left(r^{3}-R_{1}^{3}\right)
\\ 
\phi_{e}\left(r,\,\theta\right)=\dfrac{M\cos{\left(\theta\right)}}{r^{2}}\displaystyle{\int\limits^{\textup{R}_{2}}_{R_{1}}} \tilde{r}^{\,2} \,d\tilde{r}=\dfrac{M\cos{\left(\theta\right)}}{3r^{2}}\left(R_{2}^{3}-R_{1}^{3}\right)
\end{cases}
\\\\
=\begin{cases}
\phi_{i}\left(r,\,\theta\right)\;=\;0\,\textup{A} 
\\ 
\phi_{M}\left(r,\,\theta\right)=\dfrac{M}{3}\left(r-\dfrac{R_{1}^{3}}{r^{2}}\right)\cos{\left(\theta\right)}
\\ 
\phi_{e}\left(r,\,\theta\right)=\dfrac{M}{3}\left(R_{2}^{3}-R_{1}^{3}\right)\dfrac{\cos{\left(\theta\right)}}{r^{2}}
\end{cases}
\end{gathered}
\end{split}
\end{equation}

\clearpage

\subsection{Evaluation of the \textit{Breit-Hartree} contribution for a hollow sphere $\Omega$ with a radially dependent magnetization density} \label{AppendixEHBHollowSphereRadial}

One uses the material equation, (\ref{magneticfield}), (\ref{defM(r)}) and the magnetic potential from (\ref{PotentialHollowM(r)}), for the purpose of the calculation of the magnetic flux density $\vec{B}$. In this regard, only the magnetized volume is considered, because the magnitude $M(r)$ of the magnetization density is zero for the internal and exterior space, generating no contribution to the $\vec{B}$ -field. Under the use of the \textit{Nabla} operator in spherical coordinates all derivatives with respect to $\varphi$ drop, because the potential is exclusively a function in $r$ and $\theta$. Consequently, one finds:
\begin{equation*}\label{BfieldhollowSphereM(r)0App}
\begin{split}
\begin{gathered}
 \vec{B}_{M}\left(r,\,\theta\right) = \mu_{0} \left[ \vec{H}_{M}+\vec{M}_{M}\right]\underset{(\ref{defM(r)})}{\overset{(\ref{magneticfield})}{=}}\mu_{0} \left[ -\vec{\nabla}\phi_{M}+M\vec{e}_{z}\right]
\\[13pt]
\overset{(\ref{PotentialHollowM(r)})}{=}  \mu_{0}\left( -\vec{\nabla}\left(\dfrac{\cos{\left(\theta\right)}}{r^{2}}\displaystyle{\int\limits^{r}_{R_{1}}} \tilde{r}^{\,2}M(\tilde{r})\,d\tilde{r}\right)+M(r)\vec{e}_{z}\right)
\\[12pt]
=\mu_{0}\left[-\vec{e}_{r}\cos{\left(\theta\right)}\dfrac{d}{dr}\left(\dfrac{1}{r^{2}} \int\limits^{r}_{R_{1}} \tilde{r}^{\,2}M(\tilde{r})\,d\tilde{r}\right) -\vec{e}_{\theta}\dfrac{1}{r^{3}}\int\limits^{r}_{R_{1}} \tilde{r}^{\,2}M(\tilde{r})\,d\tilde{r}\underbrace{\dfrac{d}{d\theta} \left(\cos{\left(\theta\right)}\right)}_{=-\sin{\left(\theta\right)}}+M(r)\vec{e}_{z}\right]
\\[14pt]
=\mu_{0}\left[ -\vec{e}_{r}\cos{\left(\theta\right)}\left(-\dfrac{2}{r^{3}} \int\limits^{r}_{R_{1}} \tilde{r}^{\,2}M(\tilde{r})\,d\tilde{r}+\dfrac{1}{\cancel{r^{2}}} \cancel{r^{2}}M(r)\right) +\vec{e}_{\theta}\dfrac{\sin{\left(\theta\right)}}{r^{3}}\int\limits^{r}_{R_{1}} \tilde{r}^{\,2}M(\tilde{r})\,d\tilde{r} +M(r)\vec{e}_{z}\right]
\end{gathered}
\end{split}
\end{equation*}\vspace{1mm}
\begin{equation}\label{BfieldhollowSphereM(r)1App}
\begin{split}
\begin{gathered}
\underline{\underline{ \vec{B}_{M}\left(r,\,\theta\right)=\mu_{0}\left[ M(r)\left(\vec{e}_{z}-\vec{e}_{r}\cos{\left(\theta\right)}\right)+\dfrac{1}{r^{3}} \int\limits^{r}_{R_{1}} \tilde{r}^{\,2}M(\tilde{r})\,d\tilde{r}\left(2\cos{\left(\theta\right)}\vec{e}_{r}+ \sin{\left(\theta\right)}\vec{e}_{\theta}\right)\right]}}
\end{gathered}
\end{split}
\end{equation}

Eventually, one inserts (\ref{BfieldhollowSphereM(r)1App}) and (\ref{defM(r)}) into (\ref{E_WW_DIP}), receiving the \textit{Breit-Hartree} contribution, by integrating over the volume of the hollow sphere $\Omega=r\,\in\,\left[R_{1},R_{2}\right]\times\theta\,\in\,\left[0,\pi\right]\times\varphi\,\in\,\left[0,2\pi\right)$. In this context, \textit{Fubini's} theorem is applicable and one has to insert the volume element in spherical coordinates $d\Omega=r^{2}\sin{\left(\theta\right)}dr d\theta d\varphi$, including the \textit{Jacobian} determinant:
\begin{equation*}\label{EHBHollowSphereM(r)App1}
\begin{split}
\begin{gathered}
 \Delta E_{BH} \overset{(\ref{E_WW_DIP})}{=}  -\int\limits_{\Omega} \vec{M}_{M}\cdot\vec{B}_{M}\,d\Omega
\\[4pt]
\underset{(\ref{defM(r)})}{\overset{(\ref{BfieldhollowSphereM(r)1App})}{=}}-\mu_{0}  \int\limits_{\Omega} \left(  \left[M(r)\right]^{2}\left(\underbrace{\vec{e}_{z}\cdot\vec{e}_{z}}_{=\vert\vv{e}_{z}\vert=1}-\underbrace{\vec{e}_{z}\cdot\vec{e}_{r}}_{=\cos{\left(\theta\right)}}\cos{\left(\theta\right)}\right)\right.
\end{gathered}
\end{split}
\end{equation*}
\begin{equation*}\label{EHBHollowSphereM(r)App2a}
\begin{split}
\begin{gathered}
\left.+\,\dfrac{M(r)}{r^{3}} \int\limits^{r}_{R_{1}} \tilde{r}^{\,2}M(\tilde{r})\,d\tilde{r}\left(2\cos{\left(\theta\right)}\underbrace{\vec{e}_{z}\cdot\vec{e}_{r}}_{=\cos{\left(\theta\right)}}+ \sin{\left(\theta\right)}\underbrace{\vec{e}_{z}\cdot\vec{e}_{\theta}}_{=-\sin{\left(\theta\right)}}\right) \right)d\Omega
\end{gathered}
\end{split}
\end{equation*}
\begin{equation}\label{EHBHollowSphereM(r)App2}
\begin{split}
\begin{gathered}
=-\mu_{0}  \int\limits_{\Omega} \left( \left[M(r)\right]^{2}\underbrace{\left(1-\cos^{2}{\left(\theta\right)}\right)}_{=\sin^{2}{\left(\theta\right)}}+\dfrac{M(r)}{r^{3}} \int\limits^{r}_{R_{1}} \tilde{r}^{\,2}M(\tilde{r})\,d\tilde{r}\left(2\cos^{2}{\left(\theta\right)} - \sin^{2}{\left(\theta\right)} \right) \right)d\Omega
\\\\
=-\mu_{0}\left[\,\underbrace{\int\limits_{\varphi=0}^{2\pi}d\varphi}_{=\,2\pi}\underbrace{\int\limits_{\theta=0}^{\pi} \sin^{3}{\left(\theta\right)} d\theta}_{=\left[\frac{1}{3}\cos^{3}{(\theta)}-\cos{(\theta)}\right]_{0}^{\pi}=\frac{4}{3}} \int\limits_{r=R_{1}}^{\textup{R}_{2}} r^{2}M^{2}(r)dr\right.
\\
\left.+\,\int\limits_{\varphi=0}^{2\pi}\int\limits_{r=R_{1}}^{\textup{R}_{2}}\left(\dfrac{M(r)r^{2}}{r^{3}} \int\limits^{r}_{R_{1}} \tilde{r}^{\,2}M(\tilde{r})\,d\tilde{r}\right)drd\varphi\underbrace{\int\limits_{\theta=0}^{\pi} \left(2\cos^{2}{(\theta)} - \sin^{2}{(\theta)} \right)\sin{(\theta)} d\theta}_{=\left[\cos^{2}{(\theta)}\sin{(\theta)}\right]_{0}^{\pi}=\,0}\right]
\\\\
\scalebox{2}{\rotatebox[origin=c]{180}{$\Lsh$}}
\;\;\underline{\underline{\Delta E_{BH}}}=-\mu_{0}\left[\,2\pi\cdot\dfrac{4}{3} \cdot\int\limits_{r=R_{1}}^{\textup{R}_{2}} r^{2}\left[M(r)\right]^{2}dr\right]
=\underline{\underline{ -\dfrac{8\pi}{3} \mu_{0}\int\limits_{r=R_{1}}^{\textup{R}_{2}} r^{2}\left[M(r)\right]^{2}dr }}
\end{gathered}
\end{split}
\end{equation}

For the special case of a constant magnetization density $M(r)=M=\textup{const.}$ one receives the already known result from (\ref{EHBHollowSphereMConstApp}):
\begin{equation}\label{SpecialcaseEHBM(r)toMConst}
\Delta E_{BH}=  -\dfrac{8\pi}{3} \mu_{0}\int\limits_{r=R_{1}}^{\textup{R}_{2}} r^{2}M^{2} dr=  -\dfrac{8\pi}{3} \mu_{0}M^{2}\int\limits_{r=R_{1}}^{\textup{R}_{2}} r^{2} dr =-\dfrac{8\pi}{9}\mu_{0}M^{2}\left(R_{2}^{3}-R_{1}^{3}\right)
\end{equation}
\clearpage

\subsection{Evaluation of the \textit{Monte Carlo} approximation of the integral $I_{d}$} \label{CHECKMonteCarlo}

The \textit{Monte Carlo} approximation of the integral $I_{d}$ from (\ref{aPrefactor}) uses the values $I_{d,num.}$ from (\ref{outMonteCarloId}) in subsection \ref{MCNUMApp}:
\begin{equation}\label{CheckMC1}
I_{d}\approx\tilde{I}_{d,num.}= a\cdot I_{d,num.}=\dfrac{\mu_{0}}{8\pi}M^{2}R_{2}^{3}\cdot I_{d,num.}
\end{equation}
The analytical expression for $I_{d}$ for a full sphere from (\ref{DELTAEBHd2}) is then compared with the expression in (\ref{CheckMC1}), in order to find the exact solution for $I_{d,num.}$, which is:
\begin{equation}\label{CheckMC2}
\begin{split}
\begin{gathered}
\Delta E_{BH}^{\,d}=-\dfrac{4\pi}{9}\mu_{0}M^{2} R_{2}^{3}\equiv I_{d}\overset{!}{\approx}\tilde{I}_{d,num.}= a\cdot I_{d,num.}=\dfrac{\mu_{0}}{8\pi}M^{2}R_{2}^{3}\cdot
 I_{d,num.}
\\\\
\scalebox{2}{\rotatebox[origin=c]{180}{$\Lsh$}}
\;\; -\dfrac{4\pi}{9}\cancel{\mu_{0}}\cancel{M^{2} R_{2}^{3}} \overset{!}{\approx}\dfrac{\cancel{\mu_{0}}}{8\pi}\cancel{M^{2}R_{2}^{3}}\cdot
 I_{d,num.}
\\\\
\longleftrightarrow\;\; I_{d,num.} \overset{!}{\approx} -\dfrac{32\pi^{2}}{9} \approx -35.092
\end{gathered}
\end{split}
\end{equation}
Hence, one deduces the order $10^{1}$ from (\ref{CheckMC2}) for the numerical value $I_{d,num.}$ and compares it with the order of the results from (\ref{outMonteCarloId}). Consequently, one finds the correct order of $I_{d,num.}$ for more than $N_{\textup{tot.}}=20^{6}$ sampling points, which emphasizes the preferential treatment of the \textit{Monte Carlo} integration over the \textit{Gauss-Legendre} integration, which can be understood by additionally observing subsection \ref{GaussAppSubSec}.
\clearpage

\subsection{Derivation of the \Cpp implementation of the second order \textit{M\o{}ller-Plesset} correlation energy contribution} \label{AppMP2}

The second order \textit{M\o{}ller-Plesset} correlation energy contribution in GCGS-units according to \cite{Pellegrini_2020} can be written as:
\begin{equation}\label{AppMP2Plesset1}
E_{C,MP2}[n(r)]=-\dfrac{43-46\ln{(2)}}{525}\dfrac{e^{4}\hslash^{2}\left(k_{F}(n(r))\right)^{4}}{2m_{e}^{3}c^{4}\pi^{2}}
\end{equation}    
One finds the MP2-correlation energy contribution to the total energy of a hollow sphere $\Omega$, by inserting the definition of the \textit{Fermi} wave vector $k_{F}(n(r))=(3\pi^{2}n(r))^\frac{1}{3}$ \cite{ashcroft1976solid} and by integration over the volume $\Omega$ in spherical coordinates:
\begin{equation}\label{AppPlesset2}
\begin{split}
\begin{gathered}
\underline{\underline{\Delta E_{C,MP2}}}=\int\limits_{\Omega}E_{C,MP2}[n(r)]\,d\Omega=-\dfrac{43-46\ln{(2)}}{525}\dfrac{e^{4}\hslash^{2}}{2m_{e}^{3}c^{4}\pi^{2}}\int\limits_{\Omega}\left(3\pi^{2}n(r)\right)^{\frac{4}{3}}\,d\Omega
\\\\
=-\dfrac{43-46\ln{(2)}}{525}\dfrac{3^{\frac{4}{3}}\pi^{\frac{\cancel{8}^{\,2}}{3}}e^{4}\hslash^{2}}{2m_{e}^{3}c^{4}\cancel{\pi^{2}}}\underbrace{\int\limits_{0}^{2\pi} d\varphi}_{=\,2\pi} \underbrace{\int\limits_{0}^{\pi} \sin{\left(\theta\right)} \,d\theta}_{=\,2} \int\limits_{\textup{R}_{1}}^{\textup{R}_{2}}r^{2}\left(n(r)\right)^{\frac{4}{3}}\,dr 
\\\\
=-\dfrac{43-46\ln{(2)}}{525}\dfrac{3^{\frac{4}{3}}\cancel{4}^{2}\pi\pi^{\frac{2}{3}}e^{4}\hslash^{2}}{\cancel{2}m_{e}^{3}c^{4} } \int\limits_{\textup{R}_{1}}^{\textup{R}_{2}}r^{2}\left(n(r)\right)^{\frac{4}{3}}\,dr 
\\\\
=\underline{\underline{-\dfrac{43-46\ln{(2)}}{525}\dfrac{2\cdot3^{\frac{4}{3}}\pi^{\frac{5}{3}}e^{4}\hslash^{2}}{m_{e}^{3}c^{4} } \int\limits_{\textup{R}_{1}}^{\textup{R}_{2}}r^{2}\left(n(r)\right)^{\frac{4}{3}}\,dr}}
\end{gathered}
\end{split}
\end{equation}
In order to receive the contribution in SI-units one has to convert the result from (\ref{AppPlesset2}), which is stated in GCGS-units. All quantities and constants are therefore inserted in GCGS-units, leading to an energy contribution in $1\,\textup{erg}$. Whereat the GCGS-quantities are derived from the SI-quantities with the help of $1\,\textup{m}=10^{2}\,\textup{cm}$, $1\,\textup{kg}=10^{3}\,\textup{g}$, $1\,\textup{erg}=1\textup{g}\textup{cm}^{2}\textup{s}^{-2}=10^{-7}\,\textup{J}$ and $e=4.80320440\cdot10^{-10}\,\textup{g}^{\frac{1}{2}}\textup{cm}^{\frac{3}{2}}\textup{s}^{-1}$ \cite{FranzSchwabl2019}. One shows that the converted result of (\ref{AppPlesset2}) leads to an energy in joules, if all quantities are inserted in GCGS units:
\begin{equation*} 
\begin{split}
\begin{gathered}
\left[\Delta E_{C,MP2}\right]^{\textup{GCGS}}\,\overset{(\ref{AppPlesset2})}{=} \dfrac{\left[e\right]^{4}\left[\hslash\right]^{2}}{\left[m_{e}\right]^{3}\left[c\right]^{4} }\left[r\right]^{2}\left[n(r)\right]^{\frac{4}{3}}\left[dr\right]
=\dfrac{\left[\textup{g}^{\frac{1}{2}}\textup{cm}^{\frac{3}{2}}\textup{s}^{-1}\right]^{4}\left[\textup{erg}\cdot\textup{s}\right]^{2}}{\left[\textup{g}\right]^{3}\left[\textup{cm}\textup{s}^{-1}\right]^{4} }\left[\textup{cm}\right]^{2} \left[\textup{cm}^{-3}\right]^{\frac{4}{3}}\left[\textup{cm} \right]
\\\\
= \dfrac{ \textup{g}^{2}\textup{cm}^{6}\textup{s}^{-4}  \textup{erg}^{2}\textup{s}^{2}}{ \textup{g} ^{3} \textup{cm}^{4}\textup{s}^{-4}  } \textup{cm}^{2} \textup{cm}^{-4}  \textup{cm} = \dfrac{  \textup{cm}^{2}  \textup{erg}^{2}\textup{s}^{2}}{ \textup{g}      }= \dfrac{ \textup{erg}^{2} }{ \textup{erg}}=  \textup{erg}  
\end{gathered}
\end{split}
\end{equation*}
\begin{equation}\label{AppUNITSergJ}
\scalebox{2}{\rotatebox[origin=c]{180}{$\Lsh$}}
\;\;\underline{\underline{\Delta E_{C,MP2}^{\textup{SI}} =\left\lbrace \Delta E_{C,MP2}^{\textup{GCGS}} \right\rbrace 10^{-7}\,\textup{J} }} 
\end{equation}

The converted values of the physical constants can be found in the \Cpp\footref{CppMP2ImplementationGd3+}$^{;}$\footref{CppMP2ImplementationMn2+} and the \textsc{Matlab}\footref{MATLABMP2ImplementationGd3+}$^{;}$\footref{MATLABMP2ImplementationMn2+} implementations. They calculate $\Delta E_{C,MP2}$ in joule and additionally convert the result into electron volts by dividing with the elementary charge in SI-units. 
\\
The integration method of the \textsc{Matlab} script is a \textsc{Matlab}-intern trapezoidal rule, while for the \Cpp program the \textit{Gauss-Legendre} integration is used. For both programs the data for the electron density 
is extracted from two \small{\verb|.txt|}-files\footref{gdMp2}$^{;}$\footref{mnMp2}, which were generated with the help of the FPLO-code\footref{note0} in the same configuration as in the sections \ref{Gd3+E_HB} and \ref{Mn2+E_HB}. Hence, the numerical data\footnote{\label{gdMp2}Gd3+ion_electron_density.txt}$^{;}$\footnote{\label{mnMp2}Mn2+ion_electron_density.txt} has to be converted into SI-units, due to the fact that they contain a mesh of $2000$ values for the radius $r_{data}\in[0,\,30\,r_{B}]$ in \textit{Bohr} radii $r_{B}$ and the assigned numerical values for the electron density $n_{data}=r^{2}n(r)$ in units of $r_{B}^{2}r_{B}^{-3}=r_{B}^{-1}$:
\begin{equation}\label{MP2Ion_E_HBSIUNITS}
r_{SI}=r_{data}\cdot r_{B}\;\;\;;\;\;\;n_{SI}=\dfrac{n_{data}}{r^{2}_{data}\cdot r_{B}^{3}}
\end{equation}
The converted data can then be read to memory arrays within the \Cpp and the \textsc{Matlab} implementation, where they are used to evaluate equation (\ref{AppPlesset2}). However, one has to state the approximation of (\ref{AppPlesset2}) for the \textit{Gauss-Legendre} intergration, applied in the \Cpp code \cite{schwarz2009numerische}:
\begin{equation}\label{MP2GaussLegendre}
\Delta E_{BH} \overset{(\ref{AppPlesset2})}{=} -\dfrac{43-46\ln{(2)}}{525}\dfrac{2\cdot3^{\frac{4}{3}}\pi^{\frac{5}{3}}e^{4}\hslash^{2}}{m_{e}^{3}c^{4} } \dfrac{R_{2}-R_{1}}{2}\sum\limits_{i=1}^{N_{tot.}} w_{i}\cdot r_{i}^{2}\left[n_{i}(r_{i})\right]^{\frac{4}{3}}
\end{equation}
In (\ref{MP2GaussLegendre}) the weights $w_{i}$ are the standard \textit{Gauss–Legendre}-weights, while the sampling points $r_{i}$ are transformed \textit{Gauss–Legendre}-points $x_{i}$, whereat the $x_{i}$ are equivalent to the roots of the \textit{Legendre}-polynomials \cite{schwarz2009numerische}:
\begin{equation}\label{MP2Legendre} 
r_{i}=\frac{R_{1}+R_{2}}{2} + \frac{R_{2}-R_{1}}{2} \cdot x_{i}
\end{equation}
On that point, $R_{1}$ and $R_{2}$ are chosen as the minimum and maximum value of the mesh for the radial data $r_{data}$, in order to  include all contributions of the electron density distribution. 
The necessary sets of standard sampling points $x_{i}$ and weights $w_{i}$ on the interval $\left[-1,1\right]$ for the \textit{Gauss}-\textit{Legendre} quadrature are calculated separately with two modified \textsc{Mathematica} scripts from \cite{Legendre}, which can be found in the digital appendix\footref{note2}$^{;}$\footref{note3}. This allows to generate two \small{\verb|.txt|} files\footref{note4}$^{;}$\footref{note5} with $N_{max.}=200$ sampling points and weights with an internal precision of $44$ digits, which are used to evaluate the numerical equation (\ref{Gd3+Ion_E_HBGAUSS}). Eventually, the values $n_{i}(r_{i})$ have to be linearly interpolated from the given data values. Hence the values $r_{1,i}<r_{i}$ and $r_{2,i}>r_{i}$ are searched in the mesh for the radius for each iteration $i\in[1,N_{tot.}]$, so that the corresponding data values $n_{data}$ are $n_{1,i}:=n_{data}(r_{1,i})$ and $n_{2,i}:=n_{data}(r_{2,i})$, leading to the interpolation relation:
\begin{equation}\label{MP2n(r)Interpolation}
n_{i}(r_{i})=n_{1,i}+\dfrac{n_{2,i}-n_{1,i}}{r_{2,i}-r_{1,i}}\cdot (r_{i}-r_{1,i})
\end{equation}
In addtion, one examines the specific electron density data of the ions with the help of (\ref{ConditionElecDens}), because the total number of electrons for the $\textup{Gd}^{3+}$ ion is exactly given by $64-3=61$ electrons, due to its proton number $Z=64$ and its ionic charge of $3+$:
\begin{equation}\label{Gd3+Ion_MP2CONDITION}
\begin{split}
\begin{gathered}
N_{el}^{\textup{Gd}^{3+}}\overset{(\ref{ConditionElecDens})}{=}\int\limits_{\Omega} n_{\textup{Gd}^{3+}}(r)\,d\Omega
=4\pi\int\limits_{r=R_{1}}^{\textup{R}_{2}} r^{2}n_{\textup{Gd}^{3+}}(r)\,dr\overset{!}{=} 61
\\
\scalebox{2}{\rotatebox[origin=c]{180}{$\Lsh$}}
\;\;
N_{el}^{\textup{Gd}^{3+}}\approx 4\pi\cdot\dfrac{R_{2}-R_{1}}{2}\sum\limits_{i=1}^{N_{tot.}} w_{i}\cdot r_{i}^{2}n^{\textup{Gd}^{3+}}_{i}(r_{i})\overset{!}{\approx} 61
\end{gathered}
\end{split}
\end{equation}
One similarily finds for the $\textup{Mn}^{2+}$ ion with exactly $25-2=23$ electrons, due to its proton number of $Z=25$ and its ionic charge of $2+$:
\begin{equation}\label{Mn2+Ion_MP2CONDITION}
\begin{split}
\begin{gathered}
N_{el}^{\textup{Mn}^{2+}}\overset{(\ref{ConditionElecDens})}{=}\int\limits_{\Omega} n_{\textup{Mn}^{2+}}(r)\,d\Omega
=4\pi\int\limits_{r=R_{1}}^{\textup{R}_{2}} r^{2}n_{\textup{Gd}^{3+}}(r)\,dr\overset{!}{=} 61
\\
\scalebox{2}{\rotatebox[origin=c]{180}{$\Lsh$}}
\;\;
N_{el}^{\textup{Mn}^{2+}}\approx 4\pi\cdot\dfrac{R_{2}-R_{1}}{2}\sum\limits_{i=1}^{N_{tot.}} w_{i}\cdot r_{i}^{2}n_{i}^{\textup{Mn}^{2+}}(r_{i})\overset{!}{\approx} 61
\end{gathered}
\end{split}
\end{equation}
Note that in (\ref{Gd3+Ion_MP2CONDITION}) and (\ref{Mn2+Ion_MP2CONDITION}) the Gauss-Legendre approximation for the integration in \Cpp was stated addtionally, while for the \textsc{Matlab} implementation a provided trapezoidal rule was used. 
The conditions (\ref{Gd3+Ion_MP2CONDITION}) and (\ref{Mn2+Ion_MP2CONDITION}) are checked parallelly to the evaluations of the \textit{M\o{}ller-Plesset} correlation energy contribution. The redundant implementations\footref{CppMP2ImplementationGd3+}$^{;}$\footref{CppMP2ImplementationMn2+}$^{;}$\footref{MATLABMP2ImplementationGd3+}$^{;}$\footref{MATLABMP2ImplementationMn2+} are then used for the purpose of the verification of all results. 
Eventually, the outputs of both programms
\footnote{\label{R1MP2}output_MP2_gauss_legendre_integration[Gd3+_ion].txt}$^{;}$\footnote{\label{R2MP2}output_MATLAB_MP2_integration[Gd3+_ion].txt}$^{;}$\footnote{\label{R3MP2}output_MP2_gauss_legendre_integration[Mn2+_ion].txt}$^{;}$\footnote{\label{R4MP2}output_MATLAB_MP2_integration[Mn2+_ion].txt} are displayed in the appendix (\ref{CppGD3+AppendixMP2}, \ref{CppMN2+AppendixMP2}, \ref{MATLABMP2AppendixMP2}) and lead to the MP2-correlation energy contribution for the $\textup{Gd}^{3+}$ and $\textup{Mn}^{2+}$ ion:
\begin{equation}\label{MP2PlessetGd3+App}
\underline{\underline{\Delta E_{C,MP2}^{\textup{Gd}^{3+}}=-5.734\cdot10^{-50}\,\textup{J}=-3.580\cdot10^{-31}}}
\end{equation}
\begin{equation}\label{MP2PlessetMn2+App}
\underline{\underline{\Delta E_{C,MP2}^{\textup{Mn}^{2+}}=-1.156\cdot10^{-50}\,\textup{J}=-7.215\cdot10^{-32}\,\textup{eV}}}
\end{equation}

\clearpage

\subsection{Alternative calculation procedure for the second order \textit{M\o{}ller-Plesset} correlation energy contribution} \label{ConstAppMP2}

The second order \textit{M\o{}ller-Plesset} correlation energy contribution in GCGS-units according to \cite{Pellegrini_2020} can also be written for a constant electron density $n=\frac{N_{\textup{el}}}{\Omega_{R}}$, whereat $\Omega_{R}=\frac{4\pi}{3}R^{3}$ is the volume of a full sphere with the ion radius $R$. Hence, one finds with $k_{F}[n]=(3\pi^{2}n)^\frac{1}{3}=\textup{const.}$ and (\ref{AppMP2Plesset1}):
\begin{equation}\label{ConstAppMP2Plesset1}
\begin{split}
\begin{gathered}
\underline{\underline{\Delta E_{C,MP2}}}=\int\limits_{\Omega}E_{C,MP2}[n]\,d\Omega=-\dfrac{43-46\ln{(2)}}{525}\dfrac{e^{4}\hslash^{2}k_{F}^{4}}{2m_{e}^{3}c^{4}\pi^{2}}\int\limits_{\Omega}d\Omega
\\\\
=-\dfrac{43-46\ln{(2)}}{525}\dfrac{e^{4}\hslash^{2}(3\pi^{2}n)^\frac{4}{3}}{2m_{e}^{3}c^{4}\pi^{2}}\cdot vol\left(\Omega\right)
=-\dfrac{43-46\ln{(2)}}{525}\dfrac{e^{4}\hslash^{2}(3\pi^{2}\frac{N_{\textup{el}}}{\Omega_{R}})^\frac{4}{3}}{\cancel{2}m_{e}^{3}c^{4}\pi^{\cancel{2}}}\dfrac{\cancel{4}^{\,2}\cancel{\pi}}{3}R^{3}
\\\\
\underline{\underline{=-\dfrac{43-46\ln{(2)}}{525}\dfrac{2e^{4}\hslash^{2}(3\pi^{2}\frac{N_{\textup{el}}}{\Omega_{R}})^\frac{4}{3}R^{3}}{3m_{e}^{3}c^{4}\pi}}}
\end{gathered}
\end{split}
\end{equation}    
This can be implemented in \textsc{Matlab} in GCGS-units, using the conversion into SI-uints from (\ref{AppUNITSergJ}) for the $\textup{Gd}^{3+}$ ion\footnote{\label{P1}MP2_correlation_energy_contribution_Gd_ion_constant.m} and for the $\textup{Mn}^{2+}$ ion\footnote{\label{P2}MP2_correlation_energy_contribution_Mn_ion_constant.m}, whereat the ion radii $R_{\textup{Gd}^{3+}}=93.8\,\cdot 10^{-12}\,\textup{m}$ and $R_{\textup{Mn}^{2+}}=82\cdot 10^{-12}\,\textup{m}$ were measured by spectroscopy and can be found in \cite{Drager}. The output file\footnote{output_MATLAB_MP2_integration[constant_Gd3+_ion].txt} from the program\footref{P1} for the $\textup{Gd}^{3+}$ ion with $N_{\textup{el}}^{\textup{Gd}^{3+}}=61$ leads to the result:
\begin{equation}\label{ConstMP2PlessetGd3+App}
\underline{\underline{\Delta E_{C,MP2}^{\textup{Gd}^{3+}}=-1.513\cdot10^{-50}\,\textup{J}=-9.441\cdot10^{-32}\textup{eV}}}
\end{equation}

And the output file\footnote{output_MATLAB_MP2_integration[constant_Mn2+_ion].txt} from the program\footref{P2} for the $\textup{Mn}^{2+}$ ion with $N_{\textup{el}}^{\textup{Mn}^{2+}}=23$ leads to the result:
\begin{equation}\label{ConstMP2PlessetMn2+App}
\underline{\underline{\Delta E_{C,MP2}^{\textup{Mn}^{2+}}=-4.713\cdot10^{-51}\,\textup{J}=-2.942\cdot10^{-32}\,\textup{eV}}}
\end{equation}

The integration over a finite volume in (\ref{ConstAppMP2Plesset1}) can be questioned for the solid state model of the homogeneous electron gas and will increase the MP2 correlation energy contribution by $24$ orders of magnitude, if it is dropped. In particular, this leads to a total order of $10^{-8}\,\textup{eV}$ without the integration over the volume of the full sphere. Nevertheless, one notices that the order of the \textit{M\o{}ller-Plesset} correlation energy contribution is always smaller than the order of the \textit{Breit-Hartree} contribution from (\ref{Gd3+Ion_E_HB}) or (\ref{Mn2+Ion_E_HB}). Consequently, one should not use the calculation from (\ref{AppMP2}) for an estimation of the MP2 correlation energy. Instead, one might include the larger results from this subsection as a correction to the total ground state energy, using a constant electron density without the integration over a finite volume.
\clearpage

\section{Numerical calculations}\label{NumericalCalulations}

\subsection{\textit{Monte Carlo} integration of the dipolar interaction contribution for the homogeneously magnetized sphere}\label{MCNUMApp}

The output of the \Cpp implementation holds for a constant magnetization density of $M=1$ along the $z$-axis within a dimensionless full sphere with $R_{1}=R^{\prime}_{1}=0$ and $R_{2}=R^{\prime}_{2}=1$, so that $\Omega=r\,\in\,\left[0,1\right]\times\theta\,\in\,\left[0,\pi\right]\times\varphi\,\in\,\left[0,2\pi\right)$. It shows the numerically calculated $I_{d,num}$ of the integral $I_{d}$ from (\ref{Idv2}) with $I_{d}\approx a\cdot I_{d,num}$ and the prefactor $a$ from (\ref{prefactor}), for constantly increasing numbers of evaluations of the integrand (\ref{Idv2}) per coordinate. The computations of the integrand are carried out under the inclusion of the regularisation of singular results by replacement with a zero. For the purpose of an estimation of the convergence behaviour of the implementation concerning the integral from (\ref{Idv2}) the maximum number of sampling points per integration variable was set to $N_{max.}=35$. The listing \ref{outMonteCarloId} displays the console output of the \Cpp program\footnote{MonteCarlo_integration[I_d_regularised].cpp}, which is also stored in the digital appendix\footnote{output_MonteCarlo_integration[I_d_regularised].txt}:
\\
\begin{lstlisting}[language=C++, style=c,frame=single, caption={Output of the \textit{Monte Carlo} implementation (\Cpp) for $I_{d,num}$ with $M=1$ for a dimensionless full sphere with $R_{2}=R^{\prime}_{2}=1$\\},label={outMonteCarloId} ]
--------------------------------------------------------------------------
--------------------------------------------------------------------------
|| MIN random value: 0, MAX random value: 18446744073709551615          ||
--------------------------------------------------------------------------
--------------------------------------------------------------------------
||   N || N_tot. = N^6 ||                    I_d_num || regularisations ||
--------------------------------------------------------------------------
--------------------------------------------------------------------------
||   1 ||            1 ||        4.03507519899709779 ||               0 ||
--------------------------------------------------------------------------
||   2 ||           64 ||        38.7885552258482114 ||               0 ||
--------------------------------------------------------------------------
||   3 ||          729 ||       -13.4537391232425419 ||               0 ||
--------------------------------------------------------------------------
||   4 ||         4096 ||       -80.5222834399876123 ||               0 ||
--------------------------------------------------------------------------
||   5 ||        15625 ||        21.4678092427355394 ||               0 ||
--------------------------------------------------------------------------
||   6 ||        46656 ||        19.2428917724898646 ||               0 ||
--------------------------------------------------------------------------
||   7 ||       117649 ||       -23.6051761819461218 ||               0 ||
--------------------------------------------------------------------------
||   8 ||       262144 ||       -15.0056399402068638 ||               0 ||
--------------------------------------------------------------------------
||   9 ||       531441 ||        -43.548222154164641 ||               0 ||
--------------------------------------------------------------------------
||  10 ||      1000000 ||        15.5318372897040708 ||               0 ||
--------------------------------------------------------------------------
||  11 ||      1771561 ||      -0.951928539614792106 ||               0 ||
--------------------------------------------------------------------------
||  12 ||      2985984 ||        -375.37853520112168 ||               0 ||
--------------------------------------------------------------------------
||  13 ||      4826809 ||        1202.35136178463205 ||               0 ||
--------------------------------------------------------------------------
||  14 ||      7529536 ||       -558.038432561019067 ||               0 ||
--------------------------------------------------------------------------
||  15 ||     11390625 ||        23.0221610808263354 ||               0 ||
--------------------------------------------------------------------------
||  16 ||     16777216 ||        96.9801336092075547 ||               0 ||
--------------------------------------------------------------------------
||  17 ||     24137569 ||        28.5551641841373334 ||               0 ||
--------------------------------------------------------------------------
||  18 ||     34012224 ||        13.9288822887318629 ||               0 ||
--------------------------------------------------------------------------
||  19 ||     47045881 ||       -124.382254826585826 ||               0 ||
--------------------------------------------------------------------------
||  20 ||     64000000 ||         21.235147014404369 ||               0 ||
--------------------------------------------------------------------------
||  21 ||     85766121 ||        26.4026178911079444 ||               0 ||
--------------------------------------------------------------------------
||  22 ||    113379904 ||       -24.5269399192752949 ||               0 ||
--------------------------------------------------------------------------
||  23 ||    148035889 ||        7.75233551087561883 ||               0 ||
--------------------------------------------------------------------------
||  24 ||    191102976 ||       0.749245987734423448 ||               0 ||
--------------------------------------------------------------------------
||  25 ||    244140625 ||        92.2493408371053123 ||               0 ||
--------------------------------------------------------------------------
||  26 ||    308915776 ||       -21.0929482944977889 ||               0 ||
--------------------------------------------------------------------------
||  27 ||    387420489 ||       -3.84491046205120798 ||               0 ||
--------------------------------------------------------------------------
||  28 ||    481890304 ||        17.9918506849874063 ||               0 ||
--------------------------------------------------------------------------
||  29 ||    594823321 ||       -16.3220387554660987 ||               0 ||
--------------------------------------------------------------------------
||  30 ||    729000000 ||       -2.75323995209555775 ||               0 ||
--------------------------------------------------------------------------
||  31 ||    887503681 ||         46.861718282042822 ||               0 ||
--------------------------------------------------------------------------
||  32 ||   1073741824 ||        1.28763736766273793 ||               0 ||
--------------------------------------------------------------------------
||  33 ||   1291467969 ||        42.6531466206686268 ||               0 ||
--------------------------------------------------------------------------
||  34 ||   1544804416 ||         23.814772199659828 ||               0 ||
--------------------------------------------------------------------------
||  35 ||   1838265625 ||       -8.84655261423981738 ||               0 ||
--------------------------------------------------------------------------
--------------------------------------------------------------------------

Process returned 0 (0x0)   execution time : 21165.830 s
\end{lstlisting}

\clearpage

\subsection{\textit{Gauss-Legendre} integration of the dipolar interaction contribution for the homogeneously magnetized sphere}\label{GaussAppSubSec}

The output of the \Cpp implementation holds for a constant magnetization density of $M=1$ along the $z$-axis within a dimensionless full sphere with $R_{1}=R^{\prime}_{1}=0$ and $R_{2}=R^{\prime}_{2}=1$, so that $\Omega=r\,\in\,\left[0,1\right]\times\theta\,\in\,\left[0,\pi\right]\times\varphi\,\in\,\left[0,2\pi\right)$. It shows the numerically calculated $I_{d,num}$ of the integral $I_{d}$ from (\ref{Idv2}) with $I_{d}\approx a\cdot I_{d,num}$ and the prefactor $a$ from (\ref{prefactor}), for constantly increasing numbers of evaluations of the integrand (\ref{Idv2}) per coordinate. The computations of the integrand are carried out under the inclusion of the regularisation of singular results by replacement with a zero. For the purpose of an estimation of the convergence behaviour of the implementation concerning the integral from (\ref{Idv2}) the maximum number of sampling points per integration variable was set to $N_{max.}=40$. The console output of the \Cpp program\footnote{gauss_legendre_integration[I_d_regularised].cpp} is displayed in listing \ref{outGaussLegendreId}. In addition, the generated output is stored in the digital appendix\footnote{output_gauss_legendre_integration[I_d_regularised].txt} as a \small{\verb|.txt|}-file:
\\
\begin{lstlisting}[language=C++, style=c,frame=single, caption={Output of the \textit{Gauss-Legendre} implementation (\Cpp) for $I_{d,num.}$ with $M=1$ for a dimensionless full sphere with $R_{2}=R^{\prime}_{2}=1$\\},label={outGaussLegendreId} ]
--------------------------------------------------------------------------
--------------------------------------------------------------------------
||   N || N_tot. = N^6 ||                    I_d_num || regularisations ||
--------------------------------------------------------------------------
--------------------------------------------------------------------------
||   1 ||            1 ||                          0 ||               1 ||
--------------------------------------------------------------------------
||   2 ||           64 ||       -4.46038688207645547 ||               8 ||
--------------------------------------------------------------------------
||   3 ||          729 ||   1.08970846827863199e+028 ||               9 ||
--------------------------------------------------------------------------
||   4 ||         4096 ||   2.64984581152214894e+028 ||              48 ||
--------------------------------------------------------------------------
||   5 ||        15625 ||   6.10914809652650429e+025 ||             100 ||
--------------------------------------------------------------------------
||   6 ||        46656 ||   5.76733848725963214e+027 ||             144 ||
--------------------------------------------------------------------------
||   7 ||       117649 ||    3.3265787320116687e+024 ||             294 ||
--------------------------------------------------------------------------
||   8 ||       262144 ||   1.35816215726815946e+024 ||             384 ||
--------------------------------------------------------------------------
||   9 ||       531441 ||   3.96744613172616513e+026 ||             567 ||
--------------------------------------------------------------------------
||  10 ||      1000000 ||   7.81918047548289366e+022 ||             900 ||
--------------------------------------------------------------------------
||  11 ||      1771561 ||   6.13331653692164436e+026 ||            1210 ||
--------------------------------------------------------------------------
||  12 ||      2985984 ||   5.33152828722989162e+026 ||            1440 ||
--------------------------------------------------------------------------
||  13 ||      4826809 ||   1.78227284324415311e+026 ||            1859 ||
--------------------------------------------------------------------------
||  14 ||      7529536 ||   7.06880746372279801e+024 ||            2548 ||
--------------------------------------------------------------------------
||  15 ||     11390625 ||        72.3028295763693796 ||            3375 ||
--------------------------------------------------------------------------
||  16 ||     16777216 ||   1.76217578312046545e+024 ||            3584 ||
--------------------------------------------------------------------------
||  17 ||     24137569 ||        78.7725007926078985 ||            4913 ||
--------------------------------------------------------------------------
||  18 ||     34012224 ||   9.35391439828736841e+025 ||            3888 ||
--------------------------------------------------------------------------
||  19 ||     47045881 ||   2.30998749144771713e+024 ||            6137 ||
--------------------------------------------------------------------------
||  20 ||     64000000 ||   1.44926361876166361e+025 ||            6800 ||
--------------------------------------------------------------------------
||  21 ||     85766121 ||   8.50876865019942314e+023 ||            8820 ||
--------------------------------------------------------------------------
||  22 ||    113379904 ||   4.89489899730216369e+025 ||            8228 ||
--------------------------------------------------------------------------
||  23 ||    148035889 ||   5.33332208939237658e+025 ||            9522 ||
--------------------------------------------------------------------------
||  24 ||    191102976 ||   5.90283557864769887e+025 ||           10944 ||
--------------------------------------------------------------------------
||  25 ||    244140625 ||   1.47363165320993061e+025 ||           13125 ||
--------------------------------------------------------------------------
||  26 ||    308915776 ||   5.75912462842413628e+025 ||           14196 ||
--------------------------------------------------------------------------
||  27 ||    387420489 ||   3.58189276645997988e+025 ||           15309 ||
--------------------------------------------------------------------------
||  28 ||    481890304 ||   1.92896424179940193e+025 ||           18816 ||
--------------------------------------------------------------------------
||  29 ||    594823321 ||   3.11678775147635201e+025 ||           19343 ||
--------------------------------------------------------------------------
||  30 ||    729000000 ||   2.89470387747501548e+025 ||           23400 ||
--------------------------------------------------------------------------
||  31 ||    887503681 ||    2.2665476706420919e+025 ||           24025 ||
--------------------------------------------------------------------------
||  32 ||   1073741824 ||   2.38756418568201474e+025 ||           23552 ||
--------------------------------------------------------------------------
||  33 ||   1291467969 ||   1.30673708701666005e+025 ||           29403 ||
--------------------------------------------------------------------------
||  34 ||   1544804416 ||   2.67260113119786037e+025 ||           28900 ||
--------------------------------------------------------------------------
||  35 ||   1838265625 ||   3.81869288633299845e+024 ||           36750 ||
--------------------------------------------------------------------------
||  36 ||   2176782336 ||   2.12462040360686068e+025 ||           34992 ||
--------------------------------------------------------------------------
||  37 ||   2565726409 ||   2.47673613748213751e+025 ||           41070 ||
--------------------------------------------------------------------------
||  38 ||   3010936384 ||    1.7760733609622043e+025 ||           40432 ||
--------------------------------------------------------------------------
||  39 ||   3518743761 ||     6.405620226245139e+024 ||           54756 ||
--------------------------------------------------------------------------
||  40 ||   4096000000 ||   6.31720348245553144e+024 ||           56000 ||
--------------------------------------------------------------------------
--------------------------------------------------------------------------

Process returned 0 (0x0)   execution time : 29634.631 s
\end{lstlisting}

\clearpage

\subsection{\textit{Monte Carlo} integration of the benchmark problem}

The output of the \Cpp implementation for the benchmark problem shows the numerically calculated $I_{a,num.}$ of the integral $I_{a}$ from (\ref{benchmark2}) for constantly increasing numbers of evaluations of the integrand (\ref{benchmark1}) per coordinate. The computations of the integrand are carried out under the inclusion of the regularisation of singular results by replacement with a zero. For reason of comparability the maximum number of sampling points per integration variable was set to $N_{max.}=35$. Moreover, the absolute deviation from the analytical result $I_{\textup{a,analyt.}}$ from (\ref{benchmark2}) was computed using the equation $\vert I_{\textup{a,analyt.}}-I_{\textup{a,num.}}\vert$. The console output of the \Cpp program\footnote{montecarlo_integration[benchmark_regularised].cpp} is displayed in listing \ref{outGaussLegendreBenchmark}. In addition, the generated output is stored in the digital appendix\footnote{output_montecarlo_integration[benchmark_regularised].txt} a \small{\verb|.txt|}-file:
\\
\begin{lstlisting}[language=C++, style=c,frame=single, caption={Output of the \textit{Monte Carlo} implementation (\Cpp) for the benchmark problem $I_{a,num.}$ from (\ref{benchmark2})},label={outMonteCarloBenchmark} ]
-------------------------------------------------------------------------------------
-------------------------------------------------------------------------------------
||   N || N_tot. = N^6 ||              I_a_num || reg. || abs_err = abs(I_a-I_num) ||
-------------------------------------------------------------------------------------
-------------------------------------------------------------------------------------
||   1 ||            1 ||  786.771336254804917 ||    0 ||      476.538027084632888 ||
-------------------------------------------------------------------------------------
||   2 ||           64 ||  753.092049430385472 ||    0 ||      510.217313909052333 ||
-------------------------------------------------------------------------------------
||   3 ||          729 ||  1298.34838791192797 ||    0 ||       35.039024572490166 ||
-------------------------------------------------------------------------------------
||   4 ||         4096 ||  1394.46153780546584 ||    0 ||      131.152174466028035 ||
-------------------------------------------------------------------------------------
||   5 ||        15625 ||  1245.39235236687967 ||    0 ||      17.9170109725581299 ||
-------------------------------------------------------------------------------------
||   6 ||        46656 ||  1231.07981434669906 ||    0 ||      32.2295489927387456 ||
-------------------------------------------------------------------------------------
||   7 ||       117649 ||  1219.34981912713187 ||    0 ||      43.9595442123059306 ||
-------------------------------------------------------------------------------------
||   8 ||       262144 ||  1219.67587624510975 ||    0 ||      43.6334870943280563 ||
-------------------------------------------------------------------------------------
||   9 ||       531441 ||  1228.57193286296384 ||    0 ||      34.7374304764739613 ||
-------------------------------------------------------------------------------------
||  10 ||      1000000 ||  1244.13577764202485 ||    0 ||      19.1735856974129576 ||
-------------------------------------------------------------------------------------
||  11 ||      1771561 ||  1259.40824980074797 ||    0 ||      3.90111353868983335 ||
-------------------------------------------------------------------------------------
||  12 ||      2985984 ||  1246.37399209444246 ||    0 ||      16.9353712449953477 ||
-------------------------------------------------------------------------------------
||  13 ||      4826809 ||  1262.88395381887194 ||    0 ||     0.425409520565861232 ||
-------------------------------------------------------------------------------------
||  14 ||      7529536 ||   1261.1273915249698 ||    0 ||      2.18197181446800403 ||
-------------------------------------------------------------------------------------
||  15 ||     11390625 ||  1250.37997881925876 ||    0 ||      12.9293845201790479 ||
-------------------------------------------------------------------------------------
||  16 ||     16777216 ||  1254.02004633916136 ||    0 ||      9.28931700027644058 ||
-------------------------------------------------------------------------------------
||  17 ||     24137569 ||  1255.14667904402556 ||    0 ||       8.1626842954122466 ||
-------------------------------------------------------------------------------------
||  18 ||     34012224 ||  1265.11367123891672 ||    1 ||      1.80430789947891768 ||
-------------------------------------------------------------------------------------
||  19 ||     47045881 ||  1257.82828854080639 ||    0 ||        5.481074798631414 ||
-------------------------------------------------------------------------------------
||  20 ||     64000000 ||   1260.4282485389225 ||    0 ||      2.88111480051530899 ||
-------------------------------------------------------------------------------------
||  21 ||     85766121 ||  1261.47706646467139 ||    0 ||      1.83229687476641789 ||
-------------------------------------------------------------------------------------
||  22 ||    113379904 ||   1262.5678766777916 ||    1 ||     0.741486661646200362 ||
-------------------------------------------------------------------------------------
||  23 ||    148035889 ||  1261.83808703690143 ||    0 ||       1.4712763025363792 ||
-------------------------------------------------------------------------------------
||  24 ||    191102976 ||  1257.75265331876225 ||    0 ||      5.55671002067555297 ||
-------------------------------------------------------------------------------------
||  25 ||    244140625 ||  1262.73534190814878 ||    0 ||     0.574021431289021389 ||
-------------------------------------------------------------------------------------
||  26 ||    308915776 ||  1265.22055069718692 ||    1 ||      1.91118735774911885 ||
-------------------------------------------------------------------------------------
||  27 ||    387420489 ||  1262.30185618634251 ||    1 ||      1.00750715309529637 ||
-------------------------------------------------------------------------------------
||  28 ||    481890304 ||  1262.63012421153653 ||    0 ||     0.679239127901271056 ||
-------------------------------------------------------------------------------------
||  29 ||    594823321 ||  1263.57273340678894 ||    1 ||     0.263370067351130888 ||
-------------------------------------------------------------------------------------
||  30 ||    729000000 ||  1262.89879197406728 ||    4 ||     0.410571365370520747 ||
-------------------------------------------------------------------------------------
||  31 ||    887503681 ||  1263.13197727285249 ||    2 ||     0.177386066585317193 ||
-------------------------------------------------------------------------------------
||  32 ||   1073741824 ||   1264.5531312162842 ||    2 ||      1.24376787684639156 ||
-------------------------------------------------------------------------------------
||  33 ||   1291467969 ||  1262.97846267997818 ||    1 ||     0.330900659459620217 ||
-------------------------------------------------------------------------------------
||  34 ||   1544804416 ||  1261.97736038840288 ||    1 ||      1.33200295103492294 ||
-------------------------------------------------------------------------------------
||  35 ||   1838265625 ||  1263.66681434548013 ||    1 ||     0.357451006042326158 ||
-------------------------------------------------------------------------------------
-------------------------------------------------------------------------------------

Process returned 0 (0x0)   execution time : 3665.112 s
\end{lstlisting}

\clearpage

\subsection{\textit{Gauss-Legendre} integration of the benchmark problem}

The output of the \Cpp implementation of the \textit{Gauss-Legendre} integration for the benchmark problem shows the numerically calculated value $I_{a,num.}$ of the integral $I_{a}$ from (\ref{benchmark2}) for constantly increasing numbers of evaluations of the integrand (\ref{benchmark1}) per coordinate. The computations of the integrand are carried out under the inclusion of the regularisation of singular results by replacement with a zero. For reason of comparability the maximum number of sampling points per integration variable was set to $N_{max.}=50$. Moreover, the absolute deviation from the analytical result $I_{\textup{a,analyt.}}$ from (\ref{benchmark2}) was computed using the equation $\vert I_{\textup{analyt.}}-I_{\textup{a,num.}}\vert$. The console output of the \Cpp program\footnote{gauss_legendre_integration[benchmark_regularised].cpp} is displayed in listing \ref{outGaussLegendreBenchmark}. In addition, the generated output is stored in the digital appendix\footnote{output_gauss_legendre_integration[benchmark_regularised].txt} as a \small{\verb|.txt|}-file:
\\
\begin{lstlisting}[language=C++, style=c,frame=single, caption={Output $I_{a,num.}$ of the \textit{Gauss-Legendre} implementation (\Cpp) for the benchmark problem from (\ref{benchmark2})},label={outGaussLegendreBenchmark} ]
-------------------------------------------------------------------------------------
-------------------------------------------------------------------------------------
||   N || N_tot. = N^6 ||              I_a_num || reg. || abs_err = abs(I_a-I_num) ||
-------------------------------------------------------------------------------------
-------------------------------------------------------------------------------------
||   1 ||            1 ||  157.913670417429726 ||    0 ||      1105.39569292200808 ||
-------------------------------------------------------------------------------------
||   2 ||           64 ||  399.307320774877269 ||    0 ||      864.002042564560536 ||
-------------------------------------------------------------------------------------
||   3 ||          729 ||  568.647319931676065 ||    0 ||       694.66204340776174 ||
-------------------------------------------------------------------------------------
||   4 ||         4096 ||  685.694081072820934 ||    0 ||       577.61528226661687 ||
-------------------------------------------------------------------------------------
||   5 ||        15625 ||  770.018943390814722 ||    0 ||      493.290419948623082 ||
-------------------------------------------------------------------------------------
||   6 ||        46656 ||  833.274263686112592 ||    0 ||      430.035099653325213 ||
-------------------------------------------------------------------------------------
||   7 ||       117649 ||  882.343298764252375 ||    0 ||       380.96606457518543 ||
-------------------------------------------------------------------------------------
||   8 ||       262144 ||  921.460412482786419 ||    0 ||      341.848950856651385 ||
-------------------------------------------------------------------------------------
||   9 ||       531441 ||  953.347875819192687 ||    0 ||      309.961487520245118 ||
-------------------------------------------------------------------------------------
||  10 ||      1000000 ||   979.82684255769228 ||    0 ||      283.482520781745524 ||
-------------------------------------------------------------------------------------
||  11 ||      1771561 ||  1002.15831921064878 ||    0 ||      261.151044128789026 ||
-------------------------------------------------------------------------------------
||  12 ||      2985984 ||  1021.24167412579395 ||    0 ||      242.067689213643858 ||
-------------------------------------------------------------------------------------
||  13 ||      4826809 ||  1037.73490774937634 ||    0 ||      225.574455590061467 ||
-------------------------------------------------------------------------------------
||  14 ||      7529536 ||  1052.13018005693734 ||    0 ||      211.179183282500463 ||
-------------------------------------------------------------------------------------
||  15 ||     11390625 ||  1064.80277231248393 ||    0 ||       198.50659102695387 ||
-------------------------------------------------------------------------------------
||  16 ||     16777216 ||  1076.04373243508239 ||    0 ||      187.265630904355412 ||
-------------------------------------------------------------------------------------
||  17 ||     24137569 ||  1086.08219070905883 ||    0 ||      177.227172630378976 ||
-------------------------------------------------------------------------------------
||  18 ||     34012224 ||  1095.10095645512345 ||    0 ||      168.208406884314355 ||
-------------------------------------------------------------------------------------
||  19 ||     47045881 ||  1103.24763696709885 ||    0 ||       160.06172637233895 ||
-------------------------------------------------------------------------------------
||  20 ||     64000000 ||  1110.64270666158672 ||    0 ||      152.666656677851084 ||
-------------------------------------------------------------------------------------
||  21 ||     85766121 ||  1117.38545780612511 ||    0 ||      145.923905533312698 ||
-------------------------------------------------------------------------------------
||  22 ||    113379904 ||  1123.55845335552451 ||    0 ||      139.750909983913293 ||
-------------------------------------------------------------------------------------
||  23 ||    148035889 ||  1129.23090339387635 ||    0 ||      134.078459945561455 ||
-------------------------------------------------------------------------------------
||  24 ||    191102976 ||  1134.46125657473542 ||    0 ||      128.848106764702387 ||
-------------------------------------------------------------------------------------
||  25 ||    244140625 ||   1139.2992112789256 ||    0 ||      124.010152060512208 ||
-------------------------------------------------------------------------------------
||  26 ||    308915776 ||  1143.78729246084189 ||    0 ||      119.522070878595918 ||
-------------------------------------------------------------------------------------
||  27 ||    387420489 ||  1147.96209969169663 ||    0 ||      115.347263647741171 ||
-------------------------------------------------------------------------------------
||  28 ||    481890304 ||  1151.85530362693667 ||    0 ||      111.454059712501138 ||
-------------------------------------------------------------------------------------
||  29 ||    594823321 ||  1155.49444808805141 ||    0 ||      107.814915251386399 ||
-------------------------------------------------------------------------------------
||  30 ||    729000000 ||  1158.90360056995521 ||    0 ||      104.405762769482595 ||
-------------------------------------------------------------------------------------
||  31 ||    887503681 ||  1162.10388354739571 ||    0 ||      101.205479792042093 ||
-------------------------------------------------------------------------------------
||  32 ||   1073741824 ||  1165.11391129056252 ||    0 ||      98.1954520488752892 ||
-------------------------------------------------------------------------------------
||  33 ||   1291467969 ||  1167.95015121954685 ||    0 ||      95.3592121198909582 ||
-------------------------------------------------------------------------------------
||  34 ||   1544804416 ||  1170.62722457086172 ||    0 ||      92.6821387685760835 ||
-------------------------------------------------------------------------------------
||  35 ||   1838265625 ||   1173.1581579374437 ||    0 ||      90.1512054019941091 ||
-------------------------------------------------------------------------------------
||  36 ||   2176782336 ||  1175.55459479345229 ||    0 ||      87.7547685459855132 ||
-------------------------------------------------------------------------------------
||  37 ||   2565726409 ||  1177.82697423492198 ||    0 ||      85.4823891045158233 ||
-------------------------------------------------------------------------------------
||  38 ||   3010936384 ||  1179.98468271247042 ||    0 ||      83.3246806269673864 ||
-------------------------------------------------------------------------------------
||  39 ||   3518743761 ||  1182.03618339818197 ||    0 ||      81.2731799412558354 ||
-------------------------------------------------------------------------------------
||  40 ||   4096000000 ||  1183.98912693908615 ||    0 ||      79.3202364003516528 ||
-------------------------------------------------------------------------------------
||  41 ||   4750104241 ||  1185.85044664716311 ||    0 ||      77.4589166922746921 ||
-------------------------------------------------------------------------------------
||  42 ||   5489031744 ||  1187.62644061862978 ||    0 ||      75.6829227208080226 ||
-------------------------------------------------------------------------------------
||  43 ||   6321363049 ||  1189.32284282754328 ||    0 ||      73.9865205118945294 ||
-------------------------------------------------------------------------------------
||  44 ||   7256313856 ||   1190.9448848847129 ||    0 ||      72.3644784547249051 ||
-------------------------------------------------------------------------------------
||  45 ||   8303765625 ||  1192.49734985781917 ||    0 ||      70.8120134816186373 ||
-------------------------------------------------------------------------------------
||  46 ||   9474296896 ||  1193.98461932063234 ||    0 ||      69.3247440188054626 ||
-------------------------------------------------------------------------------------
||  47 ||  10779215329 ||  1195.41071460329421 ||    0 ||      67.8986487361435972 ||
-------------------------------------------------------------------------------------
||  48 ||  12230590464 ||  1196.77933305981683 ||    0 ||      66.5300302796209698 ||
-------------------------------------------------------------------------------------
||  49 ||  13841287201 ||  1198.09388004244195 ||    0 ||      65.2154832969958552 ||
-------------------------------------------------------------------------------------
||  50 ||  15625000000 ||  1199.35749716312763 ||    0 ||      63.9518661763101749 ||
-------------------------------------------------------------------------------------
-------------------------------------------------------------------------------------

Process returned 0 (0x0)   execution time : 16328.706 s
\end{lstlisting}

\clearpage

\subsection{\textit{Gauss-Legendre} integration of the \textit{Breit-Hartree} contribution for a $\textup{Gd}^{3+}$ ion}\label{CppGD3+Appendix}

The output of the \Cpp implementation of the \textit{Gauss-Legendre} integration for the \textit{Breit-Hartree} contribution for the $3+$-ionic state of gadolinium shows the numerically calculated value $\Delta E_{HB}$ of the expression from (\ref{EHBHollowSpheres(r)}) for the spin density of the $4f$-shell of a $\textup{Gd}^{3+}$ ion in joules and in electron volts. In addition, the number of $4f$-electrons is calculated, in order to show that the given data for the spin density are physically valid for the $\textup{Gd}^{3+}$ ion, whereat seven $4f$-electrons are expected. The computations of the integrand are carried out under the inclusion of the regularisation of singular results by replacement with a zero. The console output of the \Cpp program\footnote{EHB_gauss_legendre_integration[Gd3+_ion].cpp} is displayed in listing \ref{outGaussLegendreGd3+} for increasing total numbers of sampling points $N_{tot.}$, whereat only an excerpt of the generated output is shown. The complete output is stored in the digital appendix\footnote{output_gauss_legendre_integration[Gd3+_ion].txt} as a \small{\verb|.txt|}-file:
\\
\begin{lstlisting}[language=C++, style=c,frame=single, caption={Output of the \Cpp implementation for the calculation of the \textit{Breit-Hartree} contribution for a $\textup{Gd}^{3+}$ ion  },label={outGaussLegendreGd3+} ]
------------------------------------------------------------------------------
------------------------------------------------------------------------------
|| R1 = 0 m, R2 = 1.5875316e-009 m                                          ||
------------------------------------------------------------------------------
|| M(R1) = 0 A/m, M(R2) = -8.8347438e-038 A/m                               ||
------------------------------------------------------------------------------
------------------------------------------------------------------------------
|| N_tot. ||      N_el_4f ||      E_HB [Joule] ||         E_HB [eV] || reg. ||
------------------------------------------------------------------------------
------------------------------------------------------------------------------
||     10 ||      7.11395 || -1.043384873e-020 ||    -0.06512296153 ||    0 ||
------------------------------------------------------------------------------
||     15 ||      6.08678 || -1.529674374e-021 ||   -0.009547476489 ||    0 ||
------------------------------------------------------------------------------
||     20 ||      7.38917 || -5.773475084e-021 ||    -0.03603519713 ||    0 ||
------------------------------------------------------------------------------
||     25 ||      6.90262 || -4.770297935e-021 ||    -0.02977385785 ||    0 ||
------------------------------------------------------------------------------
||     30 ||      6.92894 || -4.317666726e-021 ||    -0.02694875605 ||    0 ||
------------------------------------------------------------------------------
||     35 ||      6.96121 || -4.770353881e-021 ||    -0.02977420703 ||    0 ||
------------------------------------------------------------------------------
||     40 ||      6.95036 ||  -4.64803826e-021 ||    -0.02901077298 ||    0 ||
------------------------------------------------------------------------------
||     45 ||      6.95444 || -4.633663897e-021 ||    -0.02892105526 ||    0 ||
------------------------------------------------------------------------------
||     50 ||      6.95373 || -4.650738182e-021 ||    -0.02902762456 ||    0 ||
------------------------------------------------------------------------------
||     55 ||      6.95237 || -4.643682444e-021 ||    -0.02898358611 ||    0 ||
------------------------------------------------------------------------------
||     60 ||      6.95337 || -4.648134954e-021 ||     -0.0290113765 ||    0 ||
------------------------------------------------------------------------------
||     65 ||      6.95341 || -4.647224281e-021 ||    -0.02900569252 ||    0 ||
------------------------------------------------------------------------------
||     70 ||       6.9528 || -4.645360545e-021 ||    -0.02899405999 ||    0 ||
------------------------------------------------------------------------------
||     75 ||      6.95343 || -4.647243888e-021 ||     -0.0290058149 ||    0 ||
------------------------------------------------------------------------------
||     80 ||      6.95323 || -4.646675923e-021 ||    -0.02900226994 ||    0 ||
------------------------------------------------------------------------------
||     85 ||       6.9531 || -4.646194406e-021 ||    -0.02899926455 ||    0 ||
------------------------------------------------------------------------------
||     90 ||      6.95316 || -4.646819558e-021 ||    -0.02900316644 ||    0 ||
------------------------------------------------------------------------------
||     95 ||      6.95326 || -4.646607948e-021 ||    -0.02900184567 ||    0 ||
------------------------------------------------------------------------------
||    100 ||      6.95313 ||  -4.64644607e-021 ||    -0.02900083531 ||    0 ||
------------------------------------------------------------------------------
||    105 ||      6.95323 || -4.646726235e-021 ||    -0.02900258396 ||    0 ||
------------------------------------------------------------------------------
||    110 ||      6.95313 ||   -4.6465428e-021 ||    -0.02900143905 ||    0 ||
------------------------------------------------------------------------------
||    115 ||      6.95314 || -4.646525018e-021 ||    -0.02900132807 ||    0 ||
------------------------------------------------------------------------------
||    120 ||       6.9532 || -4.646654094e-021 ||    -0.02900213369 ||    0 ||
------------------------------------------------------------------------------
||    125 ||      6.95317 || -4.646577697e-021 ||    -0.02900165686 ||    0 ||
------------------------------------------------------------------------------
||    130 ||      6.95317 || -4.646592362e-021 ||    -0.02900174839 ||    0 ||
------------------------------------------------------------------------------
||    135 ||       6.9531 || -4.646596636e-021 ||    -0.02900177507 ||    0 ||
------------------------------------------------------------------------------
||    140 ||      6.95323 || -4.646584516e-021 ||    -0.02900169942 ||    0 ||
------------------------------------------------------------------------------
||    145 ||      6.95313 || -4.646588424e-021 ||    -0.02900172381 ||    0 ||
------------------------------------------------------------------------------
||    150 ||      6.95318 || -4.646614593e-021 ||    -0.02900188715 ||    0 ||
------------------------------------------------------------------------------
||    155 ||      6.95313 || -4.646592691e-021 ||    -0.02900175044 ||    0 ||
------------------------------------------------------------------------------
||    160 ||      6.95321 || -4.646581027e-021 ||    -0.02900167765 ||    0 ||
------------------------------------------------------------------------------
||    165 ||      6.95312 || -4.646586367e-021 ||    -0.02900171098 ||    0 ||
------------------------------------------------------------------------------
||    170 ||      6.95311 || -4.646571714e-021 ||    -0.02900161952 ||    0 ||
------------------------------------------------------------------------------
||    175 ||      6.95316 || -4.646602912e-021 ||    -0.02900181424 ||    0 ||
------------------------------------------------------------------------------
||    180 ||      6.95322 || -4.646599945e-021 ||    -0.02900179572 ||    0 ||
------------------------------------------------------------------------------
||    185 ||      6.95315 || -4.646584526e-021 ||    -0.02900169948 ||    0 ||
------------------------------------------------------------------------------
||    190 ||      6.95309 || -4.646531101e-021 ||    -0.02900136603 ||    0 ||
------------------------------------------------------------------------------
||    195 ||       6.9531 || -4.646533446e-021 ||    -0.02900138067 ||    0 ||
------------------------------------------------------------------------------
||    200 ||      6.95315 || -4.646540121e-021 ||    -0.02900142233 ||    0 ||
------------------------------------------------------------------------------
------------------------------------------------------------------------------
Process returned 0 (0x0)   execution time : 1.771 s
\end{lstlisting}

\clearpage

\subsection{\textit{Gauss-Legendre} integration of the \textit{Breit-Hartree} contribution for an $\textup{Mn}^{2+}$ ion}\label{CppMN2+Appendix}

The output of the \Cpp implementation of the \textit{Gauss-Legendre} integration for the \textit{Breit-Hartree} contribution for the $2+$-ionic state of manganese shows the numerically calculated value $\Delta E_{HB}$ of the expression from (\ref{EHBHollowSpheres(r)}) for the spin density of the $3d$-shell of an $\textup{Mn}^{2+}$ ion in joules and in electron volts. In addition, the number of $3d$-electrons is calculated, in order to show that the given data for the spin density are physically valid for the $\textup{Mn}^{2+}$ ion, whereat five $3d$-electrons are expected. The computations of the integrand are carried out under the inclusion of the regularisation of singular results by replacement with a zero. The console output of the \Cpp program\footnote{EHB_gauss_legendre_integration[Mn2+_ion].cpp} is displayed in listing \ref{outGaussLegendreMn2+} for increasing total numbers of sampling points $N_{tot.}$, whereat only an excerpt of the generated output is shown. The complete output is stored in the digital appendix\footnote{output_gauss_legendre_integration[Mn2+_ion].txt} as a \small{\verb|.txt|}-file:
\\
\begin{lstlisting}[language=C++, style=c,frame=single, caption={Output of the \Cpp implementation for the calculation of the \textit{Breit-Hartree} contribution for an $\textup{Mn}^{2+}$ ion  },label={outGaussLegendreMn2+} ]
------------------------------------------------------------------------------
------------------------------------------------------------------------------
|| R1 = 0 m, R2 = 1.5875316e-009 m                                          ||
------------------------------------------------------------------------------
|| M(R1) = 0 A/m, M(R2) = -3.2329273e-032 A/m                               ||
------------------------------------------------------------------------------
------------------------------------------------------------------------------
|| N_tot. ||      N_el_3d ||      E_HB [Joule] ||         E_HB [eV] || reg. ||
------------------------------------------------------------------------------
------------------------------------------------------------------------------
||     10 ||      3.90362 || -1.142627222e-021 ||   -0.007131718173 ||    0 ||
------------------------------------------------------------------------------
||     15 ||       5.0501 || -7.297851698e-022 ||    -0.00455496076 ||    0 ||
------------------------------------------------------------------------------
||     20 ||       5.0532 || -1.137681437e-021 ||    -0.00710084901 ||    0 ||
------------------------------------------------------------------------------
||     25 ||      4.98607 || -9.372795402e-022 ||   -0.005850038756 ||    0 ||
------------------------------------------------------------------------------
||     30 ||      5.00011 || -9.852670518e-022 ||   -0.006149553244 ||    0 ||
------------------------------------------------------------------------------
||     35 ||      4.99591 || -9.829405945e-022 ||    -0.00613503264 ||    0 ||
------------------------------------------------------------------------------
||     40 ||      4.99966 || -9.834252988e-022 ||   -0.006138057927 ||    0 ||
------------------------------------------------------------------------------
||     45 ||      4.99889 || -9.824162078e-022 ||   -0.006131759676 ||    0 ||
------------------------------------------------------------------------------
||     50 ||      4.99827 || -9.817526231e-022 ||   -0.006127617906 ||    0 ||
------------------------------------------------------------------------------
||     55 ||      4.99909 || -9.829683455e-022 ||   -0.006135205848 ||    0 ||
------------------------------------------------------------------------------
||     60 ||      4.99837 || -9.821924612e-022 ||   -0.006130363159 ||    0 ||
------------------------------------------------------------------------------
||     65 ||      4.99879 || -9.824627569e-022 ||   -0.006132050212 ||    0 ||
------------------------------------------------------------------------------
||     70 ||      4.99868 || -9.825070521e-022 ||   -0.006132326681 ||    0 ||
------------------------------------------------------------------------------
||     75 ||       4.9987 || -9.823410751e-022 ||   -0.006131290734 ||    0 ||
------------------------------------------------------------------------------
||     80 ||       4.9988 || -9.824910951e-022 ||   -0.006132227086 ||    0 ||
------------------------------------------------------------------------------
||     85 ||      4.99868 || -9.824079187e-022 ||    -0.00613170794 ||    0 ||
------------------------------------------------------------------------------
||     90 ||      4.99865 ||  -9.82427897e-022 ||   -0.006131832634 ||    0 ||
------------------------------------------------------------------------------
||     95 ||      4.99876 || -9.824483033e-022 ||       -0.00613196 ||    0 ||
------------------------------------------------------------------------------
||    100 ||      4.99865 || -9.824186318e-022 ||   -0.006131774805 ||    0 ||
------------------------------------------------------------------------------
||    105 ||      4.99869 || -9.824427119e-022 ||   -0.006131925102 ||    0 ||
------------------------------------------------------------------------------
||    110 ||      4.99869 || -9.824240449e-022 ||   -0.006131808591 ||    0 ||
------------------------------------------------------------------------------
||    115 ||      4.99868 || -9.824284218e-022 ||   -0.006131835909 ||    0 ||
------------------------------------------------------------------------------
||    120 ||      4.99872 || -9.824319613e-022 ||   -0.006131858001 ||    0 ||
------------------------------------------------------------------------------
||    125 ||      4.99871 || -9.824280927e-022 ||   -0.006131833855 ||    0 ||
------------------------------------------------------------------------------
||    130 ||      4.99866 || -9.824328261e-022 ||   -0.006131863399 ||    0 ||
------------------------------------------------------------------------------
||    135 ||      4.99863 || -9.824241675e-022 ||   -0.006131809356 ||    0 ||
------------------------------------------------------------------------------
||    140 ||      4.99875 || -9.824326239e-022 ||   -0.006131862137 ||    0 ||
------------------------------------------------------------------------------
||    145 ||      4.99867 || -9.824300682e-022 ||   -0.006131846186 ||    0 ||
------------------------------------------------------------------------------
||    150 ||      4.99869 || -9.824329878e-022 ||   -0.006131864408 ||    0 ||
------------------------------------------------------------------------------
||    155 ||      4.99868 || -9.824272992e-022 ||   -0.006131828903 ||    0 ||
------------------------------------------------------------------------------
||    160 ||      4.99871 || -9.824360546e-022 ||    -0.00613188355 ||    0 ||
------------------------------------------------------------------------------
||    165 ||      4.99867 || -9.824266696e-022 ||   -0.006131824973 ||    0 ||
------------------------------------------------------------------------------
||    170 ||      4.99867 || -9.824233443e-022 ||   -0.006131804218 ||    0 ||
------------------------------------------------------------------------------
||    175 ||      4.99868 || -9.824320446e-022 ||   -0.006131858522 ||    0 ||
------------------------------------------------------------------------------
||    180 ||      4.99872 || -9.824357117e-022 ||    -0.00613188141 ||    0 ||
------------------------------------------------------------------------------
||    185 ||      4.99868 || -9.824296966e-022 ||   -0.006131843866 ||    0 ||
------------------------------------------------------------------------------
||    190 ||      4.99865 || -9.824227728e-022 ||   -0.006131800651 ||    0 ||
------------------------------------------------------------------------------
||    195 ||      4.99866 ||  -9.82423317e-022 ||   -0.006131804048 ||    0 ||
------------------------------------------------------------------------------
||    200 ||      4.99871 || -9.824279112e-022 ||   -0.006131832723 ||    0 ||
------------------------------------------------------------------------------
------------------------------------------------------------------------------
Process returned 0 (0x0)   execution time : 1.600 s
\end{lstlisting}

\clearpage

\subsection{\textsc{MATLAB} implementations for the calculation of the \textit{Breit-Hartree} contribution for a $\textup{Gd}^{3+}$ and an $\textup{Mn}^{2+}$ ion}\label{MATLABEHBAppendix}

The output of the \textsc{Matlab} implementation for the calculation of the \textit{Breit-Hartree} contribution for the $3+$-ionic state of gadolinium shows the numerically calculated value $\Delta E_{HB}$ of the expression from (\ref{EHBHollowSpheres(r)}) for the spin density of the $4f$-shell of a $\textup{Gd}^{3+}$ ion in joules and in electron volts. In addition, the number of $4f$-electrons is calculated, in order to show that the given data for the spin density are physically valid for the $\textup{Gd}^{3+}$ ion, whereat seven $4f$-electrons are expected. Within the implementation a \textsc{Matlab}-intern trapezoidal rule is used for the numerical integration. The console output of the \textsc{Matlab} script\footnote{Breit_Hartree_contribution_Gd_ion.m} is displayed in listing \ref{outMATLABGd3+}. In addition, the generated output is stored in the digital appendix\footnote{output_MATLAB_integration[Gd3+_ion].txt} as a \small{\verb|.txt|}-file:
\vspace{3mm}
\begin{lstlisting}[language=Matlab, style=Matlab,frame=single, caption={Output of the \textsc{Matlab} implementation for the calculation of the \textit{Breit-Hartree} contribution for a $\textup{Gd}^{3+}$ ion},label={outMATLABGd3+} ]
N_el_4f = 6.952863762418840

E_HB_J = -4.646550756982213e-21

E_HB_eV = -0.029001488714647
\end{lstlisting}
\vspace{4mm}
The output of the \textsc{Matlab} implementation for the calculation of the \textit{Breit-Hartree} contribution for the $2+$-ionic state of maganese shows the numerically calculated value $\Delta E_{HB}$ of the expression from (\ref{EHBHollowSpheres(r)}) for the spin density of the $3d$-shell of an $\textup{Mn}^{2+}$ ion in joules and in electron volts. In addition, the number of $3d$-electrons is calculated, in order to show that the given data for the spin density are physically valid for the $\textup{Mn}^{2+}$ ion, whereat five $3d$-electrons are expected. Within the implementation a \textsc{Matlab}-intern trapezoidal rule is used for the numerical integration. The console output of the \textsc{Matlab} script\footnote{Breit_Hartree_contribution_Mn_ion.m} is displayed in listing \ref{outMATLABMn2+}. In addition, the generated output is stored in the digital appendix\footnote{output_MATLAB_integration[Mn2+_ion].txt} as a \small{\verb|.txt|}-file:
\vspace{3mm}
\begin{lstlisting}[language=Matlab, style=Matlab,frame=single, caption={Output of the \textsc{Matlab} implementation for the calculation of the \textit{Breit-Hartree} contribution for an $\textup{Mn}^{2+}$ ion},label={outMATLABMn2+} ]
N_el_3d = 4.998474532044643
 
E_HB_J = -9.824211036421589e-22
 
E_HB_eV = -0.006131790233324
\end{lstlisting}
\clearpage

\subsection{\Cpp implementation of the second order \textit{M\o{}ller-Plesset} correlation energy correction for a $\textup{Gd}^{3+}$ ion}\label{CppGD3+AppendixMP2}

The output of the \Cpp implementation, which uses the \textit{Gauss-Legendre} integration, for the second order \textit{M\o{}ller-Plesset} correlation energy contribution for the $3+$-ionic state of gadolinium shows the numerically calculated value $\Delta E_{C,MP2}^{\textup{Gd}^{3+}}$ of the expression from (\ref{EHBHollowSpheres(r)}) for the electron density of a $\textup{Gd}^{3+}$ ion in joules and in electron volts. In addition, the total number of electrons is calculated, in order to show that the given data for the electron density are physically valid for the $\textup{Gd}^{3+}$ ion, whereat $61$ electrons are expected. The computations of the integrand are carried out under the inclusion of the regularisation of singular results by replacement with a zero. The console output of the \Cpp program\footnote{MP2_gauss_legendre_integration[Gd3+_ion].cpp} is displayed in listing \ref{outGaussLegendreGd3+} for increasing total numbers of sampling points $N_{tot.}$, whereat only an excerpt of the generated output is shown. The complete output is stored in the digital appendix\footnote{output_MP2_gauss_legendre_integration[Gd3+_ion].txt} as a \small{\verb|.txt|}-file:
\\
\begin{lstlisting}[language=C++, style=c,frame=single, caption={Output of the \Cpp implementation for the calculation of the second order \textit{M\o{}ller-Plesset} correlation energy correction for a $\textup{Gd}^{3+}$ ion  },label={outGaussLegendreGd3+MP2} ]
------------------------------------------------------------------------------
------------------------------------------------------------------------------
|| N_tot. ||    N_el_Gd3+ ||   E_MP2_c [Joule] ||      E_MP2_c [eV] || reg. ||
------------------------------------------------------------------------------
------------------------------------------------------------------------------
||     10 ||      49.3663 || -2.181272894e-050 || -1.361443456e-031 ||    0 ||
------------------------------------------------------------------------------
||     15 ||      67.1681 ||  -5.56671862e-050 || -3.474472478e-031 ||    0 ||
------------------------------------------------------------------------------
||     20 ||      65.6261 ||   -5.2483654e-050 || -3.275772027e-031 ||    0 ||
------------------------------------------------------------------------------
||     25 ||      56.6607 || -5.703628689e-050 || -3.559925022e-031 ||    0 ||
------------------------------------------------------------------------------
||     30 ||      66.0256 || -5.980484365e-050 || -3.732724743e-031 ||    0 ||
------------------------------------------------------------------------------
||     35 ||      61.7258 || -5.569364456e-050 || -3.476123879e-031 ||    0 ||
------------------------------------------------------------------------------
||     40 ||      57.2417 || -5.046719011e-050 || -3.149914251e-031 ||    0 ||
------------------------------------------------------------------------------
||     45 ||      60.3737 || -5.549147904e-050 ||   -3.4635057e-031 ||    0 ||
------------------------------------------------------------------------------
||     50 ||      62.9599 || -6.299013294e-050 || -3.931534863e-031 ||    0 ||
------------------------------------------------------------------------------
||     55 ||      62.8527 ||  -6.56128284e-050 || -4.095230639e-031 ||    0 ||
------------------------------------------------------------------------------
||     60 ||       61.857 || -6.388976752e-050 || -3.987685637e-031 ||    0 ||
------------------------------------------------------------------------------
||     65 ||       61.135 || -6.128147907e-050 || -3.824889077e-031 ||    0 ||
------------------------------------------------------------------------------
||     70 ||      60.8357 || -5.920285023e-050 ||  -3.69515127e-031 ||    0 ||
------------------------------------------------------------------------------
||     75 ||      60.7811 || -5.781719914e-050 || -3.608665731e-031 ||    0 ||
------------------------------------------------------------------------------
||     80 ||       60.816 || -5.714328238e-050 || -3.566603155e-031 ||    0 ||
------------------------------------------------------------------------------
||     85 ||       60.869 || -5.693216282e-050 || -3.553426109e-031 ||    0 ||
------------------------------------------------------------------------------
||     90 ||      60.9144 || -5.688405515e-050 || -3.550423464e-031 ||    0 ||
------------------------------------------------------------------------------
||     95 ||      60.9472 || -5.686848292e-050 || -3.549451522e-031 ||    0 ||
------------------------------------------------------------------------------
||    100 ||      60.9685 || -5.687972431e-050 || -3.550153154e-031 ||    0 ||
------------------------------------------------------------------------------
||    105 ||      60.9824 || -5.693496008e-050 ||   -3.5536007e-031 ||    0 ||
------------------------------------------------------------------------------
||    110 ||      60.9892 || -5.702553758e-050 || -3.559254103e-031 ||    0 ||
------------------------------------------------------------------------------
||    115 ||      60.9943 || -5.712513678e-050 || -3.565470596e-031 ||    0 ||
------------------------------------------------------------------------------
||    120 ||      60.9973 || -5.720751772e-050 ||  -3.57061241e-031 ||    0 ||
------------------------------------------------------------------------------
||    125 ||      60.9988 || -5.726510341e-050 || -3.574206626e-031 ||    0 ||
------------------------------------------------------------------------------
||    130 ||      61.0007 || -5.730286096e-050 || -3.576563267e-031 ||    0 ||
------------------------------------------------------------------------------
||    135 ||      61.0005 ||  -5.73233451e-050 || -3.577841786e-031 ||    0 ||
------------------------------------------------------------------------------
||    140 ||       61.002 || -5.733509117e-050 || -3.578574918e-031 ||    0 ||
------------------------------------------------------------------------------
||    145 ||      61.0019 || -5.733997647e-050 || -3.578879835e-031 ||    0 ||
------------------------------------------------------------------------------
||    150 ||      61.0023 || -5.734302758e-050 ||  -3.57907027e-031 ||    0 ||
------------------------------------------------------------------------------
||    155 ||      61.0023 || -5.734520684e-050 || -3.579206289e-031 ||    0 ||
------------------------------------------------------------------------------
||    160 ||      61.0024 ||  -5.73474579e-050 || -3.579346789e-031 ||    0 ||
------------------------------------------------------------------------------
||    165 ||      61.0023 ||  -5.73504197e-050 ||  -3.57953165e-031 ||    0 ||
------------------------------------------------------------------------------
||    170 ||      61.0025 ||  -5.73530116e-050 || -3.579693424e-031 ||    0 ||
------------------------------------------------------------------------------
||    175 ||      61.0027 ||  -5.73547835e-050 || -3.579804017e-031 ||    0 ||
------------------------------------------------------------------------------
||    180 ||      61.0031 || -5.735608448e-050 || -3.579885217e-031 ||    0 ||
------------------------------------------------------------------------------
||    185 ||      61.0029 || -5.735726323e-050 || -3.579958789e-031 ||    0 ||
------------------------------------------------------------------------------
||    190 ||      61.0027 || -5.735755587e-050 || -3.579977055e-031 ||    0 ||
------------------------------------------------------------------------------
||    195 ||      61.0028 || -5.735802712e-050 || -3.580006467e-031 ||    0 ||
------------------------------------------------------------------------------
||    200 ||       61.003 || -5.735833377e-050 || -3.580025607e-031 ||    0 ||
------------------------------------------------------------------------------
------------------------------------------------------------------------------

Process returned 0 (0x0)   execution time : 0.933 s
\end{lstlisting}

\clearpage

\subsection{\Cpp implementation of the second order \textit{M\o{}ller-Plesset} correlation energy correction for an $\textup{Mn}^{2+}$ ion}\label{CppMN2+AppendixMP2}

The output of the \Cpp implementation, which uses the \textit{Gauss-Legendre} integration, for the second order \textit{M\o{}ller-Plesset} correlation energy contribution for the $2+$-ionic state of manganese shows the numerically calculated value $\Delta E_{C,MP2}^{\textup{Mn}^{2+}}$ of the expression from (\ref{MP2Plesset2}) for the electron density of an $\textup{Mn}^{2+}$ ion in joules and in electron volts. In addition, the total number of electrons is calculated, in order to show that the given data for the electron density are physically valid for the $\textup{Mn}^{2+}$ ion, whereat $23$ electrons are expected. The computations of the integrand are carried out under the inclusion of the regularisation of singular results by replacement with a zero. The console output of the \Cpp program\footnote{MP2_gauss_legendre_integration[Mn2+_ion].cpp} is displayed in listing \ref{outGaussLegendreMn2+MP2} for increasing total numbers of sampling points $N_{tot.}$, whereat only an excerpt of the generated output is shown. The complete output is stored in the digital appendix\footnote{output_MP2_gauss_legendre_integration[Mn2+_ion].txt} as a \small{\verb|.txt|}-file:
\\
\begin{lstlisting}[language=C++, style=c,frame=single, caption={Output of the \Cpp implementation for the calculation of the second order \textit{M\o{}ller-Plesset} correlation energy correction for an $\textup{Mn}^{2+}$ ion  },label={outGaussLegendreMn2+MP2} ]
------------------------------------------------------------------------------
------------------------------------------------------------------------------
|| N_tot. ||    N_el_Mn2+ ||   E_MP2_c [Joule] ||      E_MP2_c [eV] || reg. ||
------------------------------------------------------------------------------
------------------------------------------------------------------------------
||     10 ||      15.9818 || -4.797954491e-051 || -2.994647649e-032 ||    0 ||
------------------------------------------------------------------------------
||     15 ||      29.4591 || -1.429115183e-050 || -8.919835386e-032 ||    0 ||
------------------------------------------------------------------------------
||     20 ||       19.901 || -8.977799936e-051 || -5.603501977e-032 ||    0 ||
------------------------------------------------------------------------------
||     25 ||      22.7102 || -1.088720563e-050 || -6.795259276e-032 ||    0 ||
------------------------------------------------------------------------------
||     30 ||      24.9647 || -1.417316123e-050 || -8.846191443e-032 ||    0 ||
------------------------------------------------------------------------------
||     35 ||      23.7918 || -1.393891481e-050 || -8.699986327e-032 ||    0 ||
------------------------------------------------------------------------------
||     40 ||       22.924 || -1.240778925e-050 || -7.744332922e-032 ||    0 ||
------------------------------------------------------------------------------
||     45 ||      22.7572 || -1.157189722e-050 || -7.222610148e-032 ||    0 ||
------------------------------------------------------------------------------
||     50 ||      22.8324 || -1.133124061e-050 || -7.072404107e-032 ||    0 ||
------------------------------------------------------------------------------
||     55 ||      22.9182 || -1.128284964e-050 || -7.042200844e-032 ||    0 ||
------------------------------------------------------------------------------
||     60 ||      22.9668 || -1.133489916e-050 || -7.074687595e-032 ||    0 ||
------------------------------------------------------------------------------
||     65 ||      22.9893 || -1.142311671e-050 || -7.129748663e-032 ||    0 ||
------------------------------------------------------------------------------
||     70 ||      22.9968 || -1.149660647e-050 || -7.175617363e-032 ||    0 ||
------------------------------------------------------------------------------
||     75 ||      22.9999 || -1.153877215e-050 || -7.201935108e-032 ||    0 ||
------------------------------------------------------------------------------
||     80 ||      23.0011 ||  -1.15545326e-050 || -7.211772006e-032 ||    0 ||
------------------------------------------------------------------------------
||     85 ||      23.0009 || -1.155735293e-050 || -7.213532319e-032 ||    0 ||
------------------------------------------------------------------------------
||     90 ||      23.0011 || -1.155787849e-050 || -7.213860347e-032 ||    0 ||
------------------------------------------------------------------------------
||     95 ||      23.0013 || -1.155835247e-050 || -7.214156185e-032 ||    0 ||
------------------------------------------------------------------------------
||    100 ||      23.0011 || -1.155917537e-050 || -7.214669796e-032 ||    0 ||
------------------------------------------------------------------------------
||    105 ||      23.0013 || -1.155975071e-050 || -7.215028898e-032 ||    0 ||
------------------------------------------------------------------------------
||    110 ||      23.0011 || -1.155966154e-050 || -7.214973242e-032 ||    0 ||
------------------------------------------------------------------------------
||    115 ||      23.0012 || -1.155967342e-050 || -7.214980655e-032 ||    0 ||
------------------------------------------------------------------------------
||    120 ||      23.0011 ||  -1.15595717e-050 || -7.214917166e-032 ||    0 ||
------------------------------------------------------------------------------
||    125 ||      23.0011 ||   -1.1559599e-050 || -7.214934203e-032 ||    0 ||
------------------------------------------------------------------------------
||    130 ||      23.0012 || -1.155977234e-050 || -7.215042393e-032 ||    0 ||
------------------------------------------------------------------------------
||    135 ||       23.001 || -1.155971858e-050 || -7.215008842e-032 ||    0 ||
------------------------------------------------------------------------------
||    140 ||      23.0014 || -1.155982987e-050 || -7.215078301e-032 ||    0 ||
------------------------------------------------------------------------------
||    145 ||      23.0011 || -1.155973273e-050 || -7.215017672e-032 ||    0 ||
------------------------------------------------------------------------------
||    150 ||      23.0012 || -1.155975009e-050 || -7.215028505e-032 ||    0 ||
------------------------------------------------------------------------------
||    155 ||       23.001 ||  -1.15597271e-050 || -7.215014156e-032 ||    0 ||
------------------------------------------------------------------------------
||    160 ||      23.0013 ||  -1.15597593e-050 || -7.215034258e-032 ||    0 ||
------------------------------------------------------------------------------
||    165 ||      23.0011 ||  -1.15597352e-050 || -7.215019216e-032 ||    0 ||
------------------------------------------------------------------------------
||    170 ||      23.0011 || -1.155978881e-050 || -7.215052678e-032 ||    0 ||
------------------------------------------------------------------------------
||    175 ||      23.0011 || -1.155976059e-050 || -7.215035059e-032 ||    0 ||
------------------------------------------------------------------------------
||    180 ||      23.0013 || -1.155975604e-050 ||  -7.21503222e-032 ||    0 ||
------------------------------------------------------------------------------
||    185 ||      23.0012 || -1.155983303e-050 || -7.215080276e-032 ||    0 ||
------------------------------------------------------------------------------
||    190 ||      23.0012 || -1.155982064e-050 || -7.215072544e-032 ||    0 ||
------------------------------------------------------------------------------
||    195 ||      23.0012 || -1.155979407e-050 || -7.215055962e-032 ||    0 ||
------------------------------------------------------------------------------
||    200 ||      23.0013 || -1.155982424e-050 || -7.215074787e-032 ||    0 ||
------------------------------------------------------------------------------
------------------------------------------------------------------------------

Process returned 0 (0x0)   execution time : 0.945 s
\end{lstlisting}

\clearpage

\subsection{\textsc{MATLAB} implementations for the calculation of the second order \textit{M\o{}ller-Plesset} correlation energy correction for a $\textup{Gd}^{3+}$ and an $\textup{Mn}^{2+}$ ion}\label{MATLABMP2AppendixMP2}

The output of the \textsc{Matlab} implementation for the calculation of the second order \textit{M\o{}ller-Plesset} correlation energy correction for the $3+$-ionic state of gadolinium shows the numerically calculated value $\Delta E_{C,MP2}^{\textup{Gd}^{3+}}$ of the expression from (\ref{MP2Plesset2}) for the electron density of a $\textup{Gd}^{3+}$ ion in joules and in electron volts. In addition, the total number of electrons is calculated, in order to show that the given data for the electron density are physically valid for the $\textup{Gd}^{3+}$ ion, whereat $61$ electrons are expected. Within the implementation a \textsc{Matlab}-intern trapezoidal rule is used for the numerical integration. The console output of the \textsc{Matlab} script\footnote{MP2_correlation_energy_contribution_Gd_ion.m} is displayed in listing \ref{outMATLABGd3+MP2}. In addition, the generated output is stored in the digital appendix\footnote{output_MATLAB_MP2_integration[Gd3+_ion].txt} as a \small{\verb|.txt|}-file:
\vspace{2.5mm}
\begin{lstlisting}[language=Matlab, style=Matlab,frame=single, caption={Output of the \textsc{Matlab} implementation for the calculation of the second order \textit{M\o{}ller-Plesset} correlation energy correction for a $\textup{Gd}^{3+}$ ion},label={outMATLABGd3+MP2} ]
N_el_Gd = 61.000542609873257
 
E_MP2_J = -5.735954175918290e-50
 
E_MP2_eV = -3.580101003968511e-31
\end{lstlisting}
\vspace{4mm}
The output of the \textsc{Matlab} implementation for the calculation of the second order \textit{M\o{}ller-Plesset} correlation energy contribution for the $2+$-ionic state of maganese shows the numerically calculated value $\Delta E_{C,MP2}^{\textup{Mn}^{2+}}$ of the expression from (\ref{MP2Plesset2}) for the electron density of an $\textup{Mn}^{2+}$ ion in joules and in electron volts. In addition, the total number of electrons is calculated, in order to show that the given data for the electron density are physically valid for the $\textup{Mn}^{2+}$ ion, whereat $23$ electrons are expected. Within the implementation a \textsc{Matlab}-intern trapezoidal rule is used for the numerical integration. The console output of the \textsc{Matlab} script\footnote{MP2_correlation_energy_contribution_Mn_ion.m} is displayed in listing \ref{outMATLABMn2+MP2}. In addition, the generated output is stored in the digital appendix\footnote{output_MATLAB_MP2_integration[Mn2+_ion].txt} as a \small{\verb|.txt|}-file:
\vspace{2.5mm}
\begin{lstlisting}[language=Matlab, style=Matlab,frame=single, caption={Output of the \textsc{Matlab} implementation for the calculation of the second order \textit{M\o{}ller-Plesset} correlation energy correction for an $\textup{Mn}^{2+}$ ion},label={outMATLABMn2+MP2} ]
N_el_Mn = 23.000204590607947
 
E_MP2_J = -1.155942651526840e-50
 
E_MP2_eV = -7.214826549061007e-32
\end{lstlisting}
\clearpage